\documentclass[12pt]{article}
\pdfoutput=1

\usepackage{putex,placeins}
\usepackage{amsmath,amssymb,amsfonts}
\usepackage{psfrag}
\usepackage{footmisc}
\usepackage{url}
\usepackage{mathtools}
\usepackage[dvipsnames]{xcolor}

\usepackage{tikz}
    \usepackage{amssymb,amsfonts,amsmath}
    \usepackage{tkz-euclide}
        \usetikzlibrary{arrows,calc,patterns}
\usepackage{pgfplots}

\definecolor{darkblue}{rgb}{0.1,0.1,.7}
\definecolor{purple}{rgb}{0.6,0,0.6}
\definecolor{orange}{rgb}{1,0.2,0}
\usepackage[colorlinks, linkcolor=darkblue, citecolor=darkblue, urlcolor=darkblue, linktocpage]{hyperref} 
\usepackage[square, comma, compress,numbers]{natbib}
\usepackage[]{amsmath}
\usepackage[]{graphicx}
\usepackage[]{latexsym}
\usepackage[utf8]{inputenc}
\usepackage{geometry}
\usepackage{amscd}
\usepackage[all,cmtip]{xy}
\usepackage{mathrsfs}
\usepackage{float}

\usepackage[margin=10pt,font=small,labelfont=bf]{caption}
\usepackage{dsdshorthand}
\usepackage{changepage}
\usepackage{setspace}
\usepackage{tikz}
\usepackage{subcaption}

\usepackage[titles]{tocloft}
\usetikzlibrary{arrows,positioning}

\def\SL2{\widetilde{SL}(2,\mathbb R)}

\def\mC{\mathcal C}

\def\inst{\text{inst}}
\def\betainst{{\beta}_\inst}

\newcommand\mR{\mathbb{R}}

\numberwithin{equation}{section}

\newcommand{\grp}[1]{\mathrm{#1}}

\newcommand{\Ebrk}{E_\text{brk.}}
\def\initCharge{10^{44} |q|}

\newcommand {\bes} {\begin {equation*}}
\newcommand {\ees} {\end {equation*}}
\newcommand {\beq} {\begin {equation}}
\newcommand {\eeq} {\end {equation}}
\newcommand {\bea} {\begin {eqnarray}}
\newcommand {\ea} {\end {eqnarray}}
\newcommand {\eea} {\end {eqnarray}}

\def\rb{\rangle}
\def\lb{\langle}

\newcommand{\lrm}[1]{\left( #1 \right)}

\def\t{\text}

\numberwithin{equation}{section}

\def\<{\langle}
\def\>{\rangle}

\tikzset{
    >=stealth',
    punkt/.style={
           rectangle,
           rounded corners,
           draw=black, very thick,
           text width=15em,
           minimum height=2em,
           text centered},
    pil/.style={
           ->,
           thick,
           shorten <=2pt,
           shorten >=2pt,}
}

 \def\ie{\begin{equation}\begin{aligned}}
\def\fe{\end{aligned}\end{equation}}

\begin{document}

\preprint{}

\title{
\textbf{The evaporation of charged black holes}}

\authors{Adam R.~Brown${}^{1,2}$, Luca V.~Iliesiu${}^{2,3}$, Geoff Penington${}^3$, Mykhaylo Usatyuk${}^{3,4}$}

\begin{center}
\institution{GDM}{${}^1$ Google DeepMind, Mountain View, CA, 94043, USA}

\institution{SU}{${}^2$ Stanford Institute for Theoretical Physics, Stanford University, Stanford, CA 94305, USA}

\institution{BCTP}{${}^3$ Center for Theoretical Physics and Department of Physics, Berkeley, CA, 94720, USA}
\institution{KITP}{${}^4$ Kavli Institute for Theoretical Physics, Santa Barbara, CA 93106, USA}
\end{center}

\abstract{  
Charged particle emission from black holes with sufficiently large charge is exponentially suppressed. As a result, such black holes are driven towards extremality by the emission of neutral Hawking radiation. Eventually, an isolated black hole gets close enough to extremality that the gravitational backreaction of a single Hawking photon becomes important, and the quantum field theory in curved spacetime approximation breaks down. To proceed further, we need to use a quantum theory of gravity. We make use of recent progress in our understanding of the quantum-gravitational thermodynamics of near-extremal black holes to compute the corrected spectrum for both neutral and charged Hawking radiation, including the effects of backreaction, as well as greybody factors and metric fluctuations. At low temperatures, large fluctuations in a set of quantum-gravitational (almost) zero modes lead to drastic modifications to neutral particle emission that -- in contrast to the semiclassical prediction -- ensure the black hole remains subextremal. Relatedly, angular momentum constraints mean that,  close enough to extremality, black holes with zero angular momentum can no longer emit individual photons and gravitons; instead, the dominant radiation channel consists of entangled pairs of photons in angular-momentum singlet states. This causes a sudden slowdown in the evaporation rate by a factor of at least $10^{700}$. We also compute the effects of backreaction and metric fluctuations on the emission of charged particles. Somewhat surprisingly, we find that the semiclassical Schwinger emission rate is essentially unchanged, despite the fact that the emission process leads to large changes in the geometry and thermodynamics of the throat. Our results allow us to present, for the first time, the full history of the evaporation of a large charged black hole. A notable feature of this history is that the black hole alternates between exponentially long epochs of integer and half-integer spin that have radically different cooling rates. This corrects the semiclassical calculation, which  gives completely wrong predictions for almost the entire evaporation history, even for the crudest observables like the temperature seen by a thermometer.

}

\maketitle

\tableofcontents

\section{Introduction} \label{sec:introduction}

The no-hair theorem says that an equilibrium black hole in classical general relativity is uniquely characterized by its mass, electromagnetic charge, and angular momentum.\footnote{In Einstein-Maxwell theory, black holes can, in general, have not just electric but also magnetic charge. Magnetic monopoles have never been observed, but are a very natural feature of Beyond Standard Model physics. For most of this paper, we will assume that our black hole formed from the collapse of conventional matter and, hence, only carries an electric charge. But we will briefly comment on magnetically charged black holes in Sections \ref{sec:the-story-of-a-charged-black-hole} and \ref{sec:discussion}.} Dynamical black holes quickly settle down to a stationary solution described solely by these parameters. Thereafter, their evolution is driven by quantum effects \cite{Hawking:1974rv,Hawking:1975vcx}. These quantum effects cause the black hole to lose mass, charge, and angular momentum, but not in equal measure. The first to go is (most of) the angular momentum.  A large isolated Kerr-Newman black hole will lose angular momentum through Hawking radiation faster than it loses energy or charge \cite{Page:1976ki}, and so the black hole will approach a Reissner-Nordstr\"{o}m (or Schwarzschild) metric while still retaining much of its initial mass. 

This paper is about what happens next. There is a crucial difference between angular momentum and charge, which is that although angular momentum (and energy) may be carried away by massless particles, the lightest known charged particles in our universe are massive. This means that—unlike for angular momentum or energy—the emission of charge from a sufficiently large black hole is exponentially suppressed \cite{Gibbons:1975kk}. As a result, any black hole left in isolation that starts with a sufficiently large charge will tend to radiate energy without losing much charge, regardless of its initial mass. Such black holes will thereby be driven towards the extremal limit $M \approx Q$. As this limit approaches, the na\"{i}ve semiclassical picture of Hawking radiation breaks down \cite{Preskill:1991tb,Maldacena:1998uz,Page:2000dk} with a single Hawking quantum seemingly containing more energy than the total available energy above extremality of the black hole. As we will see, this breakdown happens when the energy above extremality drops below 
\begin{align}
    M - Q \, \lesssim \,  \Ebrk \equiv \frac{M_{Pl}}{Q^3}.
\end{align}

What happens to the black hole after this point has generally been regarded as a mystery: speculations in the literature include the suggestion that the spectrum of near-extremal black holes might be gapped, with no microstates in the energy range $(Q, Q + \Ebrk)$ \cite{Callan:1996dv,Maldacena:1996ds,Maldacena:1997ih} or alternatively that the spectrum has an approximately uniform density of states that is exponentially large but nondegenerate \cite{Page:2000dk}. In recent years, however, our understanding of the thermodynamics of near-extremal black holes has been dramatically improved by the study of a particular set of near-horizon modes that, in the near-extremal limit, have a very small classical action and hence have large fluctuations \cite{Iliesiu:2020qvm, Heydeman:2020hhw,  Ghosh:2019rcj, Iliesiu:2022onk, Boruch:2022tno, Kapec:2023ruw, Rakic:2023vhv, Kolanowski:2024zrq}. These modes, known as the Schwarzian modes because their action takes the form of a Schwarzian derivative, had originally been identified in the context of JT gravity, which forms part of the dimensional reduction to two dimensions of the near-horizon region of a Reissner-Nordstr\"om (RN) black hole. With care, the Schwarzian modes can be integrated out exactly, while treating all other modes semiclassically. When the energy above extremality drops below $\Ebrk$, the Schwarzian modes become strongly coupled, and the statistical mechanics of the black hole are significantly altered. There is no spectral gap; however, the density of states at low energies is drastically modified so that, in contrast to the semiclassical prediction, the density of states goes to zero as extremality is approached.

While the effects of the Schwarzian modes on the black hole energy spectrum (in the absence of light matter fields) are now well understood, understanding the \emph{dynamics} of the black hole after the breakdown of semiclassical thermodynamics 
\cite{Preskill:1991tb,Maldacena:1998uz,Page:2000dk} requires us to also understand the effects of the Schwarzian modes on the spectrum of Hawking quanta emitted near extremality. 
The goal of the present paper is
to fully understand the evaporation of large charged black holes by finding this spectrum, arbitrarily close to extremality, for both neutral and charged radiation. 

As we will see, a black hole with a large enough charge spends almost all its evaporation history outside the semiclassical regime---it has $M - Q \gsim \Ebrk$ for only an exponentially small fraction of its lifetime. This means that the standard QFT-in-curved-spacetime prediction for the Hawking radiation is almost always wrong. It's not just wrong for delicate fine-grained exceptionally-hard-to-measure $N$-point functions, like those involved in the information paradox \cite{Hawking:1976ra}. It's also wrong for coarse quantities like the energy and emission rates of photons, quantities that comparatively unsophisticated observers could measure with a spectrograph. 
Thus, even for these crude quantities, the quantum-field-theory-in-curved spacetime answer is almost always wrong. In this paper, we will derive an answer that is almost always right. It will be right for almost the entire evaporation history of the black hole, starting from when the black hole settles down to a static Reissner-Nordstr\"om solution and ending only when the temperature approaches the Planck scale in the last fraction of a second of its life. 

To the best of our knowledge, our results mark the first example of a controlled calculation of Lorentzian dynamics due to large, strongly coupled quantum metric fluctuations in Einstein gravity coupled to the Standard Model. They enable us to do for black holes with very large initial charge what we have long been able to do for black holes with little or no charge -- to tell almost the complete story of the evaporation of charged black holes.

\subsection*{Plan for paper} 

In \textbf{Section}~\ref{sec:background}, we give a pedagogical introduction to neutral and charged Hawking radiation for black holes with large charge. We explain how the effect of this radiation is to drive the black hole towards extremality and out of the semiclassical regime and how a Schwarzian analysis can reveal what happens next. We then review the near-extremal limit of RN black holes and the importance of the Schwarzian modes, and provide useful formulas for matter correlators in the Schwarzian theory.  This section can be safely skipped by expert readers.

In \textbf{Section}~\ref{sec:neutral-particle-emission-main}, we incorporate Schwarzian corrections into computations of the spectrum of neutral Hawking radiation and show that the spectrum is dramatically altered near extremality in a way that avoids the emission of quanta that would leave the black hole superextremal. We first consider the case of a massless scalar field, for which the semiclassical energy flux is
\begin{align}
    \frac{d E}{d t} \sim r_+^2 \Ebrk^2 (M-Q)^2\,, \qquad (M-Q) \gg \Ebrk\,.
\end{align}
This only gives the correct answer above the breakdown scale, $M - Q \, \gg \,  \Ebrk$. When $M - Q \, \ll \,  \Ebrk$, we find that Schwarzian corrections lead instead to
\begin{align}\label{eq:scalarSchwarzianrate}
    \frac{d E}{d t} \sim r_+^2 \Ebrk^\frac{1}{2} (M-Q)^{\frac{7}{2}}\,, \qquad (M-Q) \ll \Ebrk\,.
\end{align}

In our universe, there are no known massless scalar fields. As a result, the Hawking radiation of sufficiently low-temperature black holes (assuming that no massless neutrino exists\footnote{There is strong experimental evidence that some of the neutrinos have mass, but it cannot currently be excluded that one of the neutrinos is extremely light or even massless. However, if the mass of the lightest neutrino is within a few orders of magnitude of the (known) mass differences between neutrinos, neutrino emission will be exponentially suppressed for all black holes satisfying $Q \gtrsim \initCharge$.}) is dominated by photons and gravitons. 
\begin{figure}[t]
    \centering
    \includegraphics[width=.7\textwidth]{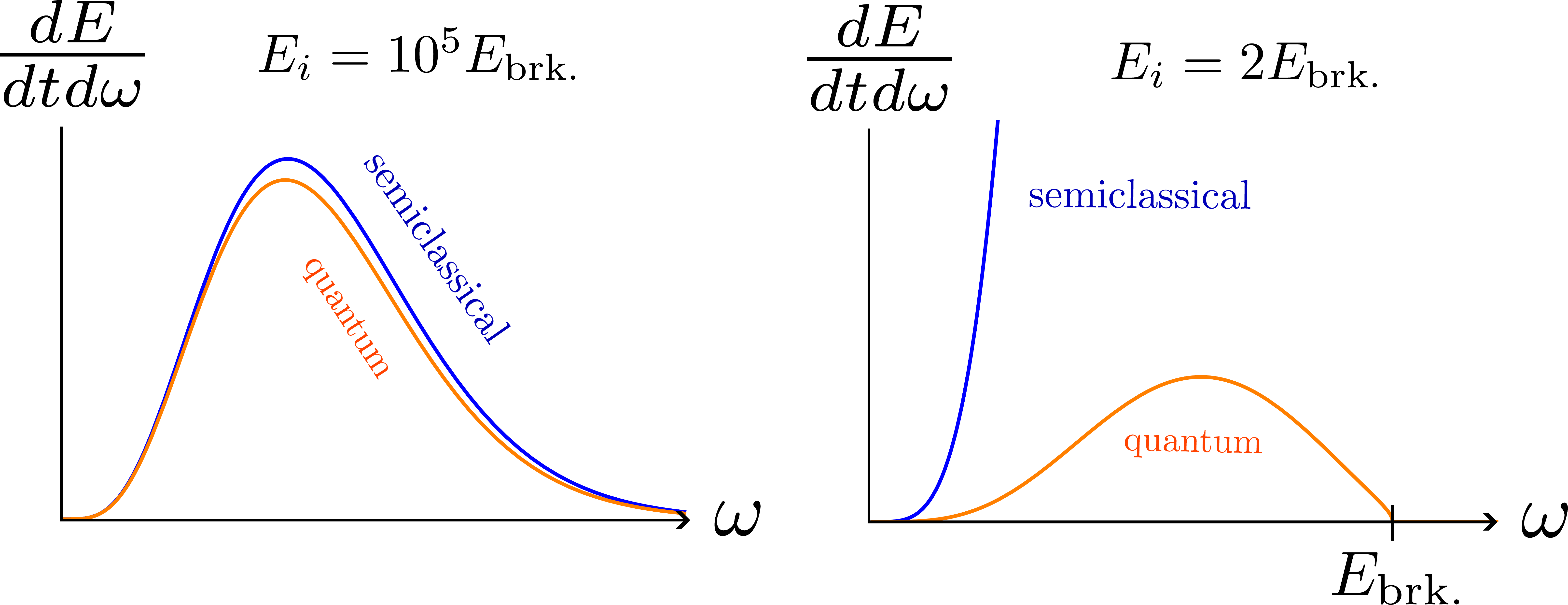}
    \caption{Comparison of the \textcolor{blue}{semiclassical prediction} vs.~\textcolor{orange}{quantum corrected} Hawking radiation into the $\ell=1$ photon mode. (The quantum curve is the integrand from \eqref{eqn:photon_fullflux_kerrnewman}; the semiclassical curve is the integrand from  \eqref{eqn:photon_semiclassicalflux}.) 
    The energy flux per unit frequency is plotted for black holes in the microcanonical ensemble at initial energy $E_i=M-Q$ above extremality with fixed charge $Q$ and initial spin $j=0$. \textit{\textbf{Left:}} A black hole far above the thermodynamic breakdown scale $(E_i = 10^5 \Ebrk)$. Quantum corrections are not important and the \textcolor{orange}{exact energy flux} approaches the \textcolor{blue}{semiclassical prediction}. \textit{\textbf{Right:}} For a BH near the breakdown scale $(E_i = 2 \Ebrk)$, quantum corrections are important. The exact quantum answer cuts off the spectrum since the BH can't transition into states that don't exist. For photon emission from $j=0$ BHs the spectrum is cut off at $E_i-\Ebrk$ due to angular momentum selection rules. Note that the scales of the two plots are very different due to the difference in energy above extremality.}
    \label{fig:hawkingradphotonintro}
\end{figure}
Because photon modes exist with angular momentum $\ell = 1$, while all graviton modes have angular momentum $\ell \geq 2$, one would na\"ively expect that at low temperatures graviton emission would be highly suppressed relative to photon emission by the centrifugal potential barrier that prevents tunneling in and out of the black hole.  This would indeed be true if photons were not excitations of the same electromagnetic field that the black hole is charged under. Instead, however, metric and electromagnetic field modes mix in the near-horizon region such that the mode with smallest scaling dimension is an $\ell = 2$ ``graviphoton'' mode that becomes a superposition of a photon and a graviton far from the black hole. Semiclassically, the lighter scaling dimension of this mode relative to the $\ell = 1$ photon mode cancels out the effect of the larger centrifugal barrier and means that both modes contribute an $O(1)$ fraction of the Hawking radiation, even very close to extremality. Schwarzian effects, however, mean that the near-horizon scaling dimension becomes essentially irrelevant at low enough temperatures, and photon emission dominates. Examples of semiclassical and quantum-corrected single-photon Hawking radiation spectra are shown in Figure \ref{fig:hawkingradphotonintro}.

Because photons necessarily carry nonzero angular momentum, a black hole with zero angular momentum cannot emit a single photon or graviton and remain at zero angular momentum. However, all black holes with angular momentum $j = 1$  have mass 
\begin{align}\label{eq:J=1extremalitybound}
     M \geq Q +  \Ebrk .
\end{align}
A black hole with angular momentum $j = 0$ and mass $M$ that violates the bound \eqref{eq:J=1extremalitybound} cannot emit a single photon and remain on-shell. Instead, the dominant radiation channel consists of entangled pairs of photons with total angular momentum zero. This is the same channel that dominates the ``forbidden'' $2s \rightarrow 1s$ transition in the hydrogen atom. We show that the resulting energy flux is
\begin{align}
     \frac{d E}{d t} \sim r_+^{16} \Ebrk^{\frac{17}{2}}  (M-Q)^{\frac{19}{2}}, \qquad (M-Q) \ll \Ebrk \ . 
\end{align}
As with forbidden transitions in atomic physics, this is considerably slower than transitions where single-photon emission is possible. 
As a result, the mass $M(t)$ of a black hole with charge  $Q\gg \initCharge$ approaches extremality as
\begin{align}
\label{eq:intro-mass-evolution}
\t{Bosonic Black Holes:} \qquad M(t)-Q \sim \begin{cases}
t^{-1/4}\qquad \qquad \qquad &\Ebrk \ll M-Q \ll M\,, \\
t^{-2/17} \qquad & \hspace{11mm}  M-Q \ll \Ebrk\, .
\end{cases}
\end{align}
In \eqref{eq:intro-mass-evolution}, we assumed that the initial black hole state was bosonic (i.e.,~it had integer spin). In such cases, angular momentum can and will eventually be driven to zero by photon and graviton emission, and at low temperatures, the evaporation rate will be controlled by di-photon emission. On the other hand, if the black hole is initially fermionic, it can never reach zero angular momentum by emitting particles with integer spin. Instead, at low temperatures, it will have angular momentum $j = 1/2$, and will flip back and forth in orientation as it emits one photon at a time. Specifically, we find that
\be 
\t{Fermionic Black Holes:} \qquad M(t) - M_0  \sim t^{-\frac{2}{9}}\,, \qquad   M-M_0 \ll \Ebrk \ , 
\ee
where $M_0 = Q +  3\Ebrk/8$ is the ground-state energy for a black hole with charge $Q$ and angular momentum $j = 1/2$. This means that fermionic black holes can emit single photons and, as a result, will cool much more quickly than bosonic black holes.

 The approach to extremality
will be interrupted if a charged particle is emitted.  
In \textbf{Section}~\ref{sec:background-charged}, we turn our attention from neutral to charged particles. For convenience, we assume that the black hole is positively charged and hence that the emitted charged particles are positrons; the behaviour of negatively charged black holes is identical except that the emitted particles are electrons. Semiclassically, the positron emission rate can be computed by integrating the Schwinger formula for pair production per unit spacetime volume over the RN spacetime.  One finds that the rate per unit time is
\begin{align}\label{eq:schwingerrateintro}
    \frac{d \langle \textrm{pairs} \rangle}{dt} = \frac{q^3 Q^3}{2 \pi^3 m^2 r_+^3} \exp \lrm{-\frac{ r_+^2}{Q_* Q}} + \ldots\,.
\end{align}
Charged particle emission is thus exponentially suppressed when $Q \gg Q_* \equiv \frac{q}{\pi m^2}$. Restoring units, $Q_* =  \frac{1}{\pi} \frac{M_{\t{pl}}^2}{m^2} q \approx  1.8 \times 10^{44} q$, where $q$ is the charge of the positron. This is an enormous charge, with an extremal BH of this charge having mass $M_* \approx 10^5 M_\odot$ with $M_\odot$ the solar mass. We also compute the energy distribution for the emitted positron: the spectrum is highly nonthermal and the typical positron is ultrarelativistic with near-Planckian energy. After emission, the black hole has an energy above extremality
\begin{align}
    \bar{M} - \bar{Q} \sim \frac{q Q_*}{Q}\,,
\end{align}
which is small compared to the Planck scale but still well above the breakdown scale.

 Since the emission of neutral radiation was drastically altered at low temperatures from semiclassical expectations, one might expect the same to occur for charged particle emission. 
In \textbf{Section} \ref{sec:charged-particle-emission-main} we provide a detailed analysis of gravitational corrections to the semiclassical emission process. We take into account both the backreaction of the particle on the semiclassical black hole spacetime and the quantum fluctuations of that spacetime geometry. The particle backreaction can be found by solving for a gravitational instanton that includes the energy-momentum of the charged particle worldline loop present in the usual Schwinger effect instanton. Outside the particle worldline, the gravitational instanton looks like an ordinary Euclidean black hole, except that the periodicity in Euclidean time is altered to ensure that the solution inside the particle worldline is smooth. As a result, the instanton acts as an effective conical defect in spacetime. To compute the effects of gravitational fluctuations, we integrate over the set of almost-zero modes associated to this gravitational instanton solution. The presence of a defect leads to two additional such modes, whose integral is proportional to the inverse temperature $\beta$ of the black hole. 
To obtain the full pair-production rate, we resum multi-instanton configurations by integrating over the moduli space of multi-defect spacetime geometries. The end result of the resummation, which yields the final production rate for positrons, gives the emission rate
\begin{align}\label{eqn:Gamma_Schwinger_final_result-intro}
\Gamma &= \sum_{k=1}^\infty \overbrace{\frac{1 }{k^3} { \frac{q^3}{2\pi^3 m^2}} e^{-2\pi k Q (q-\sqrt{q^2-m^2})} }^{\substack{\text{Answer including }\\ \text{perturbative quantum gravity corrections}}} \nn \nn \\ 
&- \underbrace{\frac{m^2 q^5 Q^2}{\pi^5} \left(\sum_{k=1}^\infty \frac{1}{k^2}e^{-2\pi k Q (q-\sqrt{q^2-m^2})} \right)^2 \left(\sum_{k=1}^\infty \frac{1}{k} e^{-2\pi k Q (q-\sqrt{q^2-m^2})} \right)}_{\substack{\text{Leading quantum correction}\\ \text{from topological expansion }}}  + \dots  \, . 
\end{align}

The first line of \eqref{eqn:Gamma_Schwinger_final_result-intro}, gives the dominant contribution to the decay rate and exactly matches the semiclassical answer of Schwinger \cite{schwinger1951gauge}. Somewhat surprisingly, the dramatic modifications to the black hole thermodynamics found at low energy do not significantly change the charged particle production rate. Essentially, this is because Schwinger pair production is a local process: it is therefore insensitive to the presence of the Schwarzian fluctuations, which are locally pure gauge.

The second line of \eqref{eqn:Gamma_Schwinger_final_result-intro} describes the corrections from gravitational interactions between an arbitrary number of instantons. These corrections are subleading compared to self-interactions of the positron loop, and potentially also compared to other multi-instanton interaction effects, such as those mediated by electromagnetism, that were not included in \eqref{eqn:Gamma_Schwinger_final_result-intro}. We included them in \eqref{eqn:Gamma_Schwinger_final_result-intro} because they demonstrate the power of the formalism we are using. Additionally, such effects might become important as the black hole charge approaches $Q_*$ and the instanton action becomes small. To obtain the result \eqref{eqn:Gamma_Schwinger_final_result-intro}, we took advantage of the fact that the instanton moduli space can be identified as a natural deformation of a $(2,\, p)$ minimal string theory. Thus, even though the corrections in \eqref{eqn:Gamma_Schwinger_final_result-intro} are not stringy, our results put the mathematical apparatus of the topological expansion to use in understanding black holes in our own universe. 

In \textbf{Section} \ref{sec:the-story-of-a-charged-black-hole}, we combine the results of Sections \ref{sec:neutral-particle-emission-main} and \ref{sec:charged-particle-emission-main} to give a complete description of the evaporation of charged black holes until they reach Planckian size. We consider an asymptotically flat universe described by Einstein gravity coupled to the Standard Model and containing an isolated single black hole.\footnote{In particular, we will generally ignore cosmological features of our own universe including dark matter and dark energy. This qualification is somewhat important since our analysis will involve timescales that are very long compared to the age of the universe and temperatures that are far below the de Sitter temperature associated to the positive cosmological constant present in the $\Lambda \text{CDM}$ model.}

\begin{figure}[t!]
\centering
\hspace*{-2.5cm}
    \includegraphics[width=1\textwidth]{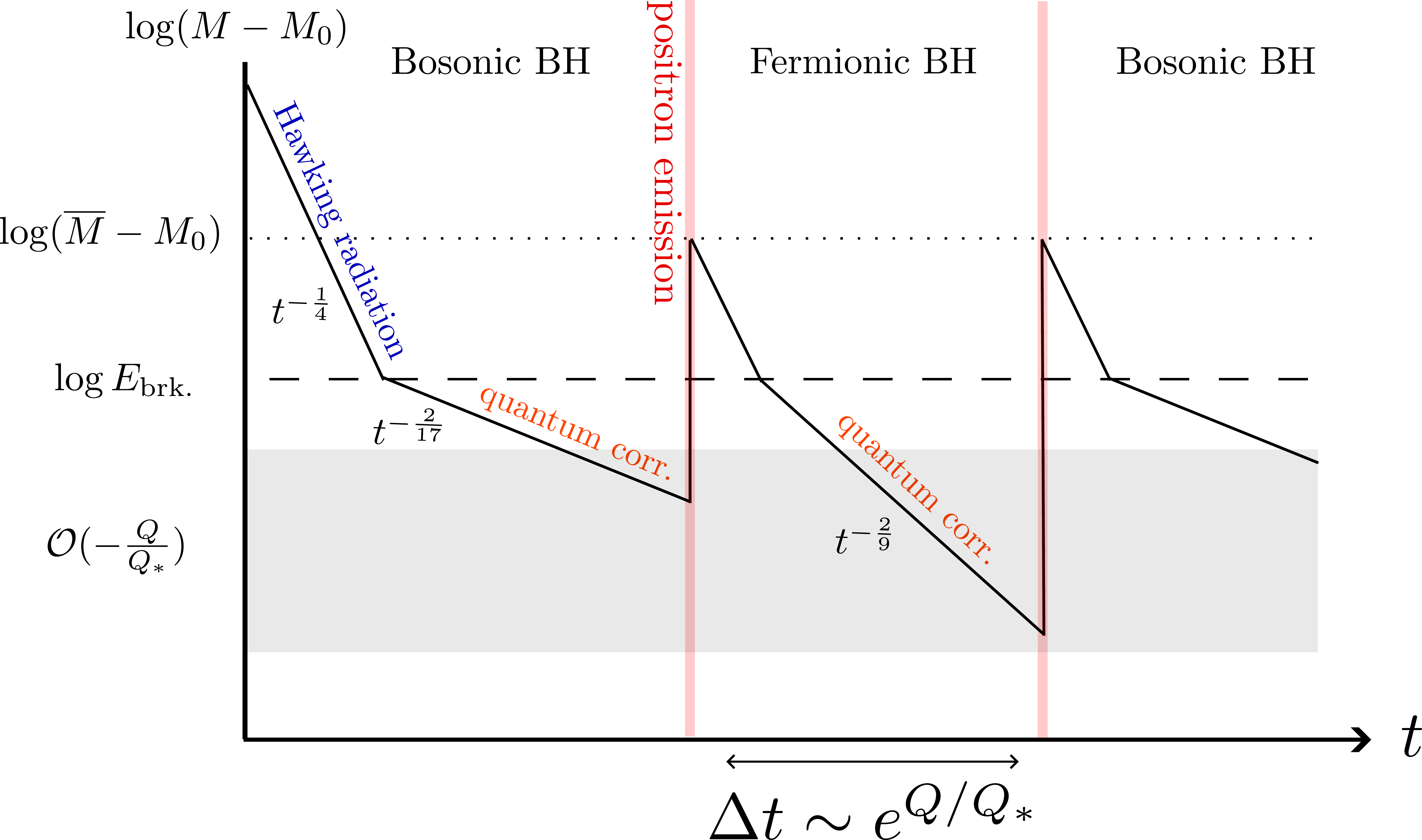}
    \caption{The evolution of the mass above extremality $M-M_0$ of a near-extremal black hole with initial charge $Q \gg Q_* \equiv 1.8 \times 10^{44} q$, as a function of time. Within the plot, time is scaled logarithmically within each straight-line segment. The emission of photons continuously decreases the mass, taking it below the breakdown scale $M-M_0 \sim \Ebrk$ and reaching exponentially close to extremality. Below the breakdown scale, the quantum effects slow down the Hawking emission process. Pair production is exponentially suppressed, with positrons appearing with Poisson statistics on timescales $\delta t  \sim e^{Q/Q_*}$. When a positron is made, it decreases the charge, increases the mass above extremality far above the breakdown scale, and causes the BH to transition from fermionic to bosonic and vice versa. 
    The quantum-corrected Hawking radiation is different for bosonic and fermionic black holes below the breakdown scale $\Ebrk$: bosonic black hole evaporation is dominated by di-photon decay, while fermionic black hole evaporation is dominated by single photon decay.  }
    \label{fig:largeBhEvapProcess1}
\end{figure}

If the black hole is sufficiently large, it will initially be driven very close to extremality by semiclassical photon and graviphoton emission. When $(M-Q) < \Ebrk$, only diphoton emission is possible from a black hole with $j=0$ and so the black hole cools very slowly relative to semiclassical expectations. Eventually -- on timescales of order $\exp(Q/Q_*)$  -- the black hole emits a positron. The positron carries away slightly more charge than energy, and so causes $(M-Q)$ to jump far above $\Ebrk$, up to about \be
\bar{M} - \bar{Q} = \bar M - (Q-q) \sim \frac{q Q_*}{Q} \gg \Ebrk
\ee
above extremality. After charged particle emission, neutral radiation begins to drive the black hole back towards extremality. However, the emission of a fermionic particle means that the black hole has now switched from being bosonic to fermionic (or vice versa). Eventually, another charged particle is emitted, and the ``tick-tock'' cycle between a slow-cooling bosonic black hole and a comparatively fast-cooling fermionic black hole repeats. This evolution is shown in figure \ref{fig:largeBhEvapProcess1}.

Through this slow process of punctuated quasi-equilibrium, eventually the black hole's charge will fall to $Q \sim Q_* = 1.8 \times 10^{44} q$, whereupon the emission of positrons is no longer exponentially suppressed. From this point onwards, the horizon electric field remains close to the critical Schwinger value $|\vec{E}| \sim \pi m^2/q$ so that  Schwinger pair production, like neutral Hawking radiation, is slow but not exponentially suppressed. This leads to the black hole approaching an approximately Schwarzschild metric, with $Q \sim M^2/ Q_*$. The neutral Hawking radiation and Schwinger pair production are well described semiclassically until the black hole reaches Planckian size. 

Finally, in \textbf{Section} \ref{sec:discussion}, we discuss possible future directions and speculate about potential observational signatures of the described quantum gravity corrections.


\section{Black hole thermodynamics and the Schwarzian} \label{sec:background}

In this section, we give a pedagogical introduction to the Hawking radiation of large charged black holes. We then review some standard results and recent developments regarding the dynamics and thermodynamics of near-extremal black holes. 

\subsection{Pedagogical overview: large black holes `self-tune' towards extremality and away from the semiclassical regime} \label{subsec:pedagogicaloverview}

Hawking radiation takes black holes that are well-described by the semiclassical approximation and drives them towards a regime where the semiclassical description breaks. 
If the black hole radiates long enough, eventually the gravitational backreaction of a single Hawking quantum becomes important and to make further quantitative predictions we must move beyond the semiclassical approximation and  use a quantum theory of gravity. 

For uncharged black holes, the reason the semiclassical approximation eventually breaks is that the black hole becomes too hot. Because temperature is inversely proportional to mass, $T \sim \frac{\hbar}{GM}$, a black hole that starts large and cold gets progressively smaller and hotter as it radiates, driving itself towards an `explosion' \cite{Hawking:1974rv} of notionally infinite temperature. At least that's what the semiclassical approximation says. But on the precipice of the explosion, as the temperature is approaching the Planck scale, the energy in a single quantum becomes comparable to the energy left in the black hole, the semiclassical approximation breaks down, and to know what happens next we need quantum gravity.

For the large charge black holes considered in this paper, we will also see a breakdown of the semiclassical approximation but in the opposite temperature regime. Hawking radiation makes black holes of extremely large charge exit the semiclassical regime by making them too \emph{cold}. The black holes `self-tune' to extremality, and when they get too close to extremality, the semiclassical approximation breaks. 
This phenomenon was first noted by Preskill \emph{et al.}~\cite{Preskill:1991tb}, building on earlier work of Gibbons \cite{Gibbons:1975kk}. 
 Let's review the basic physics 
behind this phenomenon, 
and how, even after the semiclassical analysis fails, a Schwarzian analysis maintains full control and allows us to describe what happens next.

\begin{enumerate}
\item The lightest uncharged particle is massless; the lightest charged particle is massive. 

The photon and the graviton are both massless, whereas the lightest known charged particles---the positron or the electron---are massive. This is the fact that is ultimately responsible for the self-tuning phenomenon. (If we eventually discover either a massless charged particle or a mass for both the photon and graviton, the results of this paper would need major revision.)

\item Though massive, the lightest charged particle is very light (in Planck units).

Two positrons electromagnetically repel but gravitationally attract. These two forces, though opposite, are not equal. Instead, the electromagnetic force is stronger than the gravitational force by a ratio
\begin{equation}
\frac{|\vec{F}_\textrm{electric}|}{|\vec{F}_\textrm{gravitational}|} = \frac{q^2}{4 \pi \epsilon_o r^2} \div \frac{Gm^2}{ r^2}  = \frac{G}{4 \pi \epsilon_0} \frac{q^2}{ m^2} = 
4.17 \ldots \times 10^{42} \ . \label{eq:largeratio}
\end{equation}
This is a stupendously large number. Ultimately all the large numbers in this paper (and many of the large numbers in daily life) trace back to the hugeness of this ratio.

We can ask whether it is the electromagnetic force that is very strong, or the gravitational force that is very weak. In classical physics, this question is meaningless: there is only a single dimensionless ratio. But in quantum physics we can use $\hbar$ to formulate two independent dimensionless numbers,
\begin{equation}
\frac{G m^2}{\hbar c} \equiv  \frac{m^2}{M_\textrm{Pl}^2} = 1.749 \ldots \times 10^{-45}
\ \ \textrm{ and } \ \ 
\frac{q^2}{4 \pi \epsilon_0 \hbar c} \equiv  \frac{q^2}{q_\textrm{Pl}^2} \equiv \alpha_\textrm{QED} = \frac{1}{137.06 \ldots} \ , 
\end{equation}
that characterize the strength of each force. The ratio of these two numbers is $4.17 \ldots \times 10^{42}$. Viewed this way, we see that the largeness of this ratio is caused not by the electromagnetic force being strong (indeed, electromagnetism is so `weakly coupled' that just the first dozen terms in the electromagnetic perturbation series for the anomalous magnetic moment of the electron are sufficient to give one of the most accurate experimental predictions that humanity has ever confirmed \cite{Fan:2022eto}), instead, it is caused by gravity being so very weak. Or, said another way, the ratio  \eqref{eq:largeratio} is huge because the positron ($m \sim 10^{-27}$grams) weighs so much less than the Planck mass ($ M_\textrm{Pl} \equiv  \sqrt{\hbar c/G  }  \sim 2 \times 10^{-5}$grams).

The `near masslessness' of the lightest charged particle means the self-tuning phenomenon only becomes operative for black holes with very large charge.

\item For black holes with very large charge, the electric field at the horizon is very small. 

For a Reissner-Nordstr\"om black hole of mass $M$ and charge $Q$ (reviewed in Sec.~\ref{subsec:semiclassicalRNsolution}), the electric field at the horizon is given by Gauss's Law as $|\vec{E}| = {Q}/{r_+^2}$,  where the horizon radius is $r_+ 
\equiv M + \sqrt{M^2 - Q^2}$, and where here and henceforth in this subsection we put $G= \hbar = c = 4 \pi \epsilon_0 = 1$. The inverse-square law means $|\vec{E}|$ becomes weak for large black holes (the horizon is `Rindler like'). Because all black holes must have $M \geq Q$,  the electric field at the horizon is thus upperbounded by 
\begin{equation}
|\vec{E}| \biggl|_{\textrm{horizon}} \leq \frac{1}{Q} \ . \label{eq:lowerboundonEhorizonfromQ}
\end{equation}
This inequality is saturated by extremal black holes, which have $M = Q$. 

(To be clear, \eqref{eq:lowerboundonEhorizonfromQ} is small for large charge $Q$ relative to the Planck scale  and to the Schwinger limit but not necessarily relative to the electric fields found in everyday life. Indeed the electric field around an extremal black hole with $M = Q = Q_*$ is still several orders of magnitude stronger than the strongest sustained fields ever produced by humanity.)

\item The emission of uncharged particles from a large RN black hole is only polynomially suppressed. 

The  temperature of an RN black hole is given, in the semiclassical approximation  \eqref{eq:tempandentorpyofRNblackhole}, by 
\begin{equation}
T  \sim \frac{ \sqrt{M^2 - Q^2} }{ \left( M + \sqrt{M^2 - Q^2} \right)^2} \ .  \label{eq:temperatureforRNintro}
\end{equation}
As for uncharged black holes, big means cold, but now there is an additional suppression near extremality (from the additional redshift as radiation climbs out of the gravitational potential caused by the electric field). The Stefan-Boltzmann formula says that the power in massless radiation from a blackbody of area $4 \pi r_+^2$ is $P \sim (r_+)^2 T^4$. As we will see in Sections~\ref{sec:semiclassical-result-neutral-hawking-radiation} and \ref{app:greybody} this is further suppressed by `greybody' diffraction, but crucially the greybody factors are merely powers in $r_+$ and ${(M - Q)/M}$, and so don't change the fact that the emission of massless particles is only polynomially suppressed.

\item The emission of charged particles from a large RN  black hole is exponentially suppressed. 

Because the lightest charged particle is massive,  emission becomes exponentially suppressed when the black hole gets too large. As we will explore in detail below, the semiclassical rate of charged-particle emission can be calculated by considering the rate of Schwinger pair production  $\exp[ - \pi m^2/(q |\vec{E}|)]$ near the horizon \cite{Gibbons:1975kk}. This is exponentially suppressed for small electric fields, $|\vec{E}| \ll m^2 /q$. Equation \eqref{eq:lowerboundonEhorizonfromQ} tells us that large black holes have small electric fields, and so charged particle emission is exponentially suppressed for any black hole with 
\begin{equation}
Q \gg Q_{*} \equiv \frac{q}{\pi m^2} \ .
\end{equation}
For positrons, this threshold evaluates to $Q_{*} \sim  2.9 \times 10^{25} \textrm{ Coulomb}$, which for an extremal black hole would require a mass a few percent that of Sagittarius A*. 
\item The emission of uncharged particles drives the black hole towards extremality; the emission of charged particles pushes the black hole away from extremality. 

The proximity to extremality is controlled by $M - Q$. 
Uncharged radiation carries away energy but not charge, and so reduces $M - Q$ and makes the black hole closer to extremality. Charged radiation carries away both charge and energy, but as we will discuss in Sec.~\ref{sec:charged-particle-emission-main} it carries away slightly more charge than energy, and so increases $M-Q$ and makes the black hole further from extremality.

When the black hole emits a positron, the change in the charge is $\Delta Q = -q$. Naively, you might think that $\Delta M = - m$, which since $q \gg m$ would imply that $\Delta (M - Q) = -m + q \sim q$. This is wrong. As will be explored in Sec.~\ref{sec:charged-particle-emission-main} the positron is expelled by the huge electrostatic potential of the black hole, and acquires a relativistic boost given by the squareroot of the ratio in  \eqref{eq:largeratio}, $\gamma \sim 2 \times 10^{21}$. This corresponds to a near-Planckian kinetic energy. At leading order in $Q_*/Q$, we have $\Delta M = \frac{q Q}{r_+} \rightarrow_{M \rightarrow Q} q$, meaning that extremal black holes remain extremal after emitting positrons. Subleading corrections make the black hole non-extremal, as we will show in Section \ref{sec:background-charged}.

\item Black holes with large enough charge are driven exponentially close to extremality. 

The approach to extremality is governed by the competition between the emission of mass and the emission of charge. Because, for large enough black holes, the emission of charged particles is exponentially rare, whereas the emission of uncharged particles is only polynomially rare, large black holes get driven exponentially close to extremality.

\item For black holes with large enough charge, Hawking radiation drives the black hole out of the semiclassical regime. 

We have seen that Hawking radiation forces black holes of very large charge to approach extremality. Now let's see that for black  holes that are close enough to extremality, the semiclassical approximation breaks. 

Wien's Law says that the typical energy of a blackbody quantum is given by the temperature. 
Equation \eqref{eq:temperatureforRNintro} says that as extremality is approached the semiclassical temperature is given by $r_+^{-3/2} \sqrt{M - Q}$. Because the greybody suppression  is a decreasing function of the quantum's energy, the typical energy of a photon emitted by a near-extremal semiclassical black hole is therefore slightly larger than $r_+^{-3/2}\sqrt{M - Q}$.  Close enough to extremality, this must be larger than the total energy above extremality $M - Q$. Since the black hole cannot possibly emit more energy than it has available to it, this is a sign than something has gone wrong. What has gone wrong is that the semiclassical approximation has broken. Comparing  \eqref{eq:temperatureforRNintro} with $M - Q$ we see that the breakdown must occur when the energy above extremality (and the semiclassical temperature) is no less than 
\begin{equation}
E_\textrm{brk.} \equiv \frac{M_\textrm{Pl}}{Q^3} \ . 
\end{equation}
Below this scale, the energy of a single Hawking quantum causes significant gravitational backreaction, and to calculate what happens next we need a quantum theory of gravity \cite{Preskill:1991tb}.

\item For near-extremal black holes, the relevant backreaction effects are `universal' and fully captured by a Schwarzian analysis. 

The inevitable breakdown of the semiclassical approximation for isolated black holes of large enough charge was first pointed out by Preskill \emph{et al.}, who noted that the semiclassical regime ``must be replaced by a significantly different quantum-mechanical description” \cite{Preskill:1991tb}. But what this description should be, they were unable to say. Thankfully, there has recently been great progress in understanding near-extremal black holes, and we now know the correct quantum-mechanical description: the Schwarzian \cite{Sachdev:2015efa, Almheiri:2016fws, Maldacena:2016upp, Nayak:2018qej, Moitra:2018jqs, Castro:2018ffi, Larsen:2018cts,  Moitra:2019bub, Sachdev:2019bjn,Maldacena:2019cbz, Charles:2019tiu, Larsen:2020lhg, Castro:2021wzn, Iliesiu:2020qvm, Heydeman:2020hhw,  Ghosh:2019rcj, Iliesiu:2022onk, Boruch:2022tno, Kapec:2023ruw, Rakic:2023vhv, Kolanowski:2024zrq}.

Though including the gravitational backreaction of the Hawking radiation requires quantum gravity, we will see that it does not require a full non-perturbative theory of quantum gravity and does not depend in detail on which quantum theory of gravity we choose. Rather, our results are valid in any theory where the low-energy effective theory is correctly described by Einstein-Maxwell theory coupled to matter fields.  For such theories, our answers will be correct for temperatures both above and below $E_\textrm{brk.}$.

(The details of the quantum theory of gravity only become important for temperatures exponentially small in the entropy of the black hole, which are sensitive to the discreteness of the energy spectrum.) 
\end{enumerate}


\subsection{The semiclassical RN solution} \label{subsec:semiclassicalRNsolution}
The Reissner-Nordstr\"om solution describes a black hole of mass $M$ and charge $Q$ in an otherwise empty and flat spacetime.

 The Euclidean action of Einstein gravity coupled to a $U(1)$ gauge field is given by
\be \label{eqn:I_Einstein_Maxwell}
I_{E M}=  -\frac{1}{16 \pi G_N}\left[\int d^4 x \sqrt{g}R + 2 \int_{\partial M} \sqrt{h} K\right]+\frac{1}{16\pi} \int d^4 x \sqrt{g} F_{\mu \nu} F^{\mu \nu}-\frac{1}{4\pi} \int_{\partial M} \sqrt{h} n_i F^{i j} A_j ,
\ee
where Newton's constant and the Planck length are related by $G_N = \ell_{\textrm{pl}}^2$.\footnote{ 
In our conventions, we have set $4\pi \varepsilon_0= \hbar = c = 1$. In these units, the total charge is quantized as $Q = q \cdot \mathbb{Z} = \alpha^{1/2} \cdot \mathbb{Z} $ where $q = \alpha^{1/2}$ is the charge of a positron in Planck units and the electric potential of a point charge is $V=Q/r$.} In this section, we will keep factors of $G_N$ (though set $4\pi \varepsilon_0= \hbar = c = 1$) to emphasize the scaling of various quantities. The last term is added to make the variational principle well-defined when fixing the electric charge at infinity, and $n_i$ is an outward pointing unit vector at the boundary. 

We will consider boundary conditions where we fix the asymptotic charge $Q$ to be large and the inverse temperature to be $\beta$. The temperature is fixed by demanding that the asymptotic thermal circle has periodicity $\tau \sim \tau + \beta$, while fixing the field strength $F_{i j}$ at the boundary fixes the charge. The Euclidean Reissner-Nordstr\"om solution with these boundary conditions is given by
\be \label{eqn:RN_metric}
d s^2=f(r) d \tau^2+\frac{d r^2}{f(r)}+r^2 d \Omega_2^2, \quad f(r)=1-\frac{2 G_N M}{r}+\frac{G_N Q^2}{r^2}=\frac{(r-r_+)(r-r_-)}{r^2},
\ee
where we have introduced the inner and outer horizons $r_{\pm} = G_N M \pm \sqrt{(G_N M)^2-G_N Q^2}$ given by solving $f(r_{\pm})=0$; and the 2-sphere has the standard metric $d\Omega_2^2 = d \theta^2 + \sin^2 \theta d \phi^2$. When we refer to the `horizon' we will always mean the outer horizon, $r_+$, which is the closest you can approach with a rocket if you hope to get out again. In Euclidean signature the radial coordinate runs from $r \in [r_+, \infty)$, and the time coordinate is periodic $\tau \sim \tau + \beta$ with $\beta$ being the inverse of the temperature specified below. The solution for the field strength $F=d A$ is given by
\be   
F= i \frac{Q}{r^2} dr \wedge dt, \quad A = i Q\lrm{\frac{1}{r_+}-\frac{1}{r}} dt. \label{eq:electricfieldforRN}
\ee

\begin{figure}[t!]
    \centering
    \includegraphics[width=0.85\textwidth]{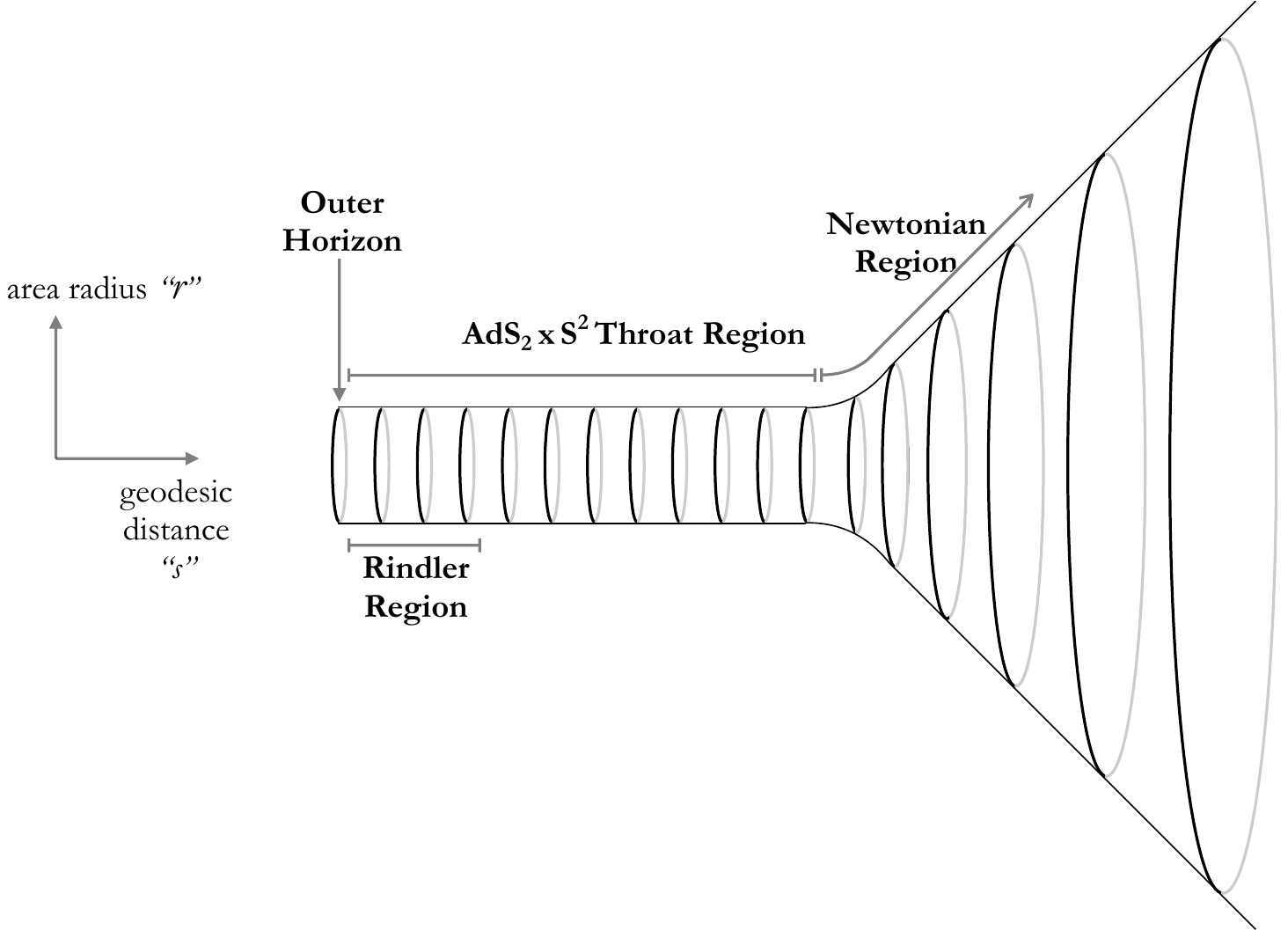}
    \caption{A snapshot of the exterior of a near-extremal RN black hole. As for the Schwarzschild solution, there is a region far from the horizon where the Newtonian approximation holds (the `Newtonian' region), and a region very close to the horizon where the gravitational field grows much more strongly than inverse-square (the `Rindler' region). What's new for the RN solution is that the Rindler region forms only part of a long `throat'. In the throat, the geometry is approximately $AdS_2 \times S^2$, so the area-radius $r$ is approximately constant, and, away from the Rindler region, the gravitational field is approximately constant. The closer the black hole is to extremality, the longer the throat.} 
    \label{fig:throat}
\end{figure}

\paragraph{The throat region.} The proper distance $\Delta s$ from $r_+$ out to an area-radius $R$ is
\begin{equation}
\Delta s = \int_{r_+}^{R} \frac{dr}{\sqrt{f(r})} = \sqrt{(R- r_+)(R- r_-)} + (r_+ + r_-) \, \textrm{arcsinh} \left[ \sqrt{ \frac{R - r_+}{r_+ - r_-}} \, \right] . \label{eq:distancefromrplus}
\end{equation}
At large $R$ the term on the left dominates and we have $\Delta s \sim R$ as in flat space. But close to the horizon the term on the right dominates and there is a long region---what below we will call the `throat'---that has approximately constant area. This is shown in Figure \ref{fig:throat}. The length of the throat ending at $r\approx 2 r_+$ is
\be \label{eqn:length_throat}
\Delta s \sim r_+ \log \lrm{ \frac{4 r_+}{r_+ - r_-}}\,,
\ee
and so gets long in the near-extremal limit $r_- \rightarrow r_+$. The Rindler region, which in our terminology is a subset of the throat, is defined by $r\lesssim 2r_+ - r_-$. The proper distance from $r_+$ to radius $r$ in this region is
\be \label{eqn:length_rindler}
\Delta s \sim 2 r_+ \sqrt{\frac{r-r_+}{r_+-r_-}}\,.
\ee
It is in the Rindler region that the proper force required to keep a particle at fixed radius $r$ deviates from Newton's law. 

\paragraph{Semiclassical thermodynamics.} The temperature and Bekenstein-Hawking entropy of the black hole are
\be
T = \frac{|f'(r_+)|}{4\pi} = \frac{r_+ - r_-}{4\pi r_+^2} = \frac{1}{2 \pi} \frac{ \sqrt{G_N^2 M^2 - G_N Q^2}}{r_+^2} , \qquad S = \frac{A}{4 G_N} = \frac{\pi r_+^2}{G_N}. \label{eq:tempandentorpyofRNblackhole}
\ee

At extremality the temperature is zero and the outer and inner horizons become degenerate $r_+ = r_-$. We will denote the horizon radius at extremality by $r_0$. At fixed charge $Q$ the extremal radius, mass, and entropy are given by
\be
r_0 = \sqrt{G_N} Q, \qquad M = \frac{r_0}{G_N} = \frac{Q}{\ell_{\t{pl}}}, \qquad S_0 = \frac{\pi r_0^2}{G_N} = \pi Q^2,
\ee
where we have also given their scaling with the large parameter $Q$. If we turn on a slightly non-zero temperature we find the outer horizon radius to be $r_+ = r_0 + 2\pi r_0^2 T+10\pi^2 G_N^{3/2} Q^3 T^2$, and the energy and entropy above extremality are
\be \label{eqn:M(Q,T)_semiclassical}
M(Q,T) = Q + \frac{2\pi^2}{\Ebrk} T^2 + 16\pi^3 G_N Q^4 T^3 +\ldots,\qquad S(Q,T) =  S_0 + \frac{4\pi^2}{\Ebrk} T + \ldots,
\ee
where we have included subleading corrections important for recovering the AdS$_2$ geometry and introduced an emergent energy scale $\Ebrk$ at small temperature given by
\be
\Ebrk = \frac{G_N}{r_0^3} = \frac{1}{\ell_{\t{pl}}Q^3 }\,. 
\ee 
We will shortly review how the thermodynamics of near extremal Reissner-Nordstr\"om break down for temperatures below the breakdown scale $T \lesssim \Ebrk$.

\subsection{The $AdS_2 \times S^2$ throat}
We have seen that when the black hole is close to extremality, the near-horizon region has approximately constant area-radius. We can can study this region---the `throat'---more closely by changing coordinates,
\be
r \to r_+(T) + 2 \pi G_N Q^2 T \lrm{\cosh(\rho) -1}, \qquad \tau \to \frac{1}{2\pi T}\tau.
\ee
With this change of coordinates the metric \eqref{eqn:RN_metric} can be expanded in the near-horizon region using \eqref{eqn:M(Q,T)_semiclassical} and taking the small temperature limit $T\to 0$ to find
\begin{align} \label{eqn:4d_NHR_metric}
ds^2 &= \underbrace{r_0^2 \lrm{d \rho^2 + \sinh^2 \rho~ d \tau^2} + r_0^2 d \Omega_2^2}_{\t{exact AdS$_2 \times S^2$}} \nn \\ &+  \underbrace{4 \pi G_N^{\frac{3}{2}} Q^3 T \cosh \rho ~ d \Omega_2^2 + \pi G_N^{\frac{3}{2}} Q^3 T (2 + \cosh \rho) \tanh^2 \frac{\rho}{2} (d \rho^2 - \sinh^2 \rho d \tau^2)}_{\t{finite temperature corrections}}\,.
\end{align}
The first line is an exact AdS$_2 \times S^2$ spacetime with identical AdS and sphere radii
\be
\ell_{\t{AdS}_2} = r_0 = \ell_{\t{pl}} Q,  \qquad R_{S^2} = r_0.
\ee
This is the near-horizon geometry for an extremal black hole with zero temperature. The finite temperature corrections capture deviations away from extremality for a near extremal black hole, where we have only kept terms up to linear order in temperature $T$. The metric is only valid up to a cutoff $\rho_c$ where higher temperature corrections become important. Note that after including finite temperature corrections, the transverse sphere grows slightly as we move out along the throat (which, from the JT gravity perspective, will correspond to the dilaton increasing). 

The same coordinate change can be applied to the gauge field. Ignoring finite temperature corrections, the leading order form of the vector potential is
\be
F = i Q \sinh \rho ~ d \rho \wedge d \tau, \qquad A= i Q \lrm{\cosh \rho -1} d \tau\,,
\ee
and so, at extremality, the throat is supported by a constant flux.\footnote{Finite-temperature corrections to the above formulas can be found through the same coordinate transformation used for the metric, but these corrections will not be important for us, see \cite{Iliesiu:2022onk} for a discussion.}

\subsection{The one-loop partition function and the Schwarzian}

\label{sec:one-loop-part-function-and-the-Schwarzian}

We now explain how to evaluate the one-loop partition function around the RN background by dimensionally reducing the four-dimensional theory to two-dimensional JT gravity. The key point is that at sufficiently low temperatures, certain modes become light, and this drastically modifies the thermodynamics of RN black holes.
We give a brief overview with all technical details readily contained in \cite{Iliesiu:2020qvm,Iliesiu:2022onk}.

The problem amounts to expanding the metric $g+\delta g$ and gauge field $A + \delta A$ around the classical solutions \eqref{eqn:RN_metric} and integrating over fluctuations $\delta g, \delta A$ to obtain the relevant one-loop determinant. The modes that become important at low temperatures can be identified by looking at fluctuations around the four-dimensional near-horizon geometry \eqref{eqn:4d_NHR_metric}, but it is easier to instead reduce to two dimensions and study the resulting theory of JT gravity. We take the ansatz for the 4d metric to be
\be
ds^2 = \frac{r_0}{\chi^{1/2}} g_{\mu \nu} dx^\mu dx^\nu + \chi d \Omega_2^2, 
\ee
where $x^\mu = (\rho, \tau)$ are AdS$_2$ coordinates, $g_{\mu \nu}(x)$ is the two dimensional metric that will become AdS$_2$, and $\chi(x)=r_0^2 + G_N \Phi(x)$ contains the two dimensional dilaton $\Phi$. The size of the transverse sphere is captured by the dilaton, and slowly grows along the AdS$_2$ throat as expected from the form of the finite temperature solution \eqref{eqn:4d_NHR_metric}. We have ignored fluctuations of the metric on the sphere since such modes are not important in the low temperature limit we will consider. A similar expansion can be applied to the gauge field, with the final result that only the zero mode of the gauge field is important at low temperatures. 

After carefully carrying out the dimensional reduction of the 4d Einstein-Maxwell action around the Reissner-Nordstr\"om black hole solution with fixed charge $Q$ and temperature $\beta$ the action is found to be controlled by 2d JT gravity \cite{Iliesiu:2020qvm}
\begin{align} \label{eqn:I_EM_reduced}
&I_{\t{EM}}^Q[g_{\t{RN}}+\delta g, A_{\t{RN}} + \delta A] = \nn \\ &\underbrace{-S_0(Q) + \beta Q}_{\t{extremal entropy, mass}} \underbrace{- \frac{1}{2} \int d^2 x \sqrt{g}  \Phi \left( R+\frac{2}{\ell^2_{\t{AdS$_2$}}} \right) - \Phi_{b,Q} \int_{\partial_M} d u \sqrt{h} \left(K-\frac{1}{\ell_{\t{AdS$_2$}}} \right) }_{\t{JT gravity action in AdS$_2$ throat}}\,,
\end{align}
where in the above we have already integrated out the gauge field fluctuations $\delta A$ to get the right-hand side. In writing the above action we have only kept track of fluctuations that become important in the low temperature limit $\beta \Ebrk \gg 1$. The first two terms are the extremal entropy $S_0$ and extremal mass $Q$ of a BH of charge $Q$. The last two terms are the 2d JT gravity action with appropriate boundary term in the AdS$_2$ throat. The contribution of the asymptotically flat region far from the BH has also been taken into account in the above formula. The correct boundary conditions fix the dilaton to the value $\Phi_{b,Q} = (\Ebrk \epsilon)^{-1}$ where $\epsilon$ is a cutoff in the AdS$_2$ throat. The four-dimensional gravity path integral is thus reduced to a two dimensional path integral
\be
Z_{\t{RN}}(\beta, Q) = e^{S_0(Q) - \beta Q}\int \mathcal{D} g_{\mu \nu} \mathcal{D} \Phi \hspace{.05cm} e^{-I_{\t{JT}}[g_{\mu \nu},\Phi]},
\ee
where the JT action can be identified from \eqref{eqn:I_EM_reduced}. The JT path integral can be evaluated exactly by integrating over the dilaton $\Phi$ along an imaginary contour to enforce a delta function constraint on the curvature. The remaining integral is over all boundary fluctuations weighted by the Schwarzian action, which comes from evaluating the boundary extrinsic curvature. Since we will refer again to the perturbative expansion of the boundary modes in later section, the  fluctuations on the near-horizon boundary are weighted by 
\be
\label{eq:Schw-quadratic-fluct}
I_\text{Schw}^\text{quad} \sim \sum_{|n|\geq 2} \frac{T}{\Ebrk} n^2(n^2-1) \epsilon_n \bar \epsilon_n\,,
\ee 
at quadratic order in the boundary fluctuations determined by the complex modes $\epsilon_n$. The integral should not include the modes $n=-1,\,0,\, 1$, which correspond to the generators of the $SL(2,\mathbb R)$ isometry, which are not physical modes since they do not change the geometry of the near-horizon region and are quotiented out in the gravitational path integral. As $T\to 0$, the quantum fluctuations over all physical modes become larger, and such modes need to be integrated out exactly. Integrating-out these fluctuations exactly, the partition function is given by 
\be \label{eqn:thermalZ}
Z_{\t{RN}}(\beta, Q) = \frac{1}{4\pi^2}\underbrace{\left( \frac{2\pi}{\beta \Ebrk } \right)^{\frac{3}{2}}}_{\t{one-loop correction}} \exp \left(\underbrace{S_0(Q) - \beta Q}_{\t{extremal entropy, mass}}  + \underbrace{\frac{2\pi^2}{\beta \Ebrk} }_{\substack{\t{semiclassical correction} \\ \t{to entropy+mass}}}\right).
\ee
The first term is the one-loop correction from the JT mode that becomes important at low temperatures $\beta \Ebrk \gg 1$, where it can be noticed that $Z_{\t{RN}} \to 0$ as $\beta \Ebrk \to \infty$. The exponential contains the extremal entropy, extremal mass, and leading semiclassical correction to the entropy and mass terms away from extremality. The partition function can be written in a way that makes it easy to extract the density of states through an inverse Laplace transform
\be
Z_{\t{RN}}(\beta, Q) = \int_{Q}^\infty d M \frac{e^{S_0 (Q)}}{2\pi^2 \Ebrk } \sinh \left(2\pi \sqrt{2} \sqrt{\frac{M-Q}{\Ebrk} } \right) e^{-\beta E}\,.
\ee
The density of states for fixed charge $Q$ and energy above the extremality bound with zero angular momentum,
\be 
E \equiv M - Q\,,
\ee 
is therefore given by 
\be \label{eqn:schwarziandos}
\rho(E,Q) =\frac{e^{S_0 (Q)}}{2\pi^2 \Ebrk } \sinh \left(2\pi \sqrt{2} \sqrt{\frac{E}{\Ebrk} } \right) \Theta \left(E\right).
\ee
(The above formula should be trusted up to exponentially small (in $Q$) temperatures, where other gravitational configurations may become important.) We see that the density of states smoothly approaches zero as the energy above extremality is lowered.  Thus at sufficiently small temperatures $T \lesssim \Ebrk$ there are many fewer states than the semiclassical answer \eqref{eqn:M(Q,T)_semiclassical} would predict. From the above we can derive the correction to the energy above extremality
\be\label{eqn:M(Q,T)_quantum}
M(Q,T) = Q + \frac{2 \pi^2}{\Ebrk}T^2 + \frac{3}{2} T + \ldots,
\ee
which should be compared with \eqref{eqn:M(Q,T)_semiclassical}. At temperatures below the breakdown scale $T < \Ebrk$ the energy above extremality begins to grow linearly with temperature. 

Note that for $T \lesssim \Ebrk$ the growth in the density of states is only a power-law (rather than exponential) in energy above extremality. As a result, the fluctuations in energy above extremality in the canonical ensemble are comparable to the average energy above extremality (given in \eqref{eqn:M(Q,T)_quantum}). Consequently, the canonical and microcanonical ensembles begin to behave quite differently, and one has to be careful when associating a single energy to the canonical ensemble or a single temperature to the microcanonical ensemble.

\begin{figure}
    \centering
    \includegraphics[width=.4\textwidth]{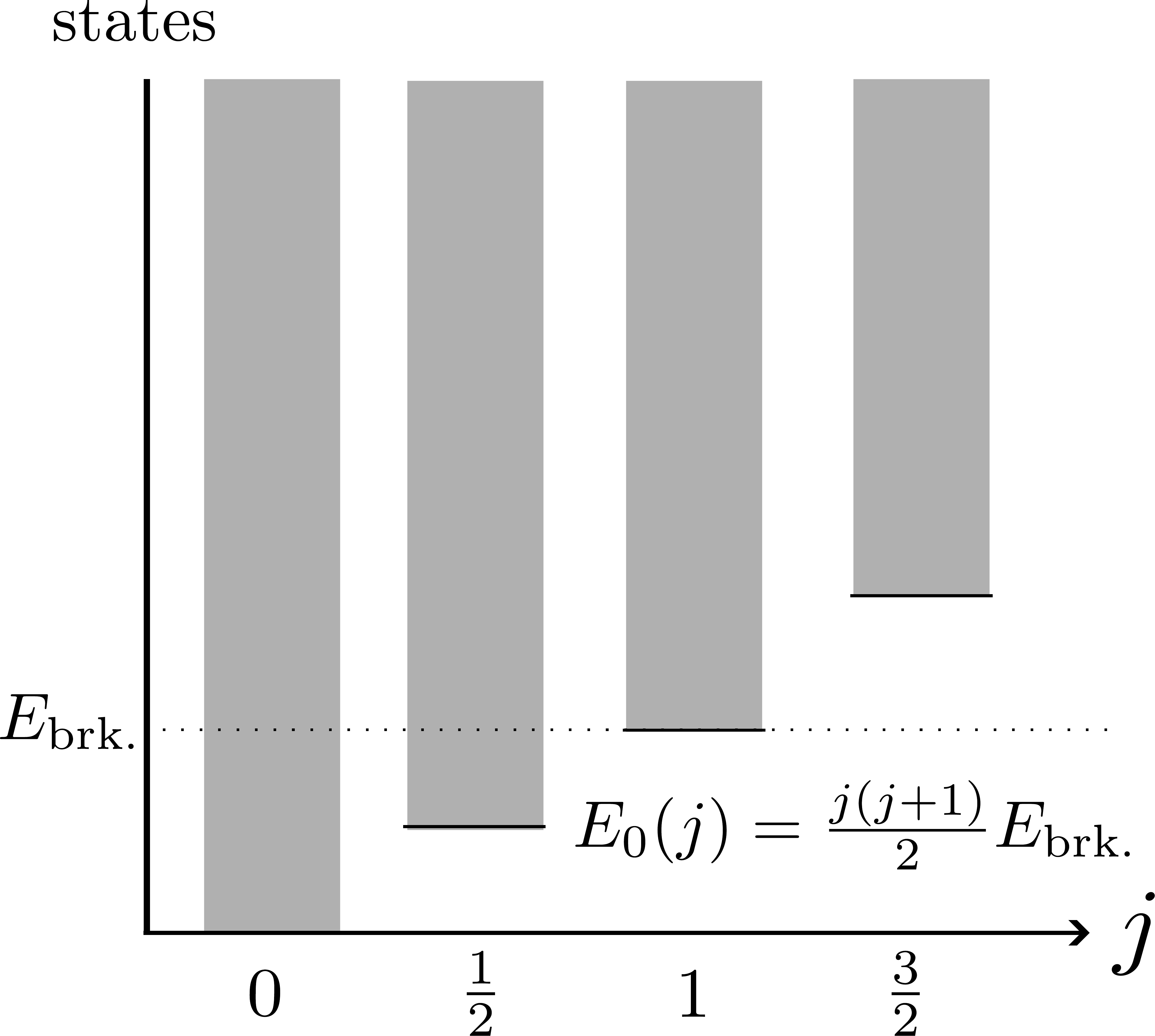}
    \caption{For any integer or half-integer angular momentum $j$, there exists a quasicontinuous density of black hole microstates for energies $E > Q+ E_{0}^j = Q + \frac{j (j+1)}{2} \Ebrk$.}
    \label{fig:spectra}
\end{figure}

\paragraph{Adding Angular momentum.}
The preceding discussion considered Reissner-Nordstr\"om black holes without angular momentum. However, a similar analysis can be done for black holes with fixed charge and fixed integer or half-integer angular momentum $j$. An extra $SU(2)$ mode capturing metric fluctuations on the $S^2$ must be included in the Schwarzian analysis, with final result for the density of states with fixed $j,m$\cite{Heydeman:2020hhw}
\be \label{eqn:rho_angularmomentum}
\rho_j(E,Q) =\frac{(2j+1) e^{S_0} }{2\pi^2 \Ebrk } \sinh \left(2\pi \sqrt{2} \sqrt{\frac{E-E_{0,j}}{\Ebrk} } \right) \Theta \left(E-E_{0,j} \right)\,.
\ee
In total, there are actually $2j+1$ times this number of states with angular momentum $j$ due to a sum over the quantum number $m$. The spectrum in each sector begins at
\be \label{eqn:gap_angularmomentum}
M = Q + E_{0}^j \equiv Q + \frac{j (j+1)}{2} \Ebrk \,,
\ee
as shown in Figure \ref{fig:spectra}. For $j=0$ the spectrum starts at $M=Q$, while for higher $j$ the spectrum is shifted by an amount proportional to $\Ebrk$. For $j=1$ the spectrum begins at $Q+ E_{0}^{j=1} = Q + \Ebrk$, and so there are no states below this energy for this value of charge, while for $j=0$ there are states in this energy window. This will be important when we consider the evaporation of such black holes in our universe since photons carry angular momentum and thus, due to selection rules, induce transitions between black holes in different angular momentum sectors.

A simple way to understand the formula \eqref{eqn:rho_angularmomentum} is that the $SU(2)$ almost-zero modes decouple from the JT gravity modes and act as an $SU(2)$ quantum rigid rotor. This rotor enforces the constraint that the angular momentum $j$ must be integer or half-integer and also contributes an energy  $j (j+1) \Ebrk/2$ to the total energy of the black hole. This energy shifts the JT gravity spectrum \eqref{eqn:schwarziandos} and leads to the density of states \eqref{eqn:rho_angularmomentum}. 

\subsection{Matter correlators}
\label{sec:matter-correlators}

\paragraph{JT gravity two-point function.} For later analyses, we will need some matter correlators and operator matrix elements in JT gravity \cite{Mertens:2017mtv, Yang:2018gdb, Iliesiu:2019xuh, Mertens:2022irh}. The basic matrix element that we are interested in is $| \langle E_f| \mathcal{O}|E_i\rangle|^2 $, where 
\begin{align}
|{E}\rangle\langle E| = \frac{P_{E}}{\Tr[P_{E}]} = \frac{P_{E}}{\rho(E)}
\end{align} 
represents a projection\footnote{Really, $P_E$ is a projection-valued measure where integrating $P_E dE$ gives a projector onto a range of energies. Our conventions are $\lb E|E' \rb=\rho(E)^{-1}\delta(E-E')$.} $P_E$ onto states of the black hole with energy $E$ above extremality that has been normalised to have trace one by dividing by the density of states 
\be
\Tr[P_{E}] = \rho(E)\equiv \frac{e^{S_0 }}{2\pi^2 \Ebrk } \sinh \left(2\pi \sqrt{2} \sqrt{\frac{E}{\Ebrk} } \right).
\ee
This formula can be found e.g. by regarding the trace as an infinite-temperature limit of the Hartle-Hawking state $|HH \rb$. 

Two matter insertions with scaling dimension $\Delta$ separated by a geodesic length $\ell$ in AdS$_2$ is equivalent to an insertion of $e^{-\Delta \ell}$ in the gravitational path integral. To account for the fluctuations of the metric, one consequently has to integrate over this geodesic length $\ell$ appropriately weighted in the gravitational path integral. To compute the matrix element  $| \langle E_f| \mathcal{O}|E_i\rangle|^2 $, we can view  $e^{-\Delta \ell}$ as an operator inserted between two eigenstates of the ADM Hamiltonian $\psi_E(\ell)$, one with energy $E=E_i$ and the other with energy $E=E_f$, expressed in the geodesic basis \cite{Yang:2018gdb}.\footnote{Such eigenstates are given by $\psi_E(\ell) = 4 K_{i \sqrt{8\frac{E}{\Ebrk}}}\lrm{e^{-\ell/(2\ell_{AdS_2})}}$.} Integrating over $\ell$,  the matrix elements of a boundary matter operator with scaling dimension $\Delta$ (coupled to the Schwarzian fluctuations) between energy eigenstates of the black hole are given by
\be \label{eqn:2ptfnJT}
\frac{\Tr(P_{E_f} \cO P_{E_i} \cO)}{\Tr(P_{E_f}) \Tr(P_{E_i})} \equiv | \langle E_f| \mathcal{O}|E_i\rangle|^2 = \frac{2 e^{-S_0} \, \Gamma \lrm{\Delta \pm i \sqrt{2 \Ebrk^{-1} E_f} \pm i \sqrt{2 \Ebrk^{-1} E_i}} }{(2  \Ebrk^{-1})^{2\Delta} \Gamma(2\Delta) }  \equiv 
\frac{2 e^{-S_0}\Gamma_{E_i,E_j}^\Delta}{(2 \Ebrk^{-1})^{2\Delta} \Gamma(2\Delta)} \,,
\ee 
where we have specified that the normalization of the matrix element through the ratio of one-sided black hole Hilbert space traces on the left. Here, we are using a standard convention where there is an implicit product over all choices of sign appearing in gamma functions; in the last equation, we also defined $\Gamma_{E_i E_j}^\Delta$ for later convenience. The energies $E_i$ and $E_f$ in \eqref{eqn:2ptfnJT} (and the energies that appear below in \eqref{eqn:jt4ptfn}) represent the energy of the black hole above the extremal energy $Q + E_{0}^j$. In other words, the total energy of a black hole in the state $|E\rangle$, with charge $Q$ and angular momentum $j$, is $Q + E_{0}^j + E$. 

In addition, higher dimensional fields that carry angular momentum, after dimensional reduction, are charged under the two dimensional $SU(2)$ field and give non-trivial matrix elements between black hole states with different spin \cite{Iliesiu:2019lfc}. Once this is taken into account, two matter operator insertions separated by a geodesic of length $\ell$ and an $SU(2)$ holonomy $g$ along the geodesic slice is equivalent to inserting $e^{-\Delta \ell} D^{\ell}_{m_p m_p}(g)$ in the gravity path integral where we now have to perform an additional integral over the $SU(2)$ group element $g$ in addition to the integral over $\ell$. Here, the matter operator has spin $\ell, m_p$ and $D^j_{m_p m_p}(g)$ are Wigner-D matrix elements for the group element $g$.\footnote{In this paper, we will use the same conventions for the Wigner-D matrix elements and for Clebsch-Gordan coefficients used in Mathematica 14.}  

The squared matrix element $|\lb E_f , j',m' | \mathcal{O}_{\ell, m_p} | E , j,m \rb|^2$ is a product of a matrix element for $e^{-\Delta \ell}$, which simply gives \eqref{eqn:2ptfnJT}, together with a matrix element for the $SU(2)$ holonomy. 
 To construct the holonomy matrix element, we will use the $SU(2)$ holonomy for the infinite-temperature limit of the Hartle-Hawking state $|HH\rb$, which has wavefunction
 \be 
 \lb g|HH\rb = \delta(g)\,, \qquad \lb HH|j;m,m\rb = \sqrt{2j+1},
 \ee
 where the states
 \be\label{eq:normalizedwigner}
\lb g|j;m,m'\rb = \sqrt{2j+1}D^j_{m m'}(g)\,,
\ee
form an orthonormal eigenbasis for the angular momentum $j$ and axial angular momentum $m$. The trace of the projector
\begin{align}
    P_{j,m} = \sum_{m'}|j;m,m'\rb \lb j;m,m'|
\end{align}
is then
\begin{align}
    \Tr[P_{j,m}] = \lb HH| P_{j,m}| HH \rb = 2j+1,
\end{align}
which gives the factor of $(2j+1)$ appearing in \eqref{eqn:rho_angularmomentum}. The matrix element for the holonomy in the matter two-point function is \cite{Iliesiu:2019lfc}: \begin{align}\label{eq:spinfactor}
\frac{\lb HH | P_{j',m'} D^{\ell}_{m_p,m_p}(g) P_{j,m}|  HH \rb}{\Tr[P_{j',m'}]\Tr[P_{j,m}]} &=\int dg (D^{j'}_{m',m'}(g))^*  D^{\ell}_{m_p,m_p}(g)  D^{j}_{m,m}(g) \nn\\
&=\frac{1}{2j'+1}\left|C^{j' m'}_{j m, \ell m_p} \right|^2\,,
\end{align}
where $C^{j' m'}_{j m, \ell m_p}$ are Clebsch-Gordan coefficients of $SU(2)$.  Putting everything together, we can now obtain  the squared matrix element $|\lb E_f , j',m' | \mathcal{O}_{\ell, m_p} | E , j,m \rb|^2$ by considering the ratio of traces 
\be \label{eqn:spin_2pt_fn}
\frac{\Tr(P_{E_f,j',m'} \cO_{\ell, m_p} P_{E_i, j, m} \cO_{\ell, m_p})}{\Tr(P_{E_f,j',m'}) \Tr(P_{E_i, j, m})} &\equiv | \langle E_f, j', m'| \mathcal{O}_{\ell, m_p} |E_i, j, m\rangle|^2 \nn \\ &= \frac{e^{-S_0}}{2j'+1}\left|C^{j' m'}_{j m, \ell m_p} \right|^2  \times \frac{2\Gamma_{E_i,E_j}^\Delta}{(2 \Ebrk^{-1})^{2\Delta}\Gamma(2\Delta)}\,,
\ee
where $j',j,\ell$ are the angular momenta of the final BH, initial BH, and particle, respectively, while $m',m,m_p$ are the respective axial $J_z$ angular momenta.   The energies $E_f$ and $E_i$ are again the energies above the extremality bounds $Q + E_0^{j'},Q +E_0^{j}$ respectively. Above, $P_{E, j, m}$ are projectors onto states with fixed energy $E$, angular momentum $j$ and axial angular momentum $m$. 


\paragraph{JT gravity four-point function.} We will also need the four-point function between energy eigenstates. For a free scalar field, the four-point function is once again normalized by considering a ratio of traces and is given by \cite{Mertens:2017mtv,Jafferis:2022wez},
\begin{align} \label{eqn:jt4ptfn}
 &\frac{\Tr(P_{E_1} \cO P_{E_2} \cO P_{E_3} \cO P_{E_4} \cO)}{\Tr(P_{E_1}) \Tr(P_{E_2}) \Tr(P_{E_3}) \Tr(P_{E_4})} = \lb E_1 | \mathcal{O} | E_2 \rb \lb E_2 | \mathcal{O} | E_3 \rb \lb E_3 |  \mathcal{O} | E_4 \rb \lb E_4 | \mathcal{O} | E_1 \rb = \\
&\quad =\mathcal{N}_{4pt}\Ebrk^{4\Delta} (\Gamma_{E_1,E_2 }^\Delta \Gamma_{E_2,E_3}^\Delta \Gamma_{E_3,E_4}^\Delta \Gamma_{E_4,E_1}^\Delta)^{1/2} \times\left(\frac{\delta\left(E_1-E_3\right)}{e^{-S_0}\rho\left(E_1\right)}+\frac{\delta\left(E_2-E_4\right)}{e^{-S_0}\rho\left(E_2\right)}+\left\{\begin{array}{ccc}
\Delta & E_1 & E_2 \\
\Delta & E_3 & E_4
\end{array}\right\}\right)\,. \nn
\end{align}
The first two terms in the last line are equal to the two-point function squared times a delta function and come from Wick contractions of neighbouring pairs of operator insertions. In a slight abuse of terminology we will refer to these terms as the time-ordered correlators, because they dominate time-ordered four-point functions in the limit where there is a large time gap between early- and late-time insertions. The final term, which is proportional to a $6J$ symbol for $SL(2, \mathbb R)$, comes from Wick contractions of antipodal pairs of operators and dominates out-of-time-ordered correlators. The normalization is $\mathcal{N}_{4pt} = (2^{2\Delta-1}\Gamma(2\Delta))^{-2} e^{-3S_0}$. The 6J symbol is expressed in terms of the Wilson function as \cite{Jafferis:2022wez}
\be
\left\{\begin{array}{ccc}
\Delta & E_1 & E_2 \\
\Delta & E_3 & E_4
\end{array}\right\} &= (\Gamma_{12}\Gamma_{2 3}\Gamma_{3 4}\Gamma_{41})^{1/2} \\ 
&\qquad \times \mathcal{W}_{\sqrt{2 \Ebrk^{-1} E_4}} \lrm{\sqrt{2 \Ebrk^{-1} E_2} ; \Delta \pm i \sqrt{2 \Ebrk^{-1} E_1}, \Delta \pm i \sqrt{2 \Ebrk^{-1} E_3}} \,,\nn
\ee
where we have restored all units of $\Ebrk$ and written it in terms of energies. If the operators in \eqref{eqn:jt4ptfn} carry spin and so are charged under the $SU(2)$ gauge field, each term in \eqref{eqn:jt4ptfn} will carry additional spin-dependent factors; however, we will only be interested in the case where the initial and final state of the black hole carry spin $j = 0$. In this case, the formulas simplify and we evaluate the four-point function of interest for our application in appendix \ref{app:angularmomentum_di_photon}. \footnote{The effect of spin in the four-point function results in a Wigner-D matrix element associated with each consecutive pair of operator insertions \cite{Iliesiu:2019lfc}; the labels of this Wigner-D are given by the representation and state of the black hole between consecutive operator insertions. The group element on which each of these four Wigner-D is evaluated is given by the holonomy between the two consecutive operator insertions.  There are two additional Wigner-D matrix elements associated with the worldlines that go between the contracted pair of operator insertions. Their representation and $m$-label are given by the spin and $m$-labels of the contracted pair of operators. The group element on which each one of these two Wigner-D is evaluated is given by the holonomy between the two operator insertions connected by the worldline.  To obtain the final result, we perform an integral over all holonomies. In the case of interest for this paper, the initial and final spins of the black hole have $j=0$: in such cases, two of the Wigner-D matrix elements are equal to $1$, and the integrals over holonomies drastically simplify.} 

\section{Neutral particle emission}
\label{sec:neutral-particle-emission-main}
In this section, we will explain both the semiclassical Hawking emission rate for uncharged particles from near extremal Reissner-Nordstr\"om black holes, as well as the quantum corrected low-temperature rate. The key point will be to extract the decay rate using Fermi's golden rule and its generalization to second-order perturbation theory, which relies on the JT gravity two- and four-point functions, respectively. We will find that these calculations correctly reproduce semiclassical results at high temperatures, but receive significant corrections at low temperatures, $T \lesssim \Ebrk$, that significantly modify the evaporation rate. In particular, for black holes with angular momentum $j = 0$ one finds that at low temperatures single photons and gravitons cannot be produced on-shell and the emission is instead dominated by the emission of entangled pairs of photons.\footnote{Schwarzian corrections to neutral Hawking radiation emission rates were independently studied in \cite{Bai:2023hpd}. We believe their results to be consistent with ours for S-wave emission of scalar fields, as described in Sec.~\ref{sec:massless_scalar_emission}. For particles with nonzero spin, their formulas are incorrect because they do not take into account angular momentum conservation or photon-graviton mixing. Separately, the conclusion of \cite{Bai:2023hpd} that Hawking radiation for small near-extremal black holes is dominated by Higgs boson emission is incorrect because it ignores charged particle emission, discussed in Sections \ref{sec:background-charged} and \ref{sec:charged-particle-emission-main}.}

\subsection{Semiclassical result}
\label{sec:semiclassical-result-neutral-hawking-radiation}

We begin by reviewing the semiclassical calculation of the Hawking emission rate for neutral particles; see \cite{Arbey:2021jif,Arbey:2021yke} for a recent discussion.  The energy flux lost by the black hole per unit time is given by
\be \label{eqn:dEdt}
\frac{d E}{d t} = \sum_{\ell,m} \int_0^\infty \frac{d \omega}{2\pi}  \frac{ \omega P_{\t{emit}}(\omega,\ell)}{e^{\beta \omega}-1}.
\ee
In the above $\ell,m$ specify angular momenta and $P_{\t{emit}}(\omega,\ell)$ is known as the transmission probability/greybody factor. The greybody factor is the probability that a spherical outgoing wave is transmitted through the effective potential of the black hole and escapes to infinity, and it depends on the frequency and angular momentum of the emitted wave as well as on the black hole background.

To calculate the energy flux, we must first calculate the transmission probability for the mode of interest. The effective potential of the black hole grows stronger with increasing angular momentum, so the dominant decay channel comes from the lowest angular momentum modes available. We will be interested in considering the case of our universe, where the dominant decay channels consist of photons and graviphotons, but we will first illustrate the general idea by considering a massless scalar field. We must solve the wave equation $\nabla^2 \phi=0$ in the near-extremal black hole background. Imposing purely in-going boundary conditions at the horizon, the solution will take the general form
\be
\phi \underset{r_* \to -\infty}{\sim} a_{\t{in}}e^{- i r_* \omega}, \qquad \phi \underset{r_* \to +\infty}{\sim} b_{\t{in}}e^{-i r_* \omega} + b_{\t{out}}e^{i r_* \omega},
\ee
where we have introduced the tortoise coordinate $r_*$, defined by $\frac{d r_*}{d r} = f^{-1}$. The absorption probability/transmission and reflection coefficients are given by
\be
P_{\t{emit}}(\omega,\ell) \equiv \mathcal{T} = \left|\frac{a_{\t{in}}}{b_{\t{in}}} \right|^2, \qquad \mathcal{R} = \left| \frac{b_{\t{out}}}{b_{\t{in}}} \right|^2,
\ee
with the condition $\mathcal{T}+\mathcal{R}=1$. Due to exponential suppression of high-energy excitations, evaporation is dominated by modes with $\beta \omega \lesssim 1$ for which the wave equation can be solved in distinct overlapping regions to extract the greybody factor. In appendix \ref{app:greybody}, we solve the wave equation and find the absorption probability for a massless scalar to be
\be\label{eq:pemitscalar}
P_{\t{emit}}^{\t{scalar}}(\omega, \ell=0)= 4 (r_+ \omega)^2,
\ee
which matches the expected universal answer \cite{Das:1996we} for 4d black holes at small frequency. The formula \eqref{eq:pemitscalar} turns out to be independent of the temperature $T$ of the black hole; this is a peculiarity specific to the S-wave mode of massless scalar fields and will not be true in general. For near-extremal RN black holes, the  semiclassical energy flux in the S-wave of the scalar for $T r_+ \ll 1$ is thus
\be \label{eqn:semiclassicalflux_scalar}
\frac{d E}{d t} = \frac{1}{2\pi}\int_{0}^\infty d \omega \omega \frac{4 (\omega r_+)^2 }{e^{\beta \omega}-1} =\frac{2 \pi^3}{15} r_+^2 T^4.
\ee
The greybody factors for the lowest modes of the photon and graviton for the near-extremal Reissner-Nordstr\"om background are:
\begin{eqnarray}
P_{\t{emit}}^{\t{photon}}(\omega, \ell=1) &=&\frac{4}{9} r_+^8 \omega^4 \lrm{\omega^2+\frac{4\pi^2}{\beta^2}} \lrm{\omega^2+\frac{16\pi^2}{\beta^2}} \label{eq:photoellisonegreybodyresult} \,,\\
P_{\t{emit}}^{\t{photon}}(\omega, \ell=2) &=&  \frac{4}{45} r_+^8 \omega^6 \lrm{\omega^2 + \frac{4\pi^2}{\beta^2}}\,, \\
P_{\t{emit}}^{\t{graviton}}(\omega, \ell=2) &=&  \frac{16}{45} r_+^8 \omega^6 \lrm{\omega^2 + \frac{4\pi^2}{\beta^2}} \,.
\end{eqnarray}
In appendix \ref{app:greybody}, we derive greybody factors for general values of angular momentum.\footnote{These results match certain limiting cases computed by \cite{Page:2000dk,Crispino:2009zza,Oliveira:2011zz}.} The first propagating photon and graviton modes are respectively $\ell=1,2$. It is surprising that the different modes have greybody factors that scale with the same powers of frequency and temperature. This is explained by the fact that the equations of motion for propagating photons and gravitons mix in the Reissner-Nordstr\"om background, which we further explain in appendix \ref{app:greybody}.

The energy fluxes for the above modes are then given by
\begin{eqnarray}\label{eqn:semiclassicalflux_photon}
\ell=1 \t{ photon:} & \frac{d E}{d t} = \frac{6}{2\pi}\int_{0}^\infty \frac{d \omega \omega}{e^{\beta \omega}-1} \lrm{\frac{4}{9} r_+^8 \omega^4 \lrm{\omega^2+\frac{4\pi^2}{\beta^2}} \lrm{\omega^2+\frac{16\pi^2}{\beta^2}}}  & = \  \frac{62848 \pi^9}{2079} r_+^8 T^{10}\,, \nn \\
\ell=2 \t{ photon:} & \frac{d E}{d t} =\frac{5}{\pi}\int_{0}^\infty \frac{d \omega \omega}{e^{\beta \omega}-1} \lrm{\frac{4}{45} r_+^8 \omega^6 \lrm{\omega^2 + \frac{4\pi^2}{\beta^2}}} & = \ \frac{3968 \pi^9}{1485} r_+^8 T^{10}\,, \nn \\ 
\ell=2 \t{ graviton:} & \frac{d E}{d t} =\frac{5}{\pi}\int_{0}^\infty \frac{d \omega \omega}{e^{\beta \omega}-1} \lrm{\frac{16}{45} r_+^8 \omega^6 \lrm{\omega^2 + \frac{4\pi^2}{\beta^2}}} &  = \ \frac{15872 \pi^9}{1485} r_+^8 T^{10} \,. \nn \\
\end{eqnarray}
For the $\ell=1$ photon, we have taken into account two polarizations and three components of axial angular momentum. For the $\ell=2$ photon/graviton, there are two polarizations and five components of axial angular momentum. We see that for QFT on a fixed background, all three modes give comparable fluxes.

The total energy flux from a near-extremal RN black hole is the sum of the three channels in \eqref{eqn:semiclassicalflux_photon}. So long as the semiclassical approximation is valid, we thus have a rate of energy loss due to massless radiation of 
\begin{equation}
    \frac{d E}{d t} = \frac{30208 \pi^9 }{693} r_+^8 T^{10} \ . 
\end{equation}

As we will see, the above semiclassical answer is correct for temperatures above the breakdown scale, $T \gg \Ebrk$. To extract the energy flux at low temperatures, $T \lesssim \Ebrk$, where the semiclassical answer is badly wrong, we will need to use different techniques, which we now explain. We will see that at low temperatures, quantum corrections make it so the dominant mode becomes the $\ell=1$ photon.

\subsection{Hawking radiation in JT gravity}
\label{sec:Hawking-radiation-in-JT-grav}

We now explain how the Hawking emission rate can be calculated from a Schwarzian two-point function by using Fermi's Golden rule. The logic essentially follows the effective string approach for calculating greybody factors of black holes \cite{Das:1996wn,Gubser:1996xe,Gubser:1996zp,Gubser:1997cm,Maldacena:1996ix,Maldacena:1997ih,Callan:1996tv}.

In pure JT gravity plus matter, there is no spontaneous emission of Hawking radiation out of AdS$_2$ because all Hawking quanta are reflected back into the black hole before they reach the boundary. As a result, the boundary Hamiltonian is conserved. If we turn on a classical external source that oscillates at frequency $\omega$, it will stimulate the emission of radiation from the black hole at the same frequency. In a near-extremal black hole, such sources describe classical waves reflecting off the black hole. Even in the absence of such a classical source, the near-horizon region of a near-extremal black hole is weakly coupled to the quantum state of propagating fields far from the horizon. This coupling is what leads to the spontaneous emission of Hawking radiation.

\paragraph{A classical external source.} Asymptotically, solutions to the scalar wave equation in AdS$_2$ can be written as
\begin{align}\label{eq:phibdy}
    \phi(z,t) = \phi_\text{bdy}(t) z^{1-\Delta} + O(z^\Delta),
\end{align}
where $z$ is a local Poincar\'{e} coordinate that is defined relative to the location of the Schwarzian boundary particle.  The scaling dimension $\Delta$ of the primary $\mathcal{O}$ is related to the mass $m$ of the field $\phi$ by $m^2 = \Delta (\Delta - 1)$. The normalizable $O(z^\Delta)$ piece in \eqref{eq:phibdy} is a dynamical degree of freedom. However, the coefficient $\phi_\text{bdy}(t)$ of the non-normalizable mode has to be fixed as part of the boundary conditions of the theory. Physically, $\phi_\text{bdy}(t)$ describes an excitation ``tunnelling into'' the spacetime from beyond the boundary. When $\phi_\text{bdy}(t) \neq 0$, the Schwarzian action becomes
\be \label{eqn:eff_string_action}
I = I_{\t{sch.}} + \int d t\, \phi_\text{bdy}(t) \mathcal{O}(t),
\ee
where $\mathcal{O}$ is a bulk operator that describes the coefficient of the normalizable mode in \eqref{eq:phibdy}. It plays the same role as a single-trace CFT primary in higher-dimensional AdS/CFT, while $\phi_\text{bdy}(t)$ plays the role of a source $J(t)$ for that primary. The relative normalizations of $\phi_\text{bdy}(t)$ and $\mathcal{O}$ need to be chosen correctly so that the second term in \eqref{eqn:eff_string_action} appears with coefficient one. When 
\be 
\phi_\text{bdy}(t) = \phi_0 e^{-i \omega t} \,,
\ee 
oscillates at frequency $\omega$, this coupling introduces a time-dependent perturbation
$H_I(t) = \phi_\text{bdy}(t) \mathcal{O}(t)$
to the Hamiltonian. Using Fermi's golden rule, this leads to the transition rate
\begin{align}\label{eq:JTstimemission}
\Gamma_{i \to f} &= 2 \pi\, |\langle E_f| H_I|E_i \rangle|^2 \,\delta(E_i - \omega - E_f) \nn
\\&=2 \pi |\phi_0|^2 \,|\langle E_f | \mathcal{O} (0) | E_i \rangle|^2\, \delta(E_i - \omega - E_f).
\end{align}

\paragraph{Coupling to an external quantum system.}
The formula \eqref{eq:JTstimemission} can be used to compute the stimulated emission rate for a near-extremal black hole in response to an incoming classical wave. This rate is related to the spontaneous emission rate by a classic argument of Einstein. However, since the latter rate is our true object of interest, we will instead describe how to derive it directly. In the absence of an incoming wave, the classical value of $\phi_\text{bdy}(t)$ is zero. However, after quantizing matter fields far from the black hole, $\hat\phi_\text{bdy}(t)$ becomes a nontrivial quantum operator acting on the far-field Hilbert space. In particular, $\hat\phi_\text{bdy}(t)$ does not annihilate the quantum vacuum. The second term in  \eqref{eqn:eff_string_action} then becomes a perturbative interaction
\be \label{eqn:interaction_H}
H_I(t) = \cO(t) \hspace{.035cm}\hat{\phi}_0(t)\,,
\ee
coupling the near-horizon and far-field Hilbert spaces. Since the far-field Hamiltonian is free, we can use the classical equations of motion to relate $\hat{\phi}_0$ to a linear combination of creation and annihilation operators at asymptotic infinity and thereby derive the spontaneous rate of Hawking emission for the near-extremal black hole. In particular, given the interaction \eqref{eqn:interaction_H}, Fermi's golden rule leads to the spontaneous emission rate
\begin{align} \label{eqn:FGR}
\Gamma_{i \to f} &= 2 \pi\, |\langle E_f, \omega| \cO \hspace{.035cm}\hat{\phi}_0|E_i \rangle|^2 \,\delta(E_i - \omega - E_f).
\end{align}
 Here the state $|E_i \rangle$ means that the black hole has energy $E_i$ and the far-field modes are in the vacuum state, while, in the state $|E_f, \omega \rangle$, the black hole has energy $E_f$ and the far-field modes contain a single delta-function-normalized particle excitation with energy $\omega$. Similarly, the state $|E_i, \omega, \omega'\rb$ would describe a black hole with energy $E_i$ and two particles in the far-field Hilbert space with energies $\omega$ and $\omega'$.
 
 To compute the full spontaneous emission rate we need to integrate over both the final energy of the black hole and the energy $\omega$ of the radiated quanta, multiplied by their respective continuum densities of states. For the JT gravity states, $|E_f\rangle$, the correct density of states is given in \eqref{eqn:schwarziandos}. Meanwhile, if the states of the far-field modes are normalized so that $\langle \omega'|\omega\rangle = \delta(\omega - \omega')$,
then the operator
\begin{align}
    P_1 = \int d\omega\, |\omega\rangle\langle \omega|\,,
\end{align}
is a projector onto single-particle states of the far-field mode, and so the far-field continuum density of states is equal to one. It follows that  the total spontaneous emission rate is
\be
\Gamma_{\t{spon.}} = \int_0^\infty d \omega \int_0^\infty d E_f \rho(E_f) \Gamma_{i \to f}.
\ee
To calculate the expected energy flux, we insert an extra factor of the energy of the mode
\be
\frac{d E}{d t} = \int_0^\infty d \omega \omega \int_0^\infty d E_f \rho(E_f) \Gamma_{i \to f}\,.
\ee

\subsection{Massless scalar emission} \label{sec:massless_scalar_emission}
As a warm-up, we begin with the case of a massless scalar field $\phi$ and a black hole with angular momentum $j=0$. The four-dimensional action is given by
\begin{align}
I = \frac{1}{2}\int d^4 x \sqrt{g_4} \nabla^a \phi \nabla_a \phi.
\end{align}
In appendix \ref{app:greybody}, we solve the equations of motion for the scalar field in the near-extremal Reissner-Nordstr\"om black hole background. The field can be decomposed as 
\begin{align}
    \phi(t,r,\theta,\phi)=\sum_{\ell,m} \frac{1}{r_+}\int_0^\infty d\omega e^{-i \omega t} \phi_{\omega\ell m}(r) Y_{\ell,m}(\theta,\phi) + \t{h.c.}
\end{align}
where $Y_{\ell,m}$ are spherical harmonics for the angular momentum mode $\ell,m$. The factor of $1/r_+$ was included so that, after dimensionally reducing to two dimensions in the near-horizon region, 
\begin{align}
    \phi_{\ell m}(r,t) = \int_0^\infty d\omega \,e^{-i \omega t} \phi_{\omega\ell m}(r)  + \t{h.c.}
\end{align} 
becomes a canonically normalized two-dimensional scalar field. Near asymptotic infinity, we can write 
\be
\phi_{\omega\ell m}(r) = \frac{a_{\omega\ell m}\,r_+}{\sqrt{4\pi \omega }r} e^{-i \omega r} + \frac{b_{\omega\ell m}\,r_+}{\sqrt{4\pi \omega }r} e^{i \omega r}\,.
\ee
Here, the normalizations were chosen so that, e.g. the quantum operator
\begin{align}\label{eq:farfieldscalarsoln}
a_{\omega\ell m} = \sqrt{\frac{\omega}{4\pi}} \int  r dr d^2\Omega\, Y^*_{\ell,m}\, e^{i \omega r} \left[\phi(0,r,\theta,\phi) + \frac{i}{\omega} \dot\phi(0,r,\theta,\phi)\right]\,,
\end{align}
satisfies
\begin{align}
[a_{\omega' \ell'm'},a_{\omega \ell m}^\dagger] = \delta_{\ell',\ell} \delta_{m',m} \delta(\omega -\omega'), 
\end{align}
and hence acts as a conventionally normalized annihilation operator on the Fock space $\mathcal{H}_{\t{rad}}$ describing the far-field radiation. At asymptotic future null infinity, the far-field Hamiltonian can be written in terms of the outgoing modes as
\begin{align}\label{eq:H+inf}
   H_{+\infty} = \sum_{\ell,m} \int_0^\infty d \omega \hspace{.035cm} \omega\, b_{\omega\ell m}^\dagger b_{\omega\ell m},
\end{align}
while at past null infinity, it can be written in terms of the ingoing modes as
\begin{align}\label{eq:H-inf}
   H_{-\infty} = \sum_{\ell,m} \int_0^\infty d \omega \hspace{.035cm} \omega\, a_{\omega\ell m}^\dagger a_{\omega\ell m}.
\end{align}
Meanwhile, in the near-horizon region, the general solution is found in \eqref{eqn:scalar_regionII} to be
\be \label{eqn:adsfield_scalar_main}
\phi_{\omega\ell m}(r) = c_{\omega\ell m} \left(\frac{r_+^2}{r - r_+}\right)^{-\ell} + d_{\omega\ell m} \left(\frac{r_+^2}{r - r_+}\right)^{\ell + 1}.
\ee
The solution \eqref{eqn:adsfield_scalar_main} is valid so long as $\omega \ll r_+^{-2}(r -r_+)$. In other words it is valid so long as we are near the asymptotic boundary of the AdS$_2$ region from the perspective of a Rindler mode with frequency $\omega$. The first and second terms in \eqref{eqn:adsfield_scalar_main} are related to the non-normalizable and normalizable modes, respectively. In the near-extremal limit, the near-horizon and far-field regions decouple, and so the modes $d_{\omega\ell m}$, which describe quantum tunneling out of the horizon, vanish up to the perturbative corrections that we will include momentarily. We show how to match coefficients in the solutions \eqref{eq:farfieldscalarsoln} and \eqref{eqn:adsfield_scalar_main} in Appendix \ref{app:A_scalarODE}. Setting $d_{\omega\ell m} = 0$, one obtains
\begin{align} \label{eqn:mode_relations_scalar}
    b_{\omega\ell m} =- e^{-i \ell \pi} a_{\omega\ell m} ~~~~\text{and} ~~~~ c_{\omega\ell m} = a_{\omega\ell m} \frac{-i e^{-\frac{i}{2}\ell \pi} \omega^{\ell+\frac{1}{2}}r_+^{2\ell + 1} }{2^{\ell}\sqrt{2\pi} \Gamma(\ell+\frac{3}{2})}.
\end{align}
These boundary conditions fix the Hamiltonians \eqref{eq:H+inf} and \eqref{eq:H-inf} to be equal
\begin{align}
H_0 = H_{+\infty} = H_{-\infty} = \sum_{\ell,m} \int_0^\infty d \omega \hspace{.035cm} \omega\, a_{\omega\ell m}^\dagger a_{\omega\ell m}.
\end{align}
since no energy is absorbed or emitted by the black hole.

The dominant contribution to the Hawking emission process comes from the scalar mode with $\ell = m = 0$. This mode is described in the near-horizon region by a two-dimensional scalar field with scaling dimension $\Delta = 1$ in JT gravity. The far-field operator $\hat{\phi}_0$ describing the non-normalizable mode that appears in \eqref{eqn:interaction_H} is
\begin{align} \label{eqn:bdy_source_scalar}
    \hat{\phi}_0(t) \propto \lim_{(r -r_+)\gg (\omega r_+^2) }\left[\left(\frac{r_+^2}{r - r_+}\right)^{\ell}\phi_{00}(r,t)\right] \propto  \int_{0}^\infty d\omega \lrm{ c_{\omega 0 0} e^{-i\omega t} + c_{\omega 0 0}^\dag e^{i\omega t} }.
\end{align}
Plugging in \eqref{eqn:mode_relations_scalar}, we obtain
\begin{align}
 \hat{\phi}_0(t) = \mathcal{N} \int_0^\infty d \omega \hspace{.035cm}\sqrt{r_+^2 \omega} \hspace{.035cm} (a_{\omega 00} + {a^\dag_{\omega 00}})\,, 
\end{align}
where the dimensionless normalization constant $\mathcal{N}$ needs to be chosen such that \eqref{eqn:eff_string_action} holds, given our choice of normalization for the JT gravity operator $\cO$. Rather than computing $\mathcal{N}$ from first principles, we will fix it by matching the fully quantum emission rate derived here to the semiclassical rate derived in Sec.~\ref{sec:semiclassical-result-neutral-hawking-radiation} (in the high energy limit). As we shall see, the correct normalization turns out to be $\mathcal{N}^2=\frac{2}{\pi^2}$.

To leading order in perturbation theory in the low-temperature limit, the interaction Hamiltonian is
\be
H_I = \mathcal{N} \cO \int_0^\infty d \omega \hspace{.035cm}\sqrt{r_+^2 \omega} \hspace{.035cm} (a_{\omega 00} + {a^\dag_{\omega 00}}).
\ee
We expect the state $|E_i\rangle$, where the black hole has energy $E_i$ and the far-field modes, are in the vacuum state $|\Omega\rangle$ to decay to a state $|E_f ,\omega\rangle$, where the black hole has energy $E_f$ and the far-field modes, are in the delta-function normalized single-particle state  
$
|\omega\rangle = a^\dagger_{\omega0 0} |\Omega\rangle.
$
The matrix element between these states is 
\be \label{eqn:scalar_matrixelement}
\lb E_f, \omega | H_I |E_i \rb = \mathcal{N} \sqrt{r_+^2 \omega} \hspace{.05cm}\lb E_f | \mathcal{O} | E_i \rb\,,
\ee
where we are left with a one-point function between initial and final energy eigenstates of the black hole.
Using Fermi's golden rule \eqref{eqn:FGR}, we therefore find the spontaneous decay rate
\begin{align}
\Gamma_{i \to f} &= 2\pi|\lb E_f, \omega | H_I  | E_i \rb|^2 \delta(E_f + \omega - E_i) \\ 
&=2\pi\mathcal{N}^2 (r_+^2 \omega) |\lb E_f | \mathcal{O} | E_i \rb|^2 \delta(E_f + \omega - E_i)\,,  \nn 
\end{align}
To get the total spontaneous emission rate into an arbitrary final state we integrate over all possible final states, leading to
\be
\Gamma_{\t{spon.}} = \int_0^\infty d \omega \int_0^\infty d E_f \rho(E_f) \Gamma_{i \to f} = 2\pi\mathcal{N}^2 \int_0^{E_i} d \omega (r_+^2 \omega) \rho(E_i - \omega) |\lb E_i - \omega | \mathcal{O} | E_i \rb|^2 \,.
\ee
The energy flux per unit time is
\be
\frac{d E}{d t} = 2\pi \mathcal{N}^2 \int_0^{E_i} d \omega (r_+\omega)^2 \rho(E_i - \omega) |\lb E_i - \omega | \mathcal{O} | E_i \rb|^2\,.
\ee
By using the JT gravity two-point function \eqref{eqn:2ptfnJT} with $\Delta=1$ for the massless scalar and the density of states \eqref{eqn:schwarziandos}, we find that the energy flux is
\be \label{eqn:scalarfinalflux}
\frac{d E}{d t} = \frac{2}{\pi}\int_0^{E_i} d \omega \hspace{.04cm}\omega (r_+ \omega)^2 \frac{\sinh \lrm{2\pi \sqrt{2 \Ebrk^{-1}(E_i - \omega)}} }{\cosh \lrm{2\pi \sqrt{ 2 \Ebrk^{-1} E_i}} -\cosh \lrm{2\pi \sqrt{2 \Ebrk^{-1} (E_i-\omega)}} } \,,
\ee
where we have simplified the various gamma functions appearing in the two-point function and set $\mathcal{N}^2=\frac{2}{\pi^2}$ as explained above. 

\begin{figure}
    \centering
    \includegraphics[width=.9\textwidth]{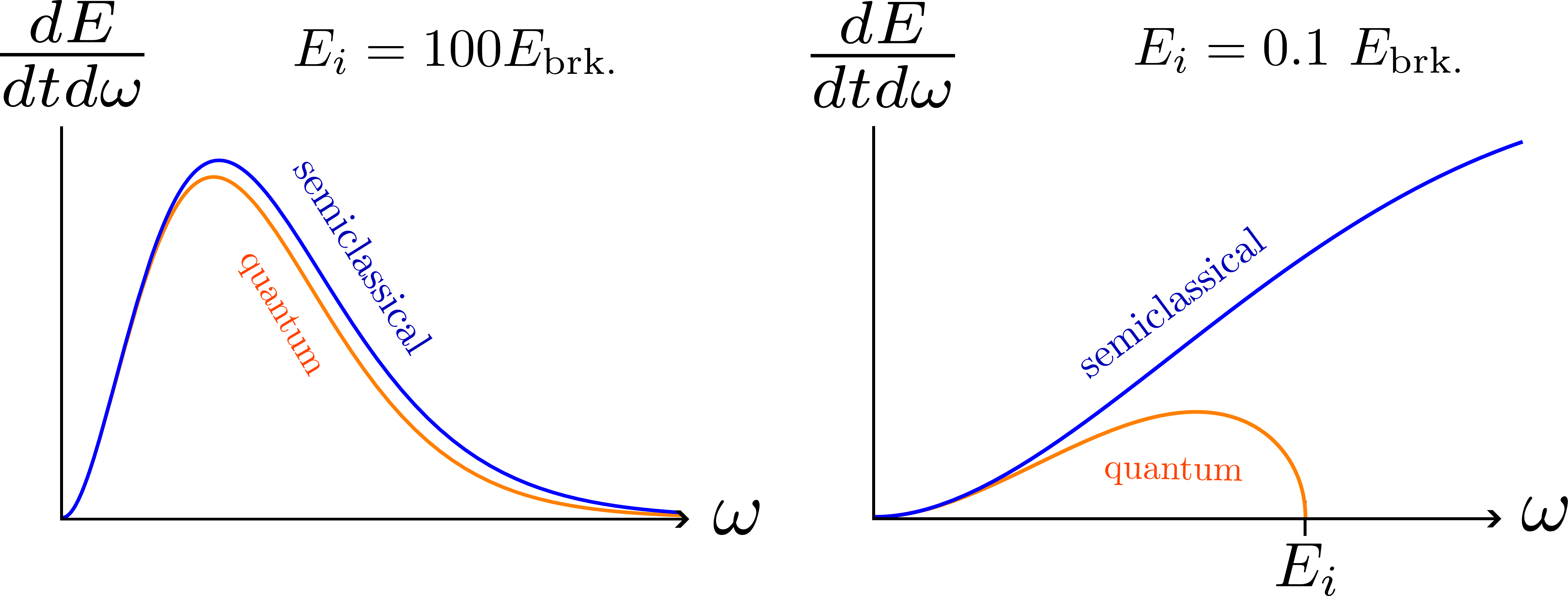}
    \caption{Comparison of the \textcolor{blue}{semiclassical prediction} vs.~\textcolor{orange}{quantum corrected} Hawking radiation into a massless scalar field. The energy flux per unit frequency is plotted for a black hole in the microcanonical ensemble at initial energy $E_i$ above extremality at fixed charge $Q$. \textit{Left:} For a BH initially far above the thermodynamic breakdown scale $(E_i = 100 \Ebrk)$, quantum corrections are not important and the \textcolor{orange}{exact energy flux} approaches the \textcolor{blue}{semiclassical prediction}. \textit{Right:} For a BH below the breakdown scale $(E_i = .1 \Ebrk)$ Schwarzian corrections are important and the distribution is no longer thermal. The exact Schwarzian answer cuts off the spectrum so particles with energy larger than the initial black hole energy, $\omega > E_i$, are not emitted. Note that the scales of the two plots are very different due to the difference in energy above extremality. The plot of the exact flux from the Schwarzian comes from the integrand of  \eqref{eqn:scalarfinalflux}, while the semiclassical flux is from  \eqref{eqn:semiclassical_flux}.}
    \label{fig:hawkingradscalar}
\end{figure}

This is the exact quantum answer for the radiated flux in the $\ell=0$ mode of a scalar field from a black hole which begins in an initial microcanonical state centered around energy $E_i$. We plot the energy flux in Figure \ref{fig:hawkingradscalar}, both at high initial energies $E_i$ where it matches the semiclassical emission spectrum, and at low energies where we see significant deviations from the semiclassical result.

\paragraph{Semiclassical limit.} 
The semiclassical answer is given by taking the limit of a black hole with initial energy far above extremality while holding the corresponding microcanonical temperature finite. This means taking $ E_i/\Ebrk \to \infty$. The full quantum energy flux \eqref{eqn:scalarfinalflux} reduces to\footnote{In this limit we are approximating the $\sinh$ \& $\cosh$ functions by exponentials and finding that the trigonometric expressions in \eqref{eqn:scalarfinalflux} simplify to $(e^{\sqrt{E_i}-\sqrt{E_i-\omega}})^{-1}\sim (e^{\beta \omega}-1)^{-1}$ once numerical prefactors are included.} 
\be \label{eqn:semiclassical_flux}
\t{semiclassical limit:} \qquad \frac{d E}{d t} = \frac{1}{2\pi} \int_0^{\infty} d \omega \omega \frac{4(r_+ \omega)^2}{e^{\beta \omega}-1} = \frac{1}{30 \pi} r_+^2 \Ebrk^2 E_i^2 \,, \qquad \beta = \sqrt{\frac{2\pi^2}{\Ebrk E_i}}
\ee
where we have identified the effective inverse temperature $\beta$ in the microcanonical ensemble at energy $E_i$. This is exactly the QFT-in-a-fixed-curved-background result for the energy flux emitted by a black hole at temperature $T$ \eqref{eqn:semiclassicalflux_scalar}. Thus, the formula for the energy flux from the JT gravity two-point function correctly reproduces the expected semiclassical energy flux with the choice of $\mathcal{N}$ given below \eqref{eqn:mode_relations_scalar}. 

\paragraph{Quantum limit.}
 In the opposite limit, $E_i, \omega \ll \Ebrk$, quantum effects become very important. We obtain
\begin{align}\label{eqn:scalar_micro_lowE}
\lim_{E_i \ll \Ebrk} \frac{d E}{d t} &= \frac{\sqrt{2}}{\pi^2} \Ebrk^{\frac{1}{2}} \int_0^{E_i} d \omega (r_+ \omega)^2 \sqrt{E_i - \omega} \\&= \frac{16 \sqrt{2}}{105 \pi^2} r_+^2 \Ebrk^{\frac{1}{2}} E_i^{\frac{7}{2}} \,.  \nn
\end{align}
We can compare the quantum corrected energy flux at low energies to the extrapolation of the semiclassical flux \eqref{eqn:semiclassical_flux} at the same energy
\be
\frac{\t{quantum corrected flux}}{\t{semiclassical flux}} \sim \lrm{\frac{E_i}{\Ebrk}}^{\frac{3}{2}}\,.
\ee
Since we are looking at $E_i \ll \Ebrk$ we find that the quantum corrected flux is much lower than the naive semiclassical flux.

\begin{figure}[t]
    \centering
    \includegraphics[width=1\textwidth]{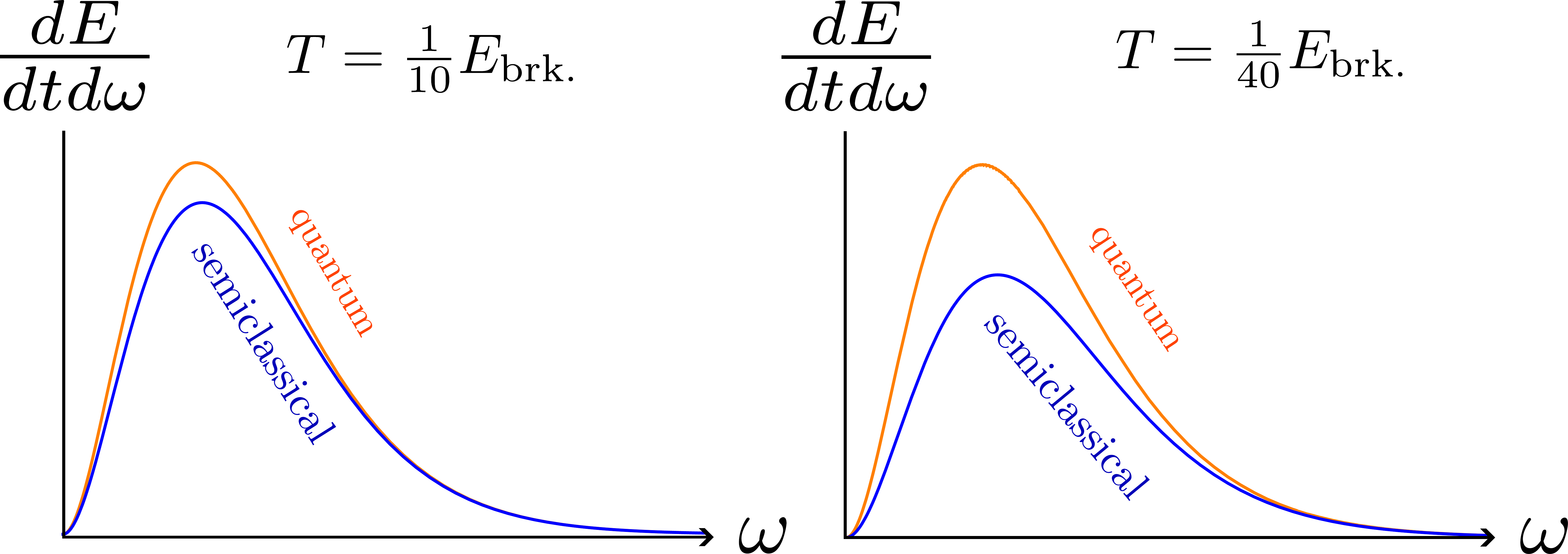}
    \caption{Comparison of the \textcolor{blue}{semiclassical prediction} vs.~\textcolor{orange}{quantum corrected} Hawking radiation into a massless scalar field. In Figure~\ref{fig:hawkingradscalar}, we plotted these quantities in the microcanonical ensemble; here, we plot them in the canonical ensemble. The energy flux per unit frequency is plotted for a black hole in the canonical ensemble at temperature $T={\Ebrk}/{10}$ (left) and $T={\Ebrk}/{40}$ (right). The semiclassical prediction is plotted using the exact expression \eqref{eqn:semiclassicalflux_scalar}, while the Schwarzian corrected answer is obtained by numerically integrating the exact expression \eqref{eqn:scalar_canonical_exact_formal}. It appears that the energy flux in the exact quantum answer is \emph{larger} than the semiclassical flux at fixed temperature $T$. Naively, this seems in contradiction with the microcanonical result, which showed \emph{reduced} flux relative to the semiclassical answer at very low energies. The resolution is that the semiclassical relation $M-Q = \frac{2\pi^2}{\Ebrk}T^2$ breaks down at low temperatures. Instead, a black hole in the canonical ensemble at temperature $T \ll \Ebrk$ is dominated by states with $M-Q \sim T$ although the distribution is no longer tightly peaked about any one value. The increase in typical energy leads to an increase in expected flux for the canonical ensemble relative to the semiclassical prediction. Furthermore, at high values of $\omega$, the quantum corrected and semiclassical spectra agree. This is because only black holes that have high enough energy above extremality can radiate such particles; such black holes are well described in the semiclassical approximation but have a large Boltzmann suppression.
    }
    \label{fig:canonicalflux}
\end{figure}

\paragraph{Converting to the canonical ensemble.} With the fluxes in the microcanonical ensemble, we can easily convert to the canonical ensemble. The initial black hole density matrix is distributed with thermal distribution $\rho_i  = Z(\beta)^{-1} \sum_i e^{-\beta E_i}|E_i\rangle \langle E_i |$. We can simply apply this distribution to the energy flux of the microcanonical ensemble to obtain the flux in the canonical ensemble using the density of states and partition function of near-extremal black holes \eqref{eqn:schwarziandos},
\be \label{eqn:scalar_canonical_exact_formal}
\left.\frac{d E}{d t}\right\rvert_{\t{canonical}} = \frac{1}{Z(\beta)}\int_0^\infty d E_i \rho(E_i) e^{-\beta E_i} \left.\frac{d E}{d t}\right\rvert_{\t{micro. $E_i$}}
\ee
Examples of the Hawking radiation spectra for a black hole in the canonical ensembles at different temperatures are shown in Figure \ref{fig:canonicalflux}. 

At large temperatures $T \gg \Ebrk$ the integral is dominated by large energies $E_i$, and the microcanonical answer matches the QFT-in-curved-spacetime result \eqref{eqn:semiclassicalflux_scalar} where we replace $\beta = \sqrt{\frac{2\pi^2}{\Ebrk E_i}}$. Performing the above integral with this answer, we find
\be
\left.\frac{d E}{d t}\right\rvert_{\t{canonical}} = \frac{2\pi^3}{15} r_+^2 T^4 + \mathcal{O}(r_+^2 \Ebrk T^3), \qquad T \gg \Ebrk\,.
\ee
This is unsurprising since the ensembles are identical at large temperatures/energies.
For very small temperatures, we must use the low energy result \eqref{eqn:scalar_micro_lowE}, finding
\begin{align}
\left.\frac{d E}{d t}\right\rvert_{\t{canonical}} &= \frac{1}{Z(\beta)}\int_0^\infty d E_i \rho(E_i) e^{-\beta E_i} \lrm{\frac{16 \sqrt{2}}{105 \pi^2}(r_+ E_i)^2 E_i^{\frac{3}{2}} \Ebrk^{\frac{1}{2}}} \nn \\ 
&= \frac{256 \sqrt{2}}{35\pi^{\frac{5}{2}}}r_+^2 \Ebrk^{\frac{1}{2}} T^{\frac{7}{2}} \lrm{1 + \mathcal{O}\lrm{\frac{T}{\Ebrk}}}, \qquad T \ll \Ebrk\,.
\end{align}
In apparent contrast with the microcanonical ensemble, we find that the quantum-corrected emission rate at low temperatures is larger than the semiclassical rate. This is because the semiclassical thermodynamic relation $M-Q = 2 \pi^2 T^2 /\Ebrk$ is dramatically modified at low temperatures such that the canonical ensemble is dominated by states with $M - Q \sim T$.

\subsection{Particles with spin} 
\subsubsection*{Photon emission}
\label{sec:singlephoton}

Since no massless scalars are known to exist in our universe, we now turn to the emission of photons. As explained in appendix \ref{app:greybody}, the photon couples to the metric in the Reissner-Nordstr\"om background, so the various angular momentum modes are composite excitations of the metric and gauge modes. The equations of motion of the composite field can be solved in overlapping regions on the black hole geometry. We leave the details to appendix \ref{app:photongreybody}, with the result that the $\ell=1$ mode couples to an operator with $\Delta=3$ in the throat. Naively, we would expect this to be the dominant decay channel in our universe since it has the smallest allowed angular momenta and hence experiences the smallest centrifugal potential far from the black hole. However it will turn out that an $\ell =2$ ``graviphoton'' mode has smaller scaling dimension  $\Delta=2$ in the throat and hence contributes an $O(1)$ fraction of the Hawking radiation in the semiclassical limit.

\paragraph{Angular momentum selection rules.} The Reissner-Nordstr\"om black hole splits up into sectors labeled by its integer or half-integer angular momentum $j$,\footnote{This is the case if fermionic states are present in the theory as is the case in our own universe. If only bosonic states are included than angular momentum may be integer quantized.} with a density of states in equation \eqref{eqn:rho_angularmomentum}.\footnote{In the massless scalar case, we restricted to the $j=0$ sector. Since we considered the $\ell=0$ mode, a BH that starts with spin $j$ can only transition to a BH with spin $j$ through the emission of $\ell=0$ massless scalars.} In the near-horizon region, this can be seen from the quantization of the near-zero modes of metric that induce rotations of the transverse $S^2$ when moving along the boundary of AdS$_2$. By computing matrix elements of spinning operators when this mode is taken into account, we find that transitions are only allowed if angular momentum is conserved. That is, the BH with initial angular momentum $j$ can only transition into states with $j'=j-1,j,j+1$ by emitting an $\ell=1$ photon since $j \subset j' \otimes 1$. An edge case is an initial state with $j=0$, which can emit a photon and transition only into a $j=1$ state. Similarly the only way to transition into a $j=0$ state is from an initial $j=1$ state. 

Another important feature is that the spectrum begins, for sectors with charge $Q$ and spin $j$, at a minimum energy \eqref{eqn:gap_angularmomentum}, which we restate here for convenience
\be
Q + E_{0}^j = Q + \frac{j (j+1)}{2} \Ebrk\,.
\ee
If the black hole has energy below this value then states with angular momentum $j$ simply do not exist, and the BH must be composed of states with smaller values of $j$. 

As the BH decays into photons and reaches energies below each successive $E_{0}^j$ it must successively lose more of its angular momentum into photons. At low energies $E \leq 3 \Ebrk$ only $j=0,1$ states remain,\footnote{We are, for the moment, ignoring fermionic black holes with $j=\frac{1}{2}$.} and the BH will transition between these two sectors, losing energy in the process. Finally, as it gets closer to extremality, it will decay to a $j=0$ state with mass $M\leq Q+\Ebrk$, at which point it can no longer decay through single photon emission to a $j=1$ state with smaller energy since such states do not exist. Thus, the BH reaches a metastable state in the $j=0$ sector at some energy $E \leq E_{0}^{j=1}$ from which it cannot decay due to single photon emission. 

However, as we shall explain below, it can decay through two-photon processes where the emitted photons are in a singlet state and do not carry away any angular momentum. This allows $j=0$ to $j=0$ transitions, which cannot occur by any other process. This allows a BH with $E\leq Q+\Ebrk$ to continue decaying towards the extremality bound $E=Q$ in our own world.

To summarize the above discussion, a BH with arbitrary initial $j$ will lose its angular momentum as it emits Hawking radiation. Eventually it will reach a state with $j=0$ below the breakdown scale $\Ebrk$. At this point it continues evaporating through di-photon emission where the emitted photons are in an entangled singlet state. We now proceed to calculate all of these emission rates.

\paragraph{The spectrum of photons.} We first calculate the emission of individual photons from the BH. Similar to the case of the scalar we can solve the equations of motion, which we do in appendix \ref{app:photongreybody}, and extract the non-normalizable mode to find the interaction Hamiltonian\footnote{There are six distinct photon modes (three axial components, two helicities) that mediate distinct transitions between BH states with different $J_z$.}
\be \label{eqn:photon_H_int}
H_I = \mathcal{N} \sum_{m_\gamma = \pm 1,0} \mathcal{O}_{\ell, m_\gamma} \int d \omega \hspace{.035cm} r_+^4 \omega^{3/2} (a_{\omega,m_\gamma} + a_{\omega,m_\gamma}^\dag), \qquad \mathcal{N}^2 = \frac{40}{3\pi^2}\,,
\ee
where $\mathcal{O}_{\ell, m_\gamma}$ is a spinning operator with dimension $\Delta=3$ for the $\ell=1$ mode, and we have chosen $\mathcal{N}$ to match the semiclassical answer in the appropriate limit. The creation and annihilation operators create photons with energy $\omega$, unit angular momentum and $J_z = m_\gamma = \pm 1,0$. We have the same free Hamiltonian and commutation relations as in the case of the scalar.

The Hilbert space is again a tensor product $\mathcal{H} = \mathcal{H}_{\t{BH}} \otimes \mathcal{H}_{\t{rad}}$, where now BH states are also labeled by angular momentum quantum numbers $|E_{i}^{j,m}\rb$. The photon creation/annihilation operator has non-trivial action on $\mathcal{H}_{\t{BH}}$ since there are non-zero matrix elements between different values of $j,m$. The matrix elements for transitions between the photon vacuum $| \Omega \rb$ and the single-photon state $|\omega^{m_\gamma} \rb = a_{\omega,m_\gamma}^\dag | \Omega \rb$ are given by
\be \label{eqn:Hint_photon_matrixele}
\lb E^{j',m'}_{f}, \omega^{m_\gamma} | H_I | E^{j,m}_{i} \rb &=  \mathcal{N}   r_+^4 \omega^{3/2} \lb E_{f} - E_{0}^{j'},j',m'| \mathcal{O}_{\ell, m_\gamma}|  E_{i} -E_{0}^j,j,m \rb (\delta_{j,j'}+\delta_{j,j'\pm 1})\,,
\ee
where $E_0^j = \frac{j(j+1)}2 E_\text{brk.}$ is the shift in the energy above extremality for a spinning black hole. The one-point function between energy eigenstates remains the same \eqref{eqn:2ptfnJT} even with $j$ indices as long as we subtract $E_0^j$ from the energy and integrate against the spin-$j$ density of states. 

We will take an initial BH state in the microcanonical ensemble at initial energy $E_i^{j,m}$ with spin $j$. Fermi's Golden rule tells us that the transition rate per unit frequency $d \omega$ to a particular final state $f$ is given by \eqref{eqn:spin_2pt_fn} 
\be
\Gamma_{i \to f} &= \frac{ 2\pi \mathcal{N}^2}{{(2j+1)(2j'+1)}}\sum_{m=-j}^j\sum_{m_\gamma = \pm 1,0} r_+^8 \omega^3 |\lb E_{f} - E_{0}^{j'}| \mathcal{O}|  E_{i} -E_{0}^j \rb|^2 \nonumber \\ &\qquad\qquad  \times (\delta_{j,j'}+\delta_{j,j'\pm 1}) \left|C^{j' m'}_{j m \hspace{.07cm} \ell m_\gamma}\right|^2 \delta(E_i - E_f - \omega)   \,,
\ee
In the above, we have averaged over initial states giving an extra prefactor of $1/(2j+1)$ relative to the matrix element \eqref{eqn:Hint_photon_matrixele} and summed over the quantum numbers $m'$ and $m_\gamma$ describing the final angular-momentum states of the black hole and photon respectively. The final black hole angular momentum $j'$ and the photon energy $\omega$ remain fixed for the moment. As explained in Section \ref{sec:matter-correlators}, the Clebsch-Gordan coefficients and spin prefactors come from a matrix element for the holonomy of the rotational modes.

To get the total decay rate we integrate and sum over these remaining degrees of freedom. We get 
\begin{align}\label{eqn:photonspontaneousrate}
\Gamma_{\t{spon.}} &= 2 \int_0^\infty d \omega \sum_{j'} \int_{E_{0}^{j'}}^\infty d E_f \rho_{j'}(E_f) \Gamma_{i \to f} \nn  \\
& =4\pi \mathcal{N}^2\sum_{j'=j,j\pm 1} \sum_{m,m',m_\gamma} \frac{\left|C^{j' m'}_{j m \hspace{.07cm} \ell m_\gamma}\right|^2}{{(2j+1)(2j'+1)}} \int_0^{E_i - E_0^{j'}} d \omega \hspace{.035cm} r_+^8 \omega^3 \rho_{j'}(E_i-\omega)\nn \\ 
&\hspace{4.0cm}\times \,\,|\lb E_{i}  - \omega - E_{0}^{j'}| \mathcal{O}|  E_{i} -E_{0}^j \rb|^2\,. 
\end{align} 
The factor of two comes from a sum over the two possible polarizations of the photon in the final state. Let us make a couple of comments about this formula. The CG coefficients ensure that selection rules are obeyed so that the tensor product representation $j \otimes \ell$ contains $j'$. Meanwhile, the restriction in the range of the integral over $\omega$ comes from the fact that we can only emit a photon when the density of states $\rho_{j'}(E_i-\omega)$ is greater than zero. This only occurs above the extremality bound for the $j'$ sector. The sum over CG coefficients gives an overall prefactor for each emission channel, giving a relative probability to emit into various final spins. A key identity is
\be
\label{eqn:CG-identity}
\sum_{m,m',m_\gamma} {\left|C^{j' m'}_{j m \hspace{.07cm} \ell, m_\gamma}\right|^2} = {2j'+1}\,,
\ee
which can be viewed as the trace within the tensor product representation $j \otimes \ell$ of a projector onto states with spin $j'$.
The energy flux is given by inserting an extra $\omega$ in the integrands. For general initial $j$ we get
\be
\frac{d E}{d t} = 4\pi \mathcal{N}^2 \sum_{j'=j,j\pm 1} {\frac{1}{2j+1}}\int_0^{E_i} d \omega \hspace{.035cm} r_+^8 \omega^4 \rho_{j'}(E_i-\omega) |\lb E_{i}  - \omega - E_{0}^{j'}| \mathcal{O}|  E_{i} -E_{0}^j \rb|^2\,,
\ee
In the special cases $j=0$ or $j = \frac{1}{2}$ the above simplifies, since then only $j'=1$ and $j'=\frac{1}{2}, \frac{3}{2}$ contribute respectively.

\paragraph{Full quantum corrected energy flux.}

\begin{figure}
    \centering
    \hspace*{-1cm}{\label{a}\includegraphics[width=1.1\linewidth]{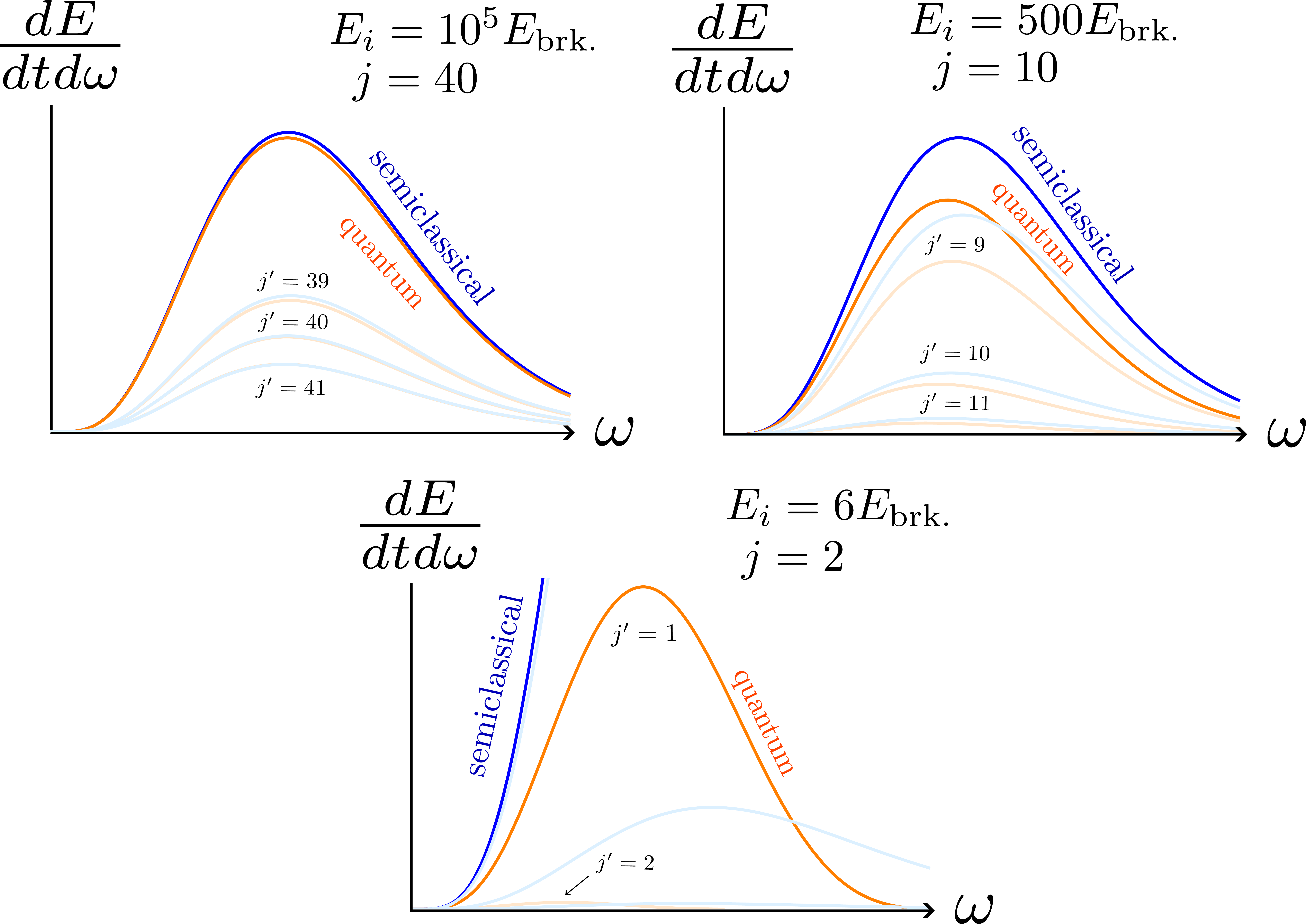}}
    \caption{Comparison of the \textcolor{blue}{semiclassical prediction} vs.~\textcolor{orange}{quantum corrected} Hawking radiation into the $\ell=1$ photon mode for various initial BH energies and spin. The energy flux per unit frequency is plotted for a black hole in the microcanonical ensemble at initial energy $E_i=M-Q$ above extremality at fixed charge $Q$ and initial angular momentum $j$. The BH state can transition into final BHs with spin $j'=j-1,j,j+1$; the semiclassical and quantum corrected predictions for each channel are shown in faint blue and faint orange respectively. The quantum-corrected flux is given by the integrand of  \eqref{eqn:photon_fullflux_kerrnewman}; the semiclassical flux is given by the integrand of \eqref{eqn:photon_semiclassicalflux_rotation} with $m_\gamma = j - j'$.  \textit{\textbf{Left:}} For a BH far above extremality $(E_i = 10^5 \Ebrk)$ with large angular momentum $(j= 40)$, the flux in each channel is close to the semiclassical prediction for a slowly rotating black hole. \textit{\textbf{Right:}} For a BH at intermediate energy $(E_i = 500 \Ebrk)$ and angular momentum $(j=10)$, quantum corrections are significant and the \textcolor{orange}{exact energy flux} deviates noticeably from the \textcolor{blue}{semiclassical prediction}. Transitions to $j' = j-1$ increasingly dominate, because this sector has the highest number of final states to transition into. \textit{\textbf{Bottom:}} At sufficiently small energies $(E_i = 6, \Ebrk, j = 2)$, angular momentum exclusion rules prevent transitions to $j' = 3$ and even transitions with $j' = 2$ are highly suppressed. The radiation is therefore almost completely dominated by transitions with $j' = 1$. The semiclassical prediction is wildly incorrect and, in particular, allows transitions to $j' = 3$ and to states with $j' = 1,2$ that are below the extremality bound.}
    \label{fig:hawkingradphoton}
\end{figure}

The full quantum corrected energy flux into the $\ell=1$ photon for a near-extremal black hole with a small amount of angular momentum $j$ and initial energy $E_i$ is given by 
\begin{align}\label{eqn:photon_fullflux_kerrnewman}
\frac{d E}{d t} = \frac{1}{144\pi^3} \sum_{j'=j,j\pm 1} {\frac{2j'+1}{2j+1}} &\int_0^{E_i-E_0^{j'}} d \omega r_+^8 \omega^4 \Ebrk^5  \sinh \lrm{2\pi \sqrt{2 \Ebrk^{-1}(E_i - \omega-E_0^{j'}}}  \\ 
&\times\Gamma\lrm{3 \pm i \sqrt{2 \Ebrk^{-1}(E_i-E_0^j)} \pm i \sqrt{2 \Ebrk^{-1}(E_i-\omega-E_0^{j'}) }} \nn \\
&\times \Theta\lrm{E_i - \omega - E_0^{j'}} \,. \nn
\end{align}
We have used the expression for the two-point function \eqref{eqn:2ptfnJT} with $\Delta=3$, the density of states for arbitrary angular momentum \eqref{eqn:rho_angularmomentum}, and our chosen normalization factor $\mathcal{N}$. The theta function in the last line enforces that when $E_i-\omega < E_0^{j'}$ there is no emission channel into $j'$ states since they don't exist at final energies $E_i -\omega$. We drop the theta function going forward, with the understanding that it is implicit in all equations. Plots of \eqref{eqn:photon_fullflux_kerrnewman} for various values of $E_i$ and $j$ are shown in Figures~\ref{fig:hawkingradphotonintro} and~\ref{fig:hawkingradphoton}.

This formula showcases that a BH with non-zero $j$ can transition into three distinct final BH states with $j'=j,j\pm 1$, and the density of states enforces the constraint that when quantum black hole states cease to exist, the flux into those states vanishes. For small values of $j$, the flux into the different final states $j'$ is slightly uneven due to the prefactor. As we will soon show, this formula interpolates between the semiclassical prediction and the full quantum answer that shows significant deviations near the various breakdown scales $E_0^{j'}$.

The full quantum corrected energy flux for a near-extremal Reissner-Nordstr\"om BH without any initial angular momentum ($j=0$) with initial energy $E_i$ is simpler and is 
\begin{align} \label{eqn:RN_quantumflux_photon}
\left.\frac{d E}{d t}\right\rvert_{j=0} &= \frac{1}{3 \pi}\int_0^{E_i - E_\text{brk.}} d \omega r_+^8 \omega^4 (\omega + \Ebrk) \left( \Ebrk^2 +8 \Ebrk E_i -2 \Ebrk \omega +\omega^2\right) \nn \\ &\times \left(\Ebrk^2+8 \Ebrk E_i+4 \Ebrk \omega +4 \omega^2\right) \nn \\
&\times \frac{ \sinh \lrm{2\pi \sqrt{2 \Ebrk^{-1}(E_i - \omega-\Ebrk)}} }{\cosh \lrm{2\pi \sqrt{ 2 \Ebrk^{-1} E_i}} -\cosh \lrm{2\pi \sqrt{2 \Ebrk^{-1} (E_i-\omega-\Ebrk}} }  \,.
\end{align}
Here we only have emission into $j'=1$ states.

\paragraph{Semiclassical limit.}
In the semiclassical limit, we again take $\frac{E_i}{\Ebrk}\to \infty$. Taking this scaling, we find the flux\footnote{In deriving \eqref{eqn:photon_semiclassicalflux_rotation}, we used the fact that either $j \gg 1$ (and hence $j' \approx j$) or $j = O(1)$.  In the former case, the sum over $j'$ matches the sum over $m_\gamma = j - j'$. In the latter case, we have $E_i - E_0^{j'} \approx E_i - E_0^{j} \approx E_i$ and $\tilde \omega \approx \omega$. The sum over $j'$ then reduces to a factor $\sum_{j'} (2j' + 1) = 3(2j+1)$, which matches the semiclassical answer where the sum over $m_\gamma$ just gives a factor of three.}
\begin{eqnarray} \label{eqn:photon_semiclassicalflux_rotation}
\frac{d E}{d t} &=& \sum_{m_\gamma = -1,0,1} \frac{6}{2\pi}\int_0^\infty d \omega \, \omega \frac{ \frac{4}{9} r_+^8 \omega^3 \tilde{\omega} \lrm{\tilde \omega^2 + \frac{4\pi^2}{\tilde\beta^2}} \lrm{\tilde \omega^2 + \frac{16\pi^2}{\tilde\beta^2}}} {e^{\tilde \beta \tilde \omega}-1}\,, \qquad \tilde \beta = \sqrt{\frac{2\pi^2}{\Ebrk (E_i-j^2/2 \Ebrk)}}\,, \nn \\
\end{eqnarray}
where  $\tilde\omega = \omega - m_\gamma j \Ebrk$ and $m_\gamma \equiv j-j'$. This matches the formula for semiclassical radiation from a slowly rotating near-extremal black hole given in Appendix \ref{app:semiclassicalangularmomentum}. If we further take the limit $\omega, T \gg j\Ebrk $, it reduces to
\begin{eqnarray} \label{eqn:photon_semiclassicalflux}
\frac{d E}{d t} &=& \frac{6}{2\pi}\int_0^\infty d \omega \, \omega \frac{ \frac{4}{9} r_+^8 \omega^4 \lrm{\omega^2 + \frac{4\pi^2}{\beta^2}} \lrm{\omega^2 + \frac{16\pi^2}{\beta^2}}} {e^{\beta \omega}-1}\,, \qquad \beta = \sqrt{\frac{2\pi^2}{\Ebrk E_i}}\label{eq:semiclassicalintegrandplotted} \\
& = & \frac{62848 \pi^9}{2079} r_+^8 T^{10} \ = \  \frac{1964}{2079 \pi} r_+^8 \Ebrk^5 E_i^5 \, , \nn
\end{eqnarray}
which is the semiclassical result an RN black hole given in \eqref{eqn:semiclassicalflux_photon}. The choice of $\mathcal{N}$ was made to match the numerical prefactor with this formula.

\paragraph{Exclusion rules.} 
When a black hole with spin $j$ has $E_i < E_0^{j+1}$, it cannot decay into a state with $j' = j + 1$ because no such states with lower energy exist. This leads to the theta function found in \eqref{eqn:photon_fullflux_kerrnewman}. In particular, if the black hole has spin $j = 0$ and $E_i \leq \Ebrk$, it cannot decay by single photon emission at all, because the angular momentum selection rules require $j' =1$.

\paragraph{Quantum limit.} Quantum effects become important occurs whenever the black hole has energy $E_i$ and a spin $j$ such that $E_i - E_0^{j} \ll \Ebrk$. As in our discussion of exclusion rules above, below these energies, states of spin $j$ cease to exist.  The transition rate will, therefore, be dominated by final states with $j'=j-1$. Expanding \eqref{eqn:photon_fullflux_kerrnewman} for $E_i \approx E_0^{j}$ we find 
\begin{align} \label{eqn:dedt_spectrumedge}
&\left.\frac{d E}{d t}\right\rvert_{j \to j'} = \frac{1}{9 \sqrt{2} \pi^2} {\frac{2j'+1}{2j+1}} \int_0^{E_i-E_0^{j'}} d \omega \omega^4 r_+^8 \sqrt{E_i- E_0^{j'} - \omega} \times \bigg((E_0^{j})^2-2 E_0^{j} (E_0^{j'}+2 \Ebrk+\omega) \nonumber \\ &+(E_0^{j'})^2-4 E_0^{j'} \Ebrk+2 E_0^{j'} \omega+4 \Ebrk^2+8 \Ebrk E_i-4 \Ebrk \omega+\omega^2\bigg) \times \bigg(4 (E_0^{j})^2 \nn \\ & -4 E_0^{j} (2 E_0^{j'}+\Ebrk+2 \omega) +4 (E_0^{j'})^2-4 E_0^{j'} \Ebrk+8 E_0^{j'} \omega+\Ebrk^2+8 \Ebrk E_i -4 \Ebrk \omega+4 \omega^2\bigg)\,.
\end{align}
The above can be evaluated but is not illuminating. The BH would emit roughly one photon in this regime before having $E_i < E_0^j$ and the above formula would no longer apply. There is one case where this is not true, and that is for $j=\frac{1}{2}$.

A state with $j=\frac{1}{2}$ can decay back into a BH with $j'=\frac{1}{2}$ by photon emission since $\frac{1}{2} \subset 1 \otimes \frac{1}{2}$, but cannot decay into a $j'=\frac{3}{2}$ state below $E_i = \frac{15}{8} \Ebrk$. We can find the flux close to the edge of the spectrum $E_i \approx E_0^{j=\frac{1}{2}} = \frac{3}{8}\Ebrk$ for these BHs by using \eqref{eqn:photon_fullflux_kerrnewman} or \eqref{eqn:dedt_spectrumedge} with $j=j'=\frac{1}{2}$ 
\begin{align}
\frac{d E}{d t} = &\frac{2 r_+^8}{9\pi} \int_0^{E_i-E_0^{j=1/2}} d \omega \omega^5 (\Ebrk^2 + 8 \Ebrk E_i -4 \Ebrk \omega + \omega^2) (4\Ebrk E_i -2 \Ebrk \omega + 2 \omega^2 -\Ebrk^2) \nn\\ 
& \times  \frac{\sinh \lrm{2\pi \sqrt{2 \Ebrk^{-1}(E_i-\omega - E_0^{j=1/2}})}}{\cosh \lrm{2\pi \sqrt{2 \Ebrk^{-1}(E_i - E_0^{j=1/2}})}-\cosh \lrm{2\pi \sqrt{2 \Ebrk^{-1}(E_i-\omega - E_0^{j=1/2}})}}\,,
\end{align}
where we have only included the $j'=\frac{1}{2}$ channel. We can expand the second line around $E_i \approx E_0^{j=\frac{1}{2}}$ to get
\begin{align}
\frac{d E}{d t} &= \frac{8}{729 \sqrt{3}\pi^2} \int_0^{E_i-E_0^{j=\frac{1}{2}}} d \omega r_+^8 \omega^4 \sqrt{E_0^{j=\frac{1}{2}}} \sqrt{E_i - \omega - E_0^{j=\frac{1}{2}}} \nn \\ 
\times & (64 E_0^{j=\frac{1}{2}} (E_0^{j=\frac{1}{2}}+3 E_i)-96 E_0^{j=\frac{1}{2}} \omega + 9 \omega^2) (9 \omega^2 - 32 (E_0^{j=\frac{1}{2}})^2 - 24 E_0^{j=\frac{1}{2}} (-2 E_i + \omega))\,.
\end{align}
We have rewritten the answer in terms of the edge of the spectrum. The total flux can be found to be 
\be
\left.\frac{d E}{d t}\right\rvert_{j=\frac{1}{2}} = \frac{8388608 }{2525985 \sqrt{3} \pi^2} r_+^8 \lrm{E_0^{j=\frac{1}{2}}}^{\frac{9}{2}} \lrm{E_i - E_0^{j=\frac{1}{2}}}^{\frac{11}{2}}, \qquad E_0^{j=\frac{1}{2}} \lesssim E_i \ll \Ebrk \,.
\ee
In the above, we have approximated that $E_i \sim E_0^{j=\frac{1}{2}}$.

\subsubsection*{Graviphoton emission.}

We briefly discuss the emission of gravitons, which is dominated by modes with $\ell=2$. These modes are actually composite graviton/photon modes which are known as graviphotons, see appendix \ref{app:photongreybody}. There are two such modes and we discuss them simultaneously since the analysis is identical up to a numerical prefactor. Similar to photons, such particles change the spin of the black from initial angular momentum $j$ to $j \otimes 2 = j,j\pm1,j\pm2$. As explained earlier these modes have semiclassical greybody factor $P \propto (r_+ \omega)^8$ which is the same scaling as the $\ell=1$ photon and thus there is a similar amount of flux emitted by $\ell=1,2$ modes when treated semiclassically. In appendix \ref{app:photongreybody} we find these modes couple to $\Delta=2$ operators in the JT region. Following previous sections the interaction Hamiltonian can be deduced to be
\be \label{eqn:graviton_HI}
H_I= \sum_{m_h = 0,\pm1,\pm2}\mathcal{N} \mathcal{O} \int d \omega r_+^4 \omega^{5/2} (a_{\omega,m_h}+a_{\omega,m_h}^\dag)\,, \qquad \mathcal{N}= \frac{8}{15\pi^2 }\,.
\ee
We can follow the same steps as for the $\ell=1$ photon, where the sum over CG coefficients for allowed transmission channels turns out to be identical as for the photon and is still given by \eqref{eqn:CG-identity}, and we again sum over the two polarizations of the graviphoton. The flux for a BH with initial spin $j$ is
\be
\frac{d E}{d t} = 4\pi\mathcal{N}^2 \sum_{j'=j,j\pm1,j\pm 2} {\frac{1}{2j+1}}\int_0^{E_i} d \omega \hspace{.035cm} r_+^8 \omega^6 \rho_{j'}(E_i-\omega) |\lb E_{i}  - \omega - E_{0}^{j'}| \mathcal{O}|  E_{i} -E_{0}^j \rb|^2\,,
\ee
where now selection rules demand that $j'=j,j\pm1,j\pm 2$. The edge cases to this set of $j'$ are $j=0,\frac{1}{2},1, \frac{3}{2}$. Using the expression for the two-point function and the density of states, we get 
\begin{align}\label{eqn:gravitonflux}
\frac{d E}{d t} = \frac{1}{45 \pi^3} \sum_{\substack{j'=j,j\pm 1,\\ j\pm2}} {\frac{2j'+1}{2j+1}} &\int_0^{E_i-E_0^{j'}} d \omega r_+^8 \omega^6 \Ebrk^3  \sinh \lrm{2\pi \sqrt{2 \Ebrk^{-1}(E_i - \omega-E_0^{j'}}}  \\ 
&\times\Gamma\lrm{2 \pm i \sqrt{2 \Ebrk^{-1}(E_i-E_0^j)} \pm i \sqrt{2 \Ebrk^{-1}(E_i-\omega-E_0^{j'}) }} \nn\,.
\end{align}
If the BH starts at $j=0$, then the only allowed transition is $j'=2$ since $0 \otimes 2 = 2$. In this case, the flux simplifies and is
\begin{align}\label{eqn:gravitonflux_j=0}
\left.\frac{d E}{d t}\right\rvert_{j=0} = &\frac{4}{9   \pi} \int_0^{E_i-3 \Ebrk} d \omega r_+^8 \omega^6 (\omega+3\Ebrk) \lrm{25 \Ebrk^2 + 8 \Ebrk E_i + 20 \Ebrk \omega + 4 \omega^2} \\ 
&\times \frac{ \sinh \lrm{2\pi \sqrt{2 \Ebrk^{-1}(E_i - \omega-3\Ebrk)}} }{\cosh \lrm{2\pi \sqrt{ 2 \Ebrk^{-1} E_i}} -\cosh \lrm{2\pi \sqrt{2 \Ebrk^{-1} (E_i-\omega-3\Ebrk}} } \nn\,.
\end{align}

\paragraph{Semiclassical limit.}

We again take the semiclassical limit of \eqref{eqn:gravitonflux} using the same logic as for the photon. We take $\frac{E_i}{\Ebrk}\to \infty$ and $j \gg 0$ (with $J\to 0$) while $\omega, T \gg j \Ebrk$. The spin-dependent prefactor simplifies and weighs each emission channel equally, with flux
\be \label{eqn:graviton_semiclassical_flux}
\frac{d E}{d t} = \frac{5}{\pi} \int_0^\infty d \omega \, \omega \frac{ \frac{16}{45} r_+^8 \omega^6 \lrm{\omega^2 + \frac{4\pi^2}{\beta^2}} } {e^{\beta \omega}-1}\,.
\ee
This agrees with the greybody factor for the $\ell=2$ graviton that we calculate in appendix \ref{app:photongreybody}. 

\paragraph{Exclusion rules.} There is no emission to states with $j' = j + 2$ when $E_i \leq E_0^{j+2}$ and no emission to states with $j' = j+ 1$ when $E_i \leq E_0^{j+1}$. In particular, since a black hole with $j= 0$ can only decay to one with $j' = 2$ and the spectrum of $j'=2$ black holes starts at $E_0^{j=2}=3\Ebrk$, the energy flux below this energy is zero for gravitons
\be
\frac{d E}{d t} = 0, \qquad E_i \leq 3\Ebrk\,.
\ee
Below this energy graviphotons can only be emitted in two graviphoton processes in an entangled singlet state. Between the range $\Ebrk \leq E_i \leq 3 \Ebrk$ emission is therefore dominated by the much faster emission of $\ell=1$ photons.

If the black hole has angular momentum $j = 1/2$, the allowed post-emission angular momenta are $j' = 3/2, 5/2$. Single-graviphoton emission is therefore excluded for $E_i < E_0^{j = 3/2} = 15 \Ebrk/8$.


\paragraph{Additional $\ell=2$ graviphoton mode.} There is another $\ell=2$ graviphoton mode that has identical formulas to the above except all fluxes must be rescaled by $\frac{1}{4}$. This is given by taking
\be
\widetilde{\mathcal{N}}_{\t{graviphoton}}=\frac{1}{2} \mathcal{N}\,,
\ee
in \eqref{eqn:graviton_HI}.

\subsection{Di-particle emission}

\subsubsection*{Di-photon emission}

\begin{figure}
    \centering
    \includegraphics[width=.8\linewidth]{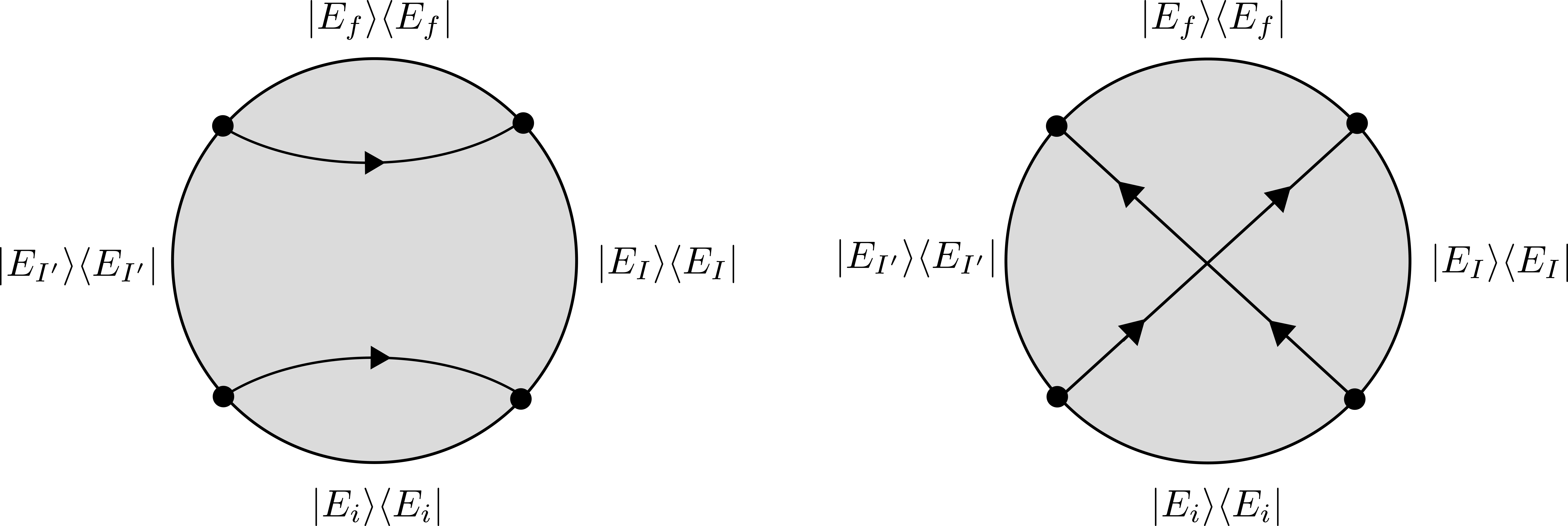}
    \caption{Pictorial representation of the four point function and the two contributing channels. The operator insertions are the black dots. On the left we have that the two intermediate energies are constrained to be equal $E_I = E_{I'}$. On the right they are not constrained, but because of the crossing there is a 6J symbol. There is a possible third channel that would set $E_i=E_f$ but this does not contribute since $E_f < E_i$.} \label{fig:4pt_fn}
\end{figure}

We now calculate the di-photon emission rate which is the dominant decay channel at energies above extremality $E_i \leq \Ebrk$, where two photons are emitted in an entangled singlet state with zero angular momentum. This is a 2nd order process in perturbation theory, with decay rate from an initial state $|\psi_A\rb$ to a final state $|\psi_B\rb$ given by \cite{Sakurai:2011zz,mizushima1970quantum,Bethe:1957ncq}
\be
\Gamma_{i \to f} = 2\pi\left| \sum_I \frac{\lb \psi_B |H_I| I\rb \lb I |H_I| \psi_A \rb}{E_I - E_A}\right|^2 \delta(E_B - E_A)\,.
\ee
The initial state will have $j=0$ and energy $E_i \leq \Ebrk$. The final state will have $j=0$ and energy $E_f < E_i$, along with two photons $\omega, \omega'$ with $E_i-E_f = \omega + \omega'$, which will necessarily be in the singlet state. We must sum over all intermediate states. Since $H_I$ creates/annihilates one photon mode the intermediate state must have a BH state $E_I^{j=1}$ and one photon $\omega_I$ of frequency either $\omega$ or $\omega'$. Taking all of this into account, the rate is
\be \label{eq:initdiphotonrate}
\Gamma_{i \to f} = 2\pi\left|\sum_I \sum_{\omega_I = \omega, \omega'} \frac{\lb E_f^{j=0}, \omega,\omega'| H_{I} | E_I^{j'=1}, \omega_I \rb \lb  E_I^{j'=1}, \omega_I | H_{I} |E_i^{j=0}\rb}{E_I + \omega_I - E_i} \right|^2 \delta(E_i-E_f-\omega-\omega')\,,
\ee
where we have integrated over intermediate photons $\omega_I$ and localized onto the terms that contribute. There are two distinct intermediate states that contribute to the rate where the energy $\omega_I$ of the single photon that has been emitted is either $\omega$ or $\omega'$. The four-point function in \eqref{eq:initdiphotonrate} can be written as a sum over the two Wick contractions that give nonzero answer, as shown in Figure~\ref{fig:4pt_fn}. Each term involves Clebsch-Gordan coefficients resulting from the integral over the holonomies between the operator insertions and a term from JT gravity four-point function \eqref{eqn:jt4ptfn}. The various integrals work to project the emitted photons into the singlet state. We work out the amplitude in appendix \ref{app:angularmomentum_di_photon}, with the final result \eqref{eqn:complicated_integrals-app} showing that, after summing over the axial angular momenta of the photon and intermediate black hole states, the transition rate \eqref{eq:initdiphotonrate} becomes 
\be\label{eq:diphotonratesinglefinalstate}
\Gamma_{i \to f} &= 2\pi\mathcal{N}^4 r_+^{16} \omega^3 \omega'^3  \delta(E_i-E_f-\omega-\omega')  \\ &\qquad \times \int_{\Ebrk}^\infty  d E_I \rho_{j=1}(E_I) \int_{\Ebrk}^\infty dE_{I'} \sum_{\omega_I, \omega_{I'} = \omega, \omega'} \frac{\mathcal{A}_{1}(E_i,E_I, E_f, E_I')}{(E_I + \omega_I - E_i)(E_{I'} + \omega_{I'} - E_i)} \,,\nn
\ee
where 
\be \label{eq:AfromappD}
\mathcal{A}_{1}(E_i,E_I, E_f, E_I')  &= \smash{\sum_{m_i, m_I, m_I'}} \lb E_i,0,0 | \mathcal{O}_{1, m_1} | E_I, 1, m_I \rb \lb E_I, 1, m_I | \mathcal{O}_{1, m_2} | E_f, 0, 0 \rb  \\& ~~~~~~~~~~~~~~~~~\times \lb E_f, 0, 0|  \mathcal{O}_{1, m_3} | E_{I'}, 1, m_{I'} \rb \lb E_{I'},1, m_{I'} | \mathcal{O}_{1, m_4} | E_i, 0,0 \rb  \,
\nonumber
\ee
is given in \eqref{eqn:complicated_integrals-app} and differs from the scalar four-point function \eqref{eqn:jt4ptfn} (with energies defined relative to extremality in the appropriate angular momentum sector) only by the constant prefactor $1/(2\ell+1)^3 = 1/27$. Recall also that the density of states $\rho_{j=1}(E) = 3 \rho(E - \Ebrk)$ for fixed angular momentum $j = 1$ and axial angular momentum $m$ differs from the Schwarzian spectrum $\rho(E)$ by a shift in the spectrum and a factor of $(2j+1)$.

The total decay rate from an initial BH state $E_i$ to any final state is given by integrating \eqref{eq:diphotonratesinglefinalstate} over all final photon and black hole states with the density of states $\rho_{j=0}(E_f)$. This gives
\begin{align}
&\Gamma_{\t{total}} = 4 \int_0^{E_i} d E_f \int_0^{E_i-E_f} d \omega \int_0^{E_i-E_f} d\omega'\, \rho_{j=0}(E_f)  \,\Gamma_{i \to f}
\end{align}\\
To be very explicit, the complete expression for the energy flux from di-photon emission is therefore 
\begin{align}
\label{eq:di-photon-energy-flux}
\frac{d E}{d t} &= 4 \int_0^{E_i} d E_f \int_0^{E_i-E_f} d \omega \int_0^{E_i-E_f} d\omega'\, \rho(E_f) (E_i - E_f) \,\Gamma_{i \to f}
\\&= \frac{8\pi\mathcal{N}^4}{3}\Ebrk^{4\Delta} r_+^{16} \int_0^{E_i} d E_f \,\rho(E_f) \int_0^{E_i-E_f} d \omega \omega^3 (E_i - E_f - \omega)^3 (E_i-E_f) \nn \\ &\,\,\,\,\qquad  \times \mathcal{N}_{4pt}\sum_{\omega_{I,I'}=\omega,\omega'}\int_{\Ebrk}^\infty \frac{d E_I  d E_{I'} \rho(E_I -\Ebrk) \rho(E_{I'} - \Ebrk) (\Gamma_{f I}\Gamma_{f I'} \Gamma_{i I}\Gamma_{i I'})^{1/2}}{(E_I + \omega_I - E_i)(E_{I'} + \omega_{I'} - E_i)} \nn \\ &\,\,\,\,\qquad\qquad \times \left(\frac{\delta\left(E_I-E_{I'}\right)}{\rho\left(E_I\right)}+\left\{\begin{array}{ccc}
\Delta & E_f & E_I - E_{0}^{j=1} \\
\Delta & E_i & E_{I'} - E_{0}^{j=1}
\end{array}\right\}\right)\,, 
\end{align}\\
where $\Delta = 3$ for the $\ell = 1$ photon mode, $E_0^{j=1} = \Ebrk$, $\mathcal{N}^2=\frac{40}{3\pi^2}$ was found using \eqref{eqn:photon_semiclassicalflux}, the factors of e.g. $\Gamma_{fI} = \Gamma^\Delta_{E_f, E_I}$ are defined as in \eqref{eqn:2ptfnJT}, and $\mathcal{N}_{4pt} = (2^{2\Delta-1}\Gamma(2\Delta))^{-2} e^{-3S_0}$ first appeared in \eqref{eqn:jt4ptfn}.

\paragraph{Approximation at low energies.} The expression \eqref{eq:di-photon-energy-flux} simplifies considerably for an initial BH energy $E_i \ll \Ebrk$. In this case we have $\omega, \omega', E_f \ll \Ebrk$ and can be dropped whenever they appear in combination with $E_I, E_I', E_0^{j=1} \geq \Ebrk$. After extracting powers of $e^{S_0}$, the only free parameter appearing in the integral over $E_I$ and $E_{I'}$ in the second and third lines of \eqref{eq:di-photon-energy-flux} is  $\Ebrk$. By dimensional analysis, those integrals then reduce to $\Ebrk^{-2}$ times an $\mathcal{O}(1)$ constant that can be found numerically. Specifically, we find that 
\be \label{eqn:numericsotoc}
e^{S_0} \mathcal{N}_{4pt}&\sum_{\omega_{I,I'}=\omega,\omega'}\int_{\Ebrk}^\infty \frac{d E_I  d E_{I'} \rho(E_I -\Ebrk) \rho(E_{I'} - \Ebrk) (\Gamma_{f I}\Gamma_{f I'} \Gamma_{i I}\Gamma_{i I'})^{1/2}}{E_I E_{I'}} \nn \\ &\,\,\,\,\qquad\qquad \times \left(\frac{\delta\left(E_I-E_{I'}\right)}{\rho\left(E_I\right)}+\left\{\begin{array}{ccc}
3 & 0 & E_I - \Ebrk \\
3 & 0 & E_{I'} - \Ebrk 
\end{array}\right\}\right)\nn
\\&\qquad\qquad= \Ebrk^{-2} \lrm{6.1 + 2.1  }\times 10^{-4}\,,
\ee
where the contribution from the delta function evaluates to the first numerical factor in the parenthesis, while the 6j symbol evaluates to the second term. The total decay rate for di-photon emission is therefore
\begin{align}
\Gamma_{\t{total}} &= 8.2\times 10^{-4} \times \frac{8\pi \mathcal{N}^4}{3} \Ebrk^{10} r_+^{16} \int_0^{E_i} d E_f e^{-S_0}\rho(E_f) \int_0^{E_i-E_f} d \omega \omega^3 (E_i - E_f - \omega)^3 \\
&= 1.8 \times 10^{-3} \times r_+^{16} (\Ebrk E_i)^{\frac{17}{2}} \,, \nn
\end{align}\\
where we have used the approximation 
\be\label{eq:lowenergyrhoE}
\rho(E_f) = \frac{e^{S_0} \sqrt{2E_f}}{\pi\Ebrk^{3/2}}
\ee
for the Schwarzian density of states when $E_f \ll \Ebrk$. Similarly, the total energy flux is
\begin{align}
\frac{d E}{d t} &= 8.2 \times 10^{-4}\times\frac{8\pi\mathcal{N}^4}{3} \Ebrk^{10} r_+^{16} \int_0^{E_i} d E_f e^{-S_0}\rho(E_f) \int_0^{E_i-E_f} d \omega \omega^3 (E_i - E_f - \omega)^3 (E_i-E_f) \nn \\
&= 1.5 \times 10^{-3} \times r_+^{16} \Ebrk^{\frac{17}{2}} E_i^{\frac{19}{2}} \, . 
\end{align}

\subsubsection*{Di-graviton emission.} We can quickly estimate two-graviton emission to show that it is suppressed relative to two-photon emission. We can adapt equation \eqref{eq:di-photon-energy-flux} where we have $\Delta=2$ for $\ell=2$ gravitons. In this case the intermediate state has $\ell=2$ so the numerical integral we must do at low energies is slightly modified. We will not compute this new integral numerically, but instead just determine the relevant scaling factors to confirm the suppression. Extra factors of $\omega$ enter the integrands from $H_I$ given in \eqref{eqn:graviton_HI}, leading to 
\be
\label{eq:di-graviton-energy-flux}
&\frac{d E}{d t} \propto \Ebrk^{4\Delta-2} r_+^{16} \int_0^{E_i} d E_f \rho_{j=0}(E_f) \int_0^{E_i-E_f} d \omega \omega^5 (E_i - E_f - \omega)^5 (E_i-E_f)\,.
\ee
Here, we have assumed we are at low energies $E_i \ll \Ebrk$ so that the sum over intermediate states becomes independent of $\omega, \omega', E_i$ and $E_f$. Evaluating the above and again using \eqref{eq:lowenergyrhoE}, we find
\be
\frac{d E}{d t} \propto r_+^{16} \Ebrk^{\frac{9}{2}} E_i^{\frac{27}{2}}\,.
\ee
Comparing the di-photon flux to di-graviton flux we find
\be
\frac{\t{di-graviton flux}}{\t{di-photon flux}} \sim \lrm{\frac{E_i}{\Ebrk}}^5\,.
\ee
Since $E_i \ll \Ebrk$ we see that most of the radiation emitted by the BH is in entangled photons rather than gravitons. This is in contrast to the semiclassical answer for $E_i \gg \Ebrk$ where the radiation is roughly evenly split between photons and gravitons.

It is useful to get some physical intuition as to why photon emission dominates over graviphoton at very low temperatures, even though the two types of radiation are comparable in the semiclassical regime. For both kinds of modes, the semiclassical rate is obtained by putting together the greybody factors in the  AdS$_2$ throat with the greybody factors between the edge of the throat and asymptotic infinity. The first greybody factor is controlled by the scaling dimension in AdS$_2$ of the propagating mode, $\Delta=3$ for the photon and $\Delta=2$ for the graviphoton. The second greybody factor is controlled by the angular momentum barrier, which is larger for the graviphoton than for the photon. In the semiclassical regime, the difference in scaling dimensions precisely compensates for the difference in the size of the angular momentum barrier to precisely yield the same power law in $\omega$ for the total photon and graviphoton greybody factors. However, in the quantum regime, the AdS$_2$ physics becomes largely independent of the scaling dimension of the associated fluctuation, only affecting the overall matrix elements by a multiplicative constant rather than changing the power of $\omega$. Therefore, at low temperatures, graviphoton emission is suppressed relative to the photon because the larger angular momentum barrier leads to a higher power of $\omega$, which is no longer compensated by a larger AdS$_2$ greybody factor. 

\subsection{Final results for energy fluxes} 

For convenience, we now list the final results for the energy fluxes. 
\paragraph{Massless scalar energy flux.} In the microcanonical ensemble for a BH with energy $E_i=M-Q$ above extremality we have 
\begin{align}
&\frac{d E}{d t} = \frac{1}{30\pi} r_+^2 \Ebrk^2 E_i^2, &&E_i \gg \Ebrk\,, \\
&\frac{d E}{d t} = \frac{16 \sqrt{2}}{105 \pi^2} r_+^2 \Ebrk^{\frac{1}{2}} E_i^{\frac{7}{2}} , &&j = 0, \, \quad E_i \ll \Ebrk\,.
\end{align}
In the canonical ensemble at temperature $T$, we have 
\begin{align}
&\frac{d E}{d t} = \frac{2 \pi^3 }{15} r_+^2 T^4\,, &&T \gg \Ebrk \\
&\frac{d E}{d t} = \frac{256 \sqrt{2}}{35\pi^{\frac{5}{2}}}r_+^2 \Ebrk^{\frac{1}{2}} T^{\frac{7}{2}}, &&j = 0, \, \quad T \ll \Ebrk\,.
\end{align}
\paragraph{Photon $\ell=1$ mode energy flux.} In the microcanonical ensemble for a BH with energy $E_i$ above extremality

\begin{align} \label{eqn:dEdt_photon_final}
&\frac{d E}{d t} = \frac{1964 }{2079 \pi} r_+^8 \Ebrk^5 E_i^5 , &&E_i \gg \Ebrk\,, \\
\t{Single photon emission} \qquad &\frac{d E}{d t} = 0, && j = 0\,,\quad E_i \leq \Ebrk\,. \\
\t{Two-photon emission} \qquad &\frac{d E}{d t}= 1.5 \times 10^{-3} \times r_+^{16} \Ebrk^{\frac{17}{2}} E_i^{\frac{19}{2}} , && j = 0\,, \quad E_i \ll \Ebrk\,.
\end{align}

\paragraph{Graviphoton $\ell=2$ mode energy flux estimate.}
For completeness, we include our estimate for two-graviphoton emission
\begin{align} \label{eqn:dEdt_photon_final2}
&\frac{d E}{d t} = \frac{7856 }{2079 \pi} r_+^8 \Ebrk^5 E_i^5 , &&E_i \gg \Ebrk\,, \\
\t{Single graviphoton emission} \qquad &\frac{d E}{d t} = 0, && j = 0 \,, \quad E_i \leq 3\Ebrk\,. \\
\t{Two-graviphoton emission} \qquad &\frac{d E}{d t} \propto r_+^{16} \Ebrk^{\frac{9}{2}} E_i^{\frac{27}{2}} , &&j= 0\,,  \quad E_i \ll \Ebrk\,.
\end{align}

\paragraph{Fermionic BHs: Photon $\ell=1$ flux.}
For a fermionic BH with energy $E_i$ and spin $j=\frac{1}{2}$ very close to extremality we have a photon flux
\begin{align} \label{eqn:dEdt_fermionic_BH}
&\frac{d E}{d t} =\frac{8388608 }{2525985 \sqrt{3} \pi^2} r_+^8 \lrm{E_0^{j=\frac{1}{2}}}^{\frac{9}{2}} \lrm{E_i - E_0^{j=\frac{1}{2}}}^{\frac{11}{2}}\,, \qquad E_i - E_0^{j=\frac{1}{2}}\ll \Ebrk \,,
\end{align}
with $E_0^{j=\frac{1}{2}}=\frac{3}{8}\Ebrk$ the edge of the spectrum for fermionic BH states. Due to exclusion rules, there is no emission of graviphotons for fermionic BHs.

\section{
Charged particle emission in a fixed background} \label{sec:background-charged}

In this section, we review the computation of charged particle emission from Reissner-Nordstr\"om black holes, in the `probe limit' in which the backreaction on the metric and electric field is small. We explain the multiple approaches to computing the exponent of the emission rate, show that these approaches agree, compute the rate including both the exponential suppression and the $O(1)$ prefactor, and discuss the properties of the post-emission black hole. 

\subsection{The motion of charged particles near  RN black holes} \label{subsec:probemassiveparticles}
To start, we review the classical motion of charged particles in the vicinity of Reissner-Nordstr\"om black holes. For simplicity we consider the `probe limit', in which the positron has sufficiently small mass and charge that it is effectively moving through a fixed spacetime and fixed electric field. (We will discuss backreaction in Section \ref{sec:charged-particle-emission-main}.) 

The motion of a like-charged particle outside a RN black hole is characterized by the competition between electromagnetic repulsion and gravitational attraction. The electrostatic repulsion follows an inverse-square law: the electric field is $|\vec{E}| = Q/r^2$ and so exerts on the positron a radial proper force
\begin{equation}
    F\Bigl|_\textrm{electric} = \frac{Q q }{r^2} \ . \label{eq:electricforcecoulumbsays}
\end{equation}
The gravitational attraction is the same as for an uncharged particle, which is inverse-square at long distance but much stronger  close to the horizon, becoming infinite at $r_+$. This means that gravity wins at short distances. But at long distances  electromagnetic  repulsion wins provided
\begin{equation}
    \textrm{repulsion at large $r$ if} : \ \ \ \ \  \ \ \  
  Q q > m M \ , \ \ \ \ \ \  \hspace{4cm} 
\end{equation}
which will be satisfied even for modestly charged black holes in light of the hugeness of $q/m$, see \eqref{eq:largeratio}.  
More quantitatively, 
to keep a positron at fixed $r$ we need to exert on it (for example with a rocket) an additional proper force  of 
\begin{equation}
    F_\textrm{rocket} 
    =  m\frac{r_+}{2r^2} \sqrt{ \frac{r - r_-}{r - r_+}}  +m \frac{r_-}{2r^2} \sqrt{ \frac{r - r_+}{r - r_-}} - \frac{Qq}{ r^2} \ \ . 
\end{equation}
At long distances (both in the Newtonian region, and less obviously also in the part of the throat region that has $r - r_+ \gg  r_+ - r_-$) this force is just the Newtonian answer $F_\textrm{rocket} = \frac{m M - q Q}{r^2}$. But in the Rindler region the required force becomes
\begin{equation}
    r  - r_+ \ll r_+ - r_- :  \ \ \ \ \  \ \ \  
   \ F_\textrm{rocket} =  \frac{m}{2r_+} \sqrt{ \frac{r_+ - r_-}{r - r_+}}  - \frac{qQ}{r_+^2} + \ldots =   \frac{m}{\Delta s } -  \frac{qQ}{r_+^2} + \ldots \ , \ \ \ \ \ \  \hspace{1cm}  \label{eq:rocketFintermsofs}
\end{equation}
where $\Delta s $ is the proper distance from $r_+$. Between the short-distance gravitational attraction and the long-distance electrostatic repulsion is the free-float radius, $r_\textrm{float}$, where the positron can sit in unstable equilibrium. 

\begin{figure}[t] 
   \centering
   \includegraphics[width=5in]{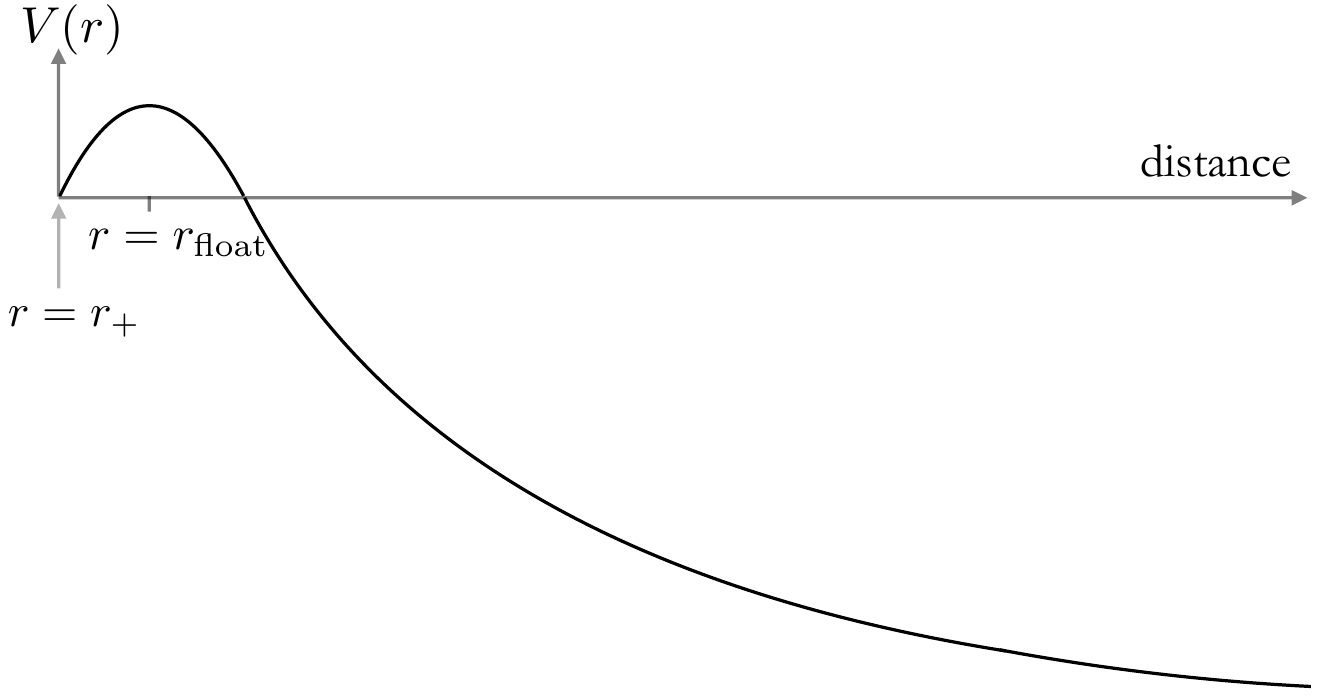} 
   \caption{A cartoon of the potential $V(r)$ seen by a positron outside a RN black hole, in the probe limit. This is the energy that must be exerted to slowly raise  a positron from $r_+$ to $r$; for large enough $r$ the potential goes negative and we are on-net extracting energy. For $r<r_\textrm{float}$ the gravitational attraction wins; for $r>r_\textrm{float}$ the electrostatic repulsion wins.  The instanton describes a decay process of thermal fluctuation of the positron from $r_+$ to $r_\textrm{float}$.} 
   \label{fig:V(r)forchargedparticle}
\end{figure}

An alternative way to visualize this is to consider the energy. The energy required to slowly extract a positron from the horizon out to $r$ is 
\begin{equation}
V(r) = m \sqrt{1 - \frac{2M}{r} + \frac{Q^2}{r^2}} + \Bigl(\frac{qQ}{r} - \frac{qQ}{r_+} \Bigl)  \ .  \label{eq:V(r)definition}
\end{equation}
This is plotted in Figure~\ref{fig:V(r)forchargedparticle}. The first term is the redshifted rest mass of the positron, which ranges from $0$ at the horizon to full $mc^2$ out at $r = \infty$; the second term is the electrostatic energy. 

  The top of the barrier gives the free-float radius, $\partial_r V |_{r_\textrm{float}}  = 0$. Any closer than $r_\textrm{float}$ and the positron falls in, any further and it is expelled.  The redshift factor at the free-float point is $\chi^2 |_{\textrm{float}} =  {    (\frac{M^2}{Q^2} - 1})/({\frac{q^2}{m^2} - 1})$, and  
\begin{eqnarray}
r_\textrm{float} &=& \frac{Q^2 \sqrt{q^2-m^2} \left(M \sqrt{q^2-m^2}+q \sqrt{M^2-Q^2}\right)}{q^2 Q^2-m^2
   M^2} =  r_+ + \frac{r_+^2 \sqrt{M^2 - Q^2} }{2 Q^2} \frac{m^2}{q^2} + \ldots \label{eq:Rforprobefreefloat} \\ 
   V(r_\textrm{float}) &=&  \frac{\sqrt{M^2 - Q^2}}{Q} \left( q - \sqrt{q^2 - m^2}  \right)  =  \frac{\sqrt{M^2 - Q^2}}{Q} \frac{m^2}{2q} + \ldots \label{eq:Vofrfloatinprobelimit}
\end{eqnarray} 

Let us make a couple of comments about these formulas. The first is that the free-float point is typically very close to the black hole. For $Qq \gg m M$,  \eqref{eq:rocketFintermsofs} tells us that the proper distance from the event horizon is 
\begin{equation}
\Delta s \Bigl|_{Qq\gg Mm} = \frac{m}{q} \frac{r_+^2}{Q} \ . \label{eq:deltasoffreefloat}
\end{equation}
Comparing to Sec.~\ref{eq:distancefromrplus}, we see that the free-float point is deep in the `Rindler' region. For $Qq \gg Mm$ gravity can only compete with electromagnetism where the inverse-square law does not apply, and the inverse-square law holds everywhere except the Rindler region.

The second comment is that $V(r = \infty) = m - \frac{q Q}{r_+}$ is enormous. The electrostatic energy of the positron is not a multiple of the rest mass of the positron $m$, it's a multiple of $M_\textrm{Pl}$ since $r_+ \approx \ell_{\t{pl}} Q$, which is stupendously larger than $m$. This means that a positron perched at $r_\textrm{float}$ that then rolls out to large $r$ arrives far from the black hole with a Planckian energy and a concomitantly large relativistic boost factor, given by the square root of the ratio in  \eqref{eq:largeratio}, 
\begin{equation}
    \gamma \sim q M_{\textrm{Pl}}/m  \sim 2 \times 10^{21}. \label{eq:gallonofgasoline}
\end{equation}

\subsection{The semiclassical tunneling exponent}
\label{subsec:probepositronemissionrate}
In the last subsection, we looked at the classical motion of massive charged particles near large RN black holes. Now, while remaining in the probe limit, let's discuss the rate at which quantum effects allow the black hole to create such particles and so spontaneously discharge. 

The electric field at the horizon of an RN black hole is given by  \eqref{eq:electricfieldforRN} as 
\begin{equation}
    |\vec{E}| = \frac{Q}{r_+^2}  \ . 
\end{equation}
As discussed in Sec.~\ref{subsec:pedagogicaloverview}, this is small for all sufficiently massive black holes. A sufficiently small electric field  will give an exponentially small emission rate for massive particles. As a first step, we will calculate that tunneling exponent in the probe limit. In the service of developing intuition, let's calculate it in four different ways. These all give the same exponent, but provide different perspectives on the same process. 
\begin{enumerate}
\item Boltzmann suppression. 

Viewed by a static observer, the immediate vicinity of the horizon is a hot plasma with abundant positrons. To reach $r = \infty$, these positrons must overcome the barrier, shown in Figure~\ref{fig:V(r)forchargedparticle}. They can do this by thermally fluctuating to the top. The temperature is given by  \eqref{eq:tempandentorpyofRNblackhole} and the height of the barrier by  \eqref{eq:Vofrfloatinprobelimit} so the Boltzmann suppression is 
\begin{equation}
\textrm{rate} \sim  \exp \left[ - \frac{V(r_\textrm{float})}{T}  \right] = \exp \Bigl[ - 2 \pi \frac{ r_+^2}{Q} \left( q -  \sqrt{q^2 - m^2}   \right)  \Big]  \  . \label{eq:boltzmannrateprobe}
\end{equation}

\item Change in horizon entropy. 

The probe limit suffices to calculate the first-order change in the charge and mass of the black hole. The new charge is $Q_\textrm{after} = Q - q$. The new mass is $M_\textrm{after} = M - (m + V(r_\textrm{float}) - V(r = \infty))$. The new horizon radius is, keeping only the leading order in the probe limit,  
\begin{equation}
\overline{r_+}  \equiv   M_\textrm{after} + \sqrt{M_\textrm{after}^2 - Q_\textrm{after}^2} =  r_+   - \frac{r_+}{Q} \left( q - \sqrt{q^2 - m^2}  \right)  \ . 
\end{equation}
The reduced area of the black hole means a reduced entropy. The reduced entropy gives an entropic suppression. To leading order in the probe limit, this suppression is 
\begin{equation}
\textrm{rate} \sim   \exp \Bigl[  \frac{\Delta \textrm{Area}}{4}  \Bigl]   \sim \exp \Bigl[ - 2 \pi \frac{ r_+^2}{Q} \left( q - \sqrt{q^2 - m^2}   \right)  \Big] \ .
\label{eq:leading-entropy-difference}
\end{equation}
   
\item Euclidean Action. 

We can also calculate the decay exponent by calculating the Euclidean action of the decay instanton. This instanton has U(1) symmetry in the Euclidean time direction, with the positron static at $r_\textrm{float}$. Euclidean time is compact with asymptotic period equal to the asymptotic inverse temperature $\beta$, given by  \eqref{eq:tempandentorpyofRNblackhole}. We calculate the fully backreacted version of this instanton in Sec.~\ref{sec:including-gravitational-backreaction}, and taking the probe limit of that answer gives 
\begin{equation}
\label{eq:instanton-expression}
\textrm{rate} \sim   \exp \Bigl[ -  S_E( \textrm{instanton}) + S_E (\textrm{false vacuum}) \Bigl]   = \exp \Bigl[ - 2 \pi \frac{ r_+^2}{Q} \left( q - \sqrt{q^2 - m^2}   \right)  \Big] \ . 
\end{equation}

\item Schwinger pair production. 

In the presence of an electric field, QED predicts the spontaneous production of electron-positron pairs. For a uniform electric field, the exponential part of the pair-creation rate  is \cite{schwinger1951gauge}
\begin{equation} 
\textrm{pair-production rate} \sim \exp \Bigl[ - \pi \frac{m^2}{q|\vec{E}|} \Bigl] . \label{eq:sppr}
\end{equation}
 During Schwinger pair production, the electron and positron tunnel until they are far enough apart that the energy recovered by discharging flux compensates for the $2mc^2$ cost of manufacturing the pair. For a uniform field, this is $\Delta s = 2m/q |\vec{E}|$.

In a non-uniform electric field, pair production is fastest where $\vec{E}$ is largest, which for an RN black hole means small $r$. However, pair production only discharges the black hole if the positron is created sufficiently far away that it does not fall back in. Thus the exponential part of the rate of emission can be calculated by considering pair production events that end with the positron perched at $r_\textrm{float}$.

Let's ask about pair-creation events centered exactly at  $r_+$. The electric field at the horizon is $|\vec{E}| = Q/r_+^2$. Approximating the electric field as being uniform, the positron is created a proper distance $\Delta s = m/q |\vec{E}| = m r_+^2/qQ$ outside the horizon, and the electron the same distance inside. Comparing to  \eqref{eq:deltasoffreefloat}, we see that this is exactly the process we want to leave the positron perched at $r_\textrm{float}$. 

Plugging $|\vec{E}| = Q/r_+^2$ into  \eqref{eq:sppr} gives $\exp[ - \pi m^2 r_+^2/ (q Q)]$. This is same as  \eqref{eq:boltzmannrateprobe} at leading order in $q/m$. Including the effect of the small non-uniformities in the electric field would precisely recover  \eqref{eq:boltzmannrateprobe}.

\end{enumerate}

\subsection{Schwinger pair production in the throat }
\label{sec:Schwinger-pair-production}

To calculate not just the exponent of the emission rate but also the non-exponential prefactor, it is most convenient to adopt the formalism of Schwinger pair production. Schwinger calculated the rate of pair production for fermions in a uniform electric field \cite{schwinger1951gauge}, 
\be 
\label{eq:QED-pair-production-rate}
\textrm{rate per spacetime volume for uniform $\vec{E}$} \ \   = \ \  \frac{q^2 |\vec{E}|^2}{4\pi^3} \sum_{k=1}^\infty \frac{1}{k^2} \exp\biggl[-\frac{\pi k m^2 }{q |\vec{E}|} \biggl] . 
\ee
This gives the rate for a constant electric field. The electric field of an RN black hole is not constant---it falls off like an inverse square-law $|\vec{E}| = \frac{Q}{r^2}$. But in the vicinity of the black hole the characteristic length scale over which the electric field varies, $r$,  is much longer than the characteristic separation of the electron-positron pair at nucleation, $\Delta s = \frac{2m}{q }|\vec{E}| \sim \frac{m r^2}{q Q}$. This means that to leading order in $q/m$ we can use the formula for constant-field pair production to calculate the contribution to the rate of pair production from each point in spacetime,
\begin{eqnarray} \label{eqn:Schwinger_rate_final0}
\textrm{rate per spacetime volume for RN} & = &   \frac{q^2 Q^2}{4\pi^3 r^4} \sum_{k = 1}^{\infty} \frac{1}{k^2} \exp \Bigl[ \frac{- \pi k m^2}{q Q} r^2 \Bigl] \ . 
\end{eqnarray}
Each unit of spacetime volume contributes the same to the expected number of pairs, so the larger the volume and the longer you wait the more total fermions are produced. In an interval $dt$ and thickness $dr$ the total spacetime volume is, using the line element  \eqref{eqn:RN_metric}, 
\begin{equation} \label{eq:spacetimevolumeelementRN}
    d(\textrm{spacetime volume}) = \sqrt{f(r)} dt  \frac{dr}{\sqrt{f(r)}} \int r^2 d \Omega = 4 \pi r^2 dt dr \ . 
\end{equation}
For $Q \gg Q_{*}$ the rate per unit time is dominated by $k=1$, so combining \eqref{eqn:Schwinger_rate_final0} and \eqref{eq:spacetimevolumeelementRN} gives 
 \cite{Hiscock:1990ex}
\begin{equation} \label{eqn:Schwinger_rate_final}
\frac{d \langle \textrm{pairs} \rangle }{dt} = \int_{r_+}^\infty d r \frac{q^2 Q^2}{\pi^2 r^2} \exp \Bigl[ {-\frac{\pi m^2}{q Q} r^2} \Bigl]  \ = \  \frac{q^3 Q^3}{2 \pi^3 m^2 r_+^3} \exp  \Bigl[ {-\frac{\pi m^2 r_+^2}{q Q}} \Bigl] \lrm{1+ \mathcal{O}\lrm{\frac{q^2}{m^2 Q}}} .
\end{equation}

\subsubsection*{Probability($r$)} 

The electric field is strongest at the horizon, but not that much weaker farther out in the throat. On the other hand because $\vec{E}$ appears in the exponent, the tunneling rate is extremely sensitive to the value of the electric field. Equation~\ref{eqn:Schwinger_rate_final} tells us that the probability for a pair to be created in a given range $dr$ is  
\begin{eqnarray}
\frac{d \langle \textrm{pairs} \rangle}{dr dt}  & = &  \frac{q^2 Q^2}{\pi^2 r^2} \exp\Bigl[ - \pi \frac{m^2 r^2}{q Q} \Bigl] \  = \  \frac{Q_*^2 Q^2}{m^4 r^2} \exp\Bigl[ - \frac{r^2}{Q Q_*} \Bigl] \ .   \label{eq:differentialprobabilitydensity}
\end{eqnarray}
(In this formula, $r$ is the center of the electron-positron pair, so the positron appears a tiny bit farther out. Only for pairs centered at $r \geq r_+$ does the positron escape the black hole.) 
\begin{figure}[t] 
   \centering
   \includegraphics[width=\textwidth]{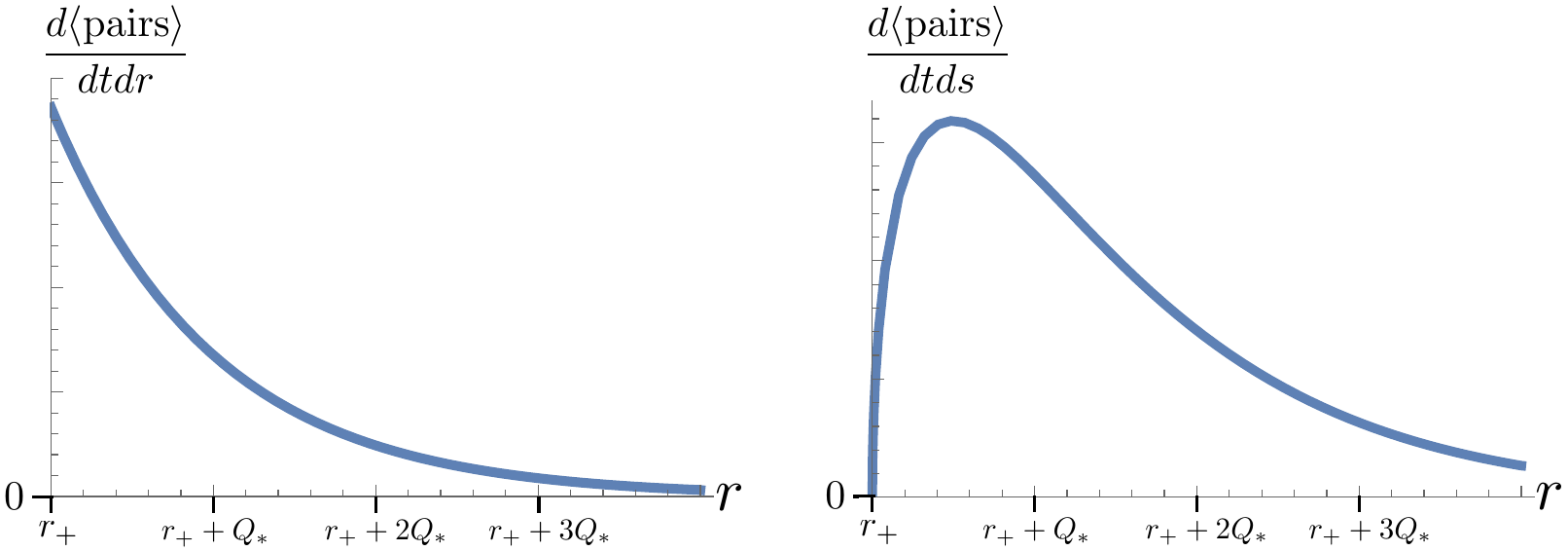} 
      \caption{The typical pair-production event is at $r - r_+ \sim Q_+ $. Left: the probability density with respect to the area-radius $r$, given by  \eqref{eq:differentialprobabilitydensity}, peaks next to the horizon. Right: the probability density with respect to the geodesic distance $s$, given by  \eqref{eq:dpdsdt}, peaks at $r = r_+ + \frac{1}{2} Q_*$.} 
   \label{fig-poslocation}
\end{figure}

The tunneling exponent is minimized (and the integrand in  \eqref{eq:differentialprobabilitydensity} is maximized) at $r = r_+$, corresponding to the positron appearing at $r_\textrm{float}$. 
However, the vast majority of positrons are made farther out than this. The mean location is  
\begin{equation}
\textrm{\bf {mean}: }  \ r = r_+  + \frac{Q Q_*}{2 r_+} + \ldots  \label{eq:meanlocationofpair}
\end{equation} 
 At leading order in $Q/Q_*$, the $r$-dependence of the integrand in  \eqref{eq:differentialprobabilitydensity} is driven by the exponent  (which kills the probability density for $r- r_+ \gg Q_*$), rather than the prefactor $r^{-2}$ (which only kills the probability density for $r - r_+ \gg r_+$), so for near-extremal black holes with $Q \gg Q_*$,  \eqref{eq:differentialprobabilitydensity} simplifies to
\begin{equation}
\frac{d \langle \textrm{pairs} \rangle}{dr dt} \biggl|_{r > Q \gg Q_*}  \sim  \frac{Q_*^2}{m^4 Q^2} \exp\Bigl[ - \frac{Q}{Q_*} \Bigl]    \exp\Bigl[ - \frac{r-Q}{\frac{1}{2}Q_*} \Bigl]  \ . \label{eq:probabilitydistributionforrandtsimplified}
\end{equation}
The probability density  is a decaying exponential with characteristic length $\frac{1}{2} Q_*$. 

\subsubsection*{Probability(s)} 
Let's calculate the probability density with respect to $s$, the proper distance from the horizon. 
The line element,  \eqref{eq:distancefromrplus}, gives 
\begin{equation}
ds = \frac{dr}{\sqrt{(1 - \frac{r_+ }{r})(1 - \frac{r_- }{r})}} \sim  \frac{dr}{\sqrt{(1 - \frac{r_+ }{r})(1 - \frac{r_- }{r_+})}}  \ ,
\end{equation}
where the $\sim$ is a good approximation inside the throat. Combining this with  \eqref{eq:probabilitydistributionforrandtsimplified} gives 
\begin{equation}
\frac{d \langle \textrm{pairs} \rangle}{ds dt} \biggl|_{r > Q \gg Q_*}  \  \sim \ 
 \sqrt{r - Q}  \exp[ - \frac{r-Q}{\frac{1}{2}Q_*} ] 
\ \sim \   \exp \Bigl[ \frac{s  - 2Q Q_*^{-1}  \sqrt{M^2- Q^2} e^{s/Q}}{{2Q} } \Bigl] 
   \label{eq:dpdsdt} 
\end{equation}
for a near-extremal black hole.  The maximum probability density is at 
\begin{equation}
\textrm{peak of $\frac{d \langle \textrm{pairs} \rangle}{ds}$ : } \ \  \ r = Q + \frac{1}{2} Q_* + \ldots  \ \ \leftrightarrow \ \ 
s = \frac{1}{2} Q \log \frac{Q_*^2}{ M^2- Q^2} + \ldots .
\end{equation} 
This is much farther out than   \eqref{eq:probabilitydistributionforrandtsimplified}, which peaks at the horizon. It peaks farther out because, even though $r$ doesn't change much in the throat, the extent to which it does change is increasing exponentially with $s$:   \eqref{eq:distancefromrplus} tells us that in the throat region $r =r_+ + (r_+ - r_-) \sinh^2 \frac{s}{2Q} \sim  Q + \frac{1}{2} \sqrt{M^2- Q^2} e^{s/Q}$. This means a step in $s$ corresponds to a bigger step in $r$ the farther out you go. 
 
(From one perspective, the reason that the Schwinger pair production rate per unit proper spatial volume does not peak at the horizon is that, as one approaches the horizon, the electric field increases, which enhances the pair production rate, but so does the gravitational time dilation, which decreases the rate. At the horizon itself, the gravitational time dilation becomes infinite while the electric field remains finite, and so the emission rate per proper spatial volume goes to zero. It is only as a function of the radial coordinate $r$ that this gravitational time dilation is cancelled by the divergence in the spatial volume associated to a small change in $r$, leading to the finite spacetime volume $4\pi r^2 dt dr$ found in \eqref{eq:spacetimevolumeelementRN}.)

\begin{figure}[t!] 
   \centering
   \includegraphics[width=4.5in]{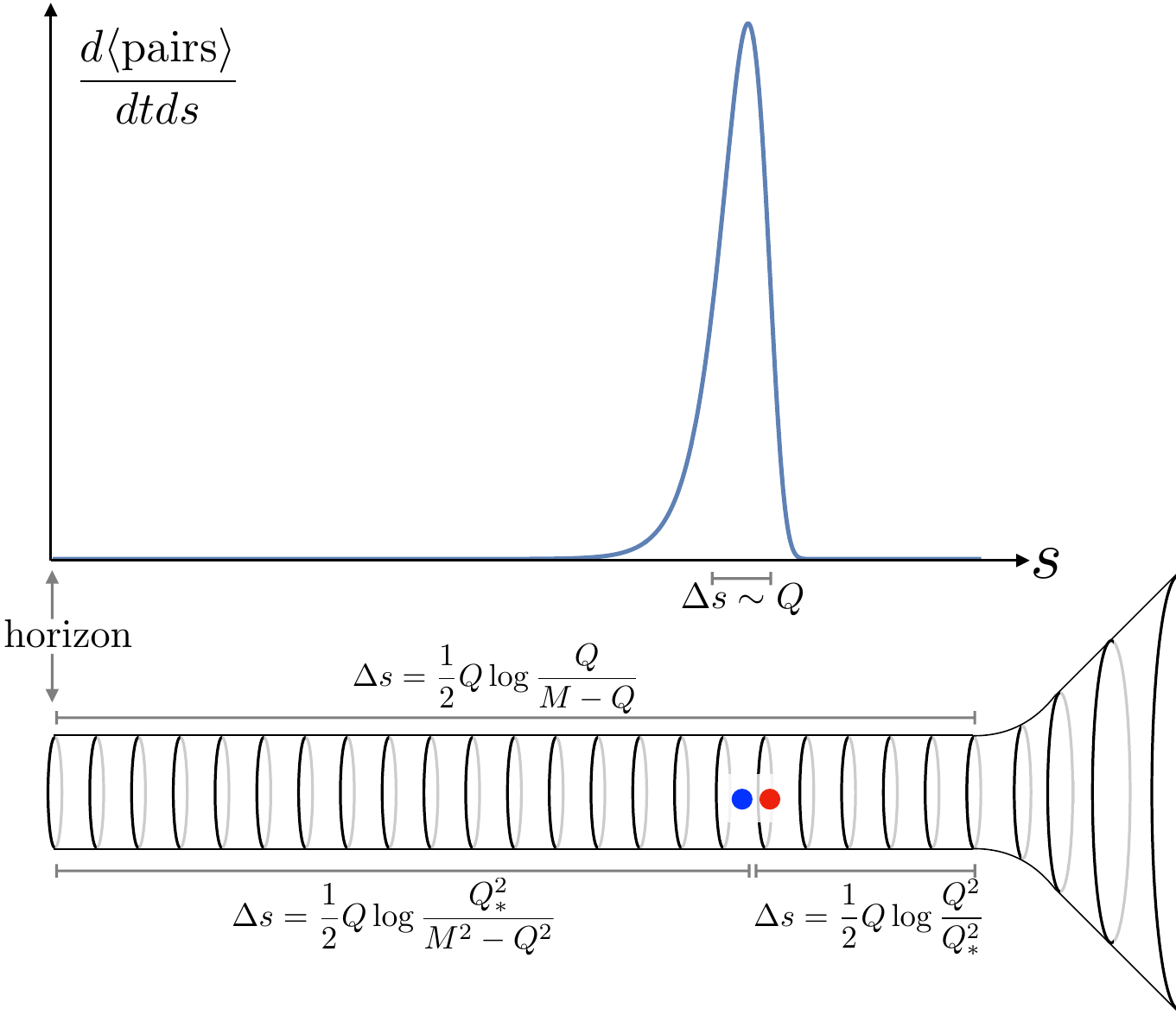} 
   \caption{The pair is typically produced a distance $\frac{1}{2} Q \log \frac{Q_*^2}{M^2 - Q^2}$ from the horizon and a distance $\frac{1}{2} Q \log \frac{Q^2}{Q_*^2}$ from the outer end of the throat. The distribution is sharply peaked, $\Delta s \sim Q$. .} 
   \label{fig-posofrlocation}
\end{figure}

The standard deviation of the distribution \eqref{eq:dpdsdt} scales like 
\begin{equation}
\textrm{spread around peak: } \ \ \ \ \Delta r \sim Q_* \ \ \ \ \ \ \Delta s \sim Q.
\end{equation} 
There is an O(1) multiplicative spread in $r - Q$ and an O(1) additive spread in $s/Q$. Since the mean value of $s/Q$ is much bigger than one for large near-extremal black holes, this means the spread in $s$ is much smaller than the typical distance from the horizon (or from the end of the throat), as shown in Figure \ref{fig-posofrlocation}.

\subsubsection*{Energy of positron} 

In the probe approximation, the final energy of a positron created at $r$ is 
\begin{eqnarray}
E_\textrm{positron} = m \sqrt{ 1 - \frac{2 M}{r} + \frac{Q^2}{r^2} } + \frac{q Q}{r} \ . 
\end{eqnarray} 
For any of the processes we will consider, the dominant contribution to $E_\textrm{positron}$ comes from the second term---not from the rest mass the positron is born with, but from the acceleration it receives from the electrostatic force later in its life. The energy is therefore 
\begin{eqnarray}
E_\textrm{positron}&=& \frac{q Q}{ r}  + \textrm{O}\Bigl(\frac{m}{Q_*}\Bigl)  \\
& = &  q - q \frac{r_+ - Q}{r_+}   - \frac{q Q (r - r_+) }{ r r_+}  + \textrm{O}\Bigl(\frac{m}{Q_*}\Bigl) \\
&=&  q - q \frac{\sqrt{M^2 - Q^2}}{Q}   - \frac{q (r - r_+) }{ Q}  + \cdots 
\end{eqnarray}
So at leading order in $q/m$, leading order at extremality, and leading order in $Q_*/r_+$, this is just $E_\textrm{positron} = q$. At next order, it is slightly less than $q$ both because the black hole is less than extremal and because the positron is produced farther out in the throat.

\subsubsection*{Energy above extremality} 
In the probe limit, 
\begin{equation}
(M - Q)_\textrm{after} = (M - Q)_\textrm{before}  - E_\textrm{positron} + q \ .
\end{equation}
(Away from the probe limit we also need to include the energy in the radiation, etc.) We've already calculated $E_\textrm{positron}$ above, and found that at leading order it is just $q$. This leading-order term exactly cancels off the change in charge. Consequently, the \emph{leading} change in $M - Q$ is given by the \emph{subleading} terms for $E_\textrm{positron}$, 
\begin{equation}
(M - Q)_\textrm{after} = M - Q + q \frac{\sqrt{M^2 - Q^2}}{Q}   + \frac{q (r - r_+) }{ Q}  + \cdots \label{eq:energyafterextremalityasfunctionofr}
\end{equation}
There are three `limits' we wish to take. These limits are $Q \gg Q_* \equiv \frac{q}{\pi m^2}$ (`big black hole limit'), and $Q_* \gg q$ (which is the same thing as $m \ll 1 $, `probe limit'), and $M  \gg M - Q$ (`near-extremal limit'). The subtlety is that these three limits do not commute. 

 Let's consider a case of physical interest. Let's take the probe limit, and then the near-extremal limit, and only \emph{then} the big black hole limit. For example, in Sec.~\ref{subsec:observationalprospects} we'll consider $Q = 2500 Q_*$, and ask what happens to positron emission when such black holes get close to extremality.  This is a reasonable thing to do \emph{even if we start with a ginormous $Q \gg 10^{100} Q_*$ black hole that is far from extremality}, because ginormous black holes emit positrons so rarely that even an \emph{arbitrarily} large black hole will almost \emph{never} emit a positron when not extremely extremal. Instead it'll get exponentially close to extremality first (well into the `closer to extremality than you are big' regime $Q_* \gg \sqrt{M^2 - Q^2}$ that makes the calculation we are about to do valid) and only then emit a positron, so calculating what happens outside this regime is irrelevant. The first time in its entire life that it'll emit a positron when not in the Schwarzian regime is once  $Q  \, \lsim \, 2500 Q_*$.  

  Specifically, if we take the `closer to extremality than you are big' limit, $Q_* \gg \sqrt{M^2 - Q^2}$, then  the last term in  \eqref{eq:energyafterextremalityasfunctionofr} dominates. Using $\langle r - r_+\rangle \sim \frac{1}{2} Q_*$ from  \eqref{eq:meanlocationofpair} gives 
\begin{equation}
\langle (M - Q)_\textrm{after} \rangle = \frac{1}{2} \frac{q Q_*}{Q} + \ldots \label{eq:expectedenergyafterdecayclosetoextremality}
\end{equation}
Indeed, we can get the whole probability distribution using  \eqref{eq:probabilitydistributionforrandtsimplified}. Conditional on having just emitted a positron, 
\begin{equation}
\textrm{Probability}[(M - Q)_\textrm{after}] =  \frac{1}{\frac{1}{2} \frac{q Q_*}{Q} }  \exp\Bigl[ - \frac{(M - Q)_\textrm{after}}{\frac{1}{2} \frac{q Q_*}{Q} } \Bigl] \ . \label{eq:decayingexponential}
\end{equation}
(This formula suggests that if the black holes starts extremal, then the mode energy above extremality after decay is zero. But this is not quite right---to correctly calculate the mode, you need to include the subleading term in the probe limit. We will show when we construct the gravitational instanton that the actual mode, and the lowerbound on $(M - Q)_\textrm{after}$, is $\frac{q^2}{2r_+} $. But in any event it doesn't matter what the `mode' says because the mode event happens vanishingly rarely.)

  An energy distribution that is a decaying exponential, like in \eqref{eq:decayingexponential},  is characteristic of a Boltzmann distribution, giving an $\textrm{`effective temperature'} =  \frac{1}{2} {q Q_*}/{Q}.$ This is much bigger than the Hawking temperature of the black hole at the end of the emission process (after the positron has been emitted and escaped to $r = \infty$), which for a black hole with typical final energy,  \eqref{eq:expectedenergyafterdecayclosetoextremality},  is given by  \eqref{eq:tempandentorpyofRNblackhole} as 
\begin{equation}
T_\textrm{after typical} = \frac{1}{2 \pi} \frac{ \sqrt{M^2 - Q^2} }{Q^2}  = \frac{1}{2 \pi} \frac{ \sqrt{2Q} \sqrt{\frac{1}{2} \frac{q Q_*}{Q} }}{Q^{2}} = \frac{1}{2 \pi}  \frac{\sqrt{q Q_*}}{Q^2} \ . 
\end{equation}
The black holes have a huge dispersion in final energy compared to the Hawking scale.

\subsubsection*{Summary of post-emission properties for typical emission} 
\begin{figure}[t] 
   \centering
   \includegraphics[width=\textwidth]{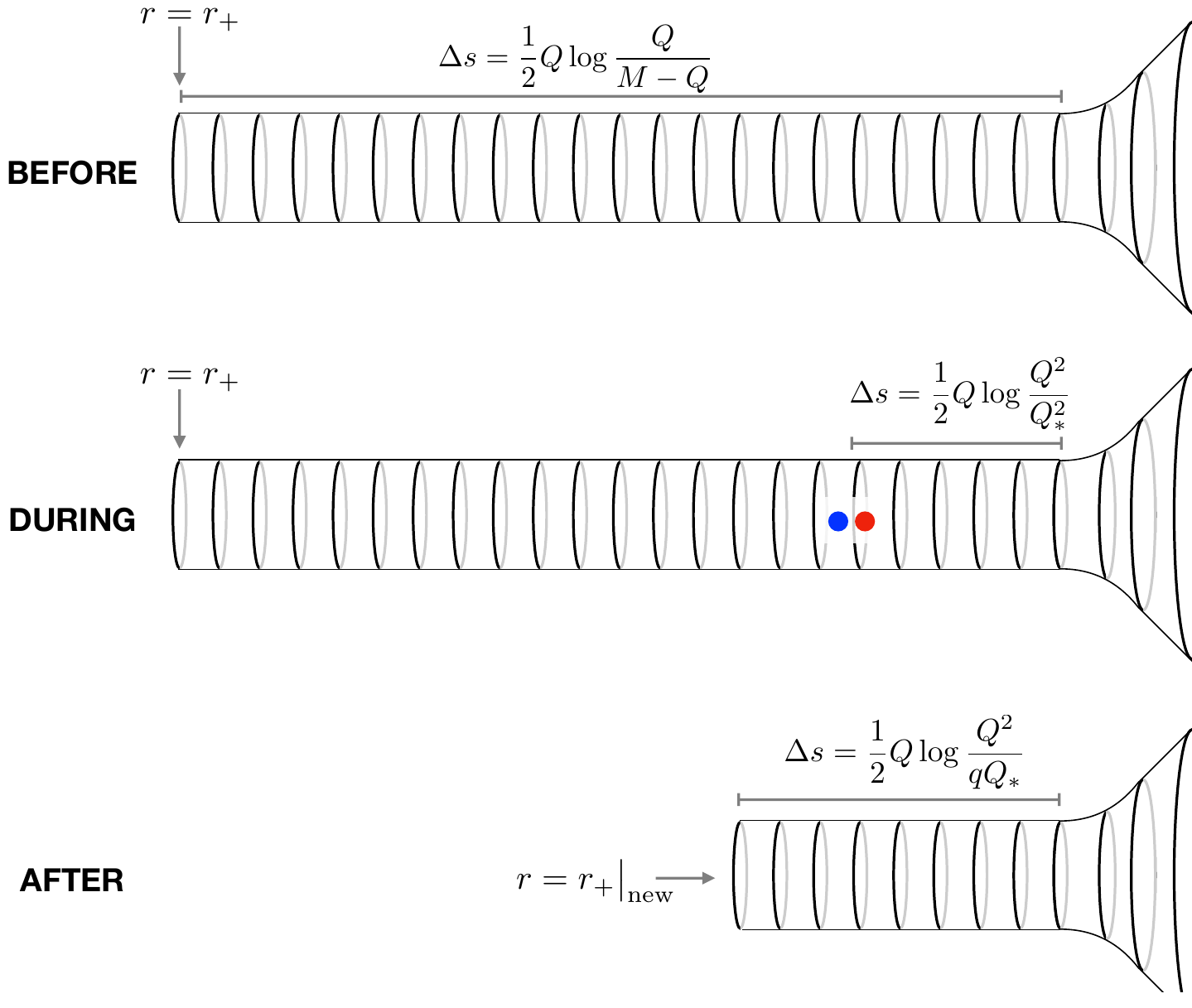} 
   \caption{Top: the throat of a near-extremal black hole before emission. Middle: the black hole immediately after pair creation, but before any subsequent classical motion of the particles, assuming the pair is produced at $r = r_+ + Q_*$ (which is a typical location, but much farther out than the location described by the instanton). The area of the black hole is unchanged. The \textcolor{blue}{electron} is made closer to the black hole than the \textcolor{red}{positron}, separated by $\Delta s \sim 2Q \frac{m}{q}$ which is much smaller than any of the other distance scales (including the size of the two-sphere). The distance from the end of the throat is independent of how nearly extremal the pre-decay black hole was (provided it was close enough to extremality that $\sqrt{M^2 - Q^2} < Q_*$). Bottom: the \textcolor{red}{positron} is accelerated away from the black hole by the electrostatic force and achieves Planckian momentum; the \textcolor{blue}{electron} is sucked into the black hole, crashing into it with large momentum and non-adiabatically increasing its area. This shortens the throat and increases the area-radius from $r_+ = Q$ out to $r_+|_\textrm{new} \sim Q + \sqrt{q Q_*}$} 
   \label{fig:throatdestruction}
\end{figure}

We have seen that typically the pair is created near $r = r_+ + Q_*$, corresponding to a distance $\Delta s = Q \log \frac{Q}{Q_*}$ from the outer end of the throat. After creation the positron is then expelled to  $r = \infty$ and the electron is sucked into the black hole. At the end of this process, at leading non-trivial order in the probe approximation, we have 
\begin{eqnarray} \label{eqn:Adams_eqns}
M - Q & = & \frac{q Q_*}{2Q} \label{eq:refertome}\\
\sqrt{\frac{ M - Q}{Q} } & \sim & \frac{\sqrt{q Q_*}}{Q}  \\
T  & \sim & \frac{\sqrt{q Q_*}}{Q^2} \\
\textrm{length of throat} & = &  Q \log \frac{Q}{\sqrt{q Q_*}} \\
\frac{1}{4} \textrm{area}   & = &  \pi Q^2 + 2 \pi Q (\sqrt{q Q_*}  - q)  + \ldots 
\end{eqnarray}
We see that the entropy has increased by a lot---by about $2 \pi Q \sqrt{q Q_*}$. As we will discuss in more detail in Section \ref{sec:whysemiclassical}, all of this entropy was made during the classical motion \emph{after} pair production, because immediately after pair production the change in area of the black hole is zero. The entropy of the black hole only goes up when the electron crashes into the horizon, as shown in Figure \ref{fig:throatdestruction}. 

The distance from where the pair is created to where the horizon will be at the end of the process is $\Delta s = Q \log [\sqrt{Q_*/q}] = Q \log [M_\textrm{Pl}/{m_e}] = Q \log [ 2.4 \times 10^{22}] = 51.5 Q$. (By comparison if the decay proceeded as in the instanton we would have had $M - Q = \frac{q^2}{2Q}$, $\sqrt{\frac{ M - Q}{Q} } \sim \frac{q}{Q}$, $T   \sim  \frac{q}{Q^2}$ and  final $\textrm{length of throat} =   Q \log \frac{Q}{q}$; and for reference, the breakdown scale is $M - Q \sim  \frac{1}{Q^3}$, $\sqrt{\frac{ M - Q}{Q} }  \sim  \frac{1}{Q^2}$, $T  \sim \frac{1}{Q^3}$, corresponding to a $ \textrm{length of throat} =   2 Q \log Q $.) 

\subsubsection*{Instanton vs.~typical}
We have seen that there is a significant difference between the discharge process described by the instanton, and the \emph{typical} discharge process. This is the difference between the mode and the median of a skewed distribution. Ultimately this is an artifact of the almost-zero-mode describing the radial location of the pair-production event. 


The quantity that is the same is the tunneling exponent, for which all processes agree. The quantities that differ the most are the properties of the residual black hole after the tunneling process is complete. The process described by the instanton manufactures the positron much closer to the horizon, and so leaves the residual black hole much closer to extremality. 

One way exemplify the difference is to ask if there is ever an electron made outside the horizon. Recall that, for a positively charged black hole, the electron is made closer to the hole than the positron, separated by $\Delta s = \frac{2m}{q|\vec{E}|} \sim \frac{2m Q}{q}$. If the positron is created sufficiently close to $r_+$, the electron will be created \emph{inside} the horizon and will proceed directly to the singularity without ever making an appearance outside. This is what happens in the process described by the instanton, for which the positron is produced at the free-float point a distance $\frac{m Q}{q}$ outside the horizon. By contrast, we saw that the typical pair-production event happens farther out, resulting in both the positron and the electron being made outside the horizon (and then later the electron being sucked back into the black hole). We can ask for what fraction of decays the picture described by the instanton is valid. There is no electron produced outside the horizon if the center of nucleation is anywhere between the horizon at $s=0$ (corresponding to $r = r_+$) and $s = \frac{m Q}{q}$ (corresponding to $r = r_+ + \frac{m^2}{q^2} \sqrt{M^2 - Q^2}$), so using  \eqref{eq:differentialprobabilitydensity}, 
\begin{equation}
\textrm{Probability[no electron outside horizon]} \ \sim \  \frac{m^2}{q^2} \frac{\sqrt{M^2 - Q^2}}{Q_*} \ . 
\end{equation}
This is minuscule for near-extremal black holes. The instanton almost always gives a misleading picture of at least this aspect of the emission process.

\FloatBarrier

\section{Charged particle emission in quantum gravity}
\label{sec:charged-particle-emission-main}

 Given the drastic effects of backreaction and quantum gravity corrections on the emission of neutral particles close to extremality, we need to revisit the analysis of charged particle emission. At first sight, the calculations presented in Section \ref{sec:background-charged} provide additional reason to believe that such corrections should exist. In particular, we saw that, at least for positrons emitted sufficiently close to the horizon, the exponent appearing in the pair-production rate was simply the difference in entropies between the pre-emission and post-emission black holes 
\begin{equation}\label{eq:gammaentropies}
\Gamma \sim \exp\left[S_\text{after} - S_\text{before}\right] \sim \exp\Bigl[  \frac{\textrm{Area}_\textrm{after}  - \textrm{Area}_\textrm{before}  }{4 G_N }   \Bigl] \,.
\end{equation}
Both the entropies appearing on the right-hand side of \eqref{eq:gammaentropies} could in principle be corrected due to quantum gravity effects. It will turn out however that backreaction in the gravitational instanton means that the post-emission black hole is never close enough to extremality to receive large Schwarzian corrections. On the other hand, pre-emission, the black hole can be arbitrarily cold and the entropy $S_\text{before}$ will be reduced by a large polynomial factor if the pre-emission  energy above extremality satisfies $M-Q \ll\Ebrk$. Assuming \eqref{eq:gammaentropies} continues to hold, we would then expect the charged particle emission rate for black hole sufficiently close to extremality to be polynomially enhanced relative to semiclassical expectations.

However, by explicitly finding the backreaction of the charged particle worldline on the instanton metric and then computing the effect of quantum gravity corrections from large fluctuations in almost-zero modes around such backgrounds, we will see that the reasoning above is flawed. Quantum gravity leaves the semiclassical charged particle emission rate largely unchanged, even when the initial temperature of the black hole is very small.

\subsection{Charged particle emission including gravitational backreaction} \label{sec:including-gravitational-backreaction}

\subsubsection*{The backreacted black hole solution }
The formalism for incorporating gravitational backreaction into the calculation of the decay rate of a false vacuum  was pioneered by Coleman and de Luccia \cite{Coleman:1980aw,Brown:1988kg}. To calculate the decay rate, Coleman and de Luccia tell us to calculate the instanton, which in this case just means the $t \rightarrow i \tau$ Euclidean version of the solution after decay.  This instanton solves the Euclidean equations of motion---the Euclidean version of Einstein's and Maxwell's equations. In this section, we will find this new instanton solution, including the backreaction on the spacetime, in terms of the initial mass $M$ and charge $Q$ of the black hole that emits the charged particle.

\begin{figure}[t!] 
   \centering
   \includegraphics[width=4in]{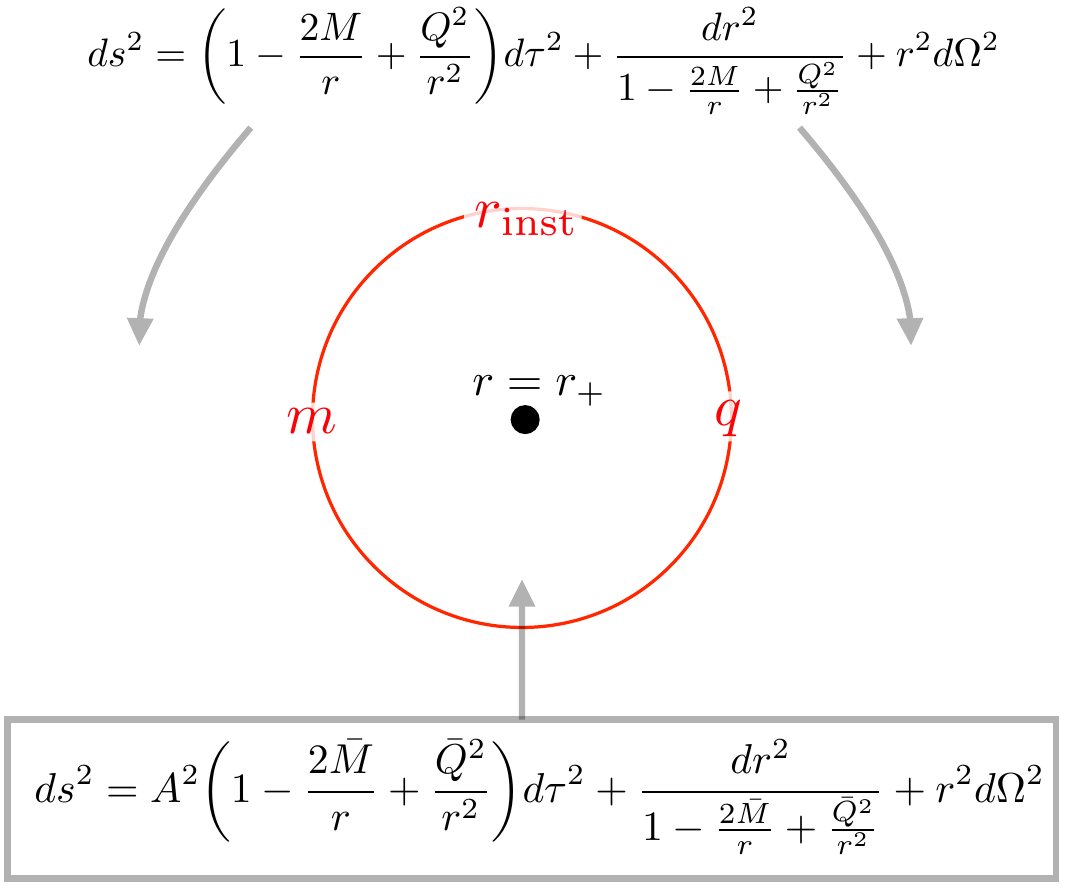} 
   \caption{
   The Euclidean ($r, \tau)$-plane in the presence of the instanton in the S-wave approximation. Inside of the instanton ($r<r_{\textrm{inst}}$), the metric is that of a black hole of mass $\bar M$ and charge $\bar Q$, surrounded by a charged particle at the free-float radius $r=r_{\textrm{inst}}$.  Outside of the instanton ($r>r_\textrm{inst}$), the metric is that of a black hole of mass $ M$ and charge $ Q$.  
   }
   \label{fig:entombedblackhole}
\end{figure}

A positron is a point particle, but at large distances the dominant contribution to gravitational backreaction comes from its S-wave sector. Consequently, we will expand the metric and electromagnetic field in spherical harmonics and only focus on their S-wave sectors. 
Within this approximation, the gravitational instanton can be described as follows (and as shown in Figure~\ref{fig:entombedblackhole}). Outside the location of the positron, the solution looks like an Euclidean black hole with mass $M$ and charge $Q$ given by their pre-emission values. Inside the location of the particle, the S-wave solution describes a new black hole, with reduced mass $\bar{M}$ and charge $\bar{Q}$. We will be able to solve the S-wave sector equations to find the instanton solution exactly to all orders in $q$ and $m$. However, other sectors will also affect the solution at subleading orders in $m/M$ and $q/Q$, so the S-wave calculation is incomplete at those orders. Nevertheless, the leading order calculation suffices to show that backreaction alone does not greatly affect the decay rate, even at very low temperatures.

The action of the particle in the S-wave sector is given by
\be \label{eqn:Action_Maxwell_Shell}
I_{\t{inst + BH}}= I_{EM} +m \int_\text{worldline} d\tau  \sqrt{\gamma} + i q \int_\text{worldline} A \,,
\ee
where $I_{EM}$ is the Einstein-Maxwell action \eqref{eqn:I_Einstein_Maxwell}, and we are taking the particle wordline in the $\tau$ axis and coupling it to the gauge field. Above, the worldline will capture the trajectory of the charged particle in the $(r,\tau)$-plane defined below. The instanton solution is found by solving the equations of motion inside and outside the particle worldline and then using the junction conditions to fix the location of the worldline. Solving the solutions inside and outside we find
\begin{align} \label{eqn:EOM_Solutions_instanton_4d}
&\text{Inside}: \quad F_{\mu \nu} F^{\mu \nu} = -\frac{2 \bar Q^2}{r^4} \quad ds^2 = A^2 \bar\chi^2(r) d \tau^2 + \frac{d  r^2}{\bar \chi(r)^2} + r^2 d \Omega_2^2 \,, \quad \bar \chi(r)^2 =1+ \frac{\bar Q^2}{r^2} - \frac{ 2 \bar M}{r}, \nn \\
&\text{Outside}: \quad F_{\mu \nu} F^{\mu \nu} = -\frac{2  Q^2}{r^4} \quad ds^2 = \chi^2(r) d \tau^2 + \frac{d  r^2}{ \chi( r)^2}+ r^2 d \Omega_2^2\,,\quad \chi(r)^2 =1+ \frac{Q^2}{r^2} - \frac{2 M}{r}\, , 
\end{align}
where $A$ is an as yet undetermined constant that needs to be included if we want to identify the coordinate $\tau$ inside and outside the particle worldline.
The gauge field is continuous across the worldline, and smoothness at the Euclidean horizon $r = \bar{r}_+$ gives
\be
A_{\t{int}} = i A \bar Q \lrm{\frac{1}{\bar r_+} - \frac{1}{r}} d\tau, \qquad  A_{\t{out}} = i Q \lrm{\frac{A \bar Q (r_\inst-\bar r_+) + Q \bar r_+}{Q \bar r_+ r_\inst } - \frac{1}{r}} d\tau \,.
\ee

We must solve for the unknowns $\bar M,\, \bar Q, \,A,\, r_{\inst}$ and the periodicity of the thermal circle $\tau \sim \tau + \betainst$. The latter is related to the periodicity of the thermal circle on the inside of the worldline, $\bar \tau \sim \bar \tau+ \bar \beta$. These can be solved by using the junction conditions across the worldline location and requiring a smooth Euclidean solution inside the worldline. The charge is given by $\bar Q = Q - q$ since the particle carries charge $q$. 

The Israel junction conditions demand continuity of the induced metric and a jump in the extrinsic curvature $K$ across the worldline location
\begin{gather} \label{eqn:Israel_Junction_conditions}
    \sqrt{h}_{\t{int}} = \sqrt{h}_{\t{out}} \Rightarrow A \bar \chi = \chi , \\
    \Delta K_{ i j} = 8 \pi \left(T_{i j} -\frac{1}{2}h_{i j} h^{k l} T_{k l}\right),
\end{gather}
where we have written the continuity equation in terms of our ansatz, and $T_{i j}$ is the stress tensor of the S-wave component of the particle. Since the worldline has no pressure the only non-zero component is $T_{\tau \tau} = \sigma g_{\tau \tau}$. The extrinsic curvature is given by $K_{\mu \nu} = \frac{1}{2} \sqrt{g^{r r}} \partial_r g_{\mu \nu}$ for $\mu,\nu \neq r$. The time and angular components give
\begin{align} \label{eqn:Israel_Junction_conditions_explicit}
    &\Delta K_{\tau \tau} = 4\pi T g_{\tau \tau} \Rightarrow \frac{1}{\chi} \left (M - \frac{Q^2}{r_\inst} \right) - \frac{1}{\bar \chi} \left(\bar M - \frac{\bar Q^2}{r_\inst} \right) = m,\\
    &\Delta K_{\theta \theta} = -4\pi T g_{\theta \theta} \Rightarrow r_\inst (\chi - \bar \chi) = m. 
\end{align}
The final condition is smoothness of the Euclidean spacetime inside the worldline. This requires the coordinate $\tau$ to have periodicity
\be 
\label{eq:smoothness-inside-inst} 
\betainst = \frac{\bar \beta}A = \frac{2\pi}{A} \frac{2\bar r^2_+}{\bar r_+ - \bar r_-} , 
\ee
where $\bar r_\pm = \bar M \pm \sqrt{\bar M^2 - \bar Q^2}$ are the locations of the inner and outer horizon of the black hole inside the worldline.
 Using the junction conditions \eqref{eqn:Israel_Junction_conditions}, \eqref{eqn:Israel_Junction_conditions_explicit} we can solve for the other unknowns $A, r_{\inst}, \bar M$ to find 
\begin{gather} 
    A = \sqrt{\frac{M^2-Q^2}{\bar M^2- \bar Q^2}}\,, \nonumber \\
    r_{\inst} = \frac{M \left(q^2-m^2\right) \left((2 Q-q)^2-m^2\right)+\left(m^2-q (q-2 Q)\right) \sqrt{\left(q^2-m^2\right) \left(M^2-Q^2\right) \left((2 Q-q)^2-m^2\right)}}{m^4-2 m^2
   \left(2 M^2+q (q-2 Q)\right)+q^2 (2 Q-q)^2}\,,  \nonumber \\
    \bar M = M - \frac{q (2Q -q)+m \left(m+2 \sqrt{Q^2-2 M r_{\inst}+r_{\inst}^2}\right)}{2 r_{\inst}}\,. \label{eq:barM-exact-formula}
\end{gather}
The radius of the new horizon and the redshift at the location where the particle is pair-produced are given by
\begin{gather} \label{eqn:rplus_after_instanton}
\bar r_+ = r_+  \frac{ \left( \sqrt{ (2Q-q)^2 - m^2} +  \sqrt{ q^2 - m^2}  \right)^2}{4 Q^2}\,, \\     
\chi^2(r_{\inst}) = \frac{4 m^2 (M^2-Q^2)}{((2 Q -q)^2 - m^2)(q^2 - m^2) }\,. \nonumber
\end{gather}

Having found the instanton solution, we can now compute the on-shell Euclidean action in the microcanonical ensemble to obtain the decay rate. We find (see Appendix \ref{app:Euclidean_instanton} for details) that the action is simply given by
\be 
I_{\t{inst + BH}} (M, Q) = -\pi \bar{r}_+^2 .
\ee 
where the value of $\bar r_+$ can be written as a function of $M$ and $Q$ from the above equations. This should be compared to the false vacuum action   in the microcanonical ensemble $I_{\t{BH}} (M, Q) = -\pi {r}_+^2$ (that of the black hole solution without the instanton). Thus, the overall decay rate given by the saddle-point approximation is
\be 
\text{rate} \sim {\exp\left[I_\text{BH} - I_\text{BH+inst}\right]} = \exp \left[-{ \pi  {\bar{r}_+}^2 + \pi r_+^2 } \right]\,. \label{eq:leading-order-decay-rate}
\ee
This completely agrees with the entropy difference between the final black hole state and the initial black hole state. To leading order in $Q$, the difference \eqref{eq:leading-order-decay-rate} completely agrees with our estimate \eqref{eq:leading-entropy-difference} for the entropy difference where we did not fully compute the backreacted solution but only estimated the final mass of the black hole $\bar M$. All the subleading terms in $Q$ that we find are finite in the low-temperature limit, and therefore, we find that backreaction does not greatly affect the decay rate. 
Of course, as we saw in Section \ref{sec:Schwinger-pair-production}, the emission of a positron typically looks qualitatively quite different from the ``modal'' emission described by the gravitational instanton and in particular occurs comparatively far from the black hole horizon. Typical emission events will soon be included as part of the one-loop determinant of quantum-gravitational fluctuations around this instanton.

\subsubsection*{Properties of the gravitational instanton}

First, however, we provide a qualitative discussion of some properties of the instanton solution that we have just found, or equivalently (taking $\tau \rightarrow i t$) of the static Lorentzian solution found immediately after the charged particle emission.

\begin{itemize}

\item When $q Q \gg m M$, which since $q \gg m$ for positrons in our universe will apply even for black holes of intermediate charge, the radius $r_\text{inst}$ sits very close to the outer horizon
\begin{align} \label{eqn:R_inst_q_greater_m}
r_\inst &= r_+ + \frac{m^2\sqrt{M^2-Q^2}}{2q^2} + \mathcal{O}\left( (M^2-Q^2)^{\frac{3}{2}} ,\,\frac{m^4}{q^4}\right)\, . 
\end{align}
We were able to see this already in the probe limit,  \eqref{eq:deltasoffreefloat}, and it continues to hold when including the effects of gravitational backreaction. Thus, the instanton sits very close to the horizon, in a region which is well approximated by Rindler space. 
Notably the throat can in principle contain a ``dilute gas'' of a large number of instantons separated by large distances compared to their own size.

\begin{figure}[t!]
    \centering
    \includegraphics[width=0.45\textwidth]{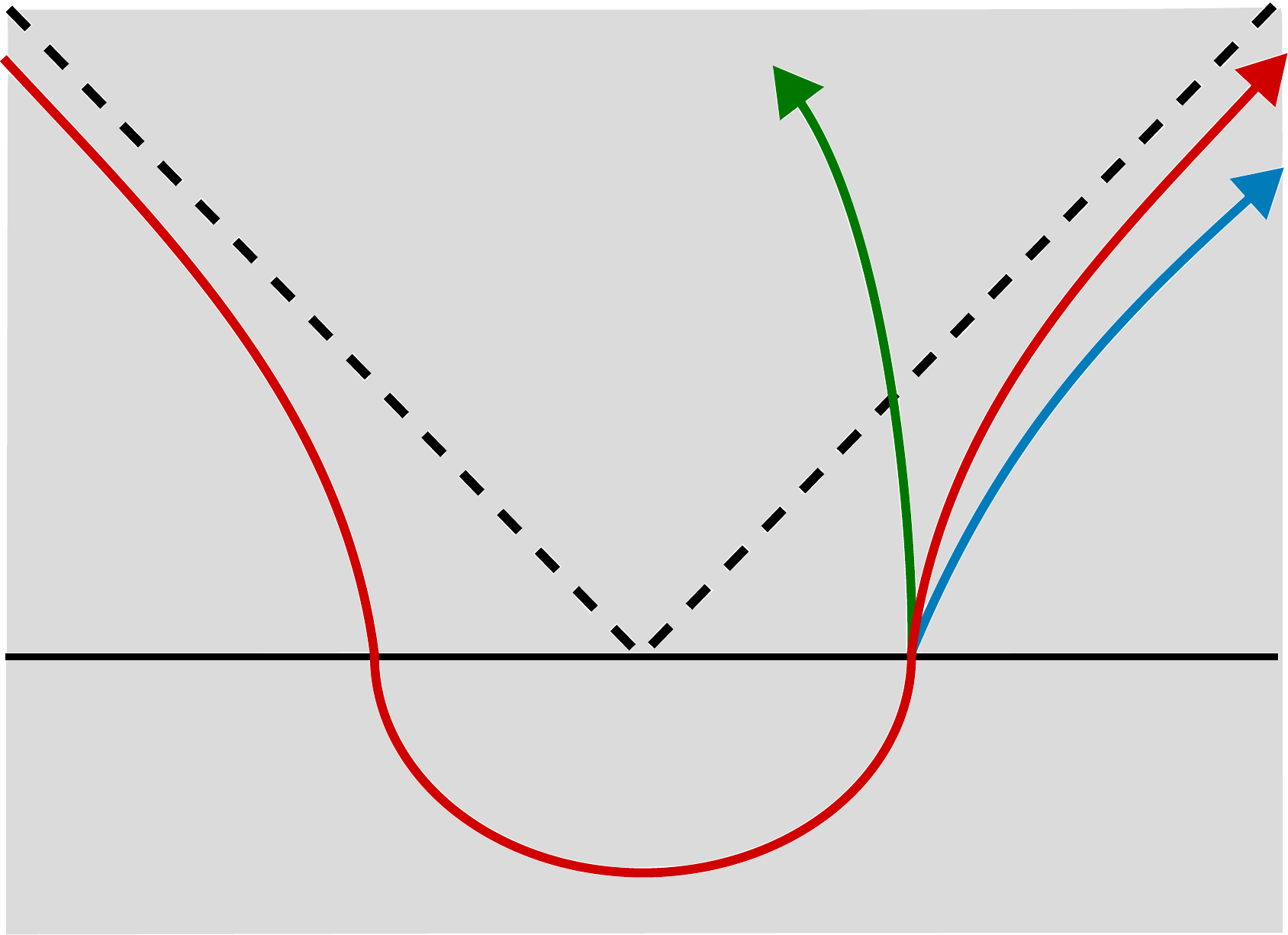}
    \caption{Instanton configuration when analytically continuing the geometry to Lorentzian signature. The bottom segment is half of the Euclidean instanton solution cut at Euclidean time $\tau=0$ and $\beta/2$, and the top segment is Lorentzian time evolution using the $\tau=0$ slice for initial data. The dashed line is the black hole horizon. When going from Euclidean to Lorentzian signature, the particle is in an unstable equilibrium. If not perturbed, the instanton will sit at constant radius outside the horizon for all time (red). With a small quantum fluctuation, the particle will be perturbed away from its unstable equilibrium and either fall into the black hole (green), or escape out to infinity (blue). We depict all of these scenarios in the figure. Only a portion of the asymptotically flat spacetime is depicted, with the interior being a portion of the Reissner-Nordstr\"om Penrose diagram excluding the singularity and inner horizon.}
    \label{fig:instantonDiagram}
\end{figure}
\item Like any instanton, the solution \eqref{eqn:EOM_Solutions_instanton_4d} contains a single unstable direction that will eventually lead to the partition function having a nonzero imaginary part describing the decay of the unstable vacuum. In this case, the unstable direction describes the radius of the worldline. In particular, in the continuation of the instanton solution to Lorentzian times, shown in Figure \ref{fig:instantonDiagram}, the positron wants to fall off the top of the effective potential and either fall into the black hole or roll out to $r = \infty$.

\item The ADM mass of the static solution in the outside region is still $M$. This is because we are in the microcanonical ensemble and so energy is conserved. However, if the positron falls off the top of $V(r)$ and rolls out to $r = \infty$, it will carry some of that energy with it. The ADM energy of the remaining smaller black hole will be $\bar{M}$, given by  \eqref{eq:barM-exact-formula}. As we saw in the probe limit, in Section \ref{sec:Schwinger-pair-production}, the energy carried away by the positron is much larger than the rest mass of the positron, because the positron is hyper-relativistic after it has been accelerated by the electrostatic potential, 
\begin{gather}
   M-\bar{M} = \frac{qM}{Q} + \ldots \\
E_{\t{positron}} \approx \sqrt{\alpha} M_{\t{pl}} = \gamma m \implies \gamma = 2 \times 10^{21}\,.  \label{eq:particle_energy}
\end{gather}

\item An extremal black hole that lost charge $q$ and mass $m$ would be far from extremal, since $q \gg m$. However, as we have just seen, the black hole loses much more energy than that. Indeed, in Sec.~\ref{subsec:probemassiveparticles} we saw that in the probe limit---i.e.~at leading order in the size of the black hole---an extremal black hole that emits a positron via the process described by the instanton remains exactly extremal. With the benefit of the calculations above we can now see that in at next order in $G_N$ the emission of a positron moves the black hole slightly away from extremality, 
\begin{equation} \label{eqn:barM}
    \bar{M} - \bar{Q}|_{M = Q} = \frac{q^2 - m^2}{2r_+} + \ldots 
\end{equation}
Since this is subleading in $G_N$, this means that even after a large near-extremal black hole has emitted half its charge, it's still parametrically close to extremality. On the other hand, emitting a single positron makes the black hole far enough away from extremality that it is above the breakdown scale, since $\frac{q^2 - m^2}{2r_+} \gg E_\textrm{brk.} \sim Q^{-3}$. Thus, the evaporation of the black hole into neutral particles is well approximation by the standard QFT in curved spacetime approximation immediately after the charged particle emission. On the other hand, the energy above extremality given in \eqref{eqn:barM} is very small compared to the typical energy above extremality found in Section \ref{sec:Schwinger-pair-production} due to semiclassical fluctuations in the emission location.

\item That $\Delta M = \frac{qM}{Q} + \ldots \gg m$  also explains why the gravitational corrections to the probe calculation of the rate,  \eqref{eq:boltzmannrateprobe}, can be so modest even for black holes that are close to extremal. 

One might be concerned that the probe calculation would have limited validity because near-extremal black holes are fragile---the energy density in the electric field nearly causes the entire throat to undergo gravitational collapse. And indeed, if a near-extremal black hole were to have its charge and mass reduced by $\Delta M = m$ and  $\Delta Q = q$ then the outer horizon would leap out so far as to engulf the entire Rindler region where decay is meant to be happening, and in many cases increase the temperature so much that  \eqref{eq:boltzmannrateprobe} would be completely unjustified.  Because $\Delta M = q {M}_\textrm{Pl}$ the reduction in mass almost completely balances the reduction in charge, cancelling the leading order correction to the temperature and horizon location. This is part of the story of how the probe limit can remain reliable even when the energy above extremality is less than $M_\textrm{Pl}$.

\item There are three relevant temperatures. Let's distinguish them. There's $T_0$, the temperature of the initial black hole, given by  \eqref{eq:tempandentorpyofRNblackhole} as
\begin{equation}
     T_0 = \frac{1}{2\pi} \frac{ r_+ -  r_-}{2 r^2_+}  \ . 
\end{equation}There's $T_\textrm{inst}$, the temperature of the static solution outside the region where the instanton is located. This is the same as the temperature of the smaller black hole, redshifted by the gravitational potential of the positron at $r_\textrm{inst}$ and is given by  \eqref{eq:smoothness-inside-inst}. Finally, there's ${T}_\textrm{final}$, the temperature of the smaller black hole once the positron has been expelled to large $r$, given by 
\begin{equation}
     T_\textrm{final} = \frac{1}{2\pi} \frac{\bar r_+ - \bar r_-}{2\bar r^2_+} \ . 
\end{equation}
Because the Hawking quanta no longer need to climb out of the gravitational well of the positron, we have $T_\textrm{final} > T_\textrm{inst}$. Similarly, we can show 
\begin{equation}
    T_\textrm{inst} = T_0 \frac{r_+^2}{\bar{r}_+^2} \ .
\end{equation}
Because $\bar{r}_{+} < r_+$ this implies $T_0 > T_\textrm{inst}$. Finally, it is straightforward algebra to show that $T_\textrm{final} > T_0$. 
Altogether, we have 
\begin{equation}
    T_\textrm{final} > T_0 > T_\textrm{inst}\ . 
\end{equation}

    \item As we have seen in Section \ref{sec:Schwinger-pair-production}, as long as we start with a sufficiently cold black hole ($\frac{M-Q}{Q} \ll \frac{1}{Q^2}$ which even applies for black holes that are well above the thermodynamic breakdown scale $\Ebrk$),  the resulting temperature after emission will be largely independent of the temperature before emission,
\be 
\label{eq:barT-leading-in-T}
 T_\textrm{final}  = \frac{\sqrt{q^2-m^2}}{2\pi Q^2} \left(1 + \mathcal{O}\left(\frac{1}{Q^2}\right) \right)\,.
\ee

\end{itemize}
These are properties of the emission process described by the gravitational instanton. Just as we saw in the non-gravitational case in Sec.~\ref{sec:Schwinger-pair-production}, the \emph{typical} emission process will involve the production of positrons farther out than is predicted by the instanton, and so leave the residual black hole further from extremality. 

\subsubsection*{Contributions from sectors with nonzero angular momentum}

In the effective JT gravity description of the near-horizon region, metric and gauge modes with nonzero angular momentum are massive, with a mass that is of order the inverse AdS$_2$ radius $r_+^{-1}$. As a result, interactions mediated via such modes are exponentially suppressed on scales that are much larger than the AdS$_2$ radius. At low temperatures, the throat is parametrically long, so the effects of the charged particle on the boundary conditions at infinity are captured to very high accuracy by the S-wave sector.

On the other hand, the proper radius of the charged particle loop is $\sim \frac{m}{q} r_+$, which is much smaller than the AdS$_2$ radius when $m \ll q$. Self-interactions within the charged particle loop (including the electromagnetic attraction between the electron and the positron) are therefore described not by an effective two-dimensional description of the S-wave sector but by an approximate description using four-dimensional flat space.

For example, the leading self-interaction is mediated by the electric field and was computed for the Schwinger effect in \cite{Affleck:1981bma}. It was found to be
\begin{align} \label{eq:selfintflat}
    \Delta I_{\rm inst} = -\frac{1}{4} q^2\,.
\end{align}
 No such term appears, however, at this order in the S-wave approximation.
\subsection{Including quantum gravity fluctuations }
\label{sec:low-T-pair-production-rate}

We will now calculate the modifications to Schwinger pair production close to extremality induced by both quantum fluctuations of the particle path and quantum gravity effects at low temperatures by using the dimensional reduction to JT gravity. Surprisingly, we will find that even when such quantum corrections are included, the Schwinger pair production rate remains unchanged. The strategy that we will follow is to compute the partition function in the presence of the instanton in the canonical ensemble, where quantum corrections are easier to study. We will then express the partition function as, 
\be 
\label{eq:partition-function-with-instanton-in-terms-of-DOS}
Z_\text{inst+BH}(\beta, Q) = \int dE \rho_\text{inst+BH}(E) e^{-\beta(\Re E + i \Im E)}\,,
\ee 
where $\Re E$ is the real part of the energy above extremality, $ \Gamma(E) = \Im E$ is its imaginary part, and $\rho_\text{inst+BH}(E) $ is the density of states in the presence of the instanton, which is purely real. Thus, $\Gamma(E)$ can be seen as the decay rate for the states with real energy $E$. If we work in a regime where $\beta \Gamma(E) \ll 1$ but still close to extremality, $\beta \gg Q$, then we can expand \eqref{eq:partition-function-with-instanton-in-terms-of-DOS}, 
\be 
\label{eq:partition-function-with-instanton-in-terms-of-DOS-expanded}
Z_\text{inst+BH} (\beta, Q) = i \beta   \int dE \Gamma(E) \rho_\text{BH}(E) e^{-\beta E} = i \beta  Z_\text{BH} (\beta, Q) \Gamma\,,
\ee
where $\Gamma(\beta)$ is the average decay rate in the thermal ensemble. In this limit, $\Gamma$ can be computed using a single instanton calculation. 

However, if we want to go beyond the approximation $\beta \Gamma(E) \ll 1$ and understand charged particle emission at the very low temperatures where it typically occurs,  
we cannot assume that the density of states $\rho_\text{inst+BH}(E)$ is unchanged by the presence of the instanton (as was done in \eqref{eq:partition-function-with-instanton-in-terms-of-DOS-expanded}). Instead, we need to resum solutions containing arbitrary numbers of instantons in order to explicitly reproduce \eqref{eq:partition-function-with-instanton-in-terms-of-DOS}, including corrections to the density of states.
We will turn to the question of how to implement this resummation later in the section after analyzing the one-loop determinant of the single instanton solution.

To compute quantum gravity corrections by using the JT description, we need the instanton to live in the AdS throat region \eqref{eqn:R_inst_q_greater_m}. This requires particles with $\frac{q}{m} \gg 1$, which is indeed true for the mass and charge of the positron. We also need fluctuations of the instanton location to remain in the throat, which requires the black hole to have charge $Q \gg Q_*$. The relevant effects we will take into account are metric fluctuations, which are influenced by the backreaction of the instanton, which creates the appearance of a conical defect in the geometry. The presence of such defects changes the boundary conditions that we have to impose on the boundary modes and, as we shall see shortly, lifts the two zero-modes $\epsilon_{\pm 1}$ that are present in the Schwarzian action when backreaction is neglected.

To then obtain the full Schwinger decay rate, we must sum over any number of instantons present in the geometry. Normally, in the dilute gas approximation, the position of each instanton is a zero mode, and distinct instantons do not ``interact" with each other in any way. As we have just explained, when the instanton backreacts on spacetime, this is no longer true, and the position of each instanton is no longer a zero mode, so we need to understand how to account for this in the presence of multiple instantons.\footnote{In our work, we will solely focus on the gravitational interaction between the instantons. It would be interesting to understand how such effects combine with other interactions between the different instantons, such as those mediated by electromagnetism. } Additionally, at the level of the gravitational path integral, understanding the moduli space of instantons brings further subtleties: we wish to count instanton configurations that are not related by large diffeomorphism since the gravitational path integral includes a quotient by the group of all diffeomorphisms. Since each instanton comes with a small conical defect, summing over distinct instanton configurations reduces to the problem of summing over conical defect geometries in JT gravity.\footnote{The integral over the defect moduli space can be written as an integral over the position of each individual defect \cite{Lin:2023wac,Stanford:2022fdt}. The integral over positions acquires a small action.} More succinctly:
\be \label{eqn:InstantonModuli_DefectModuli}
\t{Moduli space of instantons } = \t{ Moduli space of conical defects.}
\ee Therefore, the volume of the defect moduli space automatically accounts for the integral over the position of each instanton on the geometry. Since the instantons interact when gravity is turned on, this does not give a simple factor of the spacetime volume as in the case of the Schwinger effect in flat space QED and leads to subleading corrections to the decay rate. 

Performing the resummation over conical defects was solved in \cite{Witten:2020wvy,Maxfield:2020ale,Turiaci:2020fjj}, and allows us to get an answer for the left-hand-side of \eqref{eqn:InstantonModuli_DefectModuli}. We will now go through the computation in increasing levels of complexity: one instanton, one winding instanton, multiple instantons, and multiple winding instantons. The final result will be to perform the resummation over any number of instantons using \eqref{eqn:InstantonModuli_DefectModuli}, and to obtain a new density of states for BHs that takes into account the Schwinger effect.

\subsubsection*{The backreacted solution has an apparent conical defect}
\begin{figure}
    \centering
    \includegraphics[width=0.6\textwidth]{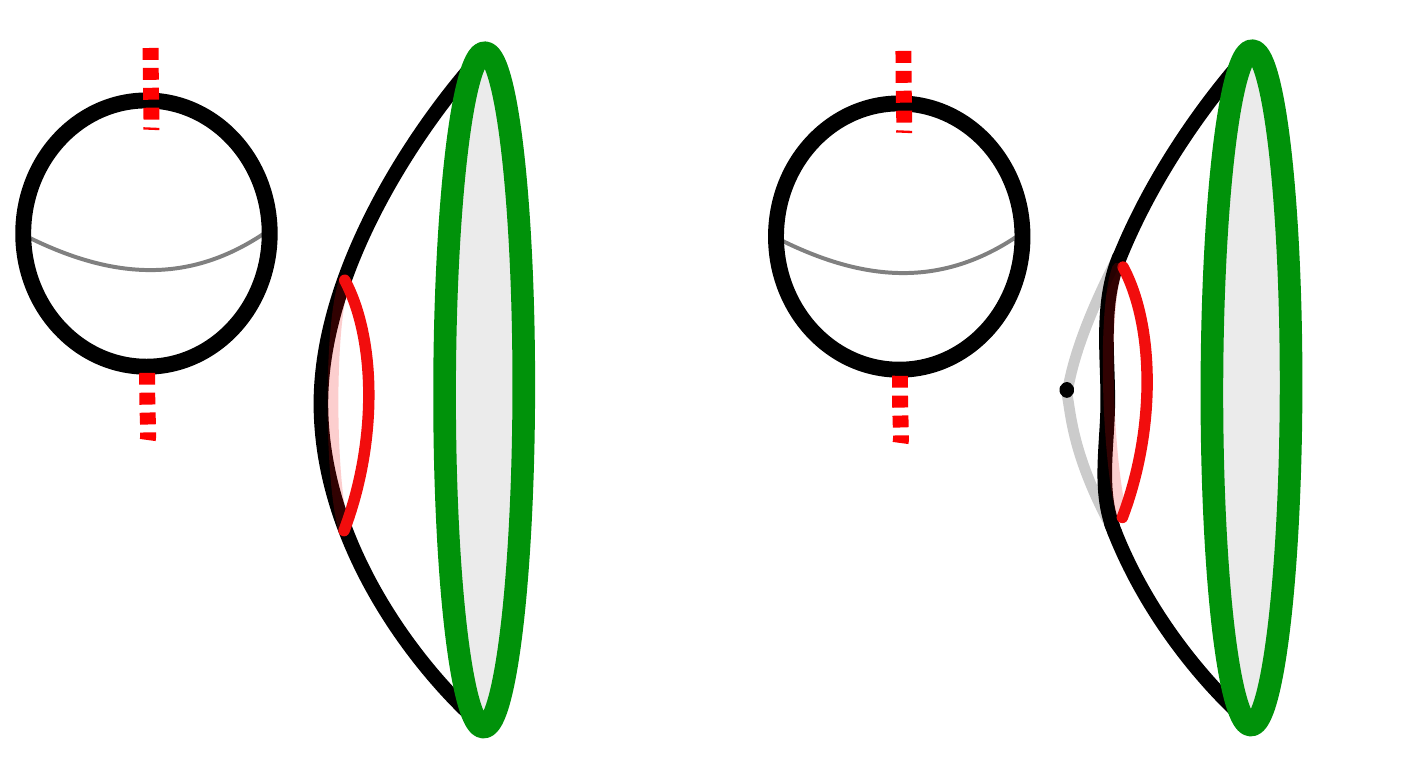}
    \caption{Instanton configuration in the near-horizon AdS$_2 \times S^2$ region of a near-extremal black hole. The effect of the backreaction is captured by the difference between the left and right figures - while both geometries are smooth, for the geometry on the right, if we hypothetically continue the spacetime past the location of the charged particle, we see that the resulting geometry has a very blunt defect whose angle is $2\pi \alpha$, with $\alpha$ given by \eqref{eq:defect-angle}. We represent the location of the defect through the dotted point. The junction conditions \eqref{eqn:Israel_Junction_conditions} are easy to understand from this figure. The first condition demands that the length of the red curve is the same both from inside and outside the worldline. The second condition implies there is a jump in extrinsic curvature as we cross the worldline, 
     which can be seen from the figure on the right.
    }
    \label{fig:Nomega}
\end{figure}
The instanton solution of Sec.~\ref{sec:including-gravitational-backreaction} is completely smooth everywhere, including at the radius $\bar{r}_+$ where the $U(1)$ Euclidean time degenerates. When imposing the periodicity $\tau \sim \tau +  \betainst $ the Euclidean solution inside the worldline is smooth. However, the black hole outside the worldline appears to have an effective conical defect at its (nonexistent) horizon $r=r_+$, as shown in Figure \ref{fig:Nomega}. 
Indeed, if the worldline didn't exist and the black hole solution with mass $M$ and charge $Q$ continued to the horizon, there would be a conical singularity with defect angle $\theta = 2\pi(1-\alpha)$ for
\be 
\label{eq:defect-angle-intro}
\alpha= \frac{  \betainst }{\beta_0} \,,\qquad \text{ where }\qquad \beta_0 = \frac{4\pi r_+^2}{r_+-r_-}\,,
\ee
because the instanton periodicity $ \betainst$ of the instanton is not equal to the BH inverse temperature $\beta_0$. Using \eqref{eq:barM-exact-formula}, one finds that
\begin{align}
\label{eq:defect-angle}
\alpha= \frac{\bar r_+^2}{r_+^2} = 1 - \frac{m^2}{q Q} + \mathcal{O}\left(\frac{1}{Q^2}, \frac{m^4}{q^4}\right) &=1 - \frac{I_\text{inst}}{\pi Q^2} + \mathcal{O}\left(\frac{1}{Q^2}, \frac{m^4}{q^4}\right) \,. 
\end{align}
where after the second equation, we have taken the limit $Q \gg q$ and $q \gg m$. In the second equality, we rewrote $\alpha$ in terms of the instanton action $I_\text{inst}$ that appears in the exponent of the final expression in \eqref{eq:instanton-expression} when taking $q \gg m$. This rewriting will be convenient for some purposes. The conical defect is therefore very blunt, with the defect angle suppressed by both the large charge of the black hole and the very light positron mass.

\subsubsection*{Metric fluctuations with a Single Instanton}

The instanton action close to extremality is given by expanding \eqref{eq:backreacted-action-with-instanton} for large $\beta_\text{inst}$,
\be 
\label{eq:1-inst-contribution}
I_\text{inst + BH} = - \pi \bar r_+^2 + \beta_\inst M  &= -\pi Q^2 + 2\pi Q \left(q - \sqrt{q^2-m^2} \right) + {\beta}_\inst Q  \nn \\ &- \frac{2\pi^2 Q^3}{\beta_\inst} \left(1 - 4\frac{q - \sqrt{q^2 - m^2}}Q \right) + \dots\,.
\ee
This can conveniently be written as,\footnote{Going forward, in our discussion of single instanton contributions, we will drop the contribution of the ground state energy $ {\beta}_\inst Q $ since that is the same with or without the instanton.}
\be 
I_\text{1 inst + BH}  = &-\pi Q^2 \alpha  + {\beta}_\inst Q  - \frac{2\pi^2 Q^3}{{\beta}_\inst} \alpha^2  + \dots\,.
\ee 
The last term can be identified as the on-shell contribution of the Schwarizan with a defect (with angle $\alpha$) \cite{Mertens:2019tcm}. In fact, in Appendix \ref{app:JT_dimensional_reduction}, we show that in the dimensional reduction on $S^2$ in the $AdS_2 \times S^2$ near-horizon region, the insertion of the instanton is equivalent to inserting an operator 
\be 
\cO_\text{inst} = \int d^2 x_\text{inst} \sqrt{g} e^{-2\pi (1-\alpha) \Phi(x_\text{inst})}\,,
\ee
in the theory of dilaton gravity arising in the dimensional reduction. Such operator insertions have been extensively studied  \cite{Mertens:2019tcm, Saad:2019lba} and are indeed equivalent to the insertion of defects whose opening angle is $2\pi(1-\alpha)$. 

The presence of the defect changes the boundary conditions for the boundary modes whose one-loop determinant caused the important quantum effects seen in Section \ref{sec:one-loop-part-function-and-the-Schwarzian} \cite{Mertens:2019tcm, Saad:2019lba}. Compared to  \eqref{eq:Schw-quadratic-fluct}, the expansion of the boundary modes to quadratic order in the presence of a defect can now be expressed as,  \be
\label{eq:Schw-quadratic-fluct-defect}
I_\text{Schw defect}^\text{quad} \sim \sum_{|n|\geq 1} \frac{T}{\Ebrk} n^2(n^2-\alpha^2) \epsilon_n \bar \epsilon_n\,,
\ee 
where the modes $\epsilon_{\pm 1}$ can be identified as generating the translation of the defect and, consequently, of the instanton, in the near-horizon geometry. As advertised, we now see that these two modes are no longer exact zero-modes once $\alpha \neq 1$, i.e.,~once backreaction is included. 

To obtain the final result, we also have to keep track of the change of measure from the 4d modes of the metric that correspond to the fluctuations of the instanton to the $\epsilon_{\pm 1}$ modes that parameterized changes in the location of the defect. This can be achieved by following \cite{Iliesiu:2022onk, Banerjee:2010qc,  Iliesiu:2022kny}, which close to extremality finds the change of measure $dg_{\mu\nu}^{\text{soft mode }\pm 1} = Q d\epsilon_{\pm 1}$ associated to each of the two modes; these two modes thus lead to a factor of $Q^2$ in the one-loop determinant. Similarly, there are two additional modes of the metric corresponding to large diffeomorphisms that induce black hole rotations that become physical once the instanton is created. Such modes thus lead to an additional factor of $Q^2$ in the one-loop determinant, which precisely corresponds to the volume of the sphere on which the charged particle is emitted. 

We can now integrate over all $\epsilon_{n}$ using the quantization of the Schwarzian theory with a defect angle $\alpha$ \cite{Mertens:2019tcm}. The partition function of the BH and instanton system can be written expressed as,\footnote{The factor of $\frac{1}{1-\alpha}$ is subtle and comes from integrating over the position of the instanton on AdS$_2$, and we explain its origin in appendix \ref{app:JT_dimensional_reduction}. }
\begin{align}
Z_{BH+1\,\inst}(\beta, Q)&=   \underbrace{\left[ i Z_{\text{1-loop}} e^{- 2\pi Q \left(q - \sqrt{q^2-m^2} \right)} \right]}_{\substack{\text{Contribution of non-interacting}\\ \text{ charged particle}}} \underbrace{\left(\frac{Q^3}{\beta}\right)^{1/2} \frac{1}{(1-\alpha)}}_{\substack{\text{Gravitational}\nn \\\text{ one-loop determinant}}} \,   \exp\underbrace{\left[{\pi Q^2 + \frac{2\pi^2 Q^3}{\beta} \alpha^2 }\right]}_{\substack{\text{Classical grav. contribution }\\ \text{including backreaction}} }\\ &\times\,  \,\underbrace{Q^\#}_{\substack{\text{Other one-loop det.'s}\\\text{also present w-out. instanton}}}\left(1+O(1/Q)\right)\,,
\label{eq:BH+1-inst-part-function}
\end{align}
where we have set $\beta = \beta_\text{inst}$ in order to compare the partition function to the false vacuum partition function with the same temperature to be able to then read off the decay rate as in \eqref{eq:partition-function-with-instanton-in-terms-of-DOS}. Above, 
$i Z_{\text{1-loop}} $ is the one-loop contribution of the particle moving in AdS$_2 \times S^2$ excluding the translation modes $\epsilon_{\pm 1}$ on AdS$_2$ already taken into account by the Schwarzian. In $Z_{\text{1-loop}}$, we will include the $Q^2$ measure associated with the two translation modes $\epsilon_{\pm 1}$, and include a factor of $Q^2$ from integrating over the position of the point particle on $S^2$. Combining the standard result for the one-loop determinant coming from the fluctuations of charged particle worldline with the change of measure for $\epsilon_{\pm 1}$ and for the two rotational modes, we find 
\be
Z_{\text{1-loop}} = \underbrace{Q^2}_{\substack{\text{Measure of the }\\\text{two lifted }SL(2, \mathbb R)\text{ modes}}} \,\,\underbrace{Q^2}_{\substack{\substack{\text{Measure of the  } \\ \text{two } SU(2) \text{ modes} \\ =\\\text{Volume of $S^2$} }}} 
\underbrace{\left( \frac{q^2}{4 \pi^3 Q^2}\right)}_{\substack{\text{Fluctuations of charged} \\ \text{particle worldline}}} =\frac{ q^2 Q^2}{4\pi^3} \,.
\ee 
The $\left(\frac{Q^3}{\beta}\right)^{1/2}$ factor in \eqref{eq:BH+1-inst-part-function} is the contribution of the graviton one-loop determinant. This is now changed from the original $\left(\frac{Q^3}{\beta}\right)^{3/2}$ seen in Section \ref{sec:one-loop-part-function-and-the-Schwarzian} in the absence of the instanton due to the fact that the modes $\epsilon_{\pm 1}$ now have to be included in the gravitational path integral.
Namely, the partition function with an instanton should be compared to the BH partition function given by
\be 
Z_{BH}(\beta, Q) = \underbrace{Q^\#}_{\text{Other one-loop det.'s}}  \underbrace{\left(\frac{Q^3}{\beta}\right)^{3/2}}_{\substack{\text{Gravitational}\\\text{ one-loop determinant}}} \,   \exp \hspace{0mm} \underbrace{\ \left[{\pi Q^2  - \beta Q + \frac{2\pi^2 Q^3}{\beta}  }\right]}_{{\text{Classical grav. contribution }} } \ . 
\label{eq:old-BH-part-function}
\ee
The $\beta$-dependence of the one-loop determinants in \eqref{eq:BH+1-inst-part-function} and {\eqref{eq:old-BH-part-function} is different because the isometries of the two geometries are different: the near-horizon of a BH has an $SL(2, \mR)$ isometry, while the BH with an instanton breaks that isometry to $U(1)$. Zero modes of the Schwarzian action associated to isometries of the spacetime geometry do not describe physical fluctuations of that geometry, and each ``missing'' mode leads to a factor of $\beta^{-1/2}$ . In particular, the modes $\epsilon_{\pm 1}$, which describe translations of the spacetime, are unphysical coordinate changes in a pure $\mathrm{AdS}_2$ geometry but become physical modes in the presence of a defect.   We thus find that the ratio of the two partition functions is given by 
\be 
\label{eq:ratio-of-two-partition-function}
\frac{Z_{BH+1\,\inst}(\beta, Q) }{Z_{BH} (\beta, Q) } =\frac{ \beta  E_\text{brk}}{ 1-\alpha} \left[ i \underbrace{Z_{\text{1-loop}} e^{- 2\pi Q \left(q - \sqrt{q^2-m^2} \right)}}_{\equiv e^{-I_{\inst \text{ eff.}}}} \right] \exp\left[{\frac{2\pi^2 Q^3}{\beta}  } (\alpha^2-1)\right]
\ee
which gives us a first estimate for the decay rate in a regime where $\frac{Z_{BH+1\,\inst}}{Z_{BH} } \ll 1$, i.e.~when $\beta/Q^3 \ll e^{I_{\inst \text{ eff.}}}$, 
\be \label{eq:gammafirstestimate}
\Gamma(\beta) = \frac{1}\beta \Im \, \log \, \left(1+\frac{Z_{BH+1\,\inst}}{Z_{BH} }\right) \sim_{\frac{Z_{BH+1\,\inst}}{Z_{BH} } \ll 1} \frac{q^3}{2\pi^3 m^2} e^{-\frac{\pi Q m^2}{q} } \left(1+ O\left(\frac{1}{Q}, \frac{m}{q}\right)\right)\,.
\ee
This is exactly the answer obtained in Section \ref{subsec:probepositronemissionrate}.\footnote{The overall $O(1)$ multiplicative factor in \eqref{eq:gammafirstestimate} is ambiguous due to the freedom to define the measure for the modes $\varepsilon_{\pm 1}$ by an arbitrary constant factor. In the above we have chosen that measure to match the standard semiclassical computation in the high-temperature limit (and, consequently, also at all other temperatures).} Even though we have included the same quantum gravity fluctuations that proved so important for the neutral radiation, we now see that for charged pair production, such corrections have no effect at leading order. The decay rate is largely $\beta$-independent, and the $Q$-dependence coming from the integration measure of the gravitational path integral together with the factors of $\Ebrk$ precisely reproduces the standard Schwinger pair-production answer. 
To fully complete the calculation for a single instanton, one also has to consider configurations where the instanton winds $k$ times around its circular trajectory. We perform this analysis in appendix \ref{sec:winding-instanton}, where we find the same answer as when including winding instantons in the standard Schwinger pair-production calculation.

\subsubsection*{Multi-instanton contribution}
\label{sec:multi-instanton-cont}

To fully understand the pair-production rate, we will thus have to re-sum all multi-instanton configurations, depicted in Figure \ref{fig:multiInst}, directly at the level of the gravitational path integral, once again taking the backreaction into account.  To perform the re-summation over any number of instantons, we need to account for the volume of moduli space of hyperbolic geometries with any number of conical defects. In particular, when summing up all possible conical defects, the partition function is given by
\be\label{eq:zsumoverdefects}
Z_{\t{sum over defects}}(\beta) = \sum_{k=0}^\infty \frac{\lambda^k}{k!} Z_{k-\t{defects}}(\beta)\,.
\ee
In the above $Z_{k-\t{defects}}$ is the gravitational path integral on the disk with $k$ conical defects, which is proportional to the volume of the moduli space for a disk with $k$ defects. Each defect has a deficit angle $\theta=2\pi(1-\alpha)$, and comes with a weight $\lambda$ that we will write in terms of the instanton parameters shortly. The factor of $k!$ is included to take into account the fact that the defects are indistinguishable. In \cite{Witten:2020wvy,Maxfield:2020ale,Turiaci:2020fjj} it was explained how to perform the above resummation. We will present the final result from \cite{Turiaci:2020fjj}, which performed the resummation for arbitrary defect angles. 

The partition function \eqref{eq:zsumoverdefects} is easiest to evaluate using the technology of the string equation $\cF(u)$, see \cite{Johnson:2022wsr,Ginsparg:1993is} for recent and classic reviews. 
One first specifies $k$ species of defects with opening angles $\alpha_i$ and weights $\lambda_i$ with $i=1, \, \dots,\, k$. The associated string equation $\cF(u)$ is then related to the density of states associated with the partition function that we are looking for through
\be 
\label{eq:density-of-states-from-F}
\rho(E) = \frac{e^{\pi Q^2}}{2\pi} \int_{2 E_0 \equiv u(0)}^{2 E}  \frac{du}{\sqrt{2 E -u}} \frac{\partial \cF}{\partial u}\,.
\ee
where the edge of the spectrum $E_0$ needs to first be determined by solving for $\cF(2 E_0) = 0$.\footnote{Restoring units, the correct equation is $\cF( \frac{2 E_0}{\Ebrk}) = 0$. The conventions used in this paper differ from the commonly used convention where $\Ebrk=2$, which would give the equation for the edge of the spectrum $\cF(E_0)=0$.} The partition function, when we include the summation over any number of defects, is given by
\be 
Z_\text{sum over defects}(\beta) = \int_{E_0}^\infty dE \rho(E) e^{-\beta E}\,. 
\ee

\begin{figure}[t!]
    \centering
    \includegraphics[width=0.3\textwidth]{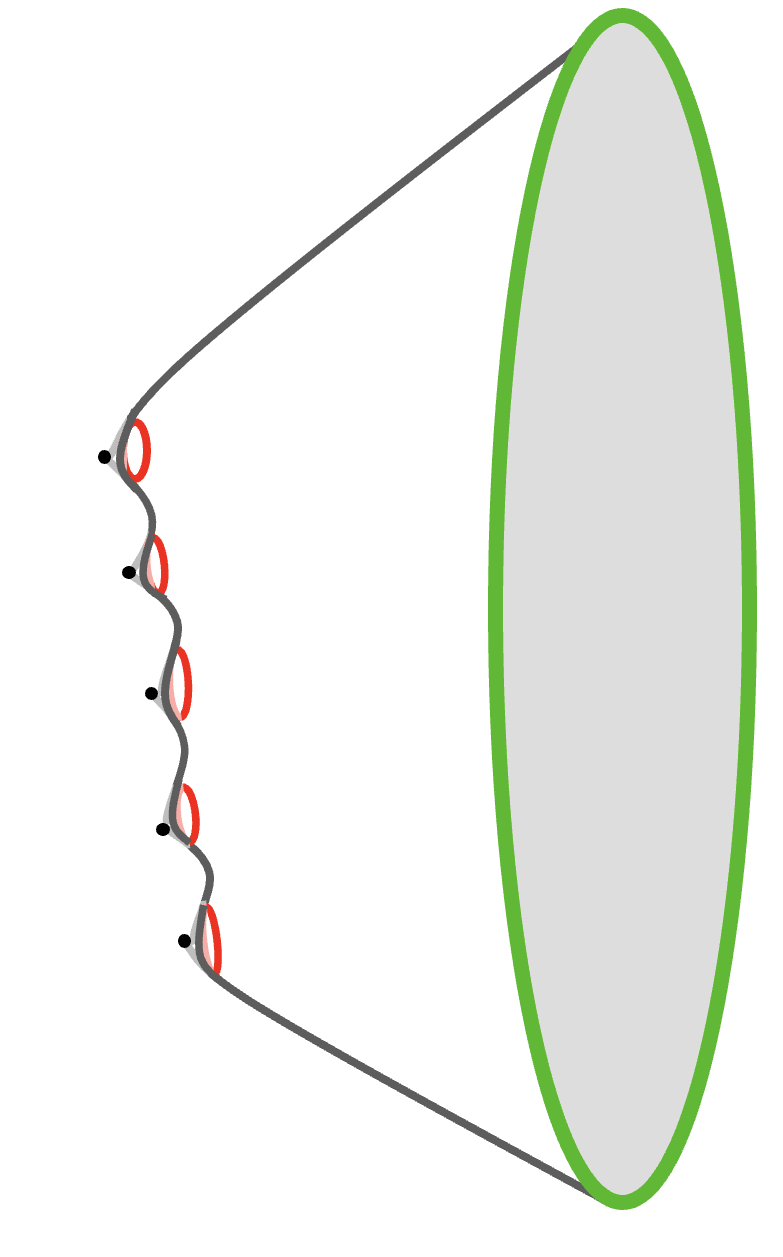}
    \caption{Cartoon of the multi-instanton configuration in the presence of backreaction in the AdS$_2$ near-horizon region for a near-extremal black hole. Each instanton has an associated defect that is seen when continuing the spacetime outside of each instanton past the instanton location. Because of the backreaction, all multi-instanton configurations are off-shell, and their re-summation is captured by the moduli space of defects in AdS$_2$.}
    \label{fig:multiInst}
\end{figure} 

\subsubsection*{String equation for instantons}
We will now demonstrate how this works with a single defect species, which is the case when ignoring winding instantons. We will assign each defect a weight given by\footnote{The prefactor $(1-\alpha)^{-1}$ comes with each defect since we are independently integrated over the position of each defect.}
\be 
\lambda \equiv i e^{-I_\text{inst eff}} = i \frac{Z_\text{1-loop}}{1-\alpha} \,e^{\frac{\pi}{16Q^2} \left[\left(\sqrt{(2Q-q)^2-m^2} +\sqrt{q^2-m^2}\right) - 16 Q^4\right] }\,,
\ee
which is the action of each instanton with deficit angle given by $\alpha$ in \eqref{eq:defect-angle} with the BH entropy subtracted off. This is the dominant instantons with winding $k=1$. In addition to this, there are also suppressed instantons that wind $k$-times whose weight will be given by $e^{-I_\text{inst eff.}^{k}} $ from \eqref{eq:inst-eff-k} with associated angle $\alpha_k$ from \eqref{eq:alpha-k}. Note that all the defect actions, as well as the defect angles, are independent from the number of defects that we put on the black hole background. 

To start, we will neglect the contribution of winding instantons. In such a case $\cF(u)$ was found to be given by a truncated sum \cite{Turiaci:2020fjj}
\be
\label{eq:F(u)-exact}
\cF(u) = \sum_{L=0}^{\left\lfloor \frac{1}{1-\alpha}\right\rfloor} \frac{i^L e^{-L I_\text{inst eff}}}{L!} \left(\frac{2\pi(1-L(1-\alpha))}{\sqrt{u}}\right)^{L-1} I_{L-1}\left(2\pi(1-L(1-\alpha)) \sqrt{u}\right) 
\ee
where $I_{L-1}(x)$ are Bessel functions. This sum can be conveniently rewritten in terms of a generating function:
\be 
\label{eq:F(u)-generating-function}
\mathcal F(u) = \int_\mC \frac{dy}{2\pi i} e^{2\pi y} \left(y-\sqrt{y^2 - u - 2 i e^{-I_\text{inst eff}} e^{-2\pi(1-\alpha)y} }\right)\,,
\ee
where the contour runs along the imaginary axis to the right of all poles in the complex $y$ plane. A similarly useful formula is that of the generating function for the density of states, which is given by 
\be 
\label{eq:density-of-states-generating-function}
\rho(E) = \frac{e^{\pi Q^2}}{2\pi} \int_\mC \frac{dy}{2\pi i} e^{2\pi y} \tanh^{-1}\left(\sqrt{\frac{2 \left(E-E_0\right)}{y^2-2  i e^{-I_\text{inst eff}} e^{-2\pi(1-\alpha)y} - 2 E_0 }}\right)\,.
\ee
The generating functions \eqref{eq:F(u)-generating-function} and \eqref{eq:density-of-states-generating-function} can also be generalized to include any number of instantons with different weights and deficit angles. For example, the $k$-times winding instantons give:
\begin{align}
\label{eq:generating-function-including-k-times-winding}
\mathcal F(u) &= \int_\mC \frac{dy}{2\pi i} e^{2\pi y} \left(y-\sqrt{y^2 - u - 2 i \sum_{k=1}^\infty e^{-I_\text{inst eff}^k} e^{-2\pi(1-\alpha_k)y} }\right)\,,\nn \\ 
\rho(E) &= \frac{e^{\pi Q^2}}{2\pi} \int_\mC \frac{dy}{2\pi i} e^{2\pi y} \tanh^{-1}\left(\sqrt{\frac{2 \left( E-E_0\right) }{y^2- 2 E_0 -2  i \sum_{k=1}^\infty e^{-I_\text{inst eff}^k} e^{-2\pi(1-\alpha_k)y} }}\right)\,.
\end{align}

\subsubsection*{A sanity check: reproducing the pair-production rate without backreaction}

As a sanity check, we can first see what happens in the complete absence of backreaction in which case we are neglecting the presence of defects, which implies we are taking $\alpha =1$ and keeping $e^{-I_\text{inst eff}
}$ as in \eqref{eq:ratio-of-two-partition-function}. The string equation for the single-wound instanton becomes
\be
\cF(u) = \sum_{L=0}^{\infty} \frac{i^L e^{-L I_\text{inst eff}}}{L!} \left(\frac{2\pi}{\sqrt{u}}\right)^{L-1} I_{L-1}\left(2\pi \sqrt{u}\right)= \frac{\sqrt{u+2i e^{-I_\text{inst eff}}}}{2\pi} I_{1}\left(2\pi \sqrt{u+2i e^{-I_\text{inst eff}}} \right)\,.
\ee
where one identifies the second expression as the infinite Taylor serier in $i e^{-I_\text{inst eff}}$ of the third expression. The edge of the spectrum in such a case is given by 
$ E_0 = -i e^{-I_\text{inst eff}} \Ebrk$, the density of states is found to be 
\be
\rho_\text{no backreaction}(E) \sim e^{\pi Q^2}\sinh\left(2\pi \sqrt{2\left( E-E_0\right)} \right).
\ee
From the above, it appears that the density of states is imaginary. Instead, we must pick a contour in the complex $E$ plane for the density of states to be real, with the resulting BH energies having a small imaginary part. This implies that our black hole states are instead metastable with a decay rate set by the imaginary part of the energy. For the above case, the correct contour is to choose $\mathcal E = E - E_0 \geq 0$ to be real and positive such that the density of states is real and positive. The BH states thus have energy and associated decay rate
\be
E = \mathcal E  -i \Gamma_\text{no backreaction}\,, \qquad \Gamma_\text{no backreaction} =Z_{\text{1-loop}} e^{- 2\pi Q \left(q - \sqrt{q^2-m^2} \right)} \Ebrk\,.
\ee
This is precisely the result in the literature that neglects backreaction but sums over the contribution of an arbitrary number of non-interacting instantons. Note that $\bar E$ is only the energy above extremality. The total energy of the BH is $E_\t{total} = Q + \bar E \gg \Gamma_\text{no backreaction}$, and so these states are metastable. 

In the case of no backreaction, we can also exactly compute the partition function using the density of states, finding
\be 
Z_\text{no backreaction} = Z_{BH} e^{-\beta E_0}  = Z_{BH} e^{\beta i e^{-I_\text{inst eff}}}\,,
\ee
from which we can again read off the complex part of the energies as the imaginary part in the exponential. 

\subsubsection*{The pair-production including back-reaction}

Coming back to \eqref{eq:F(u)-exact}, we can now determine the leading order correction coming from considering $\alpha \neq 1$. Since we are interested in the effect at small temperatures, one first observable we should understand is whether the presence of defects will affect the spectral edge $E_0$. This amounts to solving $\mathcal F(2 E_0) = 0$ order by order in $1-\alpha$.

First, let's neglect the winding solutions. To second order in $1-\alpha$ we find:
\be
\label{eq:E0-shift}
\frac{E_0}{\Ebrk} = - i e^{-I_\text{inst eff}} + 2\pi^2 e^{-2 I_\text{inst eff}} (1-\alpha)^2 + 8 i \pi^4 e^{-3 I_\text{inst eff}} (1-\alpha)^4 + \mathcal{O}\left(e^{-4 I_\text{inst eff}}(1-\alpha)^6\right)\,,
\ee
where each order in $(1-\alpha)$ is exact in terms of $e^{-I_\text{inst eff}}$. Therefore, this should not only be viewed as an expansion in terms of $(1-\alpha)$ but also in  $e^{-I_\text{inst eff}}$. Including the winding solutions studied in appendix \ref{sec:winding-instanton}, we have 
\be \label{eqn:E0_shifted}
\frac{E_0}{\Ebrk} &= - i \left(\sum_{k=1}^\infty e^{-I_\text{inst eff}^k} \right)+ 2\pi^2 (1-\alpha)^2 \left(\sum_{k=1}^\infty k e^{-I_\text{inst eff}^k} \right)^2 \nn \\ &+8i \pi^4 (1-\alpha)^4\left(\sum_{k=1}^\infty k e^{-I_\text{inst eff}^k} \right)^2 \left(\sum_{k=1}^\infty k^2 e^{-I_\text{inst eff}^k} \right) + \mathcal{O}\left(e^{-4 I_\text{inst eff}}(1-\alpha)^6\right)\,,
\ee
The above expansion should really be understood only as an expansion up to order $O\left(e^{-4 I_\text{inst eff}}\right)$, so not all the exponential terms in the parentheses should be kept. Note that $e^{-I_\text{inst eff.}^k}$ contributes at the same order as $e^{-k I_\text{inst eff.}}$.

We should also note that the partition function at low temperatures is given by the density of states around the edge of the spectrum. However, in the presence of instantons the spectrum around this edge can also change by a multiplicative factor. We shall once again start by neglecting multiple windings. In such a case, expanding \eqref{eq:density-of-states-generating-function} around $E \to E_0$, one finds 
\begin{align} \label{eqn:densityofstates_nearextremal_shifted}
\rho(E) = \sqrt{2 (E-E_0)} \frac{e^{\pi Q^2}}{2\pi} \underbrace{\int_\mC \frac{dy}{2\pi i} e^{2\pi y} \left(\frac{1}{\sqrt{y^2-2  i e^{-I_\text{inst eff}} e^{-2\pi(1-\alpha)y} - 2 E_0 }}\right)}_{\text{Energy independent const } \equiv\,\, \mC} \,,    
\end{align}
where we have defined the energy independent constant $\mC$ which can also be determined in a $1-\alpha$ expansion:
\be
\label{eq:1-loop-further-corr}
\mC = 1 - 4 i \pi^2 e^{-I_\text{inst eff}} (1-\alpha) + 2\pi^2 \left(i e^{-I_\text{inst eff}} - 4 \pi^2 e^{-2I_\text{inst eff}} \right) (1-\alpha)^2 + \mathcal{O}\left(e^{-2 I_\text{inst eff}} (1-\alpha)^3\right)\,.
\ee
Again, including the winding solutions as in appendix \ref{sec:winding-instanton}, we have 
\be \label{eqn:mC_winding}
\mC &= 1-4\pi^2 i  (1-\alpha)\left(\sum_{k=1}^\infty k\, e^{-I_\text{inst eff}^k} \right) + 2\pi^2 (1-\alpha)^2 \left[\left(\sum_{k=1}^\infty i k^2 \, e^{-I_\text{inst eff}^k} \right) - 4\pi^2 \left(\sum_{k=1}^\infty k \, e^{-I_\text{inst eff}^k}  \right)^2 \right]\nn  \\ &+8\pi^4 (1-\alpha)^3\left[3\left(\sum_{k=1}^\infty (-i k) \, e^{-I_\text{inst eff}^k} \right) \left(\sum_{k=1}^\infty (i k^2) \, e^{-I_\text{inst eff}^k} \right) + \frac{4\pi^2}{3} \left(\sum_{k=1}^\infty (-ik) \, e^{-I_\text{inst eff}^k}  \right)^3\right] + \nn \\ 
&+ \mathcal{O}\left(e^{-2I_\text{inst eff}} (1-\alpha)^4\right)\,.
\ee
The above density of states yields a low-temperature partition function given by
\be  
Z_\text{all instantons}(\beta) = \mC \left( \frac{Q^3}{\beta}\right)^{3/2} e^{\pi Q^2 - \beta E_0}\,,
\ee
which thus gives a ratio of partition functions: 
\be  
\frac{Z_\text{all instantons}(\beta)}{Z_\text{BH}(\beta)} = {\mC}  e^{- \beta E_0}\,.
\ee
Finally, the fully back-reacted decay rate close to extremality is given by 
\be 
\Gamma = \frac{1}{\beta} \Im \log\left(\frac{Z_\text{all instantons}(\beta)}{Z_\text{BH}(\beta)}\right) = -\Im E_0 + \Im \frac{1}{\beta} \log {\mC} =_{\beta \to \infty } -\Im E_0 \,.
\ee
Note that at low temperatures, the last term can be dropped, and we are left with the shift in the spectral edge as the only effect from the instantons. The rate can be easily computed from \eqref{eq:E0-shift}, \eqref{eq:1-loop-further-corr} and \eqref{eqn:mC_winding} giving
\begin{align}
\frac{\Gamma}{\Ebrk} =  &\sum_{k=1}^\infty e^{-I_\text{inst eff}^k} - 8 \pi^4 (1-\alpha)^4\left(\sum_{k=1}^\infty k e^{-I_\text{inst eff}^k} \right)^2 \left(\sum_{k=1}^\infty k^2 e^{-I_\text{inst eff}^k} \right)  \nn \\
& - \frac{4\pi^2}{\beta \Ebrk} (1-\alpha) \left(\sum_{k=1}^\infty k e^{-I_\text{inst eff}^k} \right) + \mathcal O\left(e^{-5 I_\text{inst eff}^k} \right) \,,
\end{align}
where we have included the first temperature correction for completeness. Below, we will use the explicit values of $I_\text{inst eff}^k$ and $1-\alpha$ to obtain the corrected pair-production rate for positrons in our own universe.

\subsection{Final results for the decay rate for a charged particle }
For large black holes, the instanton action is large, and the decay rate is exponentially suppressed and dominated by the $k=1$ term, with winding instantons being unimportant. Expanding the one-loop determinant and the defect angle \eqref{eq:defect-angle} at large $Q$ and $q \gg m$, we find that for charged fermions at very low temperatures, the pair-production rate is given by
\begin{align}  \label{eqn:Gamma_Schwinger_final_result}
\Gamma &= \sum_{k=1}^\infty \frac{1 }{k^3} { \frac{q^3}{2\pi^3 m^2}} e^{-2\pi k Q (q-\sqrt{q^2-m^2})} \nn \\ 
&- \frac{m^2 q^5 Q^2}{\pi^5} \left(\sum_{k=1}^\infty \frac{1}{k^2}e^{-2\pi k Q (q-\sqrt{q^2-m^2})} \right)^2 \left(\sum_{k=1}^\infty \frac{1}{k} e^{-2\pi k Q (q-\sqrt{q^2-m^2})} \right)  \nonumber \\ 
&+\mathcal O\left(e^{-10\pi k Q (q-\sqrt{q^2-m^2})} \right)   \, . 
\end{align}

The first term is precisely the expectation from the Schwinger result \eqref{eqn:Schwinger_rate_final} when including winding instantons for a charged fermion.\footnote{The above answer is not precisely correct when the pair-production rate is exponentially suppressed. There are subleading terms in the $(1-\alpha)$ expansion that are more important than highly wound instanton contributions, with the net effect that each exponential term effectively has its own $Q^{-1}$ expansion. The expression is written to contrast to the standard Schwinger rate.} The second line indeed provides a non-trivial correction coming from the fact that the instantons are weakly interacting -- nevertheless, this correction is highly suppressed in $e^{-I_\text{inst eff}}$. However, this correction becomes significant as the charge of the black hole approaches $Q_*$ and the instanton action 
becomes small. To understand how much such corrections are suppressed, it is useful to rewrite the terms coming from one-loop determinants in the second line of \eqref{eqn:Gamma_Schwinger_final_result} as $\frac{m^2 q^5 Q^2}{\pi^5} = \frac{q^3}{2\pi^3 m^2} \cdot \frac{2 I_\text{inst}^2 q^4}{\pi^4} $; in this rewriting, the first term is the one-loop determinant around the single instanton saddles while the second term corresponds to the additional suppression that appears from the one-loop determinant of the multi-instanton configuration. When $I_\text{inst} \sim O(1)$, the correction to the dilute gas approximation can thus become important since it is solely suppressed by $q^4$ compared to the leading order answer. Similarly, the kind of non-perturbative corrections seen in the second line of \eqref{eqn:Gamma_Schwinger_final_result} also lead to a non-perturbative shift of the extremal energy. Such a shift is analyzed in Appendix \ref{sec:correction-to-extremal-energy} and should be added to the already known perturbative corrections to the extremal energy. As in the second line of \eqref{eqn:Gamma_Schwinger_final_result}, such corrections once again seemingly become important when $I_\text{inst} \sim O(1)$. We plan to explore the nature of both such corrections further in future work.

To summarize, when the black hole holds a sufficiently large charge ($Q \gg Q_*$), we see that neither quantum effects nor gravitational backreaction plays a significant role in the pair-production rate. The final decay rate for a large BH is thus
\be
\label{eq:pair-production-rate-final}
\Gamma =  \frac{q^3 }{2\pi^3 m^2}  e^{-2\pi Q (q-\sqrt{q^2-m^2})} \,,
\ee
which is identical to the standard Schwinger result \eqref{eqn:Schwinger_rate_final} except for the exponent.

\subsection{Why does the semiclassical emission rate survive largely unchanged?}\label{sec:whysemiclassical}

We started this section with the 
intuition that at sufficiently low initial temperatures, the charged particle emission rate should be enhanced by a polynomial factor relative to semiclassical expectations. Instead, however, we found only exponentially small corrections to that rate. So why did the intuition from thermodynamics give an incorrect answer?

To understand why, it is helpful to first review the process by which a charged particle is emitted from the horizon of the black hole. This can be naturally decomposed into two steps, as shown in Figure \ref{fig:mitosis}. In the first step, the positron quantum tunnels out of the horizon to an unstable equilibrium just outside. This step is entropically suppressed, with an exponent equal to the difference in entropy between the pre-emission black hole and the unstable equilibrium. In the second step, it is repelled from the black hole to infinity. This second step is essentially classical in nature (or at least is allowed classically) and, in general, will be thermodynamically irreversible. We, therefore, have an inequality where the entropy of the post-emission black hole is at least as big but, in general, will not be equal to the entropy of the unstable equilibrium.

A good example of the second step increasing the entropy of the post-emission black hole occurs when the Schwinger pair production takes place outside the horizon. (As we showed in Section \ref{sec:background-charged}, this is almost always the case in practice.) After the electron-positron pair is produced, the positron will be repelled to infinity while the electron will fall into the black hole. The electron will increase the thermodynamic entropy of the black hole, making the process fundamentally irreversible. (The reverse process would involve throwing a positron at a black hole and seeing an electron spontaneously pop out of the horizon to annihilate with it. This is very unlikely to occur) As a result, even semiclassically, the Schwinger emission rate for electron-positron pairs far from the horizon cannot be related to the entropy difference between the pre- and post-emission black holes.

\begin{figure}[t] 
   \centering
   \includegraphics[width=6in]{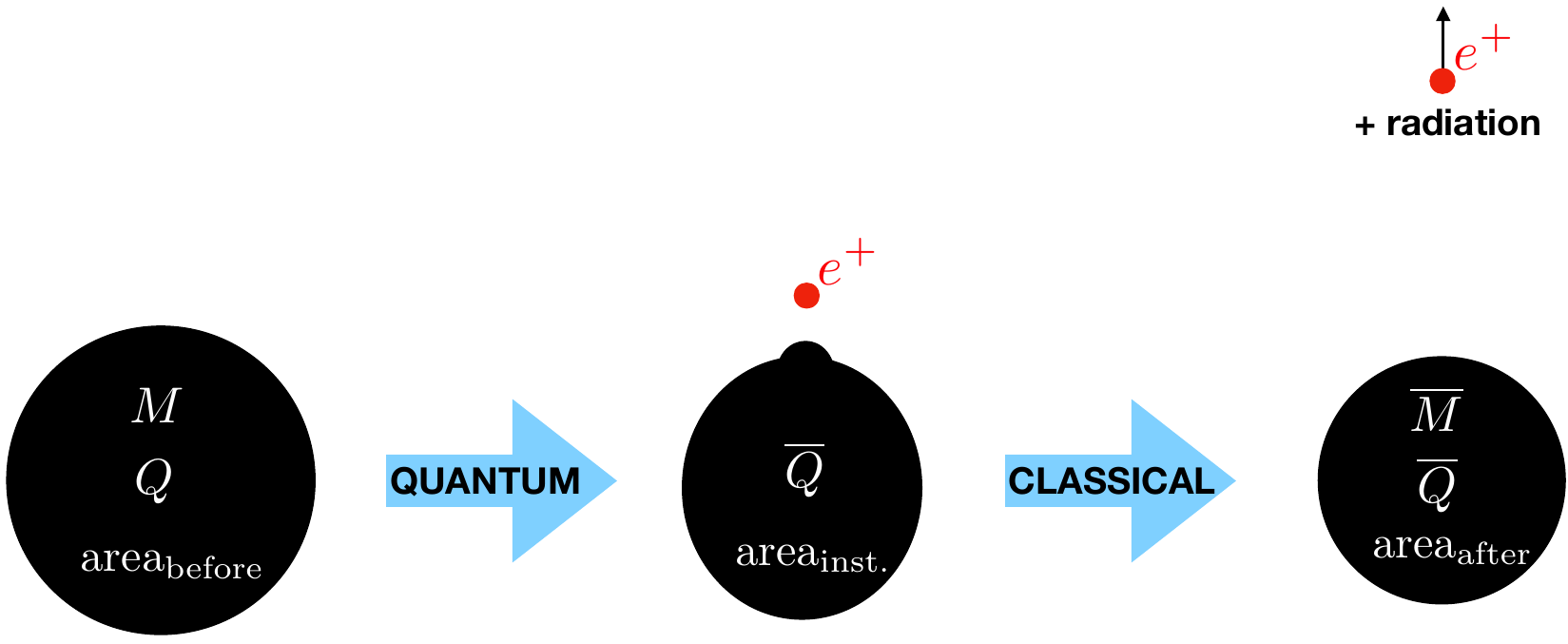} 
   \caption{\textit{Cartoon of charged particle emission from a black hole}. The black hole first quantum tunnels to a static but unstable (and non-symmetric) configuration where the positron is at the free-float point. From here, the dynamics are essentially classical as the positron is expelled with a large boost, and the black hole equilibrates to its (spherically symmetric) final state. The rate of the first step in the process is controlled by entropic suppression. The second step will, in general, be thermodynamically irreversible and will increase the entropy of the black hole.} 
   \label{fig:mitosis}
\end{figure}
Now, let's consider the case of a very low-temperature initial black hole state, with the Schwinger pair potentially being produced either near the top or the bottom of the throat. Crucially, because Schwinger pair production is a local process, it does not significantly affect the Schwarzian fluctuations of the black hole, which are locally pure gauge. This is, of course, a good reason to think that those fluctuations should not affect the local physics governing the Schwinger rate. But it also means that the unstable equilibrium solution, described by the gravitational instanton, that arises immediately after the pair is produced still has the same long throat with the same Schwarzian fluctuations as the pre-emission black hole. As a result, the entropy corrections from those fluctuations should contribute the same way to both the entropy of the pre-emission black hole and the entropy of the instanton configuration. As explained above, that is the relevant entropy difference that controls the pair production rate.

During the second step, the long throat and the fluctuations will necessarily be destroyed. Depending on where the electron-positron pair is produced in the throat, it will either be destroyed by the electron falling down or the positron flying up (or by a combination of the two). However, this destruction only affects how thermodynamically irreversible the emission process is. It does not affect the rate of charged particle emission, which is entirely controlled by the first step. We conclude that Schwarzian fluctuations make charged particle emission from a very low-temperature black hole more thermodynamically irreversible than we would expect based on semiclassical physics but do not affect its rate.

\section{The story of a charged black hole}
\label{sec:the-story-of-a-charged-black-hole}

We will now discuss the decay process of a charged black hole.\footnote{Due to magnetic fields in our universe, astrophysical black holes can carry much more charge than would be expected from balancing the electric force against gravity $F_e/F_g = 1 \implies Q/M \sim 10^{-22}$ \cite{Wald:1974np}. An estimate for astrophysical black holes is that the charge-to-mass can get as large as $\frac{Q}{M} \sim 10^{-7}$ \cite{Levin:2018mzg}. The supermassive BH in the center of our galaxy is estimated to have $\frac{Q}{M} \sim 10^{-12}$ \cite{Zajacek:2019kla}. These are very far from extremal, and our analysis is for an idealized charged black hole that does not neutralize its charge through the interstellar medium.} If the black hole starts with too small of a charge, the Schwinger process stops being exponentially suppressed well before the black hole approaches extremality, and semiclassical physics doesn't break down until the black hole becomes Planckian in size. Instead, we will be interested in the case where we start with a black hole with a large enough charge that it approaches extremality with positron emission still exponentially suppressed. As we have seen, this occurs when
$
 Q \gsim Q_* \equiv 1.8 \times 10^{44} q . 
$ 
This will be our starting point.

\paragraph{Bosonic BH Hawking radiation.} 

Let's assume that our black hole starts with integer spin and is well above extremality. It will first continuously emit neutral Hawking radiation, reducing its mass and angular momentum without losing any of its charge. In our own Universe, from \eqref{eqn:semiclassicalflux_photon} about $75\%$ of this Hawking radiation will be in the form of photons and $25\%$ in gravitons, at sufficiently high energies above extremality. 
The black hole will quickly lose most of its angular momentum until $j \ll M^2/M_\text{Pl}^2$. It will then slowly lose mass without losing charge. The mass loss due to the photons and gravitons is always polynomial in the energy above extremality, with the exact suppression dependent on the mass of the black hole above extremality. For example, we found from equation \eqref{eqn:dEdt_photon_final}
\begin{align}
\frac{d M}{d t} \sim 
\begin{cases}
-(M-Q)^5, \qquad \qquad \qquad &M-Q \gg \Ebrk\,, \\
-(M-Q)^{\frac{19}{2}} , \qquad &M-Q \ll \Ebrk.   
\end{cases}
\label{eq:mass-loss-rate}
\end{align}
The first line is valid sufficiently far above extremality that the quantum field theory in curved spacetime approximation is justified. As the black hole cools down to $(M-Q) \sim \Ebrk$, this approximation breaks down, and quantum gravity effects become important. In this regime, even the quantized spin of the black hole only takes small values since only states with $j(j+1) < \frac{2}{\Ebrk} (M-Q)$ exist.\footnote{For example, if the mass of the black hole satisfies $ \Ebrk < M-Q < 2\Ebrk$, the black hole can only have $j=1$ and $j=0$, and quantum gravity corrections will largely control the transition between any two such states due to photon emission.} After some time, photon emission will make $M-Q < \Ebrk$, at which point the black hole can only have $j=0$. At this point, single-photon emission becomes impossible!\footnote{This is because emitting a single photon forces the mass of the black hole to go down and its spin to increase by $1$.} The only way for the black hole to further shed mass is through the emission of entangled pairs of photons in angular momentum singlet states. The mass loss rate due to such emission gives the second line in \eqref{eq:mass-loss-rate}.

Ignoring positron emission, for now, the mass as a function of time can be found to go as
\begin{align} \label{eqn:M(t)_bosonicBH}
M(t) - Q \sim \begin{cases}
t^{-1/4},\qquad \qquad \qquad &M-Q \gg \Ebrk\,, \\
t^{-2/17}, \qquad &M-Q \ll \Ebrk\,,
\end{cases}
\end{align}
in the semiclassical and highly quantum regimes, respectively. Without pair production, the BH would approach extremality as $t \to \infty$. Note again that the quantum corrected decay is much slower than the QFT-in-curved-spacetime answer.

\paragraph{Fermionic BH Hawking radiation.} Black holes that start with half-integer angular momentum (fermionic states) will emit photons until they have the minimum spin $j=\frac{\hbar}{2}$. These states begin at an energy $M=Q+\frac{3}{8}\Ebrk$, and so fermionic BHs must have energies larger than this. Transitions to $j=0$ states are only allowed through fermion emission (such as positrons or neutrinos), which will take a very long time. In the meantime, they will continuously emit photons since the emission of spin one particles keeps the black hole spin to be half-integer. We found the flux into photons for such BHs to be \eqref{eqn:dEdt_fermionic_BH}
\be
\frac{d M}{d t} \sim - \lrm{M-Q-\frac{3}{8}\Ebrk}^{\frac{11}{2}}\,, \qquad  M-Q - \frac{3}{8}\Ebrk \ll \Ebrk\,,
\ee
very close to extremality in the regime where quantum gravity corrections are most important. The mass of the black hole decays as
\be \label{eqn:M(t)_fermionicBH}
M(t) - M_0 \sim t^{-\frac{2}{9}}\,,
\ee
and as $t\to \infty$ approaches the shifted extremality bound $M_0=Q+\frac{3}{8}\Ebrk$. Note that this cooling rate is somewhat slower than would be expected from a QFT-in-curved-spacetime calculation but much faster than the cooling for bosonic black holes because the photons do not need to be emitted in pairs.

\paragraph{Positron emission.} The positron emission is highly suppressed and is very well approximated by the semiclassical rate
\begin{align}
\Gamma \sim 
\frac{q^3 Q^3}{m^2 r_+^3} \exp \lrm{-\frac{\pi m^2 Q}{q}},
\end{align}
even far below the breakdown scale. The black hole will, therefore, spend most of its lifetime in the regime where $M-Q \ll \Ebrk$. The closest to extremality the BH will get can be approximated by taking the BH to emit uncharged Hawking radiation for a time $\Delta t \sim \Gamma^{-1}$ until a positron is likely to be emitted. For a bosonic BH that can only emit entangled photons, the minimum mass reached follows from \eqref{eqn:M(t)_bosonicBH} and is
\be
M_{\t{min}}^{\t{bosonic}} - Q = O \left(\exp \left(-\frac{2}{17}\frac{\pi m^2 Q}{q} \right)\right) \,,
\ee
where the numerical prefactor depends randomly on the exact time $t$ that the positron happens to be emitted. 

The BH thus gets exponentially close to extremality, well below the breakdown scale $\Ebrk \sim Q^{-3}$. The gravitational calculation remains under control since non-perturbative effects are expected to become important at energy scales $e^{-O( Q^2)}$, whereas here, we only get within $e^{-O( Q)}$.

The production of a positron after $\Delta t \sim \Gamma^{-1}$ carries away more charge than mass, greatly increases the final temperature, and changes the spin of the black hole to $j=1/2$ making the black hole fermionic. The typical mass after emission is given by \eqref{eqn:Adams_eqns}, restated here
\be
\bar M - \bar Q \sim \frac{q Q_*}{Q} \gg \Ebrk \, . 
\ee
Even though $q \gg m$, the BH loses a lot more mass than naively expected since the positron is accelerated to ultrarelativistic speeds. The BH thus remains near-extremal after emission but is now in a regime where the quantum-field-theory approximation in curved spacetime is temporarily valid again. This greatly increases the energy flux from Hawking radiation. The radiation is described semiclassically until the black hole once again reaches the breakdown scale, and quantum gravity effects become important. 

However, we are now in a different situation than the first time when the black hole approached extremality: our black hole is now fermionic with spin $j=1/2$. It must, therefore, have mass $M > Q + \frac{3}8 \Ebrk$ but will cool much faster. It will again emit radiation for a time $\Delta t = \Gamma^{-1}$, but now reaches a mass \eqref{eqn:M(t)_fermionicBH}
\be
M_{\t{min}}^{\t{fermionic}} = Q + \frac{3}8 \Ebrk + O \left(\exp \left(-\frac{2}{9}\frac{\pi m^2 Q}{q} \right) \right)\,,
\ee
where a positron is once again likely to be emitted. After the emission, the black hole once again becomes bosonic and goes far above extremality, and the process of Hawking emission described above repeats itself. The evaporation process is illustrated in Figure \ref{fig:largeBhEvapProcess1}.

When the black hole emits a positron the charge goes down by $q$. In the fullness of time, the mass also goes down by $q$ (up to O($\Ebrk$) corrections from spin) as the black hole returns to extremality. Most of this energy (about $q - \frac{q Q_*}{2Q}$, see \eqref{eq:refertome}) is carried away by the emitted positron itself, and then the remaining $\sim \frac{q Q_*}{2Q}$ energy above extremality slowly leaks out in neutral Hawking radiation. Thus for large enough black holes, $Q \gg Q_*$, most of the mass of the black hole ends up in the kinetic energy of the positron, with only a small fraction left over for neutral Hawking radiation
\begin{equation}
\textrm{fraction of mass of black hole that ends up in neutral radiation} \sim \frac{Q_*}{2Q} \ . 
\end{equation}

\paragraph{Black holes with \boldmath{$Q < Q_*$}.}
Eventually, the BH charge decays below $Q_* \equiv  1.8 \times 10^{44} q$, and the Schwinger process is no longer exponentially suppressed. Positrons are abundantly produced, and the charge $Q$ rapidly decreases, resulting in increasing temperature, which consequently increases the Hawking flux. Balancing the rate of neutral Hawking radiation with that of charged emission, we find that the equilibrium value of $M$ and $Q$ is approximately given by setting the electric field at the horizon of the black hole to be constant as $M$ and $Q$ are changed, i.e.~$\frac{m^2}q \frac{(M+\sqrt{M^2-Q^2})^2}Q \sim O(\log m/q)$. 
This will result in an attractor curve in the $M-Q$ plane that the black hole will follow once the charge becomes sufficiently small \cite{Hiscock:1990ex}. The entire process of mass and charge loss is shown in Figure \ref{fig:massChargeEvolution}.

\begin{figure}[t!]
\centering
\includegraphics[width=1\textwidth]{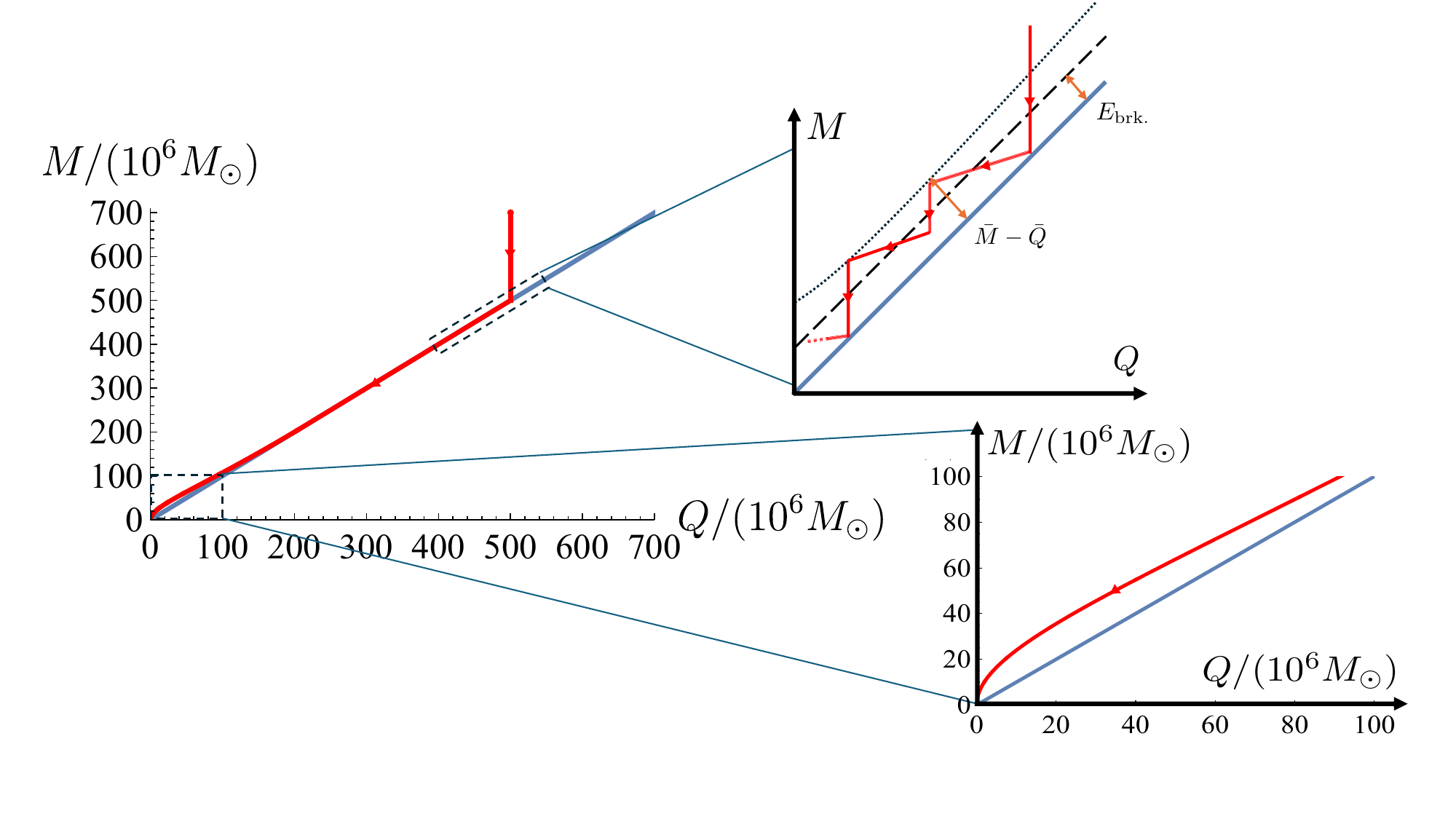}\vspace{-0.75cm}
    \caption{The evolution of the mass and charge of a black hole that starts with $M = 7 \times 10^8 M_{\odot}$, $Q = 5 \times 10^8 M_{\odot}$ and integer spin. The black hole evolves towards smaller $M$ and $Q$, in the direction indicated by the red arrows.  At first the black hole loses $M$ without losing $Q$, until it gets very close to extremality. It remains close to extremality until $Q \lesssim Q_{*}$. In the \textit{upper-right figure}, we show a zoomed-in cartoon of the evolution of $M$ and $Q$. When a positron is emitted the charge jumps down by $q$, along with a corresponding reduction in energy, and then neutral  Hawking radiation reduces $M$ without reducing $Q$ and drives the black hole back towards extremality. When the energy gets within about $\Ebrk$ of extremality the neutral Hawking radiation is strongly affected by quantum gravity corrections. With each positron emission, the black hole alternates between a bosonic era (solid red) and fermionic era (three compounded lines). Since the difference in scales between $\Ebrk$ and $\bar M-\bar Q$ is very large, this plot is not shown to scale; in particular, the energy lost in positron emission, $\sim q$, is much larger than the energy lost subsequently through neutral Hawking radiation, $\sim q Q_*/Q$, before the next positron emission. In the \textit{lower-right figure}, we zoom into the region with smaller values of $Q$ where we see large deviations from extremality. }
    \label{fig:massChargeEvolution}
\end{figure}

\paragraph{Magnetic black holes.} The above analysis applies to near-extremal magnetic black holes as well, where the magnetic field would source the production of magnetic monopoles. Since monopoles have charges $q_m \sim 1/q$ and are expected to have incredibly large masses, around the GUT scale $M_{\t{mon.}} \sim 10^{16} \t{ GeV} \sim 10^{-4} M_{\t{pl}}$, their production would be exponentially suppressed for 
\be\label{eq:magneticbound}
I = \frac{\pi M_{\t{mon.}}^2}{M^2_{\t{pl}}} \frac{Q_m}{|q_m|} \gsim 1 \implies Q_m \gsim 10^{6} |q_m| \ ,
\ee
which corresponds to a near-extremal mass
\be
M > \frac{M_{\t{pl}}^2}{M_{\t{mon.}}^2} M_{\t{pl}} \sim 10^{6}M_{\t{pl}} \sim 10^{-31} M_{\odot} \sim 10^{-2} \t{ kg}\,.
\ee
Therefore, any magnetically charged primordial black holes in our universe with a charge greater than \eqref{eq:magneticbound} would eventually approach exponentially close to extremality. It is interesting to note that quantum gravity effects (the integral over the defect moduli space) are more important for magnetic BHs since the expansion parameter for QG corrections is
\be 
(1-\alpha) \sim \frac{M_{\t{mon.}}^2}{M_{\t{pl}}^2} \frac{1}{|q_b| Q} \sim 10^{-6} \frac{1}{|q_b| Q}\,,
\ee
and is not as significantly suppressed as for positron emission.

\section{Discussion}
\label{sec:discussion}

By analyzing the quantum-corrected rates for the neutral and charged emission rates of Hawking radiation from black holes close to extremality, we are thus able to describe the full evaporation history of a charged black hole. 

\subsection{Future directions}

Let's discuss possible future directions.

The weak gravity conjecture was initially motivated by arguing that in order to avoid remnants with large entropy, charged particles satisfying $m < q$ need to exist \cite{Arkani-Hamed:2006emk, Harlow:2022ich}. Semiclassically, such charged particles would eventually take any black hole away from extremality, thus avoiding the scenario in which extremal black holes would survive indefinitely, which would lead to the existence of remnants \cite{Susskind:1995da}. Given that the analysis in this paper changes the behavior of black holes and their radiation close to extremality, it is worth reanalyzing the status of these arguments using our new results. 

Suppose that there does \textit{not} exist a particle that satisfies $m<q$. All the emitted particles, regardless of whether they are neutral or charged, would, therefore, bring the black hole closer to extremality. As the black hole approaches extremality, we have seen that the density of states and rate of Hawking radiation is modified. Nevertheless, as we approach extremality, our new results suggest no inconsistency in the theory at the perturbative level. The black hole is never at risk of becoming superextremal, the second law of thermodynamics is fully obeyed, and even the problem of remnants with large entropy at extremality \cite{Susskind:1995da} is now no longer present since the density of states vanishes at extremality. Whether the theory is still consistent once non-perturbative effects in $1/G_N$ are taken into account is, however, still an open question.

A second natural question is how nearly supersymmetric charged black holes evaporate. For simplicity, we can consider a $4d$ supergravity theory that contains supersymmetric Reissner-Nordstr\"om black holes \cite{Heydeman:2020hhw, Boruch:2022tno}. As we saw in Section \ref{sec:neutral-particle-emission-main}, the emission of neutral particles can be inferred by understanding the coarse-grained spectrum near extremality, and using Fermi's golden rule to calculate the decay rate between nearby black hole states. 

The spectrum of near-supersymmetric black holes consists of a large ground state degeneracy with $e^{S_0}$ BPS states, a mass gap, and then a continuum of near-BPS states separated from the BPS states by the mass gap. The number of near-BPS states smoothly decays to zero as the mass gap is approached from above. Using Fermi's golden rule, near-BPS states have two decay channels. The first channel is that they can decay into lower energy near-BPS states by emitting a particle with arbitrarily low energy. The second decay channel is into the $e^{S_0}$ extremal supersymmetric states by emitting a single particle with energy bounded from below by the gap scale. Since there are an exponentially large number of extremal ground states, at sufficiently small energies above the gap scale near-BPS black holes will actually entropically favor decaying directly to the extremal ground states. 

If we imagine starting with near-BPS states in the microcanonical ensemble slightly above the mass gap, the resulting spectrum of Hawking radiation will be quite unique compared to the spectrum emitted by non-supersymmetric black holes. The spectrum will consist of modes with energies far below the gap scale, corresponding to transitions between near-BPS microstates, and modes with energies bounded below by the gap scale, corresponding to transitions to extremal black holes. We leave a detailed study of this to future work \cite{susybhwip}.

\subsection{Observational prospects} 
\label{subsec:observationalprospects}

Let us discuss the use of highly charged black holes as laboratories for observing the effects of quantum gravity. 

Let us start with the advantages. Physics is an empirical science, but it is often lamented that the effects of quantum gravity do not naturally make themselves felt on everyday scales and so are very difficult to probe experimentally.
For example, a particle collider that could directly probe the Planck scale would, if built using current technology, need to be astronomically long. This is because the Planck mass $M_\textrm{Pl}$ is very much heavier than the mass of the fundamental particles, which must therefore be accelerated to enormous $\gamma$ factors in order to cram a large amount of energy $M_\textrm{Pl}$ into a small volume $\ell_\textrm{Pl}^{\, 3}$. Ultimately, this traces back to the fact that the ratio in  \eqref{eq:largeratio} is so stupendously large. By contrast, the attractive feature of highly charged black holes as laboratories for observing the effects of quantum gravity is that, having set them up with a sufficiently large charge, one does not need to then perform further difficult operations to compel them to exhibit the effects of interest. Instead, you can just wait. As we wait, and with no outside intervention required, an isolated black hole with a huge charge will tune itself towards extremality. Then, once it has tuned itself sufficiently close to extremality, the statistics of its particle emissions will deviate from those predicted by quantum field theory in a fixed curved spacetime. These deviations are due to the effects of quantum gravity and we have calculated these deviations in this paper.  To detect these deviations---these falsifiable predictions of quantum gravity---we do not need to do a phenomenally delicate interference experiment with a huge quantum computer, instead we can make a simple measurement with a spectrograph.

Now let's discuss the disadvantages. There are two we will consider. The first is how hard it is to make a black hole with a stupendously large charge. In order to tune itself towards extremality, the black hole must be large enough that Schwinger pair production near the horizon is exponentially suppressed. As discussed in Sec.~\ref{sec:background-charged}, the suppression exponent is $Q/Q_{*}$, where 
\begin{equation}
\label{eq:Q-crit}
Q_{*} \equiv \frac{q M_{\t{pl.}}^2}{\pi m^2} \sim 2.9 \times 10^{25}  \textrm{ Coulombs} \sim \textrm{the charge of }1.8 \times 10^{44} \textrm{ protons} \ . \end{equation}
The experiment we're going to be doing is slow, and in order not to be interrupted by pair production, we will see that we need at least about 
\begin{equation}
Q \geq 2500 Q_{*} . \label{eq:chargerequiredtoproceedwithoutinterruption}
\end{equation}
That is a lot of charge by laboratory standards but is less than, say, the charge in all the protons in the Earth's oceans. The hard part is separating all those protons from their partner electrons and then forcing all the protons together into a black hole. Precisely because the electric repulsion is so much stronger than the gravitational attraction, the protons will only be bound together if the corresponding black hole weighs much more than the Earth's oceans; in fact, it needs to weigh at least
\begin{eqnarray}
M \ \ \geq \  \ 2500 M_{*} & = & 8.5... \times 10^{38} kg \ = \  4.3 \times 10^8  M_\odot \ . 
\end{eqnarray}
This is very heavy. On the other hand, Nature has already provided us with a black hole in this weight class: the Sagittarius A* supermassive black hole at the center of the Milky Way is a couple dozen times heavier than $M_{*}$. The first step in our experiment will, therefore, be to extract the protons from the oceans, separate them from their electrons, and throw them into Sagittarius A*, along with enough nearby stars to bring it up to the required weight.\footnote{Alternatively, we can start with a much heavier astrophysical black hole that is naturally formed with the charge \eqref{eq:Q-crit}. However, given the estimate of \cite{Zajacek:2019kla} for the amount of charge that black holes store per unit mass ($Q/M=76.9 C/M_\odot$), no known astrophysical black hole would have a sufficiently large mass.} While doing so, we will have to be sure that the electrons do not also fall in. Indeed, we will probably be well advised, before we begin, to clean out all the other matter in our galaxy, perhaps by throwing it in too, because otherwise the huge charge of the black hole will polarize the interstellar medium, suck in electrons, and neutralize itself. Unfortunately, even in such a pristine environment, throwing that many protons into the black hole is not going to be easy. This is because the later protons, as they approach the black hole, will have to overcome the fierce electromagnetic repulsion of the earlier protons. Of course, they will be helped by the gravitational attraction of the black hole, and when the protons get close enough to the horizon, the gravitational force will dominate, and they will be sucked in, but for highly charged black holes, they will need a lot of help to get that close. The problem is illustrated in Figure \ref{fig:V(r)forchargedparticle}. This figure shows that the net force is attractive (due to gravity) very close to the horizon but repulsive (due to electromagnetism) further away. In  \eqref{eq:gallonofgasoline}, we saw that that means that a charged particle that is produced at rest near the top of the potential barrier will, by the time it makes it far from the black hole, have a tremendous kinetic energy, $\gamma  \sim 10^{21}$. Now, we are seeing the same phenomenon in reverse: in order to charge up the black hole by adding charged particles, we must throw them in with enormous $\gamma$; otherwise, they will be turned around by the electric repulsion and never make it over the barrier. Thus, to use this approach, we still need an astronomically large particle accelerator\footnote{Indeed, note that charging up a critical-mass black hole in this way needs a particle accelerator that can reach $\gamma \sim M_\textrm{Pl}/m_p$, whereas to `merely' reach the Planck energy in the center-of-mass frame of a particle collider we'd only need $\gamma \sim  (M_\textrm{Pl}/m_p)^{1/2}$ in each of the two counterrotating beams.}. We can reduce the required size of the particle accelerator either by bundling the charge we are sending in with more mass---for example, by attaching the proton to a brick---or, relatedly, by increasing the mass of the black hole. At fixed charge, larger black holes mean more gravitational attraction, and it is correspondingly easier for an ingoing charged particle to overcome the electromagnetic repulsion. But this only helps reduce the required $\gamma$ by a factor linear in the mass of the black hole, so for charge $Q_{*}$, we'd no longer need an accelerator to charge up the black hole with protons only for $M \ \gsim \  
10^{47} M_{\odot}$. This is substantially more than the mass of all the matter in the visible universe. 

But making the highly charged black hole is the easy part. The hard part---and the second and more devastating disadvantage of this scheme---is the patience required for the next step. In order for the Schwarzian prediction for the spectrum of Hawking radiation to substantially differ from the prediction of quantum field theory in curved spacetime, the black hole must be very close to extremality. It must be so close that the energy above extremality is at most a few times the energy of a single thermal Hawking quantum. Specifically, the effects of quantum gravity only become relevant for black holes with temperatures below the breakdown scale
\begin{equation}
T \lesssim \Ebrk \implies \sqrt{\frac{M - Q}{Q}} \  \lesssim \  \frac{1}{ Q^2} \ , \label{eq:extremalfactortobeinQGregime}
\end{equation}
where the temperature is given in \eqref{eq:tempandentorpyofRNblackhole}.
For a black hole of the minimum charge required to conduct our experiment, \eqref{eq:chargerequiredtoproceedwithoutinterruption}, this implies a fractional energy above extremality of 
\begin{equation}
\frac{M - Q}{Q} \ \lesssim \    \frac{1}{ (2500 Q_{*})^4}  \sim  
4.4 \times 10^{-187} \ . 
\end{equation}
This is very close to extremality. 

Because the black hole is both large and close to extremality, the photons we are trying to observe have very low energy. The effects of interest happen at (and below) the breakdown scale, 
where the characteristic temperature is
\begin{eqnarray}
 T & \sim & \frac{1}{Q^3} \sim  \frac{1}{ (2500 Q_{*})^3}  \sim  5 \times 10^{-109} \textrm{ Kelvin} \ . \label{eq:lowlowtemperature}
\end{eqnarray}
 This is well over a hundred orders of magnitude colder than the cosmic microwave background, though that is not a problem because, as we will discuss, the experiment is going to take so long that the CMB will long since have redshifted into irrelevance. More troubling, the Hawking temperature is about seventy-five orders of magnitude colder than the de Sitter temperature caused by radiation from the cosmological horizon. Alternatively, we can say that the characteristic wavelength of the black-hole radiation is seventy-five orders of magnitude longer than the visible universe. The problem is that the cosmological constant limits the precision of any experiment in our universe, and we need to be much more precise than that limit to see the effects of the Schwarzian for a black hole that is charged enough that it can tune itself into the Schwarzian regime. Our only hope of performing this experiment is if it turns out that our cosmological constant will eventually abate (or if we can force it to abate) to a value at least a hundred and fifty orders of magnitude closer to zero. That would give us the precision we need to see our signal amongst the noise.  Let us suppose that has been achieved; now, let's turn to how long we need to wait. First, we need to wait for the black hole to tune itself close enough to extremality. Then, we need to wait to measure the spectrum once it is near extremality. These take about the same length of time, and let's calculate the second of these. For a blackbody, the typical time between S-wave quanta is the inverse of the temperature, so for a blackbody with temperature given by  \eqref{eq:lowlowtemperature}, the typical time would be 
\begin{equation}
\textrm{typical time between photons if blackbody}\ \sim \ Q^3  \ \sim \ 10^{89} \textrm{years} \ . 
\end{equation}
However, the spectrum is far from that of a blackbody. As discussed in Sec.~\ref{sec:semiclassical-result-neutral-hawking-radiation}, the emission is greatly suppressed by greybody factors. Our experiment will thus require much more patience.  \eqref{eq:photoellisonegreybodyresult} tells us that the probability that a low-energy photon is not diffractively reflected by the gravitational field of a near-extremal black hole and instead escapes is 
\begin{equation}
\mathcal{P} \sim (w r_+)^8 
 \sim \left( \frac{1}{Q^2} \right)^8 \sim Q^{-16} \ . \label{eq:greybodyenormousfactor}
\end{equation}
This gives an additional greybody suppression, which stacks on top of the existing blackbody suppression, $Q^{-3}$, so the mean time between photons near the breakdown scale 
is approximately \cite{Page:2000dk} 
\begin{equation}
\textrm{time between photons} \sim Q^{19} \sim 2 \times 10^{834} \textrm{years} \ .  \label{eq:timebetweenphotons}
\end{equation}
This is how long you must wait for the first data point during the data-collection phase of the experiment. Because the first photon leaves the black hole even closer to extremality, the second and subsequent data points take even longer\footnote{By comparing to \eqref{eqn:Schwinger_rate_final}, we can now see that $Q = 2500 Q_{*}$ does indeed gives us long enough to take a few dozen data points before being interrupted by pair creation.}.  The primary disadvantage of using highly charged black holes to observe the effects of quantum gravity is the length of time we need to wait. \\

Of course, all of the above discussion assumes that single-photon emission is still possible, as is true for fermionic black holes and for bosonic black holes with $M - Q \geq \Ebrk$. To see any diphoton emission, which is the dominant emission channel for bosonic black holes with $M - Q < \Ebrk$ we need to wait considerably longer, specifically\footnote{In fact you would need to wait slightly longer than this because $Q$ needs to be somewhat larger for diphoton emission to occur before positron emission.}
\begin{equation}\label{eq:diphotontimescale}
\textrm{time between di-photons} \sim Q^{35} \sim  10^{1540} \textrm{years} \ .  
\end{equation}
How different one considers the timescales \eqref{eq:timebetweenphotons} and \eqref{eq:diphotontimescale} is somewhat subjective, but objectively their ratio is over $10^{700}$. By comparison, the allowed $2p \rightarrow 1s$ hydrogen atom transition leads to a lifetime of $1.6ns$, while the forbidden $2s \rightarrow 1s$ transition leads to a lifetime of $0.12s$, a ratio of only $10^8$. The larger ratio for black holes comes from the fact that the perturbative expansion parameter is the photon greybody factor, which is much smaller than the fine-structure constant.

  Let's discuss how we can make the wait shorter. First, it might turn out that there is a fundamental massless scalar or spin-$\frac{1}{2}$ particle (or that we could maneuver our way into a vacuum that has one). If so, this particle will radiate much faster than spin-1 photons since it will not be subject to the same punishing greybody factors,  \eqref{eq:greybodyenormousfactor}. Indeed, the spin-0 or spin-$\frac{1}{2}$ fields don't need to be actually massless; they just need to have a mass less than the characteristic temperature,  \eqref{eq:lowlowtemperature}. 

Even if Nature has not provided us with a fundamental massless scalar, we could make one. We could hasten the emission of Hawking quanta by mining the black hole, for example, by lowering cosmic or fundamental strings through the horizon \cite{Unruh:1982ic,Unruh:1983qq,Lawrence:1993sg}. For Schwarzschild black holes, the main advantage of `mining' comes from helping higher angular momentum modes radiate at the same rate as the S-wave, but in this case, the main benefit will be that the transverse modes of the string will behave as a massless scalar and so not be subject to the same greybody suppression as the spin-1 photon. However, even for Schwarzschild black holes, the gravitational back reaction of the mining equipment severely limits the achievable speed-up \cite{Brown:2012un}, and those logistical difficulties will be substantially worse for near-extremal Reissner-Nordstr\"om black holes, which are considerably more vulnerable to gravitational perturbations. 

\paragraph{Magnetic black holes.} The strongest reduction in the time we need to wait can be achieved by switching from an electric charge to a magnetic charge. Due to electric-magnetic duality, the metric of a magnetically charged black hole is still given by the Reissner-Nordstr\"om solution, and much of the analysis for electrically charged black holes directly carries over. The difference is that while there is a very light electric monopole---the positron---there are no known light magnetic monopoles. This has the disadvantage that it is difficult to give black holes a magnetic charge: we must either find a so-far unnoticed trove of magnetic monopoles, or manufacture them ourselves for example by running a particle accelerator at the magnetic-monopole-pair-production threshold.\footnote{Alternatively, if detected, primordial black holes could carry magnetic charge \cite{Stojkovic:2004hz, Bai:2019zcd, Maldacena:2020skw} and actually be in the mass range of interest.} Either way, this is much harder than harvesting electrically charged particles. On the other hand, the advantage is that magnetically charged black holes, once made, are much more useful. As discussed in Sec.~\ref{sec:the-story-of-a-charged-black-hole}, when the magnetic monopoles are heavy, even quite strong magnetic fields will not promptly decay by Schwinger pair production, and so even quite small magnetic black holes will not rapidly discharge\footnote{As discussed in Ref.~\cite{Maldacena:2020skw}, since in our universe the U(1) of electromagnetism lives within a larger gauge group, the magnetic fields of near-extremal magnetically charged black holes can restore electroweak symmetry. On the one hand this means that black holes mine themselves and emit massless radiation even faster \cite{Maldacena:2020skw} and so allow our experiment to be finished sooner, on the other hand this further complicates the analysis.}. If the lowest mass magnetic monopoles have mass $M_{\t{mon.}}$, we found that magnetic black holes must have charge
\be
Q^m \gsim \frac{|q_m|M^2_{\t{pl}}}{\pi M_{\t{mon.}}^2}\,,
\ee
in order to be driven close to extremality. The breakdown scale is then $\Ebrk = \frac{M_{\t{pl}}}{(Q^m)^3}$. For magnetic monopoles around the GUT scale we have $10^{15} \lesssim M_{\t{mon.}} \lesssim 10^{17}$ which leads to a maximum breakdown scale
\be\label{eq:magneticebrkrange}
100 \mathrm{eV} \lesssim \Ebrk  \lesssim 100 \mathrm{TeV}.
\ee
At the top end of this range, Hawking radiation has a temperature much higher than the Higgs mass, and thus, the Higgs is emitted in the same way as a massless scalar. At the breakdown scale, the radiation receives large quantum gravity corrections as described in Section \ref{sec:massless_scalar_emission}. However, unlike for particles with nonzero spin, emission of a single Higgs boson is possible even far below the breakdown scale. If the breakdown scale is low enough that Higgs production is exponentially suppressed by its nonzero mass, the dominant channel becomes single neutrino emission when $M - Q \geq 3 \Ebrk/8$ and di-neutrino emission below this energy. Emission of electrically charged particles such as quarks and charged leptons is not permitted around the breakdown scale because the extremality bound for a black hole with electric charge $q$ and magnetic charge $Q^m \gg q$ is\footnote{Dilepton and diquark emission is in principle possible if the pair of emitted particles is electrically neutral. However, it will be very slow, even compared to dineutrino emission, because the intermediate state will be parametrically further off shell.} 
\be 
M - Q^m \geq q M_{\t{pl}}/Q^m \gg \Ebrk.
\ee

For Higgs emission, and for single neutrino emission when the breakdown scale is not too low, the relevant timescales are short enough that Hawking radiation near the breakdown scale is, in principle, observable. Magnetic black holes, therefore, provide an unlikely but possible avenue by which quantum gravity effects could be directly tested in our universe.

\section*{Acknowledgments}

We thank Gustavo-Joaquin Turiaci for numerous helpful initial discussions and comments. We also thank Andreas Blommaert, Kwinten Fransen, Gary Horowitz, Maciej Kolanowski, Guanda Lin, Juan Maldacena, Mukund Rangamani, Douglas Stanford, and Leonard Susskind for illuminating discussions. MU was supported in part by grant NSF PHY-2309135 to the Kavli Institute for Theoretical Physics (KITP), and by a grant from the Simons Foundation (Grant Number 994312, DG). LVI was supported by the DOE Early Career Award DE-SC0025522. GP was supported by the Department of Energy through QuantISED Award DE-SC0019380 and an Early Career Award DE-FOA-0002563, by AFOSR award FA9550-22-1-0098 and by a Sloan Fellowship. This work was also performed in part at Aspen Center for Physics, which is supported by National Science Foundation grant PHY-2210452.
\appendix

\section{Classical greybody factors} \label{app:greybody}

In this appendix, we calculate the greybody suppression for the emission of uncharged massless radiation from near-extremal Reissner-Nordstr\"om black holes. 
These greybody factors are caused by the effects of diffraction. A common way to think about Hawking quanta is that they are `born' at the horizon and then propagate out away from the black hole. In the geometric optics approximation, in which we treat the quanta as particles that move along straight lines, all the Hawking quanta would escape to $r=\infty$ (except for quanta with large angular momentum, which all fall back in). However, the geometric optics approximation is far from valid. Even for Schwarzschild black holes, the typical wavelength of a Hawking quantum is about the same as the Schwarzschild radius; for near-extremal black holes, which have a temperature suppressed by the extremality parameter $\frac{r_+ - r_-}{r_+}$, the typical wavelength is much bigger than the size of the black hole. Small antennas find it difficult to emit long-wavelength modes, so emission is suppressed by powers of the extremality parameter. Even though ray-tracing would suggest the low-angular-momentum Hawking particles all make it to $r = \infty$, the quanta propagate like waves, diffract off the gravitational field, and are typically reflected back in. Only a small fraction escape. In this section, we will calculate this fraction. 

We will calculate the greybody factors for massless particles. We do not need to calculate the greybody suppression for positron emission because the emission of positrons is not greybody suppressed. As discussed in Sec.~\ref{subsec:probemassiveparticles}, positrons are emitted from the black hole with large momentum, and therefore, with short de Broglie wavelength, and short wavelength modes move on geodesics and do not experience diffraction or greybody reflection.

Greybody factors describe both the probability that an outgoing near-horizon Rindler mode is successfully transmitted through the near-horizon region to infinity and the probability that an ingoing wave from infinity is absorbed by the black hole. It is easy to see, using the linearity of the wave equation and the conservation of total particle flux, that these two probabilities must be equal. (In the language of quantum mechanics, this is the statement that the one-dimensional scattering S-matrix is unitary.) Our object of interest is the first process, but for practical reasons, it will be marginally more convenient to consider the second, or more precisely, to consider the probability that an ingoing wave is reflected off the black hole, which is equal to one minus this transmission probability. We will, therefore, consider scattering problems where the outgoing mode from the near-horizon region is set to zero.

The calculation of greybody factors typically requires numerics, as we must solve an ODE for the evolution of each mode \cite{Page:1976df}. However, for near-extremal RN black holes, there is a simplification. This simplification is that near-extremal RN black holes are so cold that emission is exponentially suppressed for all modes except those with a wavelength much bigger than the size of the black hole. We can thus approximate the frequency $\omega$ as zero, except very close or very far from the horizon, where the effective potential goes to zero and even long wavelength modes begin to propagate freely. This will make the calculation easier and the answer analytically tractable.

The $\omega$ dependence of the greybody factors can, therefore, be naturally decomposed into two pieces. The first involves the transmission of propagating modes in the near-horizon region into exponentially growing and decaying modes around the top of the throat. The second involves transmission between the top of the throat and propagating modes far from the black hole. Importantly, only the first term depends on the temperature (and the angular momentum) of the black hole. When $M- Q \ll Q$, the second term can be calculated in the extremal black hole background and so depends only on the frequency and angular momentum of the emitted mode.

In the quantum-corrected emission rate calculations in Section \ref{sec:neutral-particle-emission-main}, the role of these two contributions to the greybody factor is different. The first contribution, from transmission between the near-horizon region and the top of the throat, is included, together with the Bose-Einstein occupation number factor, as part of the semiclassical limit $E_i \gg \Ebrk$ of the JT gravity two-point function. At lower energies $E_i \lesssim \Ebrk$, these factors are replaced by the full quantum JT gravity two-point function. 

In contrast, the second contribution, from transmission between the top of the throat and asymptotic infinity, is not described by JT gravity. Instead, it determines the relationship between the non-normalizable mode sourcing the JT gravity dynamics and creation and annihilation operators acting on the fields far from the black hole. It, therefore, appears unchanged as part of the quantum emission rate formula, even at temperatures far below the breakdown scale $\Ebrk$.

\subsection{Scalar greybody factors}
\label{subsection:scalargreybody}

In this section, we will calculate the greybody factor for scalars incident on a near-extremal Reissner-Nordstr\"om black hole as a warm-up for the finite temperature photon/graviton calculation. For general black holes, a useful review is \cite{Harmark:2007jy}. This will be accomplished by solving the wave equation in overlapping regions and gluing together the solutions to calculate the probability for an incoming plane wave to be transmitted through the potential barrier. The metric is
\be
ds^2 = -f(r) dt^2 + f(r)^{-1} dr^2 + r^2 \lrm{d\theta^2 + \sin^2{\theta} d \phi^2}, \qquad f(r)=\frac{(r-r_{+})(r-r_-)}{r^2}\,.
\ee
The action for a massless scalar is
\be
S = \frac{1}{2} \int \sqrt{g} \nabla^a \phi \nabla_a \phi \ .
\ee
Taking the ansatz $\phi(t,r,\theta,\phi) = e^{-i \omega t} u(r) Y_{\ell, m} (\theta,\phi)$ the equation of motion is given by
\be \label{app:A_scalarODE}
r^2 \nabla^2 \phi = \partial_r \left(r^2 f u'(r) \right) + \omega^2 \frac{r^2}{f} u(r)-\ell(\ell+1) u(r)=0,
\ee
where we have used that $\nabla^2 = -\frac{1}{f} \partial_t^2 + \frac{1}{r^2} \partial_r (r^2 f \partial_r) + \frac{1}{r^2} \Delta_{S^2}$ with $\Delta_{S^2} Y_{\ell,m} = -\ell(\ell+1)$. Changing to tortoise coordinates using $\frac{\partial r_*}{\partial r} = \frac{1}{f(r)}$ we find the standard form of the scalar field equation in a black hole background
\be
\left(\frac{d^2}{d r_*^2} +  \omega^2 - \underbrace{\frac{f}{r^2} \lrm{\ell(\ell+1) + r f'}}_{V_{\t{eff}}(r)} \right) r u(r) =0.
\ee 

\paragraph{Region I: near-horizon region.}
Region I is defined to be the region where $r-r_+ \ll r_+$ which is the near-horizon region. In this region, we can simplify the differential equation \eqref{app:A_scalarODE} by replacing $r \to r_+$ except for terms containing $(r-r_+)$ and $(r-r_-)$. This gives us the equation, and solution
\be
(r-r_-)(r-r_+) u''(r)+(2r-r_+ -r_r) + \lrm{-\ell(\ell+1) + \frac{r_+^4 \omega^2}{(r-r_-)(r-r_+)}}u(r)=0
\ee
The solution is given in terms of associated Legendre functions
\be \label{eqn:nearhorizonscalarsoln}
u_I(r) = a_1 P_{\ell}^{-\frac{i \beta \omega}{2\pi}}\left(\frac{(-2r + r_+ + r_-)\beta}{4\pi r_+^2}\right) + a_2 Q_\ell^{-\frac{i \beta \omega}{2\pi}}\left(\frac{(-2r + r_+ + r_-)\beta}{4\pi r_+^2}\right)\,.
\ee
Since we are interested in greybody factors, we must set the wave to have ingoing boundary conditions at the horizon $r_+$. We thus set $a_2=0$ going forward. 

\paragraph{Region II: barrier.} The second region contains the barrier where $\omega^2 \ll V_{\t{eff}}$. We can set $\omega \to0$ and $r_- \to r_+$, since temperature corrections are not important in this region, in \eqref{app:A_scalarODE} with equation
\be
(r-r_+)^2 u''(r) + (2r-r_+) u(r) -(\ell(\ell+1))u(r)=0\,.
\ee
The solution in this region is 
\be \label{eqn:scalar_regionII}
u_{II}(r) = b_1 (r-r_+)^{\ell} + b_2 (r-r_+)^{-\ell-1}.
\ee

\paragraph{Region III: asymptotically flat space.} 
For the final region we recover asymptotically flat space where $f(r)\approx 1$. Setting $f(r)=1$ we find
\be
r^2 u''(r) + 2 r u'(r) + ( \omega^2 r^2-\ell^2-\ell) u(r)=0\, . 
\ee
The solution is given by spherical Bessel functions
\be\label{eqn:scalar_soln_infty}
u_{III} (r) &= c_1 j_l(r \omega )+c_2 y_l(r \omega )\stackrel{r \to \infty}{=}-\frac{1}{2 r \omega} \left( (i c_1 + c_2) e^{i r \omega - \frac{1}{2} i \pi \ell } +   (-i c_1 + c_2) e^{-i r \omega + \frac{1}{2} i \pi \ell }\right),
\ee
where, far from the black hole, the outgoing mode is proportional to  $c_1 -i c_2$, and the ingoing mode is proportional to $c_1 + i c_2$. 

\paragraph{Matching conditions.} Expanding the solutions $u_I, u_{II}$ and $u_{II}, u_{III}$ in overlapping regimes of validity we find the matching conditions
\begin{align} \label{eqn:scalar_matching_cond}
   &b_1 = a_1 (-1)^\ell\frac{e^{\beta \omega/4} (\beta/r_+^2)^\ell \Gamma(2\ell+1)}{(4\pi)^\ell \Gamma(\ell+1 + \frac{i \beta \omega}{2\pi})\Gamma(\ell+1)}, \qquad b_2 = a_1(-1)^{\ell+1}\frac{e^{\beta \omega/4} (\beta/r_+^2)^{-\ell-1} \Gamma(-\ell-\frac{1}{2}) \pi^{\ell+\frac{1}{2}}}{\Gamma(-\ell+\frac{i \beta \omega}{2\pi})} \\
   &c_1 =   b_1\frac{2^{\ell+1} \Gamma \left(\ell+\frac{3}{2}\right)}{\sqrt{\pi} \omega^{\ell}}, \qquad c_2=  - b_2 \frac{\sqrt{\pi }\omega^{l+1}}{2^{\ell}\Gamma \left(\ell+\frac{1}{2}\right)}.
\end{align}

\paragraph{Transmission probability.} To calculate the greybody factor, we will calculate the probability for a wave of small frequency $\omega$ to be transmitted through the effective potential. We have already imposed ingoing boundary conditions and so we can calculate the reflection/transmission probabilities. The reflection probability $R$ is given by the ratio of outgoing to ingoing amplitudes $R =|c_1-i c_2|^2 / |c_1+i c_2|^2$ in \eqref{eqn:scalar_soln_infty} at infinity. The transmission probability/greybody factor for the scalar is then $T=1-R$. We are interested in the low-temperature/low-frequency limit where $\beta \omega \sim \mathcal{O}(1)$ while $\beta \to \infty$ and $\omega \to 0$. We thus can scale both quantities simultaneously and find the leading order greybody factor
\be \label{eqn:scalargreybody_arbitrary_angularmomentum}
P_{\t{emit}}(\omega,\ell) = 1-\frac{|c_1-i c_2|^2}{|c_1+i c_2|^2} \stackrel{\omega \to 0 }{=} \frac{\pi^{2\ell+4}r_+^{4\ell+2}(\omega/\beta)^{2\ell+1}\csch \lrm{\frac{\beta \omega}{2}} }{2^{2\ell-1}\Gamma\lrm{-\ell \pm i \frac{\beta \omega}{2\pi}} \Gamma\lrm{\ell+\frac{1}{2}}^2 \Gamma\lrm{\ell+\frac{3}{2}}^2}
\ee
The dominant decay channel is the $\ell=0$ mode with greybody factor
\be
\lim_{\omega \to 0 } P_{\t{emit}}(\omega,\ell=0) = 4 (r_+ \omega)^2 \,.
\ee
Some higher greybody factors are
\begin{gather}
P_{\t{emit}}(\omega,\ell=1)=\frac{4}{9} r_+^6 \omega^4 \lrm{ \omega^2+\frac{4\pi^2}{\beta^2} }\,,\\
P_{\t{emit}}(\omega,\ell=2)=\frac{4}{2025} r_+^{10} \omega^6 \lrm{\omega^2+\frac{4\pi^2}{\beta^2}} \lrm{\omega^2+\frac{16\pi^2}{\beta^2}}\,.
\end{gather}
\subsection{Vector greybody factors} \label{app:vector_app}
We are interested in calculating the greybody factor for photons/gravitons. It turns out that in a RN background, there is an additional subtlety for photon/graviton greybody factors since propagating photons are dynamically turned into gravitons and vice versa. As a warm-up, we first calculate the greybody factor of a gauge field under which the black hole is not charged. (For simplicity, we will only consider the extremal limit where $\beta\omega \gg 1$.) In this case, there is no mixing between the gauge field and graviton, and the equations of motion remain unmodified from naive expectations. The calculation will reduce to a problem similar to that in Sec.~\ref{subsection:scalargreybody}, only now with a different effective potential because of the gauge field spin. The equations of motion for a $U(1)$ gauge field in the black hole background are given by $\nabla^\mu F_{\mu \nu}=0$, which in component form is
\be
\partial_\nu \big[\lrm{\partial_\alpha A_\sigma - \partial_\sigma A_\alpha} g^{\alpha \mu} g^{\sigma \nu} \sqrt{-g} \big]=0\,.
\ee
We emphasize again that the above equations are only correct for a gauge field in a BH background, which does not carry any charge of the gauge field (and instead, e.g.~carries the charge of a different, as yet undiscovered, U(1)). The vector field can be decomposed into vector spherical harmonics given by
\be \label{eqn:Amu}
A_{\mu}(t,r,\theta,\phi) = e^{-i \omega t}\begin{pmatrix}
\frac{u_1(r)}{r} Y_{l, m}(\theta, \phi)\\
\frac{u_2(r)}{r} Y_{l, m}(\theta, \phi) \\
\frac{1}{\sqrt{\ell(\ell+1)}} \lrm{u_3(r) \partial_\theta Y_{l,m}(\theta, \phi)+\frac{u_4(r)} {\sin \theta} \partial_\phi Y_{l,m}(\theta, \phi)}\\
\frac{1}{\sqrt{\ell(\ell+1)}} \left( u_3(r) \partial_\phi Y_{l,m}(\theta, \phi)- u_4(r) \sin \theta \partial_\phi Y_{l,m}(\theta, \phi) \right)
\end{pmatrix},
\ee
where the non-trivial radial dependence is encoded in the functions $u_i(r)$. The angular components of the vector potential have been expanded in vector spherical harmonics given by $A_a = u_3(r) \nabla_a Y_{l,m}(\theta,\phi) + u_4(r) \epsilon_{a b} \nabla^b Y_{l,m}(\theta,\phi)$ where $a$ are indices on the sphere and $\epsilon$ is the Levi-Civita density on the sphere. The equations of motion for the vector field reduce to
\begin{gather}
-r^2 f u_1''+i r^2 \omega  f u_2'+i r \omega  f u_2+\ell( \ell+1) u_1 -i \sqrt{\ell (\ell+1)} r \omega  u_3 =0, \nn\\
\ell^2 f u_2+\ell f u_2-\sqrt{\ell (\ell+1)} r f(r) u_3'+i r^2 \omega u_1'+r^2 \omega^2 u_2-i r \omega  u_1=0, \nn\\
r f f'\left(\sqrt{\ell (\ell+1)} r u_3'-\ell (\ell+1) u_2\right)+f^2 \left(\sqrt{\ell (\ell+1)} r^2 u_3''(r)-\ell (\ell+1) r u_2'+\ell (\ell+1) u_2\right)\nn\\
+\sqrt{\ell (\ell+1)} r^2 \omega ^2 u_3+i \ell (\ell+1) r \omega  u_1=0,\nn\\
r^2 f \left(f' u_4'+f u_4''\right)+u_4 \left(r^2 \omega ^2-\ell (\ell+1) f\right)=0.
\end{gather}
It is easiest to solve for $u_4(r)$ from the last equation, which can be rewritten as
\be
\lrm{\frac{d^2}{d r_*^2} + \omega^2 - \underbrace{\frac{\ell(\ell+1)f(r)}{r^2}}_{V_{\t{eff}(r)}}} u_4(r_*)=0.
\ee
with an effective potential where we use tortoise coordinates where $\frac{d r_* }{d r} = f(r)$. Using the Newman-Penrose formalism \cite{newman1962approach}, it is possible to reduce the four equations of motion to a single master equation equivalent to the above with a redefined field \cite{Arbey:2021jif}, so to solve for the greybody factor of the photon it is sufficient to solve the single equation above. We will solve the equation in three distinct regions of the black hole and match boundary conditions to extract the greybody factor. For convenience, we will define a rescaled field $u_4(r) = r \psi(r)$ with resulting equations of motion
\be \label{app:A_vectoreq}
f(r) \partial_r \lrm{r^2 f(r) \psi'(r)} + \lrm{r^2 \omega^2-f(r)\lrm{\ell(\ell+1)-r f'(r)}}\psi(r)=0\,.
\ee
\paragraph{Region I: near-horizon region.} For simplicity, we solve the equations of motion for an extremal BH with $r_+ \ r_-$. Region I is again defined to be the region where $r-r_{+} \ll r_+$, which is the near-horizon region. The differential equation \eqref{app:A_vectoreq}, setting $r=r_+$ except for terms with $r-r_+$, we find is
\be
(r-r_+)^2 \psi ''(r)+2 (r-r_+) \psi'(r)-\left(\ell (\ell+1)-\frac{r_+^4 \omega ^2}{(r-r_+)^2}\right)\psi (r) =0 \ . 
\ee
The solution in this region is 
\be \label{eqn:region1photonfield}
\psi_I(r) = \frac{i}{\sqrt{2}}\sqrt{\frac{\omega r_+^2 }{r-r_+}} \left(a_1 \Gamma\left(\frac{1}{2}-\ell\right) J_{-\ell-\frac{1}{2}}\left(\frac{\omega r_+^2 }{r_+-r}\right)+a_2 \Gamma \left(\ell+\frac{3}{2}\right) J_{\ell+\frac{1}{2}}\left(\frac{\omega r_+^2}{r_+-r}\right)\right).
\ee
\paragraph{Region II: barrier.}
Region II is defined to be the barrier region where $\omega^2 \ll V_{\t{eff}}$. Setting $\omega \to 0$ in \eqref{app:A_vectoreq} gives
\be
(r-r_+)^2 \psi ''(r)+2 (r-r_+) \psi'(r)-\left(\ell (\ell+1)-\frac{2 r_+(r-r_+)}{r^2}\right)\psi (r) =0\, .
\ee
The solution in this region is 
\be
\psi_{II}(r) = \frac{1}{r}\left( b_2 (r-r_+)^{-\ell-1}(r-r_+ + \ell r)+b_1 (r-r_+)^\ell (\ell r + r_+)\right),
\ee
which contains both a growing and a decaying piece. 

\paragraph{Region III: asymptotically flat space.}
For the final region, we recover asymptotically flat space where $f(r)\approx 1$. We set $f(r)=1$ in \eqref{app:A_vectoreq} finding
\be
r^2 \psi ''(r)+2 r \psi'(r)-\left(\ell (\ell+1)-r^2 \omega^2 \right)\psi (r) =0\, . 
\ee
The solution in this region is 
\begin{align}\label{eqn:psiIII}
\psi_{III} (r) &= c_1 j_l(r \omega )+c_2 y_l(r \omega )\stackrel{r \to \infty}{=}-\frac{1}{2 r \omega} \left( (i c_1 + c_2) e^{i r \omega - \frac{1}{2} i \pi \ell } +   (-i c_1 + c_2) e^{-i r \omega + \frac{1}{2} i \pi \ell }\right),
\end{align}
where far from the black hole, the outgoing mode is proportional to $i c_1 + c_2$, and the ingoing mode is proportional to $-i c_1 + c_2$. 

\paragraph{Matching conditions.} By expanding the above solutions in overlapping regimes of validity, we can match the coefficients. Matching $\psi_I$ at large $r$ to $\psi_{II}$ at small $r$, as well as $\psi_{II}$ at large $r$ to $\psi_{III}$ at small $r$ we find the matching conditions
\begin{align}
   &b_1 = a_1 \frac{(-2)^\ell}{(\ell+1) (\omega r_+^2)^{\ell}}, \qquad b_2 = a_2 \frac{(\omega r_+^2)^{\ell+1}  }{(-2)^{\ell+1}\ell}, \\
   &c_1 =   b_1\frac{2^{\ell+1} \ell \Gamma \left(\ell+\frac{3}{2}\right)}{\sqrt{\pi} \omega^{\ell}}, \qquad c_2=  - b_2 \frac{\sqrt{\pi } (\ell+1) \omega^{l+1}}{2^{\ell}\Gamma \left(\ell+\frac{1}{2}\right)}.
\end{align}

\paragraph{Transmission probability. } To calculate the greybody factor, we will calculate the probability for a wave of small frequency $\omega$ to be transmitted through the effective potential. To achieve this, we impose purely ingoing boundary conditions in region $I$. The purely ingoing wave is a linear combination of the contributions to $\psi_I$ with coefficients $a_{1,2}$. In the near-horizon region $r\approx r_+$ we have the solution
\be
\psi_{I}(r) \propto a_1 \Gamma (\frac{1}{2}-\ell) \cos \left(\frac{\pi  \ell}{2}+\omega  z\right)+ a_2 \Gamma (\ell+\frac{3}{2}) \sin \left(\frac{\pi  \ell}{2}-\omega z\right) ,
\ee
where we have introduced the coordinate $r-r_+ = \frac{r_+^2}{z}$. Fixing $a_{1,2}$ so that we have a purely ingoing wave, we get the condition
\be
a_1 = a_2 \frac{i e^{i \pi \ell} \Gamma(\ell+\frac{3}{2})}{\Gamma(\frac{1}{2}-\ell)}.
\ee
The reflection probability $R$ for the mode $u_4(r)$ is given by the ratio $R =|c_1-i c_2|^2 / |c_1+i c_2|^2$ of the amplitudes of the outgoing to ingoing modes in \eqref{eqn:psiIII} at infinity. The transmission probability/greybody factor for the $u_4(r)$ mode of the gauge field is therefore given by 
\be
P_{\t{emit}}(\omega,\ell) = 1-\frac{|c_1-i c_2|^2}{|c_1+i c_2|^2} \stackrel{\omega \to 0 }{=} \frac{\pi  (-1)^{-\ell}  (\ell+1)^2 \Gamma \left(\frac{1}{2}-\ell\right)}{ 2^{4 \ell-2} \ell^2 (2\ell+1)^2\Gamma \left(\ell+\frac{1}{2}\right)^3} (r_+ \omega)^{4 \ell+2} + \mathcal{O}((r_+ \omega)^{8\ell+4}),
\ee
where we have kept the leading order answer at small $r_+ \omega \ll 1$. The above is positive for all $\ell$ due to the combination $(-1)^\ell \Gamma(\frac{1}{2}-\ell)$. The dominant decay channel is given by the $\ell=1$ mode with greybody factor
\be
\lim_{\omega \to 0 } P_{\t{emit}}(\omega,\ell=1)  = \frac{16}{9} (r_+ \omega)^6 + \mathcal{O}((r_+ \omega)^{12}).
\ee
The above answer matches the low-frequency greybody factor computed by \cite{Crispino:2000jx} for vector fields in the extremal RN background. Note that the scaling differs significantly from the greybody factor computed for the photon in the previous section.

\subsection{Photon/graviton greybody factors: electromagnetic-gravitational mixing} \label{app:photongreybody}
We must solve the equations of motion for the photon in the finite temperature BH to calculate the greybody factor. In this subsection we thus work with the finite temperature background
\be
ds^2 = -f(r) dt^2 + f(r)^{-1} dr^2 + r^2 \lrm{d\theta^2 + \sin^2{\theta} d \phi^2}, \qquad f(r)=\frac{(r-r_{+})(r-r_-)}{r^2}\,.
\ee
The background electric field of the BH produces a coupling between metric and photon modes known as electromagnetic-gravitational mixing \cite{Crispino:2009zza,Gerlach:1974zz,Zerilli:1974ai,Olson:1974nk,Moncrief:1974gw,Moncrief:1974ng,Moncrief:1975sb,Matzner:1976kj}. This modifies the equations of motion of the gauge field in the BH background when the BH carries the charge of the gauge field.

We leave most technical details to \cite{Crispino:2009zza,Oliveira:2011zz}. The radial gauge field/gravitational fields can be expressed in terms of what are known as $\lambda = \t{axial, polar}$ modes, each of which come in pairs $\varphi_{\pm}^\lambda$. The fields take the form
\begin{gather} \label{eqn:photon_graviton_mixing_eqns}
F^\lambda(r) = \varphi_+^\lambda(r) \cos \psi - \varphi_-^\lambda(r) \sin \psi\,, \\
G^\lambda(r) = \varphi_+^\lambda(r) \sin \psi + \varphi_-^\lambda(r) \cos \psi\,,
\end{gather}
where $F,G$ are propagating radial photon/metric modes, and $\psi$ is an angle that takes an explicit expression depending on the parameters of the RN black hole and the angular momentum $\ell$ of the propagating modes. For a near-extremal RN BH for $\l=\t{axial}$ we have
\be
\sin (2\psi) = \frac{2 \sqrt{\ell^2 + \ell -2}}{2\ell + 1}\,, \qquad \psi < \frac{\pi}{4}\,.
\ee
The fields $\varphi_{\pm}^\lambda$ all see different effective potentials; however, it turns out that $\varphi_{+}^\lambda$ axial/polar modes have the same transmission coefficients across the potential, and the same holds for $\varphi_{-}^\lambda$ polar/axial modes. 

To calculate the transmission coefficient across the potential barrier, we must set purely ingoing boundary conditions at the horizon
\begin{align}
\varphi^\l_{\pm} &\propto a^\l_{\pm}e^{- i \omega r_*} , \qquad \qquad ~ r_* \to - \infty\,, \\
\varphi^\l_{\pm} &\propto e^{- i \omega r_*} + c_\pm^\l e^{i \omega r_*},  \quad ~ r_* \to \infty\,,
\end{align}
where $\frac{d r_*}{d r} =  f^{-1}(r)$ are tortoise coordinates. From these equations, we can tune the solutions $\varphi_{\pm}^\l$ at $r_* \to \infty$ so that we have purely ingoing electromagnetic modes at infinity with no ingoing gravitons
\begin{gather}
   \lim_{r_* \to \infty} F^\lambda = F^\lambda_\t{in} e^{- i \omega r_*}  + F^\lambda_\t{out} e^{ i \omega r_*}\,, \\
    \lim_{r_* \to \infty} G^\lambda = G^\lambda_{\t{out}} e^{ i \omega r_*}\,,
\end{gather}
as $r_* \to \infty$. The propagation of photons in the throat region excites propagating gravitational modes as well for $\ell \geq 2$. Close to the horizon, we have
\begin{gather}
   \lim_{r_* \to -\infty} F^\lambda = F^\lambda_\t{transmit} e^{- i \omega r_*} \,, \\
    \lim_{r_* \to -\infty} G^\lambda = G^\lambda_{\t{transmit}} e^{ - i \omega r_*}\,,
\end{gather}
While there are no ingoing gravitons from infinity, close to the horizon, there are graviton modes. When calculating greybody factors, we must take into account the flux of these gravitons. This is not a problem for $\ell=1$ where no gravitons are produced. The transmission coefficient/greybody factor for an incoming photon on a RN BH is thus given by
\be
T^\lambda = 1 - \left|\frac{F^\lambda_\t{out}}{F^\lambda_\t{in}}\right|^2- \left|\frac{G^\lambda_\t{out}}{F^\lambda_\t{in}}\right|^2\,.
\ee

\paragraph{Greybody factors for $\mathbf{\varphi_\pm}$.}
Before calculating greybody factors for the photon/graviton we calculate the greybody factors for the $\varphi_\pm^{\l=\t{axial}}$ modes. For $\lambda=\t{axial}$, the effective potential is quite simple, while for the polar mode, the potential is much more complicated. However, they end up giving the same transmission probability \cite{Crispino:2009zza}, and without loss of generality, we can solve the axial problem. The axial equations of motion, taking into account mixing, are given by \cite{Crispino:2009zza,Matzner:1976kj} 
\be \label{eqn:EM_Mixing_EOM}
\left(\frac{d^2}{d r_*^2} +  \omega^2 - \underbrace{f(r) \lrm{\frac{\ell (\ell+1)}{r^2} + \frac{4 r_+ r_-}{r^4} + \frac{r_+(-3 \pm (2\ell+1))}{r^3} }}_{V^{\l=\t{axial}}_{\pm}}\right) \varphi_\pm^{\lambda = \t{axial}}(r) =0.
\ee 
\be
V^{\l=\t{axial}}_{\pm, \t{ ext. RN}} = f(r) \lrm{\frac{\ell(\ell+1)}{r^2} + \frac{4 r_+ r_-}{r^4} + \frac{r_+(-3 \pm(2\ell+1))}{r^3}}
\ee
Going forward, we drop the ``axial" superscript, and we redefine the field $\varphi_\pm \to r \varphi_\pm$ to simplify the differential equation. The derivatives are simplified with the use of tortoise coordinates $\frac{d r_* }{d r} = f(r)$.

\paragraph{$\varphi_-$ Greybody factor.}
We will give a detailed derivation of the greybody factor for $\varphi_-$, since it is the dominant transmission channel for all modes $\ell \geq 2$. We will also quote the final result for $\varphi_+$ in the case of $\ell=1$ where $\varphi_-$ does not exist, and emission is dominated through $\varphi_+$.

\paragraph{Region I: near-horizon region.}
In the differential equation \eqref{eqn:EM_Mixing_EOM} we replace $r \approx r_+$ except for terms with $(r-r_+)$ or $(r-r_-)$. We get
\begin{align}
&(r-r_-)(r-r_+)\varphi''_-(r) + (2r-r_- - r_+)\varphi'_-(r) \nn\\ & + \lrm{- \ell(\ell+1) +\frac{-6r_- r_+ + r(r_- + (5+2\ell) r_+) + \frac{r^6 \omega^2}{(r-r_-)(r-r_+)}}{r^2}}\varphi_-(r)=0\,. 
\end{align} 
The solution is again given by associated Legendre Polynomials/functions\footnote{In the case of $\varphi_+$ the solutions are $P_{\ell+1}^{-i v} (x)$ and $Q_{\ell+1}^{-i v}( x)$.}
\begin{gather}
\label{app:photon_soln_NHR}
\varphi_-(r) = a_1 P_{\ell-1}^{-i v}(x) + a_2 Q_{\ell-1}^{-i v}(x)\,, \\ 
v = \frac{\beta \omega}{2\pi}, \qquad x = -1 - \frac{(r-r_+)\beta}{2\pi r_+^2}\,,  \nn
\end{gather} 
Purely ingoing boundary conditions enforce $a_2=0$, which we do from now on for simplicity.

\paragraph{Region II: barrier.}
In the barrier region, we can solve the equation exactly by setting $\omega=0$ 
\be
(r-r_+)^2 \varphi_-''(r)+2(r-r_+) \varphi_- '(r)- \frac{1}{r^2}\lrm{\ell(\ell+1)-r r_+ +6 r_+^2 - (5+2\ell)r r_+}\varphi_- (r)=0 \,. 
\ee
The solution in this region is simple but quite long, we write the explicit result for $\ell=2$ as an example
\be
\varphi_-(r) = \frac{b_1 (3 r-2 r_+)+b_2 \left(6 r^3+9 r^2 r_+ +9 r r_+^2+6 r_+^3\right) (r-r_+)^3}{r^2 (r-r_+)^2}\,.
\ee

\paragraph{Region III: asymptotically flat space.}
In the flat space region, we again find the spherical Bessel functions as in \eqref{eqn:psiIII}
\be
\varphi_{-} (r) = c_1 j_1(r \omega )+c_2 y_1(r \omega )\,.
\ee
\paragraph{Matching.}
Expanding the solution to $\varphi_-$ in the barrier region at small $r\approx r_+$, we can match to the large $r$ expansion of $\varphi_-$ in the near-horizon region. Similarly, expanding the barrier solution at large $r$ and matching to the small $r$ behavior of the spherical Bessel functions, we find the matching conditions for general $\ell$
\begin{align}\label{app:photon_matching_conds}
   &b_1 = a_1\frac{\ell (-4 \pi )^\ell r_+ \Gamma (\ell-1) e^{\frac{\beta  \omega}{4}} \left(\frac{r_+^2}{\beta }\right)^\ell}{\Gamma (2 \ell+1) \Gamma \left(-\ell+\frac{i \omega \beta }{2 \pi}+1\right)}\,, \qquad b_2 = -a_1 \frac{ (4 \pi )^{1-\ell} \beta^{\ell-1} r_+^{1-2 \ell} \Gamma (2 \ell-1) e^{\frac{\beta  \omega}{4}}}{(2 \ell +1) \Gamma (\ell+2) \Gamma \left(\ell+\frac{i \omega \beta }{2 \pi }\right)}\,, \\
   & c_1 = b_2 \frac{ 2^{\ell+1} (\ell-1) \ell (2 \ell-1) \omega^{-\ell} \Gamma \left(\ell+\frac{3}{2}\right)}{\sqrt{\pi }}\,, \qquad c_2 = -b_1\frac{\sqrt{\pi }  2^{-\ell} (\ell+1) \omega^{\ell+1}}{\Gamma \left(\ell+\frac{1}{2}\right)}\,.
\end{align}

\paragraph{Transmission probability.} 
The wave at infinity for general $\ell$ is
\be
\lim_{r\to \infty} \varphi_- \approx e^{i r \omega} \underbrace{( c_1  - i c_2 ) }_{A_{\t{out}}} + e^{-i r \omega} \underbrace{( - c_1 + ic_2)}_{A_{\t{in}}}\,.
\ee
The reflection probability is then given by $R = \frac{|A_{\t{out}}|^2}{|A_{\t{in}}|^2}$, with the transmission probability, equivalent to the greybody factor, given by $T=1-R$. We are working in the limit $\beta \omega \sim \mathcal{O}(1)$, $r_+ \omega \to 0$, $\beta r_+ \to \infty$. This can be achieved by scaling $\beta \to \infty$ while $\omega \to 0$ while keeping the product fixed. We find the greybody factor for general $\ell$
\be \label{eqn:photon_greybody_l=1_Final}
\varphi^{\l=\t{axial}}_-(\omega, \ell) ~\t{ greybody:} \qquad P_{\t{emit}}(\omega,\ell)  = \frac{\pi^{2\ell} (\ell+1)^2 (\beta \omega) \lrm{\frac{r_+^2 \omega}{\beta}}^{2\ell} \sinh \lrm{\frac{\beta \omega}{2}} \Gamma\lrm{\ell\pm i \frac{\beta \omega}{2\pi}}}{2^{2\ell-1}(\ell-1)^2 \Gamma(\ell+\frac{1}{2})^4}\,,
\ee
where the gamma function notation indicates, we multiply both signs. We also quote the final answer for $\varphi_+$ in the case of $\ell=1$
\be \label{eqn:photon_greybody_l=1_Final2}
\varphi^{\l=\t{axial}}_+(\omega, \ell=1) ~\t{ greybody:} \qquad P_{\t{emit}}(\omega,\ell=1)  = \frac{4}{9}r_+^8 \omega^4 \lrm{\omega^2 + \frac{4\pi^2}{\beta^2}} \lrm{\omega^2 + \frac{16\pi^2}{\beta^2}}\,.
\ee
This can be found by following a procedure similar to the above with an alternate effective potential. 

\paragraph{Photon greybody factors.}
We are now ready to calculate the photon greybody factor. The case $\ell=1$ is distinct from all higher values. For $\ell=1$ only $\varphi_+$ exists with $\varphi_-=0$, and we have the mixing angle $\psi(\ell=1)=0$. This gives $F^\lambda(r) = \varphi_+^\lambda(r)$ and $G^\lambda(r) = 0$. The photon greybody factor is thus equivalent to the $\varphi_{+,\ell=1}$ greybody factor. We immediately find
\be
\ell=1 \t{ photon greybody:} \qquad P_{\t{emit}}(\omega,\ell=1)  = \frac{4}{9}r_+^8 \omega^4 \lrm{\omega^2 + \frac{4\pi^2}{\beta^2}} \lrm{\omega^2 + \frac{16\pi^2}{\beta^2}}\,,
\ee
which matches the answer computed by \cite{Page:2000dk,Crispino:2009zza}. For higher $\ell \geq 2$ there is a non-trivial angle $\psi(\ell \geq 2) \neq 0$ and we have from \eqref{eqn:photon_graviton_mixing_eqns} $F^\lambda(r) = \varphi_+^\lambda(r) \cos \psi - \varphi_-^\lambda(r) \sin \psi$ and $G^\lambda(r) = \varphi_+^\lambda(r) \sin \psi + \varphi_-^\lambda(r) \cos \psi$. Due to this mixing, the photon greybody factor comes from both $\varphi_\pm$ modes. The mode $\varphi_-$ sees a smaller effective potential, and so the leading result for the greybody factor is given by taking $\varphi_+$ to be purely reflected, with transmission only through $\varphi_-$. Additionally, a propagating photon can now be turned into gravitons, which must be taken into account in the transmission probability, which we restate
\be
T^\lambda = 1 - \left|\frac{F^\lambda_\t{out}}{F^\lambda_\t{in}}\right|^2- \left|\frac{G^\lambda_\t{out}}{F^\lambda_\t{in}}\right|^2\,.
\ee
The leading result turns out to be independent of the second term above. In our conventions \cite{Crispino:2009zza}
\begin{gather}
\frac{F^\lambda_\t{out}}{F^\lambda_\t{in}} = \frac{\sqrt{1-2\sin^2(2\psi)}+1 }{2} \lrm{1-\frac{c_1-i c_2}{c_1 + i c_2}} +\frac{c_1-i c_2}{c_1 + i c_2}\,, \\
\frac{G^\lambda_\t{out}}{F^\lambda_\t{in}}=\frac{\sin(2\psi)}{2} \lrm{1-\frac{c_1-i c_2}{c_1 + i c_2}}\,.
\end{gather}
Using the value of the angle $\psi$ defined earlier, we can evaluate the leading order answer for the greybody factor, finding that it is simply a multiplicative factor multiplied by the $\varphi_-$ greybody factor
\begin{align}
&\ell \geq 2 \t{ photon greybody:} \quad P_{\t{emit}}(\omega,\ell) &=& ~\frac{\ell-1}{2\ell+1} \frac{\pi^{2\ell} (\ell+1)^2 (\beta \omega) \lrm{\frac{r_+^2 \omega}{\beta}}^{2\ell} \sinh \lrm{\frac{\beta \omega}{2}} \Gamma\lrm{\ell\pm i \frac{\beta \omega}{2\pi}}}{2^{2\ell-1}(\ell-1)^2 \Gamma(\ell+\frac{1}{2})^4} \,, \\
&\ell=2: &=& ~\frac{4}{45} r_+^8 \omega^6 \lrm{\omega^2 + \frac{4\pi^2}{\beta^2}}\,,\\
&\ell=3: &=& ~\frac{32}{354375} r_+^{12} \omega^8 \lrm{\omega^2 + \frac{4\pi^2}{\beta^2}} \lrm{\omega^2 + \frac{16\pi^2}{\beta^2}}\,.
\end{align}
In the limit $\beta \omega \gg 1$, these results match those computed by \cite{Crispino:2009zza}. The above calculation was for $\l=\t{axial}$, but the final result is identical for the polar mode. Both polarizations of the photon thus have the above greybody factors. 

\paragraph{Graviton greybody factors.}
Graviton greybody factors for the extremal Reissner-Nordstr\"om black hole were examined in \cite{Oliveira:2011zz}. Similar to the previous discussion, they considered the case of ingoing gravitons without ingoing photons
\begin{gather}
    \lim_{r_* \to \infty} G^\lambda = G^\lambda_{\t{in}} e^{ - i \omega r_*}+G^\lambda_{\t{out}} e^{ i \omega r_*}\,,\\
       \lim_{r_* \to \infty} F^\lambda = F^\lambda_\t{out} e^{ i \omega r_*}\,,
\end{gather}
where there are outgoing photons created by mixing. In this case, the greybody factor is
\be
T^\l = 1 - \left|\frac{G^\lambda_\t{out}}{G^\lambda_\t{in}}\right|^2- \left|\frac{F^\lambda_\t{out}}{G^\lambda_\t{in}}\right|^2\,.
\ee
Taking the $\varphi_+$ mode to be purely reflected, the greybody factor comes entirely from $\varphi_-$. In our conventions, we have \cite{Oliveira:2011zz}
\be
\frac{G^\lambda_\t{out}}{G^\lambda_\t{in}} = \sin^2(2\psi) - \cos^2(2\psi) \lrm{\frac{c_1-i c_2}{c_1 + i c_2}}\,, \qquad \frac{F^\lambda_\t{out}}{G^\lambda_\t{in}} = \frac{1}{2}\sin(2\psi)\lrm{1-\frac{c_1-i c_2}{c_1 + i c_2}}\,.
\ee
The conclusion is that the greybody factor for the $\ell=2$ graviton at small frequencies is given by
\begin{align}
&\ell \geq 2 \t{ graviton greybody:} \quad P_{\t{emit}}(\omega,\ell) &=& ~\frac{\ell+2}{2\ell+1} \frac{\pi^{2\ell} (\ell+1)^2 (\beta \omega) \lrm{\frac{r_+^2 \omega}{\beta}}^{2\ell} \sinh \lrm{\frac{\beta \omega}{2}} \Gamma\lrm{\ell\pm i \frac{\beta \omega}{2\pi}}}{2^{2\ell-1}(\ell-1)^2 \Gamma(\ell+\frac{1}{2})^4}\,,\\
&\ell=2  &=& ~\frac{16}{45} r_+^8 \omega^6 \lrm{\omega^2 + \frac{4\pi^2}{\beta^2}}\,.
\end{align}
We find for $\ell \geq 2$ for near-extremal RN BHs that $P_{\t{graviton}}/P_{\t{photon}} = \frac{\ell+2}{\ell-1}$\,.

\paragraph{Scaling dimensions of photon modes.} 
We point out a subtlety with the boundary scaling dimension of the operator dual to the various photon modes. This arises because of the electromagnetic-gravitational mixing, which modifies the asymptotic fall-off conditions of the field in the throat. The photon is given by a linear combination of two fields with different scaling dimensions in the AdS$_2$ throat. Recall that $F^\lambda(r) = \varphi_+^\lambda(r) \cos \psi - \varphi_-^\lambda(r) \sin \psi$ where for $\ell \geq 2$ we have that $\psi \neq 0$.

The scaling dimension of axial/polar modes $\varphi_{+}^\lambda$ turns out to be the same, and similar for $\varphi_{-}^\lambda$. We only derive the axial case. Near the horizon, we take $r\to r_+$, and the equations of motion reduce to those of a massive field in AdS. The effective potential written earlier takes the form
\begin{align}
\lim_{r \to r_p} V^{\l=\t{axial}}_{\pm, \t{ ext. RN}} \sim \frac{(r-r_+)^2}{r_+^4} \lrm{\ell(\ell+1)+1 \pm (2\ell+1)}\,.
\end{align}
The term in the parenthesis is simply the mass $m^2 = \Delta (\Delta-1)$. We find that $\Delta_{(+)} = \ell+2$ and $\Delta_{(-)} = \ell$. This can also be confirmed by solving the differential equation and examining the fall-off conditions of the field.

At low frequencies, the transmission will be dominated by the field $\varphi_{\pm}^\lambda$ that sees the smallest effective barrier to tunnel through the BH. It turns out that
\be
V_{-}^{\lambda = \t{axial/polar}} \ll V_{+}^{\lambda = \t{axial/polar}}\,,
\ee
and so $\varphi_-^\lambda$ always dominates the transmission probability. As explained earlier, for $\ell=1$ the $\varphi_{-}^\lambda$ field doesn't exist and the photon is given by $\varphi_+^\lambda$. The scaling dimension of the $\ell=1$ photon is thus $\Delta=3$. For higher values of $\ell$ both fields $\varphi_\pm$ exist, and the greybody factor will be least suppressed for the $\varphi_-^\lambda$ field with scaling dimension $\Delta = \ell$ for $\ell \geq 2$. We thus reach the surprising conclusion that the $\ell=1$ photon has a \emph{larger} scaling dimension than the $\ell=2$ photon.
\subsection{Radiation from slowly rotating black holes}\label{app:semiclassicalangularmomentum}
For completeness, we also discuss semiclassical Hawking radiation in the case of a charged black hole with a small amount of angular momentum. See \cite{Frolov:1998wf} for additional technical details.

For $j \ll M^2$, the angular momentum only changes the scattering process in region I, where its effect is to replace the term $r_+^2 \sin^2 \theta d \phi^2$ in the metric \eqref{eqn:RN_metric} by
\begin{align}
    r_+^2 \sin^2 \theta d \tilde\phi^2 = r_+^2 \sin^2 \theta \left(d \phi - \Omega_H dt\right)^2,
\end{align}
where $\Omega_H = \frac{j}{r_+^3} = j \Ebrk$ is the angular velocity of the black hole. This can be absorbed into a change of coordinates by replacing $\phi$ with $\tilde \phi$, so long as we take into account the effect of this change in coordinates on the wavefunction. In the scalar case, this is
\begin{align}
e^{-i\omega t} u(r) Y_{\ell,m}(\theta, \phi) = e^{-i(\omega - m \Omega_H) t} u(r) Y_{\ell,m}(\theta, \tilde\phi).
\end{align}
 Since the dependence on $t$ and $\phi$ is identical for the wavefunctions of the other fields considered in this section, their transformations are identical to the scalar case. (Note that we have chosen coordinates where $m$ is the angular momentum of the photon in the direction of the black hole so that (classically) $m = j - j'$ is the difference between the black hole angular momentum $j$ before emission and $j'$ after emission.)
 
 As a result, to compute greybody factors for near-extremal black holes with small angular momentum, we simply take the greybody factor for zero angular momentum and replace all factors of $\omega$ that come from region I by $\tilde\omega = \omega - m \Omega_H$. This includes, in addition to the region I contribution to the greybody factor, the thermal Bose-Einstein occupation number for the outgoing near-horizon modes, since those modes are thermally excited with respect to the Rindler-like Killing vector $\partial_t|_{\tilde\phi}$ and hence have thermal energy $\tilde \omega$. The flux is therefore given by
\be\label{eq:angularmomentumemission}
\frac{d E}{d t} = \sum_{\ell,m} \int_0^\infty \frac{d \omega}{2\pi} \frac{\omega P_{I \to II}(\tilde \omega) P_{II \to III} (\omega)}{e^{\tilde \beta \tilde \omega}-1}\,,
\ee
where $P_{I \to II}(\tilde \omega)$ is the greybody factor for transmission from region I to II and $P_{II \to III}$ is the greybody factor for transmission from region II to III. (Since the dynamics in region II are independent of both $\omega$ and $\tilde\omega$, it does not matter how the greybody factor from region II is divided between $P_{I \to II}$ and $P_{II \to III}$.) The inverse temperature $\tilde \beta$ of the black hole depends not only on its mass $M$ and charge $Q$ but also on its angular momentum $j$. Close to extremality, we have
\be
\tilde\beta = \sqrt{\frac{2\pi^2}{\Ebrk (M - Q -\frac{1}{2} j^2 \Ebrk)}}\, . 
\ee

A confusing feature of the formula \eqref{eq:angularmomentumemission} is that the integral goes from $\omega = 0$ to infinity and hence for $m > 0$ includes negative near-horizon frequencies $\tilde \omega$. This is the superradiant regime: instead of a Rindler excitation with local energy $\tilde \omega$ tunneling through the barrier and escaping to infinity as a Hawking quanta with energy $\omega$, a near-horizon particle with local energy $-\tilde\omega$ is created simultaneously with the Hawking quanta so that the total energy is conserved. 

Despite its very different physical interpretation, the formula for the superradiant emission rate is identical to the formula for Hawking radiation with $\tilde\omega > 0$. A somewhat schematic explanation for why this is the case is the following. Greybody scattering leads to the relation
\begin{align}
    b_{\text{out},\,\omega} = t \,a_{\text{out},\tilde\omega} + r\, b_{\text{in},\,\omega}
\end{align}
where $b_{\text{in/out},\,\omega}$ are annihilation operators for the ingoing and outgoing far-field modes respectively, $a_{\text{out},\,\tilde\omega}$ is the annihilation operator for the outgoing near-horizon mode and 
\begin{align}
    P_{\text{emit}}(\omega,|\tilde\omega|) = |t|^2 = 1 - |r|^2
\end{align}
is the transmission probability, which depends only on $\omega^2$ and $\tilde\omega^2$. For $\tilde\omega > 0$, we have
\begin{align}
\langle b_{\text{out},\,\omega}^\dagger b_{\text{out},\,\omega}\rangle = P_{\text{emit}}(\omega,|\tilde\omega|) \langle a_{\text{out},\,\tilde\omega}^\dagger a_{\text{out},\,\tilde\omega}\rangle = P_{\text{emit}}(\omega,|\tilde\omega|) \frac{1}{e^{\tilde\beta\tilde\omega} - 1} . 
\end{align}
When $\tilde\omega<0$, $a_{\text{up}\,,\tilde\omega} = a^\dagger_{\text{up},\,-\tilde\omega}$ becomes the creation operator for the mode with energy $-\tilde\omega$ and we have
\begin{align}
\langle b_{\text{out},\,\omega}^\dagger b_{\text{out},\,\omega}\rangle = P_{\text{emit}}(\omega,|\tilde\omega|) \langle a_{\text{out},\,-\tilde\omega} a_{\text{out},\,-\tilde\omega}^\dagger\rangle = t^2 \left[1 +\frac{1}{e^{-\tilde\beta\tilde\omega} - 1}\right] = - P_{\text{emit}}(\omega,|\tilde\omega|) \frac{1}{e^{\tilde\beta\tilde\omega} - 1},
\end{align}
which matches the formula for $\tilde\omega > 0 $ up to a sign. Since $P_{\text{emit}}(\omega,|\tilde\omega|)$ turns out to only involve odd powers of $|\tilde\omega|$, this sign is naturally absorbed into the greybody factor by simply replacing $|\tilde\omega|$ with $\tilde\omega$ to give
\begin{align}
   P_{\text{emit}} (\omega,\tilde\omega) = \text{sign}(\tilde\omega) P_{\text{emit}}(\omega,|\tilde\omega|).
\end{align}
This leads to the formula 
\begin{align}
\frac{d E}{d t} = \sum_{\ell,m} \int_0^\infty \frac{d \omega}{2\pi} \frac{P_{\text{emit}} (\omega,\tilde\omega)}{e^{\beta \tilde \omega}-1}\,, 
\end{align}
which works for both positive and negative $\tilde\omega$.

 The formula \eqref{eq:angularmomentumemission} allows us to immediately obtain semiclassical fluxes in the presence of a small amount of angular momentum from our earlier calculations in this section, so we shall just state the relevant formulas without further discussion. For the $\ell=1$ photon mode we have
\be
P_{I \to II}(\tilde\omega) = \tilde\omega \lrm{\tilde\omega^2 + \frac{4\pi^2}{\tilde\beta^2}} \lrm{\tilde\omega^2 + \frac{16\pi^2}{\tilde\beta^2}}\,, \qquad P_{II \to III}(\omega) = \frac{4}{9}r_+^8 \omega^3\,,
\ee
so that
\be
\frac{d E}{d t} = \frac{1}{\pi} \sum_{m=0,\pm 1}\int_0^\infty d \omega \, \omega \frac{ \frac{4}{9} r_+^8 \omega^3 \times \tilde \omega \lrm{\tilde \omega^2 + \frac{4\pi^2}{\tilde\beta^2}} \lrm{\tilde \omega^2 + \frac{16\pi^2}{\tilde\beta^2}}} {e^{\tilde\beta \tilde{\omega}}-1}\,.
\ee
For the $\ell=2$ photon mode we have
\be
P_{I \to II}(\tilde\omega) = \tilde\omega \lrm{\tilde\omega^2 + \frac{4\pi^2}{\tilde\beta^2}}\,, \qquad P_{II \to III}(\omega) = \frac{4}{45}r_+^8 \omega^5\,,
\ee
which gives
\be
\frac{d E}{d t} = \frac{1}{\pi} \sum_{m=0,\pm 1, \pm 2}\int_0^\infty d \omega \, \omega \frac{ \frac{4}{45}r_+^8 \omega^5 \times \tilde \omega \lrm{\tilde \omega^2 + \frac{4\pi^2}{\tilde\beta^2}}} {e^{\tilde\beta \tilde{\omega}}-1}\,.
\ee
The formula for the $\ell=2$ graviton is four times this formula.

\section{Evaluation of the Euclidean instanton action}\label{app:Euclidean_instanton}

  We evaluate the Euclidean action for the instanton solution used in Section \ref{sec:charged-particle-emission-main}. The total action is
\begin{align}
I_{\t{inst + BH}} &=  -\frac{1}{16 \pi}\left[\int d^4 x \sqrt{g}R + 2 \int_{\partial M} \sqrt{h} (K-K_0)\right]+\frac{1}{16\pi} \int d^4 x \sqrt{g} F_{\mu \nu} F^{\mu \nu}- \frac{1}{4\pi}\int_{\partial M} \sqrt{h} n_i F^{i j} A_j \nn \\ 
&+ \sigma \int_{\t{wordline}} \sqrt{\gamma}  + i q \int_{\t{wordline}} d \tau A_\tau \, ,    
\end{align}
where we have added a counterterm given by the extrinsic curvature $K_0$ in flat space. The particle contributes two terms 
\begin{gather}
I \Bigl|_{\textrm{particle}}^\text{mass} = \sigma \int \sqrt{\gamma}  =  m \chi \betainst \,,\nonumber\\
I \Bigl|_{\textrm{particle}}^\text{EM} = i q \int d \tau A_\tau = - q \lrm{\frac{A \bar Q (r_\inst-\bar r_+)}{\bar r_+ r_\inst }} \betainst \,.
\end{gather}
where $\chi \equiv \chi(r_{\inst})$ is the redshift factor at the  location where the particle is pair-produced. The scalar curvature is zero for the BH, but contains a delta-function piece supported at the location of the particle. The equations of motion\footnote{Since our matter action has a relative minus sign compared with the Ricci scalar term, the equations of motion $G_{\mu \nu} = -8 \pi T_{\mu \nu}$ have a minus sign compared with standard conventions, where $G_{\mu \nu}$ is the Einstein tensor. This is also taken into account in the junction conditions.} give $R = 8 \pi T$, where $T=\sigma \delta(r-r_\text{inst})$ is the trace of the stress tensor localized on the particle worldline found below \eqref{eqn:Israel_Junction_conditions}. Therefore,
\begin{equation}
I \Bigl|_{\textrm{Ricci}}=- \frac{1}{16 \pi} \int \sqrt{g} R = -\frac{1}{2} m \chi \betainst \ . 
\end{equation}
The field strength $F^2$ is constant in the throat and give in \eqref{eqn:EOM_Solutions_instanton_4d}, so the bulk gauge field contribution is
\begin{eqnarray}
-I \Bigl|_{\textrm{Maxwell}}  & = & \frac{A \betainst }{16 \pi}   \int_{\bar r_+ }^{r_{\inst}} dr  4 \pi  r^2 \frac{2\bar{Q}^2}{r^4} + \frac{ \betainst }{16 \pi}   \int^{\infty}_{r_{\inst}} dr  4 \pi  r^2 \frac{2{Q}^2}{r^4} \nonumber \\
& = & \frac{1}{2} \betainst \left(  A \frac{\bar{Q}^2}{\bar r_+}  -  A \frac{\bar{Q}^2}{r_{\inst}} + \frac{{Q}^2}{r_{\inst}}  \right) \ .
\end{eqnarray} 
The extrinsic curvature of a constant-$r$ sphere for $r>r_{\inst}$ is 
\begin{eqnarray}
K &=&  \frac{d \log [\textrm{area}]}{ds} =  \frac{\partial r}{\partial s}  \frac{\partial_r ( \chi r^2)} { \chi r^2} = \chi   \frac{\partial_r ( \chi^2 r^4)} { 2\chi^2 r^4}  = \frac{1}{\chi} \left( \frac{M}{r^2} - \frac{Q^2}{r^3}  \right)  + \frac{2 \chi}{r}  \ . 
\end{eqnarray}
In flat space, the same surface would have $K_0 = \frac{2}{r}$, which can be seen by taking $M, Q \to 0$. The boundary term in the action gives
\begin{eqnarray}
I \Bigl|_{K \textrm{ bdy}} = -\frac{1}{8 \pi} \int_{\partial M} \sqrt{h} (K - K_0)  &=& -\frac{1}{2} \int d \tau \lim_{r \rightarrow \infty} \chi  r^2 \left( \frac{1}{\chi} \left( \frac{M}{r^2} - \frac{Q^2}{r^3}  \right)  + \frac{2 \chi}{r}       - \frac{2}{r} \right) \nn\\
& = & \frac{1}{2} M \betainst \,.
\end{eqnarray}
The Maxwell boundary term is
\be
I \Bigl|_{F^2 \textrm{ bdy}} = -\frac{1}{4\pi} \int_{\partial M} \sqrt{h} n_i F^{i j} A_j = \betainst \frac{Q^2}{r_+} = \betainst Q \frac{A \bar Q (r_\inst-\bar r_+) + Q \bar r_+}{\bar r_+ r_\inst }\,, \quad 
\ee
after subtracting off the divergence. Putting everything together, we find that in the canonical ensemble, the action is given by
\begin{align}
I_{\t{inst + BH}} (\beta_\text{inst}, Q) &= \frac{1}{2}  \betainst \chi m -  \frac{1}{2} \betainst \left(  A \frac{\bar{Q}^2}{\bar r_+}  -  A \frac{\bar{Q}^2}{r_\inst} + \frac{{Q}^2}{r_\inst}  \right) +  \frac{1}{2} \betainst M \nonumber \\
&+ \betainst Q \frac{A \bar Q (r_\inst-\bar r_+) + Q \bar r_+}{\bar r_+ r_\inst } - \betainst q \lrm{\frac{A \bar Q (r_\inst-\bar r_+)}{ \bar r_+ r_\inst }}  \,, \nonumber \\
& = - \pi \bar{r}_+^2 + {\beta}_{\inst} M \,,
\label{eq:backreacted-action-with-instanton}
\end{align}
where we have given a few equivalent representations of the action.

\section{JT dimensional reduction} \label{app:JT_dimensional_reduction}
In this section, we include additional details on dimensionally reducing Einstein-Maxwell coupled to the particle \eqref{eqn:Action_Maxwell_Shell} to obtain a slightly modified version of 2d JT gravity. The action is again
\be
I_{\t{total}} = I_{\t{EM}} + \sigma \int_{\t{worldline}} \sqrt{\gamma} + i q \int_{\t{worldline}} d \tau A_\tau .
\ee
The classical radius of the worldline is found by extremizing the competing effects of electromagnetic repulsion and gravitational attraction. We take the metric ansatz 
\be 
\label{eq:ansatz-for-the-metric}
ds^2 = \frac{r_0}{\Phi^{1/2}} g_{\mu\nu} + \Phi d\Omega_{S^2}
\ee
where $\Phi$ is the dilaton field that parameterizes the transverse area of $S^2$ and $g_{\mu \nu}$ is the $2d$ metric. On the two dimensional space, the particle is a closed loop tracing out a path $x^\mu(u)$ with proper time $u$. 

After reducing to two dimensions, the electromagnetic coupling of the particle gives an additional Wilson loop term, $ I q \oint A$. As explained in \cite{Iliesiu:2020qvm, Iliesiu:2019lfc}, integrating out the 2d gauge field is equivalent to the problem of computing the partition function of 2d $U(1)$ Yang-Mills on the disk. Including the coupling to the S-wave sector of the particle is thus equivalent to $U(1)$ YM on a disk with a Wilson loop in the charge $q$ representation, splitting the disk into an interior disk $M_{\t{int}}$ and an outside annulus $M_{\t{out}}$. After integrating out the gauge field (in the canonical ensemble fixing the asymptotic charge $Q$) we find the action is 
\begin{align}
\label{eq:total-action}
I_{\rm total}[g,\Phi]\;\;&=\;\;-\frac{1}{4G_{N}}\int_{M_{\rm int}}d^{2}x\sqrt{g}\:[\Phi R-2U_{Q-q}(\Phi)] + m \int_{\partial M_{\rm int}} du \sqrt{\frac{r_0}{\Phi^{1/2}}g_{\mu \nu} \dot x^{\mu} \dot x^{\nu}} \nonumber \\ 
&-\frac{1}{4G_{N}}\int_{M_{\rm out}}d^{2}x\sqrt{g}\:[\Phi R-2U_{Q}(\Phi)] - \frac{1}{2G_N} \int_{\partial M_{\rm out}} du \sqrt{h} \Phi K, 
\end{align}
where $Q-q$ can be identified as the new charge of the black hole, while $q$ and $m$ are the charge and mass of the particle, respectively. $M_{\text{interior}}$ is the geometry within the charged instanton; $M_{\text{outside}}$ is the geometry outside of the instanton region whose asymptotic boundary is $\partial M_{\rm outside}$. The dilaton potential $U_{Q}(\Phi)$ is given by
\be \label{eqn:U_Q_Dilaton_Potential}
U_{Q}(\Phi)~~=~~r_{0}\left[\frac{G_N}{4\pi}\frac{Q^2}{\Phi^{3/2}}-\frac{1}{\Phi^{1/2}}\right].
\ee
The variation of such an action is given by
\begin{align}
\label{eq:total-action-variation}
    \delta I_{\rm total}[g, \Phi] &= EOM_{\t{int}}+EOM_{\t{out}} +  \frac{1}{4G_N} \int_{\partial' M_{\rm out}} du ([2(\partial_n \Phi - \Phi K)] \delta(\sqrt{g_{uu}}) - 2 \Phi \sqrt{g_{uu}} \delta K)  \nonumber \\
    & - \frac{1}{2 G_N} \int_{\partial M_{\rm out}} du \sqrt{h} \Phi \delta K  - \frac{1}{4G_N} \int_{\partial M_{\rm out}} du ([2(\partial_n \Phi - \Phi K)] \delta(\sqrt{g_{uu}}) - 2 \Phi \sqrt{g_{uu}} \delta K)  \nonumber \\ 
    & -   \frac{1}{4G_N} \int_{\partial M_{\rm int}} du ([2(\partial_n \Phi - \Phi K)] \delta(\sqrt{g_{uu}}) - 2 \Phi \sqrt{g_{uu}} \delta K) \nn \\ 
    &+  m \int_{\partial M_{\rm int}} du \frac{r_0^{1/2}}{\Phi^{1/4}} \left( \delta(\sqrt{g_{uu}}) - \sqrt{g_{uu}} \frac{\delta \Phi}{4\Phi}\right)\, ,  
\end{align}
where $\partial M_{\text{int}}$, $\partial' M_{\text{out}} $ and $\partial M_{\text{out}}$ are the boundary of the instanton region measured from the inside, from the outside or the asymptotic boundary of the entire geometry respectively. The induced boundary metrics used in \eqref{eq:total-action-variation} are given by 
\be 
g_{uu} = g_{\mu \nu} \dot x^\mu \dot x^{\nu}|_{\partial M_{\text{outside}} \text{ or } \partial' M_{\text{outside}} \text{ or } \partial M_{\text{interior}}} \,.
\ee
If imposing Dirichlet boundary conditions on the outside boundary (fixed BH temperature and fixed size of the transverse $S^2$), then the boundary terms on $\partial{M_{\text{out}}}$ in the variation of \eqref{eq:total-action-variation} vanish. We are left with solving the EOM$_{\text{int}}$ and EOM$_{\text{out}}$ as well as the variation along the boundary of the instanton. The solution for the former two are given by: 
\begin{align}
\text{Interior}: \qquad \Phi(\bar r) = \bar r^2. \qquad ds^2 = \frac{\Phi^{1/2}}{r_0} \left[\bar\chi^2(\bar r) d\bar\tau^2 + \frac{d \bar r^2}{\bar \chi(\bar r)^2}\right], \qquad \bar \chi(\Phi)^2 =1+ \frac{\bar Q^2}{\Phi} - \frac{ 2 \bar M}{\Phi^{1/2}}, \nn \\
\text{Outside}: \qquad \Phi( r) =  r^2 \qquad ds^2 = \frac{\Phi^{1/2}}{r_0} \left[\chi^2(r) d \tau^2 + \frac{d  r^2}{ \chi( r)^2}\right]\,,\qquad \chi(\Phi)^2 =1+ \frac{Q^2}{\Phi} - \frac{2 M}{\Phi^{1/2}},
\end{align}
where $\bar Q = Q-q$ and $M$ and $\bar M$ are, so far, two undetermined constants. We can also rescale time inside the instanton such that $\bar \tau=A \tau$. The continuity condition, which says that the proper length of the instanton boundary has to be the same when measured from the inside/outside is: 
\be 
\label{eq:cont-criterion}
A \bar \chi = \chi\,.
\ee
Additionally, setting the variation of the boundary terms on $\partial' M_{\text{outside}} + \partial M_{\text{interior}}$ to vanish we find:
\begin{align}
\delta K&: \qquad \Phi|_{\partial M_{\text{outside}}} = \bar \Phi|_{\partial M_{\text{inside}}} \Rightarrow r=\bar r=r_\inst\,,\nn \\ 
\delta \Phi&: \qquad \Delta K = \frac{m r_0^{1/2}}{2 \Phi^{5/4}} \Rightarrow \,\, \Delta(K_{\tau\tau} g^{\tau \tau}) = \frac{m r_0^{1/2}}{2 r^{5/2}}\,,  \nn \\ 
\delta \sqrt{g_{uu}}&:\qquad \frac{1}{4} \Delta \left(2\partial_n \Phi - 2\Phi K \right ) = -{m}\frac{r_0^{1/2}}{\Phi^{1/4}} \Rightarrow \Delta( n^\mu \partial_\mu \Phi) =  -2 m \frac{r_0^{1/2}}{r^{1/2}}\,,
\label{eq:variation-boundary}
\end{align}
where $\Delta(\dots) = \dots|_{\partial M_\text{out}'} - \dots|_{\partial M_\text{int}}$. In the second equation we have that $K_{\tau \tau} = \frac{1}{2} \sqrt{g^{r r}} \partial_r g_{\tau \tau}$.\footnote{The second equation comes from considering the variation due to $\delta \Phi$ and integrating the variation of the full action on an annulus surrounding the instanton. From the bulk equations of motion, only $\int R \delta \Phi$ contributes since $R$ is singular across the particle worldline while other terms are continuous. The variation of the particle action also contributes. Using Gauss-Bonnet, the bulk term can be rewritten as the contribution of the extrinsic curvature along the boundaries of the annulus since $\int R = - 2 \int_\partial K$, which becomes $\Delta K$ across the particle worldline as we shrink the annulus to be infinitesimally thin. Thus, the contribution of the bulk terms yields $\frac{1}{4G_N}\int R\delta \Phi = -\frac{1}{2G_N}\int (\Delta K) \delta \Phi $.} In the last line, $n_\mu$ is a unit normal vector to the particle worldline, and all quantities are evaluated at $r_\inst$. From the last two equations, we get 
\be {\Rightarrow} \frac{1}{\chi} &\left( M-
\frac{Q^2}{r_\inst}\right) - \frac{1}{\bar \chi} \left(\bar M- \frac{\bar Q^2}{r_\inst} \right) = m \,,\nonumber\\
 &r_\inst \left(\chi-\bar \chi\right)  = -m\,,
\ee
consistent with the $4d$ calculations \eqref{eqn:Israel_Junction_conditions_explicit} presented in Section \ref{sec:including-gravitational-backreaction}.

\paragraph{Modifications to dimensional reduction.} We now highlight how the introduction of the instanton modifies the dimensional reduction. There are two key points:
\begin{itemize}
    \item The AdS radii inside and outside of the instanton are different.
    \item The outside geometry, if continued past the instanton, would predict the presence of a conical defect.
\end{itemize}
The first point can be seen from the dilaton potential \eqref{eqn:U_Q_Dilaton_Potential} and action \eqref{eq:total-action}. Expanding the potentials $U_{Q-q}(\Phi)$ in the throat around a large constant dilaton value $\Phi = \frac{G_N}{4\pi}(Q-q)^2 + \frac{G_N}{\pi} \phi$, which corresponds to the horizon area $r_0^2$ after emission. This gives an interior AdS$_2$ length $\ell_{\t{in}} = \sqrt{G_N}(Q-q)$. The same expansion applied to the exterior region with potential $U_{Q}(\Phi)$ gives a slightly different AdS length.

The second point already appeared in the context of the full black hole geometry. The periodicity of the $\tau$-circle is fixed, so the interior geometry is smooth, but due to the different AdS radii, the exterior geometry, if continued past the instanton, would contain a conical defect. The defect is not actually present on the full geometry since the instanton changes the spacetime before the defect is reached. However, since the asymptotic boundary conditions appear to imply the presence of a conical defect, the boundary fluctuations that are integrated over will be modified compared with the standard disk calculation. 

\subsection*{The one-loop determinant for the particle worldline}
In this subsection, we explain the one-loop determinant that comes from integrating the position of the instanton from a two-dimensional perspective. Expanding the action \eqref{eq:total-action} as described in the previous section around the near-extremal background following \cite{Iliesiu:2020qvm} and setting $G_N=1$ we get the two-dimensional theory
\begin{align}
I[g,\phi]=&-\int_{M_{\rm int}}d^{2}x\sqrt{g} \phi \lrm{R+ \frac{2}{\ell^2_{\t{in}}}} -\int_{M_{\rm out}}d^{2}x\sqrt{g} \phi \lrm{R+ \frac{2}{\ell^2_{\t{out}}}} \nonumber \\ 
&- 2 \int_{\partial M_{\rm out}} du \sqrt{h} \phi K - \frac{m}{Q^2} \int_{\partial M_{\rm int}} du ~ \phi(x(u)) \sqrt{g_{\mu \nu} \dot x^{\mu} \dot x^{\nu}}\,,
\end{align}
where we have dropped topological terms, terms of order $\mathcal{O}((\phi/\Phi_0)^2)$, and the linear term $m L$ for simplicity since they don't modify the dilaton gravity dynamics. Integrating out the dilaton, we get a constant negative curvature surface, with different cosmological constants inside and outside the instanton, alongside a contact term along the path of the instanton coming from the last term in the above. This enforces that there is a jump in extrinsic curvature across the instanton. Solving for the instanton path, it is a circle of radius $r_{\inst} = \frac{m}{q} Q$ where $Q$ is the AdS$_2$ radius. The proper length of the instanton is $\int_{\partial M_{\t{int}}} \sqrt{h} = 2\pi Q \frac{m}{q}$ which is incredibly small in AdS units when $m\ll q$ as in the case of the positron. We must integrate over the position of the instanton since such a mode exists from the matter partition function. We can treat the loop effectively as a point, set its length to be equal to its classical value, and approximate the dilaton as constant along the small loop. The last term in the action corresponding to the instanton becomes
\be
- \frac{m}{Q^2} \int_{\partial M_{\rm int}} du ~ \phi \sqrt{g_{\mu \nu} \dot x^{\mu} \dot x^{\nu}} \to -2\pi (1-\alpha) \phi(x_\inst)\,,
\ee
where $\alpha=1-\frac{m^2}{q Q}$. This is precisely the JT gravity operator that creates a conical defect with deficit angle $2\pi(1-\alpha)$ \cite{Witten:2020wvy,Maxfield:2020ale,Turiaci:2020fjj}. Integrating over the position of the instanton amounts to evaluating the following JT gravity amplitude
\be
\left\langle \int d^2 x_\inst \sqrt{g} e^{-2\pi (1-\alpha) \phi(x_\inst)} \right \rangle. 
\ee
In \cite{Lin:2023wac,Stanford:2022fdt} it was argued that the above amplitude in the limit $\alpha \to 1$ is given by 
\be
\lim_{\alpha \to 1}\left\langle \int d^2 x_\inst \sqrt{g} e^{-2\pi (1-\alpha) \phi(x_\inst)} \right \rangle = \frac{1}{2\pi(1-\alpha)} \frac{1}{\sqrt{\beta \Ebrk}} e^{\frac{2\pi^2 (\alpha^2-1)}{\beta \Ebrk}}\,,
\ee
where the overall numerical prefactor is dependent upon the normalization of the ultralocal measure. Importantly, this is not equal to the more common JT defect amplitude \cite{Mertens:2019tcm}, which is defined by multiplying the above by $2\pi(1-\alpha)$.\footnote{The two different amplitudes arise from the origin of the integral over positions. If the integral over positions of the defect comes from integrating over the moduli space of metrics, the associated ultralocal measure contains an extra factor of $(1-\alpha)$ to give the more commonly studied JT gravity amplitude. If the integral over positions come from a matter theory, then we will obtain the amplitude above. See \cite{Lin:2023wac} for additional details.} This is the origin of the $(1-\alpha)^{-1}$ prefactor in the one-loop determinant computed in \eqref{eq:BH+1-inst-part-function}.

\section{Angular momentum contributions to di-particle emission} \label{app:angularmomentum_di_photon}
We briefly explain how to take into account the effect of angular momentum on di-photon emission at low energies. The matrix element we want to calculate is
\be \label{eq:4ptappd}
\mathcal{A}_{\ell}(E_i,E_I, E_f, E_I') &= \sum_{m_i, m_I, m_I'}\frac{\Tr(P_{E_i,0,0} \cO_{\ell, m_1} P_{E_I, \ell, m_I} \cO_{\ell, m_2} P_{E_f, 0,0} \cO_{\ell, mp} P_{E_{I'}, \ell, m_{I'}} \cO_{jp, mp})}{\Tr(P_{E_i,0,0}) \Tr(P_{E_I, \ell, m_{I}}) \Tr(P_{E_f,0,0}) \Tr(P_{E_{I'}, \ell, m_{I'}})}\nn \\ &\qquad = \sum_{m_i, m_I, m_I'} \lb E_i,0,0 | \mathcal{O}_{\ell, m_1} | E_I, \ell, m_I \rb \lb E_I, \ell, m_I | \mathcal{O}_{\ell, m_2} | E_f, 0, 0 \rb \nn \\ &\qquad \qquad \quad \times \lb E_f, 0, 0|  \mathcal{O}_{jp, m_3} | E_{I'}, \ell, m_{I'} \rb \lb E_{I'},\ell, m_{I'} | \mathcal{O}_{jp, m_4} | E_i, 0,0 \rb  \,.
\ee
where $\mathcal{O}_{\ell m_p}$ represents a particle of angular momentum $\ell$ and axial angular momentum $m_p$, and the intermediate state has angular momentum $j_I = \ell$ and axial angular momentum $m_I$. Since we are only interested in di-photon and di-graviton evaporation when they are the dominant channels in the final stages of evaporation, we have set the initial and final states to have zero angular momentum. 

The matrix element \eqref{eq:4ptappd} is a matter four-point function that in a free matter theory can be decomposed into three possible Wick contractions. The Schwarzian contribution to each Wick contraction is the same as found in \eqref{eqn:jt4ptfn}. For two of these Wick contractions, the Schwarzian contributions enforces the pairs of energies, $E_I$ and $E_{I'}$ or $E_i$ and $E_f$ to be equal. However, we are only interested in the case where $E_i \neq E_f$, i.e.~when the black hole mass changes. Thus, there are only two Wick contractions to consider: one that typically computes time-ordered correlators (TOCs) and one that computes out-of-time-ordered correlators (OTOCs). These two Wick contractions are shown in Figure \ref{fig:4pt_angularmomentum}. 

We will first consider the TOC. Since we already know the Schwarzian contribution, we still need to compute the additional contribution coming from spin modes. The TOC can be computed by integrating over holomomies in the left diagram of Figure \ref{fig:4pt_angularmomentum},
\begin{figure}
    \centering
    \includegraphics[width=.8\textwidth]{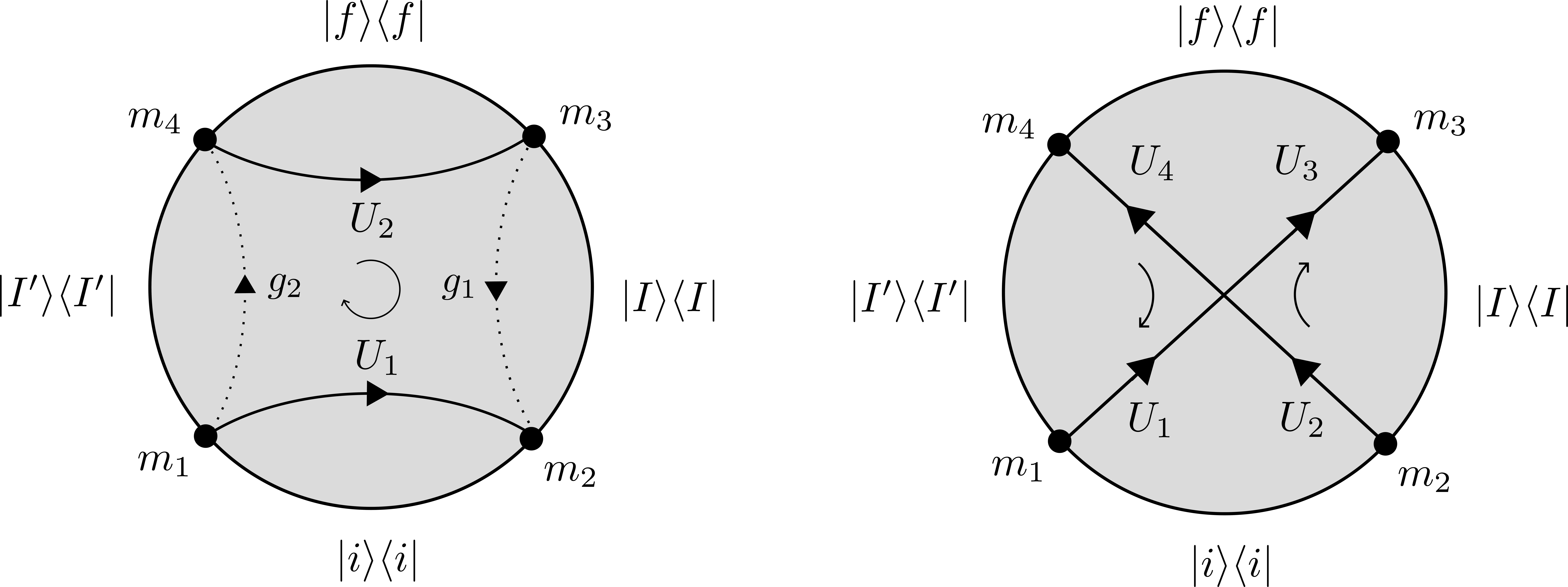}
    \caption{Spin dependent prefactor for the four-point function. On the left we have the TOC and on the right the OTOC. We have four operator insertions labelled by $m_i$, with various intermediate states between the particles. The arrows indicate how the holonomies $U_i, g_i$ are defined. The disk has a clockwise orientation and when an amplitude for a portion of the disk is evaluated it must be evaluated with respect to this orientation. The middle plaquette contributes $\delta(g_1 U_1^\dag g_2 U_2)$, while the initial and final states give no contribution since they are in the trivial representation, while the intermediate states contribute $\propto D(g_i^\dag)$ for the TOC and $\propto D(U_i U_j)$ for the OTOC.}
    \label{fig:4pt_angularmomentum}
\end{figure}
\begin{align}
&\sum_{\substack{ m_i, m_I, m_{I'}}} \int d U_1 d U_2 dg_1 dg_2 \underbrace{D^{\ell}_{m_1 m_2}(U_1) D^{\ell}_{m_4 m_3}(U_2)}_{\t{particle worldlines}} \underbrace{D_{m_I m_I}^{j_I}(g_1^\dag)  D_{m_{I'} m_{I'}}^{j_{I'}}(g_2^\dag) \delta(U_2 g_1 U_1^\dagger g_2)}_{\t{intermediate Plaquette}}\,.
\end{align}
As in \eqref{eq:spinfactor}, for each factor of $|E,j,m\rb\lb E,j,m|$ in \eqref{eq:4ptappd} we have a holonomy 
\begin{align}
\frac{P_{j,m}|HH\rb}{\lb HH|P_{j,m}|HH\rb} = \frac{1}{\sqrt{2 j+ 1}} |j;m,m\rb, 
\end{align}
which in the holonomy basis is just a Wigner D-matrix. These holonomies are glued to a plaquette, which gives a delta function setting the holonomies around the Plaquette equal to the identity. We have to be careful to define the representation wavefunction with respect to the orientation of the disk, which conjugates some matrix elements $g \to g^\dag$. The two particle Wilson lines are oriented to go from left to right, and give additional Wigner-D matrices. 

We can evaluate the above to find
\begin{align}
&\mathcal \sum_{m_i, m_{I},m_{I'}}\int d U_1 d U_2 dg_1 D^{\ell}_{m_1 m_2}(U_1) D^{\ell}_{m_4 m_3}(U_2) D_{m_I m_I}^{j_I}(g_1^\dag) D_{m_{I'} m_{I'}}^{j_{I'}}(U_2 g_1 U_1^\dagger) \nn\\
&=\sum_{m_i, m_{I}, m_{I'}} \frac{1}{2j_I+1}\int d U_1 d U_2 D^{\ell}_{m_1 m_2}(U_1) D^{\ell}_{m_4 m_3}(U_2) D^{j_I}_{m_{I'} m_I}(U_1)^* D^{j_I}_{m_{I'} m_I} (U_2) \delta_{j_I, j_I'}\nn \\
&=\sum_{m_i, m_I ,m_{I'}} \frac{(-1)^{m_I - m_{I'}}}{2j_I+1} \int d U_1 d U_2 D_{m_1 m_2}^{\ell}(U_1) D_{m_4 m_3}^{\ell}(U_2) D_{m_{I'}, m_I}^{j_I}(U_1)^*  D_{-m_{I'}, -m_I}^{j_I}(U_2)^*\nn \delta_{j_I, j_I'}\\ 
&= \sum_{m_i, m_I, m_{I'}} (-1)^{m_I - m_{I'}} \frac{\delta_{m_1,m_{I'}}\delta_{m_4,-m_{I'}} \delta_{m_2,m_I} \delta_{m_3,-m_I}\delta_{j_I, {\ell}} \delta_{j_I, j_I'}}{(2j_I+1)^3}  = \frac{\delta_{j_I, \ell} \delta_{j_{I'}, \ell}}{(2{\ell}+1)^3}\,.
\end{align}
This is equal to $1/27$ for di-photon emission ($\ell = 1$). In the first line, we have integrated over $g_2$ and enforced that $g_2=U_1 g_1 U_2^\dag$. In the second line we have used the property that $D_{m n}(g_1 \cdot g_2) = \sum_{m'} D_{m m'} (g_1) D_{m' n} (g_2)$ repeatedly and integrated over $g_1$ using the orthogonality relation
\be
\int d g D^j_{m n}(g)^* D^{j'}_{m' n'}(g) = \frac{\delta_{j' j} \delta_{m m'} \delta_{n n'}}{2j+1}\,.
\ee
Furthermore we have used that $D_{m n} (g^\dag) = D_{n m}(g^*) = D_{n m}(g)^*$. In the third line we used $D_{m n}(g) = (-1)^{m-n} D_{-m,-n}(g)^*$ and finally integrated over the remaining holonomies.

The second diagram is the OTOC and can be evaluated similarly
\be
 \sum_{\substack{j_I, j_I', \\ m_i, m_I, m_{I'}}}& \int d U_1 d U_2 d U_3 d U_4 \underbrace{D_{m_1 m_3}^{{\ell}}(U_1 U_3) D_{m_4 m_2}^{{\ell}}(U_4^\dag U_2^\dag)}_{\t{particle worldlines}} \underbrace{D_{m_{I'} m_{I'}}^{j_I'}(U_4^\dag U_1^\dag) D_{m_I m_I}^{j_I}(U_2 U_3)}_{\text{intermediate Plaquettes}}\,.
\ee
To evaluate this we apply the decomposition formula and related identities described above and evaluate each integral. We find
\begin{align}
&\sum_{\substack{m_i, m_I, m_{I'} \\  a,b,c, d}}  \int (\prod_i d U_i) D^{\ell}_{m_1 a}(U_1) D_{a m_3}^{\ell}(U_3) D_{b m_4}^{\ell}(U_4)^* D_{m_2 b}^{\ell}(U_2)^*\nn \\ &\qquad \qquad \qquad  \qquad 
 \times D_{c m_{I'}}^{j_I'}(U_4)^* D_{m_{I'} c}^{j_I'}(U_1)^* D_{m_I d}^{j_I}(U_2) D_{d m_I}^{j_I}(U_3)\nn
\\
&=  \sum_{\substack{m_i, m_I, m_{I'} \\  a,b,c, d}}   (-1)^{c-m_{I'}+d-m_{I}}\int (\prod_i d U_i)  D^{\ell}_{m_1 a}(U_1) D_{a m_3}^{\ell}(U_3) D_{b m_4}^{\ell}(U_4)^* D_{m_2 b}^{\ell}(U_2)^*\nn \\  &\qquad \qquad \qquad  \qquad \qquad   \qquad 
 \times D_{-c,-m_{I'}}^{j_I'}(U_4) D_{m_{I'} c}^{j_I'}(U_1)^* D^{j_I}_{m_I d}(U_2) D_{-d, -m_I}^{j_I}(U_3)^* \nn \\
&= \sum_{\substack{ m_I, m_{I'}}}\sum_{m_i} \sum_{a,b,c,d} (-1)^{c-m_{I'}+d-m_I} \frac{\delta_{j_I, {\ell}} \delta_{j_I',{\ell}}}{(2j_I+1)^4} \nn \\ & \qquad \qquad \qquad  \qquad \qquad   \qquad \times \delta_{m_1 m_{I'}} \delta_{m_2 m_I} \delta_{m_3, -m_I} \delta_{m_4, -m_{I'}} \delta_{a c} \delta_{b,-c} \delta_{b,d} \delta_{a,-d} \nn \\
&= \sum_{m_i} \sum_{a,b,c,d} (-1)^{c-m_1+d-m_2} \frac{\delta_{m_1,-m_4}\delta_{m_2,-m_3}}{(2{\ell}+1)^4}\delta_{a c} \delta_{b,-c} \delta_{b,d} \delta_{d, -c} \delta_{j_I, {\ell}} \delta_{j_I',{\ell}} = \frac{\delta_{j_I, {\ell}} \delta_{j_I',{\ell}}}{(2\ell+1)^3}\,.
\end{align}
For di-photon emission this is again $1/27$, so the spin dependent factors of the OTOC and TOC contribute equally. 

Thus, in total, the relevant product of matrix elements seen in \eqref{eq:4ptappd} evaluates to, 
\begin{align} \label{eqn:complicated_integrals-app}
\mathcal{A}_{\ell}(E_i,E_I, E_f, E_I')& =
\frac{\Ebrk^{4\Delta} \mathcal{N}_{4pt} (\Gamma_{f I}^\Delta \Gamma_{f I'}^\Delta \Gamma_{i I}^\Delta \Gamma_{i I'}^\Delta)^{1/2} \delta_{j_I, \ell} \delta_{j_{I'}, \ell}}{(2\ell+1)^3} \nn \\  & \qquad \qquad \qquad \times \left(\frac{\delta\left(E_I-E_{I'}\right)}{e^{-S_0}\rho\left(E_I-E_0^{\ell}\right)}+  \left\{\begin{array}{ccc}
\Delta & E_f & E_I - E_{0}^{\ell} \\
\Delta & E_i & E_{I'} - E_{0}^{\ell}
\end{array}\right\}\right)\,, 
\end{align}
where $\rho\left(E_I-E_0^{\ell}\right)$ is the Schwarzian density of states. Thus, accounting for angular momentum effectively adds a factor of $1/(2\ell+1)^3$ in front of the scalar four-point function.


\section{A winding instanton}
\label{sec:winding-instanton}

We now considering a particle that winds $k$ times around its circular trajectory. This amounts to sending $m \to k\, m$ and $q \to k \,q$. Therefore, the defect angle becomes 
\be
\label{eq:alpha-k}
\alpha_k =  \frac{\left(\sqrt{(2Q-k q)^2 - (km)^2} + k\sqrt{q^2-m^2}\right)^4}{16Q^4} = 1-2k\frac{q-\sqrt{q^2-m^2}}{Q} + \mathcal{O}\left(\frac{1}{Q^2}\right)\,,
\ee 
where the latter expression is valid as long as $k q \ll Q$. The on-shell action is given by 
\be 
I_\text{1 inst + BH}  &= -\pi Q^2 + 2\pi k Q \left(q - \sqrt{q^2-m^2} \right)  - \frac{2\pi^2 Q^3}{\beta} \alpha_k^2 + \mathcal{O}\left(\frac{1}{Q^2}, \, \frac{\pi^2 Q^6}{\beta^2}\right)\,.
\ee 
The one-loop determinant for a $k-$times wound fermion can be extracted from Schwinger's calculation in flat space \cite{schwinger1951gauge}. The integral over the position of a $k$-wound instanton comes with an additional factor of $(1-\alpha_k)$. Putting everything together, we find the sum over all winding instantons to be
\begin{align}
\label{eq:inst-eff-k}
\sum_{k=1}^\infty e^{-I_\text{inst eff.}^{k}} &= \sum_{k=1}^\infty \frac{1}{k^2} \frac{q^2 Q^2}{2\pi^3(1-\alpha_k)} e^{ -2\pi k Q \left(q - \sqrt{q^2-m^2} \right)} \\
&\approx \sum_{k=1}^\infty \frac{1}{k^3} \frac{q^3 Q^3}{2\pi^3 m^2} e^{ -2\pi k Q \left(q - \sqrt{q^2-m^2} \right)}\,,\nn
\end{align}
where we have defined the effective action for a $k$ wound instanton. For a bosonic particle, the one-loop determinant must be multiplied by $\frac{1}{2} (-1)^{k+1}$ \cite{Gelis:2015kya}. Note again that the overall $\mathcal{O}(1)$ numerical prefactor is ambiguous due to the freedom to define the normalization of the ultralocal measure.

\section{Corrections to the extremal energy due to charged particle emission}
\label{sec:correction-to-extremal-energy}

Another effect of the quantum backreaction due to the instantons is that we see a small shift of the extremal energy (i.e.~the lowest energy  in a sector of fixed charge $Q$) given by the real part of equation \eqref{eq:E0-shift}
\be 
\delta E_{\t{inst}} \equiv \Re E_0 =  2\pi^2 e^{-2 I_\text{inst eff}} (1-\alpha)^2 \Ebrk  + \mathcal{O}\left(e^{-4 I_\text{inst eff}}(1-\alpha)^6\right)\,,
\ee
where we ignored the subleading winding contribution by assuming we were working with a sufficiently large BH. This is easiest to see by writing the density of states \eqref{eqn:densityofstates_nearextremal_shifted} close to the edge of the spectrum 
\be
\rho(E) \sim \mC  e^{S_0} \sqrt{E-(E_{\t{ext}}+\delta E_{\t{inst}}) + i \Gamma} + \ldots\,,
\ee
where in the above we have shifted the energy so that $E$ is the total energy of the BH, and at extremality $E_{\t{ext}}=Q$.

However, the above is not the full shift of the spectrum. The extremality shift arises from integrating out the positron on the BH background and finding the new spectrum of near extremal RN BHs. There are two effects that must be included to properly integrate out the positron \cite{Jentschura:2001qr,Dunne:2004nc}, the first is that an infinite tower of higher dimension electromagnetic terms are generated in the effective action. The second effect is that non-perturbative Euclidean instantons must be re-summed to obtain the imaginary shift of the effective action. In this work, we have only included the Euclidean instanton contribution, and so the full shift of the extremality bound is given by
\be
E_{\substack{\t{corrected}\\ \t{ext}}} = Q+\delta E_{\t{pert.}}+\delta E_{\t{inst}}\, . 
\ee
The leading perturbative corrections $\delta E_{\t{pert.}}$ have been computed in \cite{Ruffini:2013hia}. It is perhaps noteworthy that $|\delta E_{\t{pert}}|>\delta E_{\t{inst}}>0$ for large BHs, while $\delta E_{\t{pert}}<0$ which means that the corrected extremal energy satisfies $E_{\substack{\t{corrected}\\ \t{ext}}} < Q$. The interpretation of this shift is that states below this energy in a sector of fixed charge $Q$ do not exist at all, not even as metastable states.

It is useful to explicitly write out the non-perturbative extremality shift to leading order
\be
\delta E_{\t{inst}} =\frac{q^4 Q}{2 \pi^4} e^{ -4\pi Q \left(q - \sqrt{q^2-m^2} \right)}\, . 
\ee
The only assumption in the derivation of the above is that $q \gg m$, and that we are working with black holes with sufficiently large $Q$ so that the instanton approximation is reliable.


\bibliographystyle{utphys2}
{\small \bibliography{Biblio}{}}
\end{document}